\documentclass[12pt]{article}
\usepackage{amsthm,amsmath,amsfonts,amssymb}
\usepackage{natbib}%% uncomment this for author-year citations
\usepackage[colorlinks,citecolor=blue,urlcolor=blue]{hyperref}
\usepackage{graphicx}
\usepackage[table,xcdraw]{xcolor}
\usepackage{enumerate}
\usepackage{color}
\usepackage{url}
\usepackage[space]{grffile}
\usepackage{latexsym}
\usepackage{textcomp}
\usepackage{longtable}
\usepackage{tabulary}
\usepackage{booktabs,array,multirow}
\usepackage[figuresright]{rotating}
\usepackage[margin=1in]{geometry}

\long\def\symbolfootnote[#1]#2{\begingroup%
\def\thefootnote{\fnsymbol{footnote}}\footnote[#1]{#2}\endgroup}
\usepackage{times}

\def\mE{\mathbb E}
\def\tr{\mathrm {tr}}
\def\cov{\mathrm {cov}}
\def\var{\mathrm {var}}
\def\mE{\mathbb{E}}

\def\U{{\mathbf U}}
\def\n{\nonumber}

\def\cp{\mathop{\rightarrow}\limits^{p}}
\def\mE{\mathbb E}
\def\tr{\mathrm {tr}}
\def\cov{\mathrm {cov}}
\def\var{\mathrm {var}}
\def\mE{\mathbb{E}}

\def\U{{\mathbf U}}
\def\n{\nonumber}

\def\tr{\mathrm {tr}}

\def\P{\mathbb{P}}
\def\be{\begin{eqnarray}}
\def\ee{\end{eqnarray}}
\def\bse{\begin{eqnarray*}}
\def\ese{\end{eqnarray*}}

\def\cp{\mathop{\rightarrow}\limits^{p}}
\numberwithin{equation}{section}

\newtheorem{theorem}{Theorem}[section]
\newtheorem{corollary}{Corollary}[section]

\newtheorem{assumption}{Assumption}
\newtheorem{remark}{Remark}
\allowdisplaybreaks[4]
\theoremstyle{plain}

\newtheorem{lemma}{Lemma}[section]

\begin{document}
\title{\bf Fisher's combined probability test for cross-sectional independence in panel data models with serial correlation}
\vspace{0.5in}
\author{Hongfei Wang$^{1}$, Binghui Liu$^{1}$, Long Feng$^{2}$ and Yanyuan Ma$^{3}$\\
\fontsize{10}{10}\selectfont\itshape
$^{1}$\, School of Mathematics and Statistics and KLAS, Northeast Normal University, Renmin Street, Changchun, Jilin, China.  \\
\fontsize{10}{10}\selectfont\itshape
$^{2}$\,School of Statistics and Data Science, LPMC, LEBPS and KLMDASR, Weijin Road, Tianjin, China.\\
\fontsize{10}{10}\selectfont\itshape
$^{3}$\, Department of Statistics, Pennsylvania State University, Old Main, University Park, Pennsylvania, United States of America.
}

\date{}
\maketitle

\begin{abstract}
Testing cross-sectional independence in panel data models is of
fundamental importance in econometric analysis with high-dimensional
panels. Recently, econometricians began to turn their attention to the problem in the
presence of serial dependence.
 The existing procedure for testing cross-sectional independence with
 serial correlation is based on the sum of the sample
 cross-sectional correlations, which generally performs well when
the alternative has dense cross-sectional correlations, but suffers from
 low power against sparse alternatives. To deal with sparse
 alternatives, we propose a test based on the maximum of the
 squared sample cross-sectional correlations. Furthermore, we propose
 a combined test to combine the p-values of the max based and sum based tests, which
 performs well under both dense and sparse
 alternatives.  The combined test relies on the
 asymptotic independence of the max based and sum based test
 statistics, which we show rigorously.
We  show that the proposed max based and
 combined tests have attractive theoretical
 properties and demonstrate the superior performance via extensive
 simulation results. We apply the two new tests to
   analyze the weekly returns on the
 securities in the S\&P 500 index under the Fama-French three-factor
 model, and confirm the usefulness of the proposed combined test in
 detecting cross-sectional independence.
\end{abstract}
\vfill
\newpage
\baselineskip=24.5pt
\section{Introduction}\label{sec:intro}

In this paper, we consider the problem of testing cross-sectional
independence in heterogeneous panel data models.
In statistics and econometrics, panel data occur
  frequently, which contain observations
of various types obtained over multiple time periods for
any single unit. In the study of
panel data models, the cross-sectional dependency is an important
concept, described as the
interaction between cross-sectional units, which could arise from the
behavioral interaction
between units \citep{breusch1980l,feng2020}. To make theoretical study easier,
experts often assume cross-sectional independence in the model
setups. If data across units are dependent,
inferences under the assumption of cross-sectional independence would
be inaccurate and misleading; see literature on spatial
econometrics, such as
\cite{anselin1998,kelejian1999,kapoor2007,lee2007,lee2010}
for examples of cross-sectional dependence.
Therefore, testing the existence of cross-sectional dependence is
important  and  attracts increasing attention.

A large number of literatures on testing cross-sectional dependence
are available, among which
the most widely known is likely the Lagrange Multiplier (LM)
test proposed by \cite{breusch1980l}.
The LM test is based on the sum of the squared cross-sectional
correlations of residuals, which is applicable when
the sample size $T$ is large and the dimension $N$ is finite, but
is not a valid test when $N\rightarrow \infty$.
To develop tests applicable when both $N$ and $T$ are
large, two limit schemes have been considered
in the literature.
One is the sequential limit scheme: $T\rightarrow \infty,$ followed by
$N\rightarrow \infty$, and the other is the
simultaneous limit scheme: $N$ and $T$ tend to infinity
simultaneously, i.e. $(N,T)\rightarrow \infty$.
Under the sequential limit scheme, \cite{pesaran2004} proposed a
scaled version of the LM test, as well as
 a test based on the sum of the residual correlations,
 instead of the squared sum,
to address the issue that the former test may suffer from substantial
size distortions in the case of large $N$ and small $T$;
\cite{pesaran2008} proposed a bias-adjusted LM test based on the scaled LM test.
Later, under the simultaneous limit scheme,  \cite{pesaran2015}
established the asymptotical properties of the sum based
  test by \cite{pesaran2004};
\cite{feng2020} established the asymptotical properties of the
bias-adjusted LM test proposed by \cite{pesaran2008}. In addition to
these tests,
there are also works on testing cross-sectional independence under
other related models, such as the fixed effect panel data models;
see, for instance, \cite{baltagi2012,feng2020fixed}.

All the above tests are sum based tests, i.e., they are based
on the sum of the correlations or squared correlations of the residuals.
These tests generally perform well under dense alternatives, but may
suffer from low power against sparse alternatives. To deal with sparse
alternatives,
\cite{feng2020} proposed a max based test for testing cross-sectional
dependence, and further developed a combined test that integrates the
advantages of the max based and sum based
tests, by establishing the asymptotical independence between the test statistics.

All the tests mentioned above make the common
assumption that the errors in the panel data models are independent
across time.
However, the existence of serial dependence
is likely to be the rule rather than the exception; see, for
instance, \cite{wei2006,Hong2010,box2015}.
In many applications, serial dependence may have great impact on
statistical inference, such as leading to deviation of the limiting
spectral distribution of the sample covariance matrix
\citep{gao2017high}. Hence, in analyzing panel data with serially
correlated errors, using tests for cross-sectional independence
that are based on the assumption that the errors are serially
independent may lead to wrong conclusions.

To solve this problem,  \cite{baltagi2016} proposed a test for
cross-sectional correlation under heterogeneous panel
data models with serial  correlations by adjusting the sum based test
by \cite{pesaran2004}.
Similarly, \cite{lan2017} proposed a test for cross-sectional
independence under fixed effects panel data models with serial
correlations. Both
 tests are sum based, hence the scope of their
 application is limited to dense alternatives. As far as we know,
research on max based tests for sparse alternatives or
combined tests regardless of whether alternatives are sparse or not is
not yet available for data with serial dependence.

We aim to fill this gap in this work.
To this end, for testing cross-sectional independence under
heterogeneous panel data models with serial correlations, we propose a
max based test based on the maximum of the squared cross-sectional
residual correlations to deal with sparse alternatives. The method
follows the framework of \cite{chen2018} for testing
independence of correlated samples, while relaxing their Gaussian
sample assumption.
Furthermore, we propose a Fisher's combined probability test by
combining the p-values of our proposed max based test and the sum based
test  by \cite{baltagi2016}. This combined test is applicable regardless
the alternatives are sparse or sparse. We derive the
asymptotic null distribution of the proposed combined test, by first
rigorously establishing the asymptotic independence of the max based and
sum based test statistics.
In summary, there are two main contributions of our work.
\begin{enumerate}
 \item[(1)]
 We propose a max based test for testing cross-sectional
 independence
in models with serial correlation and sub-Gaussian error. The new test
is
powerful in detecting sparse alternatives.
 \item[(2)]
 We establish the asymptotic independence between the sum based and
 max based test statistics.
We propose a combined test for testing cross-sectional independence
in models with serial correlation and sub-Gaussian error. The new test is
powerful overall.
 \end{enumerate}

The rest of the paper is organized as follows. We propose two new
tests and establish their asymptotic properties in Section
\ref{sec:statistic}.
Simulation results of the proposed tests and their comparison to some
existing methods are demonstrated in Section \ref{sec:simu}, followed
by an empirical application in Section \ref{sec:data}. We
conclusion the paper in Section \ref{sec:discuss} and relegate
the technical proofs to Supplementary Materials.

\textsc{Notations.} For any square matrix $\mathbf{A},$
$(\mathbf{A})_{ij}$ denotes the $(i,j)$-th entry of $\mathbf{A}$,
$\tr(\mathbf{A})$ denotes the trace of $\mathbf{A}$,
$\lambda_{\max}(\mathbf{A})$ and $\lambda_{\min}(\mathbf{A})$ denote the
maximum and the minimum eigenvalues of $\mathbf{A}$, respectively,
$\|\mathbf{A}\|_{\mathrm{F}}$ denotes the Frobenius norm of $\mathbf{A}$, and
$\mathbf{A}^{1/2}$  denotes the principal square root matrix of
$\mathbf{A}$ if $\mathbf{A}$ is a positive definite matrix.
$\mathbf{I}_{n}$ denotes the $n\times n$ identity matrix for each positive integer $n$.
For any two real numbers $x$ and $y$, let $ x \vee y=\max (x, y)$ and
$x \wedge y=\min (x, y).$
For any vector $v,$ $\|v\|$ denotes the Euclidean norm of $v.$
Let $\mathcal{N}(a,b)$ denote the normal distribution with mean
$a\in\mathbb{R}$ and variance $b\geq0$, $t_{v}$ denote the
$t$-distribution with degree of freedom $v$  and $\chi_{v}^2$
 denote the chi-square distribution with degree of freedom $v.$
Let $\Phi(\cdot)$ denote the cumulative distribution function of the
standard normal distribution.
 We use $(N, T) \rightarrow \infty$ to denote the joint convergence of
 $N$ and $T$ to infinity, i.e. $\min(N,T)\to\infty$.

\section{ The proposed tests}\label{sec:statistic}

\subsection{Problem description}

We consider the heterogeneous panel data model taking the form
\begin{align}\label{model}
y_{i t}=x_{i t}^{\prime} \beta_{i}+\epsilon_{it},\,\, 1\leq i\leq N,\,\,1\leq t\leq T,
\end{align}
where $i$ indexes the cross-sectional units, and $t$ indexes the time dimension.
$y_{it}\in \mathbb{R}$ is the dependent variable, $x_{i
  t}\in\mathbb{R}^{p}$ is the exogenous regressors with slope
parameters $\beta_{i}\in
\mathbb{R}^{p}$ that are allowed to vary across $i$ and
$\epsilon_{it}\in\mathbb{R}$ is the corresponding idiosyncratic error
term.
For each $1\leq i\leq N$, let
$\mathbf{x}_{i}=(x_{i1},\cdots,x_{iT})^{\prime}\in\mathbb{R}^{T \times
  p},$
${y}_{i}=(y_{i1},\cdots,y_{iT})^{\prime}\in\mathbb{R}^{T }$ and
$\epsilon_{i\cdot}=(\epsilon_{i1},\cdots,\epsilon_{iT})^{\prime}\in\mathbb{R}^{T}$.
For each $1\leq t\leq T$, let $\epsilon_{\cdot
    t}=(\epsilon_{1t},\cdots,\epsilon_{Nt})^{\prime}$.
The null hypothesis of cross-sectional independence can be written as
\begin{align}\label{null hypothesis}
H_{0}: \epsilon_{1\cdot}, \epsilon_{2\cdot}, \cdots, \epsilon_{N\cdot}
  \text{ are
  independent random vectors.}
\end{align}

\subsection{Related works}\label{secR}

Early studies on testing cross-sectional independence in \eqref{null
  hypothesis} are based on the assumption that there is  no serial
dependence in $\{\epsilon_{\cdot t}\}_{t=1}^T$, where
$\epsilon_{\cdot t}$ is assumed to be iid over time $t$.  The earliest
work is the LM test \citep{breusch1980l},
with test statistic
$$
\mathrm{L M_{B P}}=T \sum_{j=2}^{N} \sum_{i=1}^{j-1} \hat{\rho}_{i j}^{2},
$$
where $\hat{\rho}_{i j}$ is the sample correlation constructed by the
OLS residuals $\hat{\epsilon}_{i t}=y_{i t}-x_{i t}^{\prime}
\hat{\beta}_{i}$,
with
$$
\hat{\beta}_{i}=\left(\mathbf{x}_{i}^{\prime}
  \mathbf{x}_{i}\right)^{-1} \mathbf{x}_{i}^{\prime} {y}_{i}
\text{~~and~~} \hat{\rho}_{i j}=\frac{\sum_{t=1}^{T} \hat{\epsilon}_{i
    t} \hat{\epsilon}_{j t}}{\sqrt{\sum_{t=1}^{T} \hat{\epsilon}_{i
      t}^{2} \sum_{t=1}^{T} \hat{\epsilon}_{j t}^{2}}}.
$$
The asymptotic null distribution of $\mathrm{LM}_{\mathrm{BP}}$ is a
chi-squared distribution with $N(N-1) / 2$ degrees of freedom, which
is established when $N$ is fixed and $T$ diverges to infinity.  Hence,
it is not applicable to the case of large $N$. To overcome the size
distortions of the scaled version of the $\mathrm{LM_{BP}}$ test
proposed by \cite{pesaran2004} for large $N$ and small $T$,
\cite{pesaran2008} proposed a bias-adjusted test, with
  test statistic
$$
\mathrm{LM_{PUY}}=\sqrt{\frac{2}{N(N-1)}} \sum_{j=2}^{N}
\sum_{i=1}^{j-1}\frac{(T-p) \hat{\rho}_{i j}^{2}-\mu_{T i j}}{v_{T i
    j}},
$$
where
\begin{align*}
&\mu_{T i j}=\frac{1}{T-p} \tr\left[\mE\left(\mathbf{P}_{i}
  \mathbf{P}_{j}\right)\right],~~v_{T i
  j}^{2}=\left[\tr^{2}\left\{\mE\left(\mathbf{P}_{i}
  \mathbf{P}_{j}\right)\right\}\right] a_{1 T}+2
  \tr\left[\left\{\mE\left(\mathbf{P}_{i}
  \mathbf{P}_{j}\right)\right\}^{2}\right] a_{2 T},\\
&a_{1 T}=a_{2 T}-\frac{1}{(T-p)^{2}} \text{~~and~~} a_{2
  T}=3\left\{\frac{(T-p-8)(T-p+2)+24}{(T-p+2)(T-p-2)(T-p-4)}\right\}^{2}.
\end{align*}
Here, for each $1\leq i\leq N$,
$\mathbf{P}_{i}=\mathbf{I}_{T}-\mathbf{x}_{i}\left(\mathbf{x}_{i}^{\prime}
  \mathbf{x}_{i}\right)^{-1} \mathbf{x}_{i}^{\prime}$,
hence $\hat{\epsilon}_{i\cdot}=\left(\hat{\epsilon}_{i 1}, \cdots,
  \hat{\epsilon}_{i T}\right)^{\prime}=\mathbf{P}_{i} \epsilon_{i\cdot}$.
The asymptotic null distribution of $\mathrm{LM_{PUY}}$ is
$\mathcal{N}(0,1)$, which is established when $T \rightarrow \infty$
first and then $N \rightarrow \infty$.
Later, \cite{feng2020} established that $\mathrm{LM_{PUY}}\to
\mathcal{N}(0,1)$ in distribution when $\min(N,T)\to\infty$,
and also established  that
$\mathrm{LM_{FJLX}}\to \mathcal{N}(0,1)$ in distribution when $\min(N,T)\to\infty$, where
\begin{align*}
\mathrm{LM_{FJLX}}=\frac{\sum_{j=2}^{N} \sum_{i=1}^{j-1}T \hat{\rho}_{i j}^{2}-\mu_{N}}{N} \text{ and } \mu_{N}=\frac{T}{(T-p)^2} \sum_{1 \leq i<j \leq N} \tr\left(\mathbf{P}_i \mathbf{P}_j\right).
\end{align*}

These tests are all  based on the sum of squared sample correlations
$\sum\sum_{1\le i<j\le N}\hat{\rho}_{i j}^{2}$.
 In contrast, \cite{pesaran2004} proposed a test for cross-sectional
 independence directly based on the sum of sample correlations,
 with test statistic
$$
\mathrm{C D_{P}}=\sqrt{\frac{2 T}{N(N-1)}} \sum_{j=2}^{N} \sum_{i=1}^{j-1} \hat{\rho}_{i j}.
$$
\cite{pesaran2015} established the asymptotic null distribution of
$\mathrm{C D_{P}}$ to be $\mathcal{N}(0,1)$ when $\min(N,T)\to\infty$.
Recently, to test cross-sectional correlation with serially
correlated errors, \cite{baltagi2016} proposed a new test with  test
statistic
\begin{equation}\label{sumtest}
S_{N}=\sqrt{\frac{2}{N(N-1)}}\underset{1\leq i<j\leq N}{\sum\sum}\hat{\rho}_{i j},
\end{equation}
which is very similar to $\mathrm{CD_{P}}$, but has different
asymptotic variance. Under certain assumptions, they established that
under the null hypothesis,
\begin{equation}\label{sumtest0}
S_{N}/{\hat{\sigma}}_{S_{N}}\to \mathcal{N}(0,1)  \text{~~in
  distribution when~~} \min(N,T)\to\infty,
\end{equation}
hence a level-$\alpha$ test can be performed by rejecting $H_0$
when $S_N/\hat{\sigma}_{S_N}$ is larger than the $(1-\alpha)$-quantile
$z_{\alpha}= \Phi^{-1}(1-\alpha).$ Here,
\begin{align}\label{sigmasumhat}
\hat{\sigma}^2_{S_{N}}=\frac{2}{N(N-1)}\underset{1\leq i<j\leq
  N}{\sum\sum} v_{j}^{\prime}\left(v_{i}-\bar{v}_{ij}\right)
  v_{i}^{\prime}\left(v_{j}-\bar{v}_{i j}\right),
\end{align}
 $\bar{v}_{ij}=\sum_{1<k \neq i, j<N} v_{k}/(N-2)$ and
 $v_{k}=\hat{\epsilon}_{k\cdot}/\|\hat{\epsilon}_{k\cdot}\|$ for all $1\leq
 k\leq N.$

\subsection{Max based test}

All tests mentioned in Section \ref{secR} are sum based tests, which
generally perform very well under dense alternatives,
i.e. alternatives with dense cross-sectional correlation matrices,
but may suffer from low power against sparse alternatives,
i.e. alternatives with sparse cross-sectional correlation matrices. To
deal with sparse alternatives,
we now propose a max based test based on the maximum of the squares of the
sample cross-sectional correlations. The test statistic we propose is
\begin{equation}\label{maxtest}
L_{N}=\max_{1\leq i<j\leq N}\hat{\rho}^2_{ij}.
\end{equation}
Max based tests have been widely studied in testing independence
among variables, e.g.,  \cite{lx2015}, \cite{chen2018}  and
\cite{feng2020}. Specifically, \cite{feng2020} used it to test cross-sectional
independence under the assumption of no serial correlation between the
errors, and established that
$TL_{N}-4 \log N+\log\log N  \to G(y)$ in distribution, where
$G(y)=\exp \big\{- \exp \left(-y/2\right)/\sqrt{8 \pi}\big\}$.
However, this test may perform poorly if blindly used in
the situation where serial correlation exists.

To utilize the test statistic in \eqref{maxtest}
for data with serial correlations, we must reinvestigate
 the asymptotic properties of $L_{N}$.
We impose the following assumptions.

\begin{assumption}\label{assum:E distribution}
Assume  that
$\mathbf{E}=(\epsilon_{1\cdot},\cdots,\epsilon_{N\cdot})^{\prime}=\mathbf{U}^{1/2}\mathbf{Z}\mathbf{\Sigma}^{1/2},$
where $\mathbf{U}\in \mathbb{R}^{N \times N}$  and $\mathbf{\Sigma}\in
\mathbb{R}^{T \times T}$ are positive definite matrices,
all elements of $\mathbf{Z}\in \mathbb{R}^{N \times T}$ are iid
variables with mean zero and variance one, the density function of
$(\mathbf{Z})_{it}$ is symmetric and the sub-Gaussian norm of
$(\mathbf{Z})_{it}$ is bounded by $K$,
i.e. $\mE\exp{(\mathbf{Z})^2_{it}/K^2}\le 2$, for each $1\leq i\leq N$
and $1\leq t\leq T$.
\end{assumption}
\begin{assumption}\label{assum:x_it}
(i) Assume that $p >0$ is fixed, and the regressors $x_{i t}$ are strictly exogenous, such that $$
\mE\left(\epsilon_{i t} | \mathbf{x}_{i}\right)=0, \text { for all } 1\leq i\leq N \text { and } 1\leq t\leq  T
.$$ (ii)Assume that $\mathbf{x}_{i}^{\prime}\mathbf{x}_{i}/T=
\sum_{t=1}^{T} {x}_{i t} {x}_{i t}^{\prime}/T$ is non-singular and $\mathbf{x}_{i}\big(\mathbf{x}_{i}^{\prime}\mathbf{x}_i/T\big)^{-1} \mathbf{x}_{i}^{\prime}$
 is stochasticly bounded for all $1\leq i\leq N.$
\end{assumption}

\begin{assumption}\label{assum:matrix}
 (i) Assume that $C^{-1} \leq \lambda_{\min }(\mathbf{\Sigma}) \leq \lambda_{\max }(\mathbf{\Sigma})
 \leq C$ and $C^{-1} \leq \lambda_{\min }(\mathbf{U}) \leq$
 $\lambda_{\max }(\mathbf{U}) \leq C$ for some constant $C>0$.
 (ii) Assume that  $\max_{j=1, \dots T}\sum_{k=1}^{T}\left|(\mathbf{\Sigma})_{j
     k}\right|^{\tau} \leq C'$
 and $\max_{j=1, \dots
     N}\sum_{k=1}^{N}\left|(\mathbf{U})_{j k}\right|^{\tau} \leq C'$
 for some $0<\tau<2$ and $C'>0$.
  (iii) Assume that all diagonal elements of $\mathbf{U}$ are one.
 \end{assumption}

Assumption \ref{assum:E distribution} writes the error into
$\mathbf{E}=\mathbf{U}^{1/2}\mathbf{Z}\mathbf{\Sigma}^{1/2}$, which is
investigated by \cite{hornstein2018joint} in the case that the random
matrix $\mathbf{Z}$ has independent sub-Gaussian entries.
$\mathbf{U}$ represents the row covariance matrix of $\mathbf{E}$,
containing information on the cross-sectional correlation between
$\{\epsilon_{1\cdot},\cdots,\epsilon_{N\cdot}\}$,  whereas
$\mathbf{\Sigma}$
represents the column covariance matrix of $\mathbf{E}$, containing
the serial correlation information between $\{\epsilon_{\cdot 1},\cdots,\epsilon_{\cdot
  T}\}$. It is known that when $(\mathbf{Z})_{it}\overset{iid}{\sim}
\mathcal{N}(0,1)$, $\mathbf{E}$ has the matrix-variate normal
distribution $\mathcal{N}({\bf 0}, \mathbf{U}\otimes\mathbf{\Sigma})$,
where $\otimes$ denotes the Kronecker product.

Assumption \ref{assum:x_it}(i) is used in \cite{pesaran2008}, which is
a common condition for panel data model in \eqref{model}.
Assumption \ref{assum:x_it}(ii) is imposed to ensure that the
difference between the distribution of the max based statistic based on
the residuals and that based on the errors is negligible.

Assumption \ref{assum:matrix}(i) is the same as Condition (C1) in
\cite{chen2018}, which is a common eigenvalue assumption in
high-dimensional inference literature such as in \cite{crz2016} and
contains many important types of covariance matrices, including the
bandable, Toeplitz and sparse covariance matrices.
Assumption \ref{assum:matrix}(ii) is the same as Condition (C2) in
\cite{chen2018}, which assumes the sparsity of $\mathbf{U}$ and
$\mathbf{\Sigma}$.
Note that the correlation between the rows in $\mathbf{E}$ is the
correlation we are interested in testing, which corresponds to
the diagonality of $\U$. The
  sparsity of $\U$ characterizes the sparse alternative. On the other
  hand, the sparsity of $\mathbf{\Sigma}$ is imposed to ensure good
  estimation property. To eliminate the complexity
caused by the serial correlation across $t$ within each row, ideally
we hope to work with
$\mathbf{E}\mathbf{\Sigma}^{-1/2}=\mathbf{U}^{1/2}\mathbf{Z}$, and to
test if $\U$ has zero off-diagonal elements. Thus, in our test
procedure, we will need to estimate
a $\mathbf{\Sigma}$ dependent quantity
$\tr^2(\mathbf{\Sigma})/\|\mathbf{\Sigma}\|_{\mathrm{F}}^2$. Since
$\mathbf{\Sigma}$ is a high dimensional matrix, sparsity is a common
assumption to regularize the properties of the related estimation.
 Finally,
 Assumption \ref{assum:matrix}(iii) is imposed to make the
error model identifiable.

Throughout the text, we let $\gamma$ be a positive finite constant.
\begin{theorem}\label{th:max null}
Under Assumptions \ref{assum:E distribution}-\ref{assum:matrix} and
the null hypothesis in \eqref{null hypothesis},
when $\min(N,T)\to\infty$ and $N/T\to\gamma$,
for any
$y\in\mathbb{R},$
we have
$$\mathbb{P}\left(\frac{\tr^2({\mathbf{\Sigma}})}{\|{\mathbf{\Sigma}}\|_{\mathrm{F}}^{2}}
L_{N}-4 \log N+\log\log N  \leq y\right) \rightarrow G(y),$$
where $G(y)=\exp \big\{- \exp \left(-y/2\right)/\sqrt{8 \pi}\big\}$
is a type-I Gumbel distribution function.
\end{theorem}
The column covariance matrix $\mathbf{\Sigma}$, which is needed in the
construction of the max based test statistic, can be
assessed via a procedure similar to that proposed in
\cite{chen2018} for testing independence among variables using
correlated samples. First, define the column sample covariance matrix
 $\hat{\mathbf{\Sigma}}=\left(\hat{\sigma}_{i j}\right)_{1 \leq i, j \leq T}$ with
 $$\hat{\sigma}_{i j}=\frac{1}{N-1} \sum_{l=1}^{N}\left(\hat{\epsilon}_{l i}-
 \bar{\hat{\epsilon}}_{\cdot i}\right)\left(\hat{\epsilon}_{l j}-\bar{\hat{\epsilon}}_{\cdot j}
 \right),\quad \bar{\hat{\epsilon}}_{\cdot j}=\frac{1}{N}
 \sum_{l=1}^{N}{\hat{\epsilon}}_{lj} \text{~~and~~} \theta_{i
   j}=\frac{\hat{\sigma}_{i j}}{\sqrt{\hat{\sigma}_{i i}
     \hat{\sigma}_{j j}}}$$
 for each $1 \leq i, j \leq T$. Then, define
 $\tilde{\mathbf{\Sigma}}=\left(\tilde{\sigma}_{i j}\right)_{1 \leq i,
   j \leq T}$ with
\begin{equation}\label{v>1.42}
\tilde{\sigma}_{i j}=\left\{
\begin{aligned}
&\hat{\sigma}_{i j} I\bigg\{\frac{|{\theta}_{i j}|}{1-{\theta}_{i j}^{2}} \geq
\nu \sqrt{\frac{\hat{P}_{N} \log T}{N}}\bigg\}  , & i\neq j, \\
&\hat{\sigma}_{i i} , & i=j
\end{aligned}
\right.
\end{equation}
for each $1 \leq i, j \leq T$. Here, $\nu>\sqrt{2}$ and $
\hat{P}_{N}=\left[\|\hat{\mathbf{U}}\|_{\mathrm{F}}^{2}-\frac{1}{T}
\{\tr(\hat{\mathbf{U}})\}^{2}\right]/N,$
where $(\hat{\mathbf{U}})_{ij}=\hat{\epsilon}_{i\cdot}^{\prime}\hat{\epsilon}_{j\cdot}/
\tr(\hat{\mathbf{\Sigma}})$ for each $1 \leq i, j \leq N.$
Based on $\tilde{\mathbf{\Sigma}}$, $\tr^2(\mathbf{\Sigma})$ and
$\|{\mathbf{\Sigma}}\|_{\mathrm{F}}^{2}$
are estimated by $\tr^2(\tilde{\mathbf{\Sigma}})$ and
$\|{\tilde{\mathbf{\Sigma}}}\|_{\mathrm{F}}^{2}$, respectively.
Hence, ${\tr^2({\mathbf{\Sigma}})}/{\|{\mathbf{\Sigma}}\|_{\mathrm{F}}^{2}}$
is estimated by ${\tr^2(\tilde{\mathbf{\Sigma}})}/{\|\tilde{\mathbf{\Sigma}}\|_{\mathrm{F}}^{2}}$,
which is a ratio-consistent estimator of
${\tr^2({\mathbf{\Sigma}})}/{\|{\mathbf{\Sigma}}\|_{\mathrm{F}}^{2}}$
as indicated in the following theorem.

\begin{theorem}\label{th:max cons}
Under Assumptions \ref{assum:E distribution}-\ref{assum:matrix} and
the null hypothesis in \eqref{null hypothesis}, for any $\nu > \sqrt{2}$,
\begin{align}
\frac{\tr^2(\tilde{\mathbf{\Sigma}})}{\|\tilde{\mathbf{\Sigma}}\|_{\mathrm{F}}^{2}}
\frac{\|{\mathbf{\Sigma}}\|_{\mathrm{F}}^{2}}{\tr^2({\mathbf{\Sigma}})}
=1+O_{\mathbb{P}}\left\{\Big(\frac{\log T}{N}\Big) ^{ \frac{1}{2}\wedge(1-\frac{\tau}{2})}\right\},
\end{align}
when $\min(N,T)\to\infty$ and
   $N/T\to\gamma$.
\end{theorem}

Combining Theorems \ref{th:max null} and \ref{th:max cons}, we have that for any $y\in\mathbb{R},$
\begin{align}\label{maxtest0}
\mathbb{P}\left(\frac{\tr^2(\tilde{\mathbf{\Sigma}})}{\|\tilde{\mathbf{\Sigma}}\|_{\mathrm{F}}^{2}}
L_{N}-4 \log N+\log \log N  \leq y\right) \rightarrow G(y).
\end{align}
Based on \eqref{maxtest0}, for a given significance level $\alpha,$
the null hypothesis in \eqref{null hypothesis} will be rejected by the
established max based test when
$L_{N}{\tr^2(\tilde{\mathbf{\Sigma}})}/{\|\tilde{\mathbf{\Sigma}}\|_{\mathrm{F}}^{2}}
 \geq w_{\alpha}+ 4 \log N-\log \log N$,  where $w_{\alpha}$ is the $1-\alpha$ quantile of
 the type I extreme value distribution with the cumulative
 distribution function $G(y)$ and has the specific form of
 $w_{\alpha}=\log (8 \pi)-2 \log \log (1-\alpha)^{-1}.$

Next, we turn to the power analysis of the proposed max based test.

\begin{theorem}\label{th:maxpower}
 Under Assumptions \ref{assum:E distribution}-\ref{assum:matrix},
 suppose that for some $\delta>2,$ some $\nu>0,$ and sufficiently large $N$ and $T$,
 $
u_{N}=\max _{1 \leq i<j \leq N}\left|(\mathbf{U})_{i j}\right| \geq
\delta \sqrt{\|\mathbf{\Sigma}\|_{\mathrm{F}}^{2} \log
  N/\tr^{2}(\mathbf{\Sigma})},
$
then
\begin{align}\label{maxpower}
\mathbb{P}\left(\frac{\tr^2(\tilde{\mathbf{\Sigma}})}{\|\tilde{\mathbf{\Sigma}}\|_{\mathrm{F}}^{2}}
L_{N}-4 \log N+\log \log N> w_{\alpha}\right) \rightarrow 1,
\end{align}
as $\min(N,T)\to\infty$ and $N/T\to\gamma$.
\end{theorem}

Theorem \ref{th:maxpower} indicates that the proposed
  max based test is consistent under the sparse alternative in which
the maximum non diagonal  entries of $\mathbf{U}$ is
  sufficiently large. More specifically, because
  $\sqrt{\|\mathbf{\Sigma}\|_{\mathrm{F}}^{2}
    /\tr^{2}(\mathbf{\Sigma})}\asymp T^{-1/2}$ under Assumption
  \ref{assum:matrix}, the test is able to
detect the dependence as long as a single covariance is at the order
of $(\log N/T)^{1/2}$, which leads to an overall detection
  rate of $(\log N/T)^{1/2}$.  In contrast, the sum based test
can detect  the departure from null if each covariance
  reaches $T^{-1/2}N^{-1}$
  (Theorem 4 of \cite{baltagi2016}), hence the overall detection rate is $N/T^{1/2}$.
This theoretical result is consistent with the simulation results in
Section \ref{sec:simu}, where the max based test performs
better than the sum based test in terms of empirical power
under sparse alternatives.
The presence of the possible serial correlation makes it
more difficult to establish asymptotic properties of
$L_{N}$. For example, we need to reestablish the
asymptotic variance of $L_{N}$ that depends on $\mathbf{\Sigma}$. This is
the reason why we replace the test statistic $TL_{N}-4 \log N+\log\log
N$ in \cite{feng2020}, where no serial correlation is considered, by
$L_{N}{\tr^2({\mathbf{\Sigma}})}/{\|{\mathbf{\Sigma}}\|_{\mathrm{F}}^{2}}-4
\log N+\log\log N$ and
$L_{N}{\tr^2(\tilde{\mathbf{\Sigma}})}/{\|\tilde{\mathbf{\Sigma}}\|_{\mathrm{F}}^{2}}-4
\log N+\log\log N$.

On the other hand, we note that the max based test for testing
independence among normally distributed variables with correlated
samples has been studied in \cite{chen2018}, whereas in this paper we
consider the max based test under panel data models with
strictly
exogenous regressors, and we relax the Gaussian assumption to
sub-Gaussian. The proof under sub-Gaussianity
is much more challenging because many properties associated with
Gaussianity, such as rotation invariance, can no longer
  be used.
To establish the theoretical results under sub-Gaussian distributions,
more advanced technical tools, such as the
Hanson-Wright inequality \citep{rudelson2013}, need to be engaged in deriving
the asymptotical distribution of the max based test statistic
$L_{N}$. In addition, the expressions of the moments of the quadratic
forms under sub-Gaussian distributions are also much more complex than that
under Gaussian distributions, which further complicate the proof.

\subsection{Fisher's combined probability test}

It is intuitively expected that the sum based test $S_N$ performs well
under
  dense alternatives, whereas the max based test $L_N$ performs well
  under sparse alternatives. However, in practice,  it is usually
  unknown whether the correlation matrix of the errors is sparse or
  not, hence it is difficult to decide which test to
  use.  For this reason, in this subsection, we propose using Fisher's
  combined probability test by combining the sum based and max based
  tests, which is expected to take advantage of both
  tests. To construct the test, we first
  need to establish the asymptotic
  independence between the two statistics $S_N$ and $L_N$ under the
  null hypothesis.

\begin{theorem}\label{th:sum-max}
Under Assumptions \ref{assum:E distribution}-\ref{assum:matrix} and
the null hypothesis in \eqref{null hypothesis},
$L_{N}{\tr^2(\tilde{\mathbf{\Sigma}})}/{\|
 \tilde{\mathbf{\Sigma}}\|_{\mathrm{F}}^{2}}
-4 \log N+\log \log N $ and $S_{N}/\hat{\sigma}_{S_{N}}$ are
asymptotically independent, as $\min(N,T)\to\infty$ and
  $N/T\to\gamma$.
\end{theorem}

Note that in many related studies, such as \cite{lx2015} and \cite{feng2020},
the asymptotic independence between the sum based and max based
statistics is established under the assumption that all components of
the random vectors concerned are independent and identically
distributed. In contrast, we establish the asymptotic independence in
the situation where the error vectors may have serial correlation,
which makes the proof of the asymptotic independence much more
challenging and requires more complex technical treatments
  and tools.

Let $P_{L_{N}} = 1-G\Big\{L_{N}{\tr^2(\tilde{\mathbf{\Sigma}})}/
{\|\tilde{\mathbf{\Sigma}}\|_{\mathrm{F}}^{2}}
-4\log N+\log \log N \Big\}$ and
$P_{S_{N}} =1-\Phi\left(S_{N} / \hat{\sigma}_{S_{N}}\right)$. We
construct Fisher's combined probability test statistic as
\begin{align}\label{testSUMMAX}
T_{C}=-2\log(P_{L_{N}})-2\log(P_{S_{N}}).
\end{align}
By Theorem \ref{th:sum-max}, \eqref{sumtest0} and \eqref{maxtest0},
we see that $P_{L_{N}}$ and $P_{S_{N}}$
are asymptotically independent under
the null hypothesis and each has limit distribution $U[0,
1]$, the uniform distribution on $[0, 1]$. We thus obtain
  the following corollary.

\begin{corollary}\label{Tc}
Assume that the assumptions in Theorem \ref{th:sum-max} hold,
 then we have $T_{ C } \rightarrow \chi_{4}^{2}$ in distribution when
 $\min(N,T)\to\infty$ and $N/T\to\gamma$.
\end{corollary}

According to Corollary \ref{Tc}, we proposed Fisher's combined
probability tes at level $\alpha$ by rejecting the null hypothesis in \eqref{null
  hypothesis} if $T_{C} \geq q_{\alpha}$, where
$q_{\alpha}$ is the $1-\alpha$ quantile of $\chi_{4}^{2}$.

\section{Simulation studies}\label{sec:simu}

We now conduct simulations to investigate the finite sample performance of
the two tests proposed in this paper, i.e.
the test based on $L_{N}$ and the Fisher's combined probability test based
on $T_{C}$. For comparison, we also implement three
other existing tests, i.e. the  test based on $S_{N}$
proposed by \cite{baltagi2016}, the $\mathrm{LM_{PUY}}$ test
proposed by \cite{pesaran2008} and the $\mathrm{CD_{P}}$ test proposed by \cite{pesaran2004}.
Here,  the max based test is implemented by setting $\nu = 1.42$ in
\eqref{v>1.42} as in \cite{chen2018}.
For simplicity, we will abbreviate these five tests as $L_{N}$,
$T_{C}$, $S_{N}$,  $\mathrm{LM_{PUY}}$ and $\mathrm{CD_{P}}$,
respectively.

We consider the data generating process
\begin{align}\label{simu model}
y_{i t}=\alpha_{i}+\sum_{l=2}^{p}x_{li,t} \beta_{li}+\epsilon_{i t},\,\, 1\leq i\leq N,\,\,1\leq t\leq T,
\end{align}
where $x_{it}=\left(1, x_{2 i ,t}, \cdots, x_{pi,t }\right)^{\prime}
 \in \mathbb{R}^{p}$ and $\beta_{i}= \left(\alpha_{i}, \beta_{ 2i}, \cdots, \beta_{ pi }\right)^{\prime}  \in \mathbb{R}^{p}$.
We generate $\alpha_{i} \mathop{\sim}\limits^{iid} \mathcal{N}(0,1)$ and
 $\beta_{li} \mathop{\sim}\limits^{iid}\mathcal{N}(1,0.04)$ for $2 \leq l \leq p.$
The strictly exogenous regressors are generated by $x_{li ,t }=0.6 x_{li, t-1 }+v_{li ,t }$
for $1\leq i\leq N  ,$ $-50 \leq t \leq T$ and $2\leq l \leq p$,  where
$x_{ li,- 51}=0$ and $v_{li, t } \mathop{\sim}\limits^{iid}\mathcal{N}\left(0, \psi_{li  }^{2}
/\left(1-0.6^{2}\right)\right)$ with $\psi_{li  }^{2} \mathop{\sim}\limits^{iid}
 \chi_{6}^{2} / 6 $.

Consider the following two settings of serial correlation of the errors $\epsilon_{i t}$.
\begin{enumerate}[(i)]
 \item The errors follow an
auto-regressive (AR) model of order one over time, i.e. AR(1): $\epsilon_{i 1}=e_{i 1}$
 and $\epsilon_{i t}=0.6 \epsilon_{i t-1}+e_{i t}$
for $2\leq t\leq T$ and $1\leq i\leq N.$  \label{H01}
\item The errors follow an auto-regressive and
moving average (ARMA) model of order (1,1) over time, i.e. ARMA(1,1): $\epsilon_{i 1}=e_{i 1}$
 and
$\epsilon_{i t}=0.6 \epsilon_{i t-1}+e_{i t}+0.2e_{i t-1}$ for $2\leq t\leq T$ and $1\leq i\leq N.$
\label{H02}
\end{enumerate}
To produce data under the null hypothesis, $e_{it}$ are independently generated from the following three distributions:
(1) $\mathcal{N}(0,1);$
(2) $t_{6}/\sqrt{6/4};$
(3) $(\chi_{5}^{2}-5)/\sqrt{10}$.

Then, we turn to produce data under the alternative hypothesis. The
data generating process is specified as
\begin{align}\label{simu model1}
y_{i t}=\alpha_{i}+\sum_{l=2}^{p}x_{li,t} \beta_{li}+\epsilon_{i t}^{\ast},\,\, 1\leq i\leq N,\,\,1\leq t\leq T,
\end{align}
where $\epsilon_{i t}^{\ast}$ are generated from the following two settings.
\begin{enumerate}[(I)]
 \item Non-sparse case. Spatial moving average (SMA) model with order one, i.e.  SMA(1): for all $1\leq t\leq T,$ $\epsilon_{1 t}^{*}=0.5\delta\epsilon_{2 t}+\epsilon_{1 t},$
 $\epsilon_{N t}^{*}=0.5\delta\epsilon_{N-1 t}+\epsilon_{N t}$ and
$\epsilon_{i t}^{*}=\delta\left(0.5 \epsilon_{i-1 t}+0.5 \epsilon_{i+1 t}\right)
+\epsilon_{i t},$
where $2\leq i\leq N-1,$ $\delta=0.2$ and $\epsilon_{it}$ are
generated from settings (\ref{H01})-(\ref{H02}) with distributions
(1)-(3).\label{H11}
\item Sparse case. Let $(\epsilon_{1\cdot}^{\ast},\cdots,\epsilon_{N\cdot}^{\ast})^{\prime}
=\mathbf{\Psi}^{1/2}(\epsilon_{1\cdot},\cdots,\epsilon_{N\cdot})^{\prime},$
 where $\epsilon_{i\cdot}^{\ast}=(\epsilon_{i1}^{\ast},\cdots,\epsilon_{iT}^{\ast})^{\prime}$ and $\epsilon_{i\cdot}=(\epsilon_{i1},\cdots,\epsilon_{iT})^{\prime},$ for all $1\leq i\leq N.$ Here, $\epsilon_{i\cdot}$ are generated from settings (\ref{H01})-(\ref{H02}) with distributions (1)-(3).
 $\mathbf{\Psi}$ is constructed as follows.
Randomly select a subset $S \subset\{1, \cdots, N\}$ with cardinality
$\lceil N^{0.3}\rceil$,  the greatest integer grater than or equal to $N^{0.3}.$ Let $(\mathbf{\Psi})_{i j}=1$
 if $i=j$.
For $i<j$, define $(\mathbf{\Psi})_{i j}=0$ if $i \notin S$ or $j \notin S$,
and $(\mathbf{\Psi})_{i j}\mathop{\sim}\limits^{iid} U[\sqrt{4 \log N / T},
\sqrt{6\log N / T}]$ if $i, j \in S.$
\label{H12}
\end{enumerate}

In addition, for the choices of $p$, $N$ and $T$,  we set $p\in \{3,5\},$ $N \in\{100, 200\}$, $T \in \{200, 300, 400,500\}.$

\begin{figure}[http]
\centering
\includegraphics[width=6in]{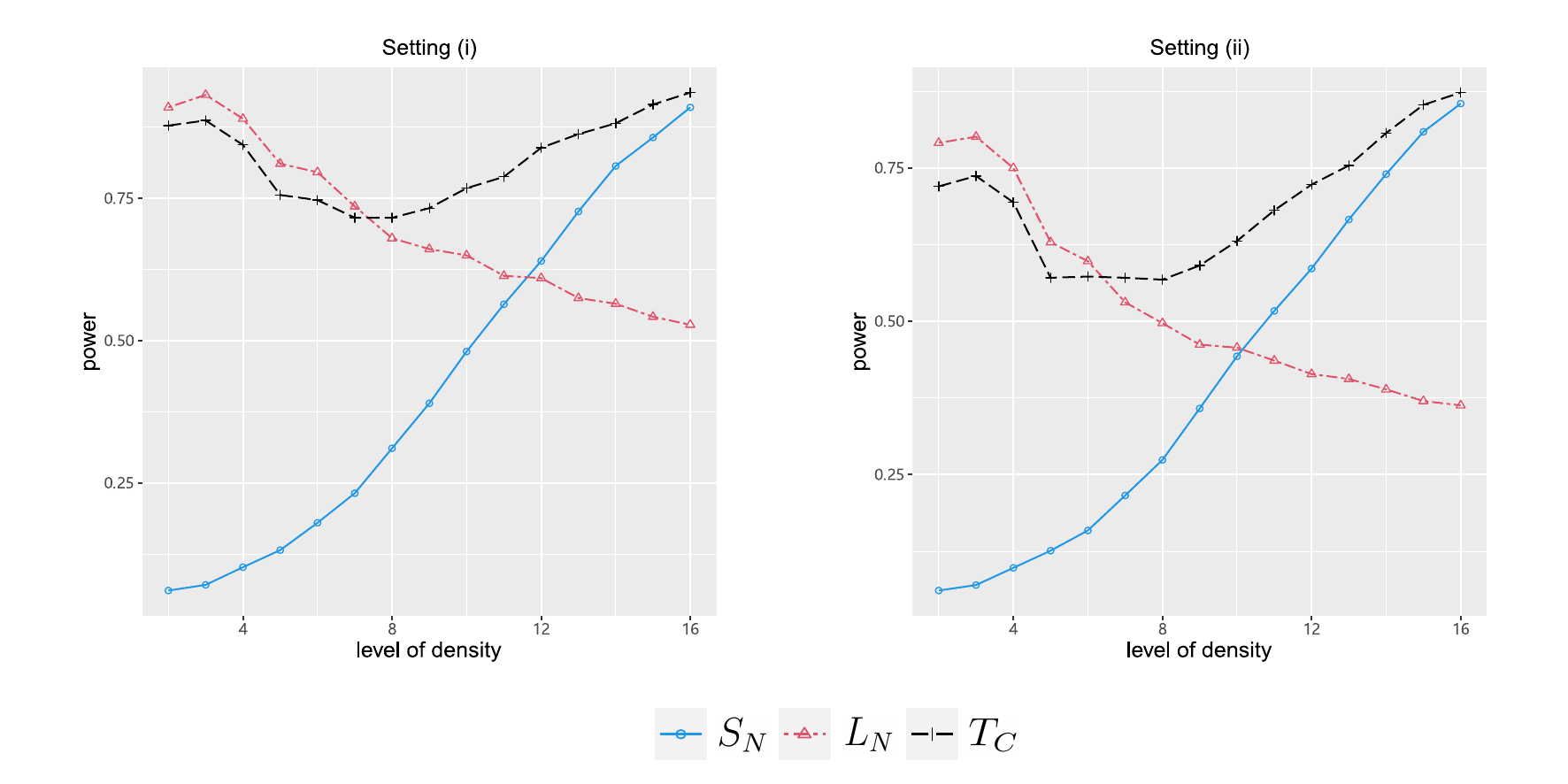}
\caption{The empirical power curves of the three tests at 5\% level under settings
 (\ref{H01}) and (\ref{H02}).}\label{fig1}
\end{figure}

The results of the empirical size and power of the five tests in the
non-sparse and sparse cases of error correlation matrices are
summarized in Tables \ref{t1}-\ref{t3}. The power curves are
plotted in Figure \ref{fig1}. All results are based on 1,000
replications. We next analyze them in detail.

Table \ref{t1} indicates that in most cases $T_{C}$, $L_{N}$ and
$S_{N}$ have empirical sizes not much larger than $5\%$. Here, the
max based test $L_{N}$ and Fisher's combined probability test $T_{C}$
tend to have smaller empirical sizes than the sum based test $S_{N}$,
especially as $T$ is relatively small. This is not surprising
and  is common for many max based tests, because the convergence rate
of the type I extreme distribution is typically slow
\citep{liu2008asymptotic}.
$\mathrm{CD_{P}}$ and $\mathrm{LM_{PUY}}$ fail to control the
empirical size because $\mE(\hat{\rho}_{ij}^2)$ may be seriously
affected by serial correlation.
This is also observed by \cite{baltagi2016}.

Tables \ref{t2} and \ref{t3} show the empirical powers in both
non-sparse and sparse cases of correlation matrices.
Since $\mathrm{CD_{P}}$ and $\mathrm{LM_{PUY}}$ fail to control the
empirical size, we exclude them from the empirical power results.
Table \ref{t2} and \ref{t3} indicate that $S_{N}$ and $T_{C}$
generally perform better than $L_{N}$ in non-sparse cases in terms of
empirical
powers, whereas $L_{N}$ and $T_{C}$ generally perform better than
$S_{N}$ in sparse cases. As expected, Fisher's combined probability
test $T_{C}$ has power
advantages regardless  the local alternative is sparse or not.

\begin{table}[!ht]
           \centering
           \caption{The empirical size of the five tests under
            settings (\ref{H01}) and (\ref{H02}) with distributions (1)-(3) at 5\% level. }\label{t1}
           \vspace{0.35cm}
      \renewcommand{\arraystretch}{0.98}
     \tabcolsep 3pt
          \scalebox{0.75}{
         \begin{tabular}{cccccccccccccccccccccccc}\hline \hline                                            \multicolumn{1}{l}{} & \multicolumn{1}{l}{} & \multicolumn{1}{l|}{} & \multicolumn{8}{c|}{Setting (\ref{H01})}                         & \multicolumn{8}{c}{Setting (\ref{H02})}                        \\\hline
                     & \multicolumn{2}{c|}{$p$}                     & \multicolumn{4}{c|}{3}     & \multicolumn{4}{c|}{5}     & \multicolumn{4}{c|}{3}     & \multicolumn{4}{c}{5}    \\\hline
                     & $N$                  & \multicolumn{1}{c|}{$T$}                  & 200  & 300  & 400  & \multicolumn{1}{c|}{500}  & 200  & 300  & 400  & \multicolumn{1}{c|}{500}  & 200  & 300  & 400  & \multicolumn{1}{c|}{500}  & 200  & 300  & 400 & 500  \\\hline
                     & \multicolumn{18}{c}{(1) Normal distribution}                                                                                                                                                                                                      \\\hline
 & 100 & \multicolumn{1}{c|}{$S_{N}$}           & 6.2  & 5.4  & 4.6  & \multicolumn{1}{c|}{5.9}  & 5.7  & 6.4  & 5.2  & \multicolumn{1}{c|}{4.8}  & 6.0    & 5.5  & 4.4  & \multicolumn{1}{c|}{5.6}  & 5.6  & 6.5  & 5.2  & 4.6  \\
 &     & \multicolumn{1}{c|}{$L_{N}$}           & 3.4  & 3.2  & 3.9  & \multicolumn{1}{c|}{4.9}  & 3.6  & 4.4  & 3.9  & \multicolumn{1}{c|}{3.9}  & 2.2  & 2.5  & 3.7  & \multicolumn{1}{c|}{4.4}  & 3.4  & 3.9  & 3.4  & 4.0    \\
 &     & \multicolumn{1}{c|}{$T_{C}$}             & 3.5  & 4.8  & 4.6  & \multicolumn{1}{c|}{4.9}  & 3.9  & 5.4  & 4.3  & \multicolumn{1}{c|}{4.9}  & 3.3  & 4.3  & 4.2  & \multicolumn{1}{c|}{4.2}  & 3.3  & 5.1  & 4.1  & 3.6  \\
 &     & \multicolumn{1}{c|}{$\mathrm{CD_{P}}$}   & 11.5 & 12.1 & 11.5 & \multicolumn{1}{c|}{13.6} & 12.5 & 14.3 & 12.2 & \multicolumn{1}{c|}{12.3} & 12.9 & 14.0   & 13.5 & \multicolumn{1}{c|}{15.5} & 14.3 & 16.3 & 14.3 & 14.0   \\
 &     & \multicolumn{1}{c|}{$\mathrm{LM_{PUY}}$} & 100  & 100  & 100  & \multicolumn{1}{c|}{100}  & 100  & 100  & 100  & \multicolumn{1}{c|}{100}  & 100  & 100  & 100  & \multicolumn{1}{c|}{100}  & 100  & 100  & 100  & 100  \\ \hline
 & 200 & \multicolumn{1}{c|}{$S_{N}$}           & 5.4  & 5.1  & 3.7  & \multicolumn{1}{c|}{6.4}  & 5.6  & 5.7  & 5.7  & \multicolumn{1}{c|}{5.4}  & 5.9  & 5.1  & 4.0    & \multicolumn{1}{c|}{6.6}  & 5.9  & 5.7  & 6.3  & 5.3  \\
 &     & \multicolumn{1}{c|}{$L_{N}$}           & 2.1  & 2.9  & 3.6  & \multicolumn{1}{c|}{3.6}  & 1.8  & 2.4  & 2.2  & \multicolumn{1}{c|}{3.1}  & 1.5  & 2.7  & 2.6  & \multicolumn{1}{c|}{3.4}  & 0.9  & 1.9  & 1.7  & 2.1  \\
 &     & \multicolumn{1}{c|}{$T_{C}$}             & 3.7  & 4.3  & 3.6  & \multicolumn{1}{c|}{5.1}  & 3.4  & 4.3  & 3.5  & \multicolumn{1}{c|}{3.7}  & 3.3  & 3.7  & 3.4  & \multicolumn{1}{c|}{4.3}  & 2.9  & 3.6  & 3.0    & 3.5  \\
 &     & \multicolumn{1}{c|}{$\mathrm{CD_{P}}$}   & 12.0   & 11.9 & 10.2 & \multicolumn{1}{c|}{13.9} & 12.3 & 12.6 & 13.0   & \multicolumn{1}{c|}{12.9} & 13.2 & 14.1 & 12.6 & \multicolumn{1}{c|}{15.3} & 13.7 & 14.8 & 15.2 & 14.9 \\
 &     & \multicolumn{1}{c|}{$\mathrm{LM_{PUY}}$} & 100  & 100  & 100  & \multicolumn{1}{c|}{100}  & 100  & 100  & 100  & \multicolumn{1}{c|}{100}  & 100  & 100  & 100  & \multicolumn{1}{c|}{100}  & 100  & 100  & 100  & 100 \\ \hline
  & \multicolumn{18}{c}{(2) $t_{6}$-distribution}                                                                                                                                                                                                      \\ \cline{2-19}
                     & 100 & \multicolumn{1}{c|}{$S_{N}$}           & 4.9  & 5.7  & 5.2  & \multicolumn{1}{c|}{5.9}  & 6.3  & 5.5  & 4.8  & \multicolumn{1}{c|}{5.2}  & 4.7  & 5.8  & 5.1  & \multicolumn{1}{c|}{5.9}  & 6.2  & 5.6  & 4.9  & 5.3  \\
 &     & \multicolumn{1}{c|}{$L_{N}$}           & 3.3  & 4.5  & 3.4  & \multicolumn{1}{c|}{5.2}  & 2.0    & 2.9  & 3.8  & \multicolumn{1}{c|}{4.3}  & 2.3  & 3.7  & 3.1  & \multicolumn{1}{c|}{5.1}  & 1.5  & 2.6  & 3.7  & 3.9  \\
 &     & \multicolumn{1}{c|}{$T_{C}$}             & 3.7  & 5.6  & 4.8  & \multicolumn{1}{c|}{6.0}    & 3.9  & 4.1  & 4.2  & \multicolumn{1}{c|}{4.6}  & 3.5  & 5.2  & 4.7  & \multicolumn{1}{c|}{5.6}  & 3.9  & 4.3  & 4.7  & 4.6  \\
 &     & \multicolumn{1}{c|}{$\mathrm{CD_{P}}$}   & 11.9 & 11.9 & 11.9 & \multicolumn{1}{c|}{12.5} & 12.3 & 13.5 & 11.5 & \multicolumn{1}{c|}{11.7} & 14.0   & 13.9 & 13.8 & \multicolumn{1}{c|}{14.0}   & 13.2 & 15.3 & 13.7 & 13.6 \\
 &     & \multicolumn{1}{c|}{$\mathrm{LM_{PUY}}$} & 100  & 100  & 100  & \multicolumn{1}{c|}{100}  & 100  & 100  & 100  & \multicolumn{1}{c|}{100}  & 100  & 100  & 100  & \multicolumn{1}{c|}{100}  & 100  & 100  & 100  & 100  \\ \hline
 & 200 & \multicolumn{1}{c|}{$S_{N}$}           & 5.1  & 4.7  & 4.6  & \multicolumn{1}{c|}{4.9}  & 5.4  & 6.7  & 6.0    & \multicolumn{1}{c|}{4.5}  & 4.9  & 5.0    & 4.7  & \multicolumn{1}{c|}{5.3}  & 5.4  & 7.1  & 5.8  & 4.8  \\
 &     & \multicolumn{1}{c|}{$L_{N}$}           & 2.4  & 3.5  & 4.2  & \multicolumn{1}{c|}{4.2}  & 3.0    & 2.5  & 3.2  & \multicolumn{1}{c|}{3.2}  & 1.4  & 2.1  & 2.9  & \multicolumn{1}{c|}{3.9}  & 2.0    & 2.4  & 3.0    & 2.8  \\
 &     & \multicolumn{1}{c|}{$T_{C}$}             & 4.2  & 4.7  & 4.4  & \multicolumn{1}{c|}{5.0}    & 4.6  & 5.3  & 4.5  & \multicolumn{1}{c|}{4.2}  & 3.8  & 4.0    & 3.8  & \multicolumn{1}{c|}{5.0}    & 4.2  & 5.4  & 4.9  & 4.1  \\
 &     & \multicolumn{1}{c|}{$\mathrm{CD_{P}}$}   & 10.5 & 11.8 & 9.3  & \multicolumn{1}{c|}{14.0}   & 11.4 & 12.0   & 13.2 & \multicolumn{1}{c|}{12.9} & 12.4 & 13.9 & 11.3 & \multicolumn{1}{c|}{15.8} & 13.3 & 14.3 & 15.0   & 15.2 \\
 &     & \multicolumn{1}{c|}{$\mathrm{LM_{PUY}}$} & 100  & 100  & 100  & \multicolumn{1}{c|}{100}  & 100  & 100  & 100  & \multicolumn{1}{c|}{100}  & 100  & 100  & 100  & \multicolumn{1}{c|}{100}  & 100  & 100  & 100  & 100  \\ \hline
                     & \multicolumn{18}{c}{(3) $\chi_{5}^2$-distribution}                                                                                                                                                                                                 \\ \cline{2-19}
 & 100 & \multicolumn{1}{c|}{$S_{N}$}           & 5.6  & 5.3  & 5.8  & \multicolumn{1}{c|}{6.4}  & 6.7  & 5.2  & 4.8  & \multicolumn{1}{c|}{5.8}  & 5.8  & 5.2  & 6.0    & \multicolumn{1}{c|}{6.6}  & 6.9  & 5.2  & 4.6  & 5.6  \\
 &     & \multicolumn{1}{c|}{$L_{N}$}           & 3.4  & 3.4  & 3.4  & \multicolumn{1}{c|}{3.1}  & 3.4  & 2.8  & 4.6  & \multicolumn{1}{c|}{3.7}  & 2.5  & 2.3  & 2.9  & \multicolumn{1}{c|}{3.5}  & 2.1  & 2.6  & 4.1  & 3.3  \\
 &     & \multicolumn{1}{c|}{$T_{C}$}             & 5.7  & 3.7  & 4.7  & \multicolumn{1}{c|}{5.3}  & 5.0    & 4.0    & 3.9  & \multicolumn{1}{c|}{5.2}  & 4.5  & 3.6  & 4.2  & \multicolumn{1}{c|}{4.8}  & 4.3  & 3.3  & 3.9  & 4.8  \\
 &     & \multicolumn{1}{c|}{$\mathrm{CD_{P}}$}   & 13.9 & 13.2 & 12.4 & \multicolumn{1}{c|}{13.7} & 12.4 & 11.7 & 11.4 & \multicolumn{1}{c|}{13.0}   & 15.6 & 15.6 & 14.6 & \multicolumn{1}{c|}{14.8} & 13.7 & 12.9 & 13.6 & 15.1 \\
 &     & \multicolumn{1}{c|}{$\mathrm{LM_{PUY}}$} & 100  & 100  & 100  & \multicolumn{1}{c|}{100}  & 100  & 100  & 100  & \multicolumn{1}{c|}{100}  & 100  & 100  & 100  & \multicolumn{1}{c|}{100}  & 100  & 100  & 100  & 100  \\ \hline
 & 200 & \multicolumn{1}{c|}{$S_{N}$}           & 6.5  & 5.2  & 5.7  & \multicolumn{1}{c|}{6.5}  & 5.4  & 5.9  & 6.3  & \multicolumn{1}{c|}{5.5}  & 6.8  & 5.7  & 5.7  & \multicolumn{1}{c|}{6.6}  & 5.7  & 5.8  & 6.7  & 5.1  \\
 &     & \multicolumn{1}{c|}{$L_{N}$}           & 1.9  & 3.3  & 3.9  & \multicolumn{1}{c|}{4.9}  & 3.1  & 3.9  & 4.1  & \multicolumn{1}{c|}{3.7}  & 1.2  & 2.7  & 3.2  & \multicolumn{1}{c|}{3.8}  & 1.7  & 3.4  & 3.6  & 3.2  \\
 &     & \multicolumn{1}{c|}{$T_{C}$}             & 4.4  & 4.5  & 5.0    & \multicolumn{1}{c|}{5.1}  & 4.7  & 4.6  & 5.2  & \multicolumn{1}{c|}{3.9}  & 4.0    & 4.1  & 4.5  & \multicolumn{1}{c|}{5.0}    & 3.8  & 4.1  & 4.4  & 3.8  \\
 &     & \multicolumn{1}{c|}{$\mathrm{CD_{P}}$}   & 13.3 & 12.8 & 13.3 & \multicolumn{1}{c|}{14.5} & 13.4 & 13.0   & 13.5 & \multicolumn{1}{c|}{12.2} & 14.6 & 14.4 & 15.6 & \multicolumn{1}{c|}{16.6} & 16.3 & 14.4 & 15.4 & 13.3 \\
 &     & \multicolumn{1}{c|}{$\mathrm{LM_{PUY}}$} & 100  & 100  & 100  & \multicolumn{1}{c|}{100}  & 100  & 100  & 100  & \multicolumn{1}{c|}{100}  & 100  & 100  & 100  & \multicolumn{1}{c|}{100}  & 100  & 100  & 100  & 100  \\ \hline
\hline
               \end{tabular}}
           \end{table}

\begin{table}[!ht]
           \centering
          \caption{The empirical power of the three tests at 5\% level under case (\ref{H11}). }\label{t2}
           \vspace{0.35cm}
      \renewcommand{\arraystretch}{1.1}
     \tabcolsep 3.8pt
     \scalebox{0.75}{
         \begin{tabular}{cccccccccccccccccccccccc}\hline \hline                                            \multicolumn{1}{l}{} & \multicolumn{1}{l}{} & \multicolumn{1}{l|}{} & \multicolumn{8}{c|}{Setting (\ref{H01})}                         & \multicolumn{8}{c}{Setting (\ref{H02})}                        \\\hline
                     & \multicolumn{2}{c|}{$p$}                     & \multicolumn{4}{c|}{3}     & \multicolumn{4}{c|}{5}     & \multicolumn{4}{c|}{3}     & \multicolumn{4}{c}{5}    \\\hline
                     & $N$                  & \multicolumn{1}{c|}{$T$}                  & 200  & 300  & 400  & \multicolumn{1}{c|}{500}  & 200  & 300  & 400  & \multicolumn{1}{c|}{500}  & 200  & 300  & 400  & \multicolumn{1}{c|}{500}  & 200  & 300  & 400 & 500  \\\hline
                     & \multicolumn{18}{c}{(1) Normal distribution}                                                                                                                                                                                                      \\\hline
 & 100 & \multicolumn{1}{c|}{$S_{N}$} & 79.3 & 89.2 & 95.5 & \multicolumn{1}{c|}{98.5} & 76.3 & 88.6 & 95.4 & \multicolumn{1}{c|}{98.5} & 73.4 & 84.7 & 92.0   & \multicolumn{1}{c|}{96.6} & 70.3 & 83.8 & 91.9 & 95.8 \\
 &     & \multicolumn{1}{c|}{$L_{N}$} & 40.9 & 83.3 & 98.7 & \multicolumn{1}{c|}{100}  & 35.6 & 82.9 & 98.9 & \multicolumn{1}{c|}{100}  & 23.5 & 59.2 & 90.4 & \multicolumn{1}{c|}{99.7} & 22.1 & 63.7 & 89.4 & 99.0   \\
 &     & \multicolumn{1}{c|}{$T_{C}$}   & 87.5 & 98.3 & 100  & \multicolumn{1}{c|}{100}  & 84.6 & 98.5 & 99.9 & \multicolumn{1}{c|}{100}  & 77.5 & 93.8 & 99.0   & \multicolumn{1}{c|}{99.9} & 74.6 & 94.9 & 99.6 & 100  \\ \hline
 & 200 & \multicolumn{1}{c|}{$S_{N}$} & 77.2 & 90.3 & 95.1 & \multicolumn{1}{c|}{98.1} & 78.1 & 89.4 & 95.3 & \multicolumn{1}{c|}{97.9} & 70.9 & 85.4 & 91.1 & \multicolumn{1}{c|}{96.4} & 71.8 & 84.8 & 92.3 & 95.7 \\
 &     & \multicolumn{1}{c|}{$L_{N}$} & 27.9 & 77.7 & 98.2 & \multicolumn{1}{c|}{100}  & 25.2 & 77.0   & 97.3 & \multicolumn{1}{c|}{100}  & 12.7 & 53.0   & 88.7 & \multicolumn{1}{c|}{98.8} & 12.6 & 52.9 & 86.5 & 99.2 \\
 &     & \multicolumn{1}{c|}{$T_{C}$}   & 82.8 & 98.8 & 99.9 & \multicolumn{1}{c|}{100}  & 81.6 & 98.0   & 100  & \multicolumn{1}{c|}{100}  & 70.3 & 94.3 & 99.1 & \multicolumn{1}{c|}{99.8} & 70.8 & 94.5 & 99.2 & 100 \\
 \hline
                     & \multicolumn{18}{c}{(2) $t_{6}$-distribution}                                                                                                                                                                                                    \\ \cline{2-19}
                    & 100 & \multicolumn{1}{c|}{$S_{N}$} & 80.1 & 91.0   & 95.5 & \multicolumn{1}{c|}{98.3} & 78.0   & 87.8 & 94.4 & \multicolumn{1}{c|}{97.8} & 74.2 & 85.1 & 92.1 & \multicolumn{1}{c|}{95.9} & 72.5 & 83.6 & 91.0   & 95.9 \\
 &     & \multicolumn{1}{c|}{$L_{N}$} & 35.9 & 79.3 & 98.4 & \multicolumn{1}{c|}{100}  & 34.9 & 81.5 & 97.9 & \multicolumn{1}{c|}{100}  & 21.9 & 57.8 & 90.3 & \multicolumn{1}{c|}{99.4} & 20.4 & 60.9 & 88.6 & 99.0   \\
 &     & \multicolumn{1}{c|}{$T_{C}$}   & 86.8 & 97.6 & 99.9 & \multicolumn{1}{c|}{100}  & 85.0   & 98.3 & 100  & \multicolumn{1}{c|}{100}  & 75.2 & 92.4 & 99.1 & \multicolumn{1}{c|}{99.9} & 75.5 & 93.9 & 99.1 & 100  \\ \hline
 & 200 & \multicolumn{1}{c|}{$S_{N}$} & 77.9 & 90.7 & 94.9 & \multicolumn{1}{c|}{98.0}   & 79.3 & 89.7 & 95.8 & \multicolumn{1}{c|}{98.3} & 72.2 & 86.4 & 92.3 & \multicolumn{1}{c|}{96.4} & 73.4 & 85.0   & 92.2 & 96.3 \\
 &     & \multicolumn{1}{c|}{$L_{N}$} & 27.3 & 78.8 & 98.6 & \multicolumn{1}{c|}{100}  & 24.3 & 75.4 & 98.4 & \multicolumn{1}{c|}{100}  & 13.0   & 54.7 & 87.3 & \multicolumn{1}{c|}{98.1} & 12.9 & 51.2 & 86.5 & 99.2 \\
 &     & \multicolumn{1}{c|}{$T_{C}$}   & 83.4 & 98.5 & 99.9 & \multicolumn{1}{c|}{100}  & 82.8 & 98.3 & 100  & \multicolumn{1}{c|}{100}  & 72.8 & 93.4 & 98.9 & \multicolumn{1}{c|}{100}  & 70.8 & 93.7 & 99.1 & 100  \\
           \hline
                     & \multicolumn{18}{c}{(3) $\chi_{5}^2$-distribution}                                                                                                                                                                                               \\ \cline{2-19}
 & 100 & \multicolumn{1}{c|}{$S_{N}$} & 79.6 & 90.5 & 95.6 & \multicolumn{1}{c|}{98.0}   & 78.8 & 90.2 & 94.8 & \multicolumn{1}{c|}{98.9} & 73.6 & 86.6 & 93.0   & \multicolumn{1}{c|}{96.1} & 74.3 & 85.9 & 91.8 & 97.3 \\
 &     & \multicolumn{1}{c|}{$L_{N}$} & 41.4 & 82.5 & 98.2 & \multicolumn{1}{c|}{100}  & 36.8 & 81.7 & 98.0   & \multicolumn{1}{c|}{100}  & 24.7 & 62.9 & 90.3 & \multicolumn{1}{c|}{98.8} & 21.3 & 60.9 & 88.4 & 99.1 \\
 &     & \multicolumn{1}{c|}{$T_{C}$}   & 87.6 & 98.9 & 99.9 & \multicolumn{1}{c|}{100}  & 86.9 & 98.1 & 99.7 & \multicolumn{1}{c|}{100}  & 77.4 & 94.6 & 99.3 & \multicolumn{1}{c|}{100}  & 76.9 & 93.0 & 99.1 & 100 \\ \hline
 & 200 & \multicolumn{1}{c|}{$S_{N}$} & 77.7 & 90.0   & 96.6 & \multicolumn{1}{c|}{99.3} & 81.7 & 90.5 & 97.2 & \multicolumn{1}{c|}{97.8} & 71.7 & 85.6 & 93.6 & \multicolumn{1}{c|}{98.0}   & 74.8 & 84.5 &94.4  & 96.1 \\
 &     & \multicolumn{1}{c|}{$L_{N}$} & 31.4 & 78.6 & 98.5 & \multicolumn{1}{c|}{100}  & 33.8 & 79.1 & 98.4 & \multicolumn{1}{c|}{100}  & 16.5 & 54.1 & 90.2 & \multicolumn{1}{c|}{99.3} & 17.4 &  54.4 &88.6 & 99.0   \\
 &     & \multicolumn{1}{c|}{$T_{C}$}   & 84.0   & 98.5 & 100  & \multicolumn{1}{c|}{100}  & 86.7 & 98.8 & 100  & \multicolumn{1}{c|}{100}  & 74.0   & 93.7 & 99.6 & \multicolumn{1}{c|}{100}  & 75.1   & 94.2  &99.9  & 100 \\
            \hline\hline
               \end{tabular}
              }
           \end{table}
\begin{table}[!ht]
           \centering
           \caption{The empirical power of the three tests at 5\% level under case (\ref{H12}). }\label{t3}
           \vspace{0.35cm}
      \renewcommand{\arraystretch}{1.1}
     \tabcolsep 3.8pt
     \scalebox{0.75}{
         \begin{tabular}{cccccccccccccccccccccccc}\hline \hline                                            \multicolumn{1}{l}{} & \multicolumn{1}{l}{} & \multicolumn{1}{l|}{} & \multicolumn{8}{c|}{Setting (\ref{H01})}                         & \multicolumn{8}{c}{Setting (\ref{H02})}                        \\\hline
                     & \multicolumn{2}{c|}{$p$}                     & \multicolumn{4}{c|}{3}     & \multicolumn{4}{c|}{5}     & \multicolumn{4}{c|}{3}     & \multicolumn{4}{c}{5}    \\\hline
                     & $N$                  & \multicolumn{1}{c|}{$T$}                  & 200  & 300  & 400  & \multicolumn{1}{c|}{500}  & 200  & 300  & 400  & \multicolumn{1}{c|}{500}  & 200  & 300  & 400  & \multicolumn{1}{c|}{500}  & 200  & 300  & 400 & 500  \\\hline
                     & \multicolumn{18}{c}{(1) Normal distribution}                                                                                                                                                                                                     \\ \cline{2-19}
& 100 & \multicolumn{1}{c|}{$S_{N}$} & 11.8 & 14.4 & 13.6 & \multicolumn{1}{c|}{18.1} & 12.8 & 16.0   & 15.7 & \multicolumn{1}{c|}{16.4} & 10.5 & 12.4 & 13.8 & \multicolumn{1}{c|}{16.8} & 11.8 & 14.7 & 13.8 & 15.2 \\
 &     & \multicolumn{1}{c|}{$L_{N}$} & 99.1 & 100  & 100  & \multicolumn{1}{c|}{100}  & 100  & 100  & 100  & \multicolumn{1}{c|}{100}  & 97.0   & 100  & 100  & \multicolumn{1}{c|}{100}  & 97.1 & 100  & 100  & 100  \\
 &     & \multicolumn{1}{c|}{$T_{C}$}   & 98.5 & 100  & 100  & \multicolumn{1}{c|}{100}  & 99.5 & 100  & 100  & \multicolumn{1}{c|}{100}  & 94.9 & 100  & 100  & \multicolumn{1}{c|}{100}  & 94.2 & 100  & 100  & 100  \\ \hline
 & 200 & \multicolumn{1}{c|}{$S_{N}$} & 8.2  & 9.5  & 8.2  & \multicolumn{1}{c|}{12.6} & 8.9  & 9.9  & 11.7 & \multicolumn{1}{c|}{11.8} & 8.4  & 9.3  & 8.0    & \multicolumn{1}{c|}{12.0}   & 8.0    & 9.8  & 11.9 & 11.1 \\
 &     & \multicolumn{1}{c|}{$L_{N}$} & 58.3 & 97.2 & 99.6 & \multicolumn{1}{c|}{100}  & 48.5 & 97.1 & 100  & \multicolumn{1}{c|}{100}  & 35.6 & 88.7 & 99.2 & \multicolumn{1}{c|}{99.9} & 32.3 & 87.5 & 98.4 & 100  \\
 &     & \multicolumn{1}{c|}{$T_{C}$}   & 48.4 & 94.5 & 98.6 & \multicolumn{1}{c|}{100}  & 41.4 & 94.1 & 100  & \multicolumn{1}{c|}{100}  & 29.7 & 81.5 & 98.6 & \multicolumn{1}{c|}{100}  & 27.6 & 81.4 & 97.2 & 100
 \\
             \hline
                     & \multicolumn{18}{c}{(2) $t_{6}$-distribution}                                                                                                                                                                                                    \\ \cline{2-19}
& 100 & \multicolumn{1}{c|}{$S_{N}$} & 12.2 & 13.5 & 15.0   & \multicolumn{1}{c|}{18.3} & 11.8 & 14.8 & 14.9 & \multicolumn{1}{c|}{16.5} & 10.6 & 12.7 & 14.5 & \multicolumn{1}{c|}{16.3} & 12.1 & 14.2 & 13.2 & 15.8 \\
 &     & \multicolumn{1}{c|}{$L_{N}$} & 99.8 & 100  & 100  & \multicolumn{1}{c|}{100}  & 99.6 & 100  & 100  & \multicolumn{1}{c|}{100}  & 96.0   & 100  & 100  & \multicolumn{1}{c|}{100}  & 97.2 & 100  & 100  & 100  \\
 &     & \multicolumn{1}{c|}{$T_{C}$}   & 99.8 & 100  & 100  & \multicolumn{1}{c|}{100}  & 98.9 & 100  & 100  & \multicolumn{1}{c|}{100}  & 93.9 & 100  & 100  & \multicolumn{1}{c|}{100}  & 93.9 & 99.9 & 100  & 100  \\ \hline
 & 200 & \multicolumn{1}{c|}{$S_{N}$} & 7.3  & 9.3  & 8.9  & \multicolumn{1}{c|}{11.9} & 8.3  & 11.0   & 11.3 & \multicolumn{1}{c|}{11.7} & 7.2  & 9.1  & 8.4  & \multicolumn{1}{c|}{11.7} & 7.7  & 9.9  & 11.3 & 10.6 \\
 &     & \multicolumn{1}{c|}{$L_{N}$} & 65.3 & 98.2 & 100  & \multicolumn{1}{c|}{100}  & 60.9 & 96.0   & 100  & \multicolumn{1}{c|}{100}  & 35.8 & 89.1 & 99.3 & \multicolumn{1}{c|}{100}  & 33.5 & 88.8 & 99.0   & 100  \\
 &     & \multicolumn{1}{c|}{$T_{C}$}   & 56.2 & 95.9 & 99.9 & \multicolumn{1}{c|}{100}  & 51.5 & 92.8 & 99.8 & \multicolumn{1}{c|}{100}  & 29.3 & 80.9 & 98.0   & \multicolumn{1}{c|}{100}  & 27.7 & 83.0   & 97.7 & 99.8\\
        \hline
                     & \multicolumn{18}{c}{(3) $\chi_{5}^2$-distribution}                                                                                                                                                                                               \\ \cline{2-19}
                     & 100 & \multicolumn{1}{c|}{$S_{N}$} & 13.6 & 14.6 & 15.5 & \multicolumn{1}{c|}{18.0}   & 12.9 & 12.6 & 14.4 & \multicolumn{1}{c|}{17.5} & 12.6 & 13.9 & 14.2 & \multicolumn{1}{c|}{15.9} & 12.0   & 12.4 & 14.2 & 16.2 \\
 &     & \multicolumn{1}{c|}{$L_{N}$} & 99.9 & 100  & 100  & \multicolumn{1}{c|}{100}  & 100  & 100  & 100  & \multicolumn{1}{c|}{100}  & 97.0   & 100  & 100  & \multicolumn{1}{c|}{100}  & 97.7 & 100  & 100  & 100  \\
 &     & \multicolumn{1}{c|}{$T_{C}$}   & 99.7 & 100  & 100  & \multicolumn{1}{c|}{100}  & 100  & 100  & 100  & \multicolumn{1}{c|}{100}  & 93.9 & 100  & 100  & \multicolumn{1}{c|}{100}  & 94.9 & 100  & 100  & 100  \\ \hline
 & 200 & \multicolumn{1}{c|}{$S_{N}$} & 10.3 & 10.3 & 11.6 & \multicolumn{1}{c|}{13.2} & 9.4  & 10.1 & 12.1 & \multicolumn{1}{c|}{11.6} & 10.2 & 10.2 & 9.7  & \multicolumn{1}{c|}{12.7} & 9.8  & 9.8  & 11.9 & 11.1 \\
 &     & \multicolumn{1}{c|}{$L_{N}$} & 56.3 & 98.1 & 99.9 & \multicolumn{1}{c|}{100}  & 59.1 & 97.9 & 100  & \multicolumn{1}{c|}{100}  & 33.9 & 87.4 & 99.4 & \multicolumn{1}{c|}{100}  & 35.4 & 85.5 & 99.1 & 100  \\
 &     & \multicolumn{1}{c|}{$T_{C}$}   & 48.9 & 96.6 & 99.8 & \multicolumn{1}{c|}{100}  & 50.9 & 95.4 & 99.9 & \multicolumn{1}{c|}{100}  & 28.0   & 82.7 & 98.5 & \multicolumn{1}{c|}{100}  & 31.4 & 79.9 & 97.7 & 99.9\\
    \hline                                                                                                                                                                                                   \hline
               \end{tabular}
              }
           \end{table}

Figure \ref{fig1} shows how the empirical powers of the
three tests change as the degree of sparsity of the correlation matrix
changes, where
 the $x$-axis  is the level of density $k$ and the $y$-axis represents the empirical power.
To generate Figure \ref{fig1}, we designed the following simulation.
We set $N=100$, $T=300$, $p=3,$ $k=2, \cdots, 16$; a subset $S \subset\{1, \cdots, N\}$ is
randomly selected with cardinality $|S|=k$; $(\mathbf{\Psi})_{i j}=1$ if $i=j$; for
$i \neq j$, $(\mathbf{\Psi})_{i j}=0$ if $i \notin S$ or $j \notin S$, and $(\mathbf{\Psi})_{i j} \mathop{\sim}\limits^{iid} U[\sqrt{7 /k\log N / T}, \sqrt{9/k\log N / T}]$ if $i
\in S$ and $j \in S$. Hence, a larger $k$ means a higher level of density.

Figure \ref{fig1} indicates that the empirical power of Fisher's
combined probability test $T_{C}$ is always very close
to the maximum power of both tests for all $k$, which suggests that it
has robust empirical power performance regardless
the alternative is sparse or not. In contrast, the empirical power
 curves of $S_N, L_N$ are both monotone, with the sum
   based test $S_{N}$ gains more power
with the increase of the level of density, while the max based test
$L_{N}$ gains more power
with the decrease of the level of density.

\section{Empirical application}\label{sec:data}

We now apply the proposed tests to analyze the securities in the S\&P 500 index. To account for the changes to the composition
of the index over time, we compiled weekly returns on all the securities that constitute the S\&P 500 index that have been listed over the
period from January 2005 to September 2018. Because the
securities that make up the index change over time,
we only consider $N=424$ securities that were included in the S\&P
500 index during the entire period. A total of $T=716$ consecutive
observations were obtained. The time series data on the safe rate of return, and the market
factors are obtained from Ken French's data library web page. The
one-week US treasury bill rate is chosen as the risk-free rate
($r_{ft}$). The value-weighted return on all NYSE, AMEX, and NASDAQ
stocks from CRSP is used as a proxy for the market return ($r_{mt}$).
The average return on the three small portfolios minus the average
return on the three big portfolios ($SMB_t$), and the average return
on two value portfolios minus the average return on two growth
portfolios ($HML_t$) are calculated based on the stocks listed on
the NYSE, AMEX and NASDAQ.

We use the Fama-French three-factor model  proposed by \cite{Fama1993Common} to describe the above panel data, which is given by
\begin{align}\label{ff3}Y_{it}=r_{it}-r_{ft}=\alpha_i+\beta_{i1}
  (r_{mt}-r_{ft})+\beta_{i2}SMB_t+\beta_{i3} HML_t+\epsilon_{it}, \end{align}
for each $1\leq i\leq N  $ and $1\leq t\leq T$, where $r_{mt}-r_{ft}$ is referred
to as the market factor. Since cross-sectional dependence among the errors may lead to misspecified inference \citep{bernard1987cross},
we are interested in testing
\begin{align}\label{nulld}
H_{0}: \epsilon_{1\cdot}, \epsilon_{2\cdot}, \cdots, \epsilon_{N\cdot} \text { are independent random vectors.}
\end{align}

\begin{figure}[ht]
\centering
\includegraphics[width=4in]{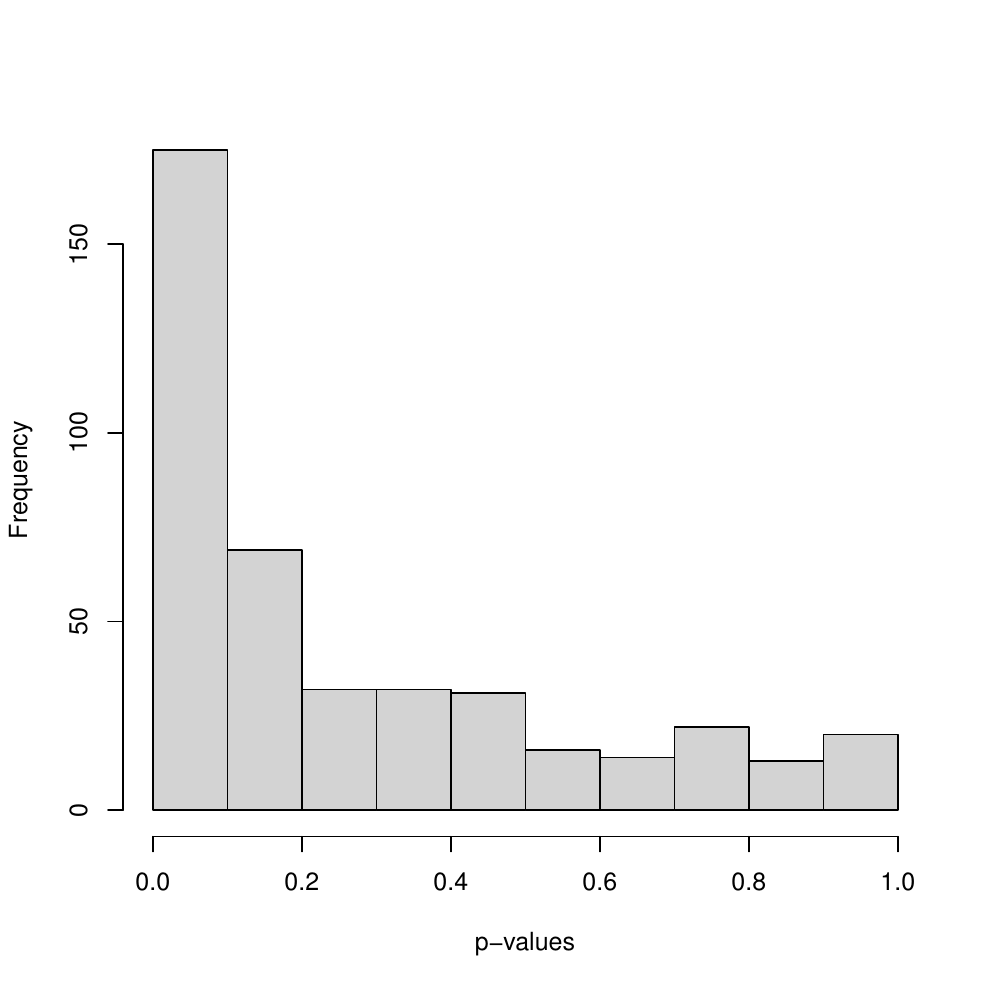}
\caption{Histogram of the p-values of the Box-Pierce test for all residual sequences under the Fama-French three-factor model.}
\label{Figp}
\end{figure}

Before applying the proposed tests to the above panel data under the
Fama-French three-factor model, we need
to investigate whether there exists serial correlation in the residuals under such model.
To this end, we applied the Box-Pierce test, a traditional test for
autocorrelation, to the residual sequence of each security under the
Fama-French three-factor model. Figure \ref{Figp} is the histogram of
the $p$-values of the residual sequences, which suggests that for most
securities the Box-Pierce tests are rejected. Furthermore, we applied
a high-dimensional white noise test proposed by \cite{lz2019} to the
sequences of the residual vector and the resulting $p$-value is 0,
which also suggests rejecting. These results indicate that there
exists serial correlation in the residuals under the Fama-French
three-factor model. This may be because the model fails to take into
account all the useful factors with serial correlation. See
\cite{schwartz1977evidence} and \cite{rosenberg1982factor} for further
discussions on causes of serial correlation in the residuals. Hence,
it is reasonable and necessary for us to use a test that allows serial
correlation for the above panel data under the Fama-French
three-factor model.

\begin{figure}[ht]
\centering
\includegraphics[width=4in]{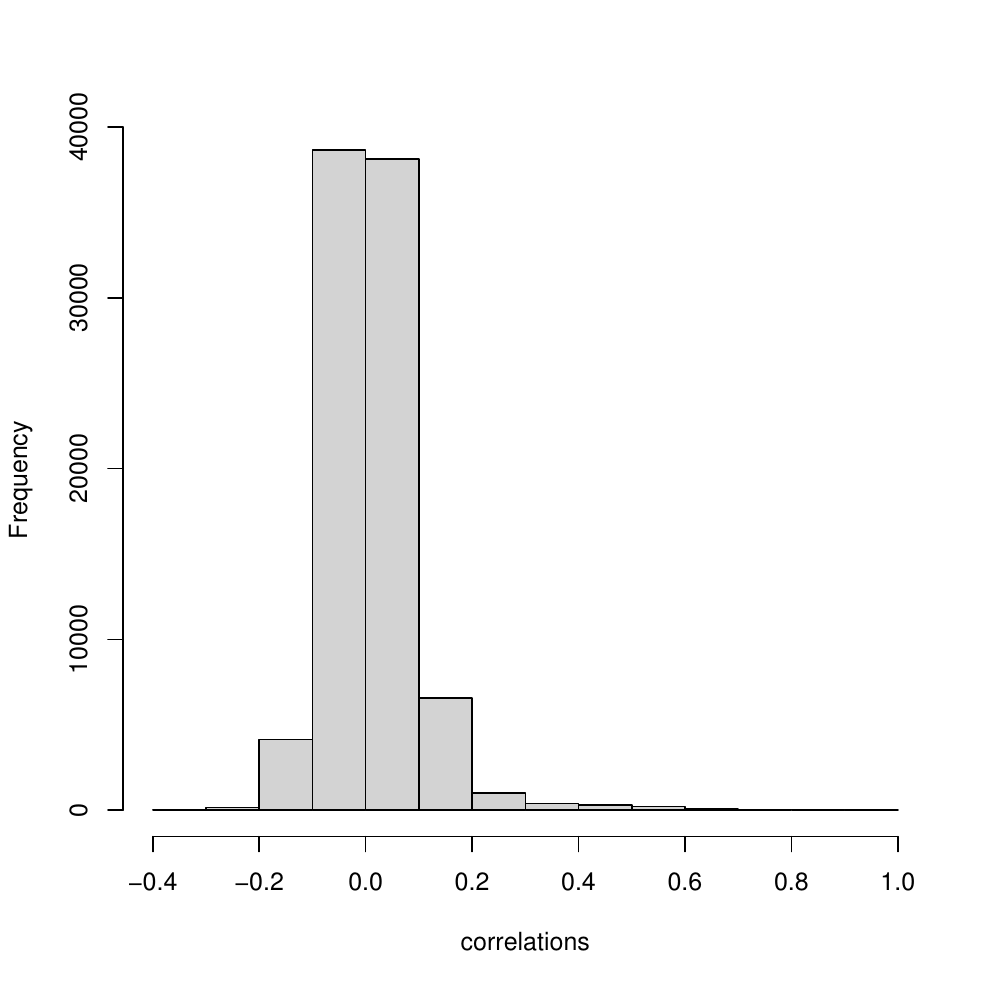}
\caption{Histogram of the correlations between all pairwise residual sequences of the securities under the Fama-French three-factor model.}
\label{Figcor}
\end{figure}

To accomodate the residual serial correlation, we used
 $S_N, L_N$ and $T_{C}$ to test the null hypothesis in (\ref{nulld}),
 and obtained
the $p$-values  $0$, $0.93$ and $0$
respectively. This seems to suggest a dense residual correlation
matrix with weak signals overall. To verify this,
we plotted the histogram of the correlations between all pairwise
residual sequences of the securities in Figure \ref{Figcor}, which
indicates that there are extensive non-zero correlations with small
absolute values.  This explains why  the sum based test $S_{N}$
and Fisher's combined probability test $T_{C}$ are able to detect the
deviation from $H_0$. In conclusion,
because both $S_N$ and $T_C$
reject the null hypothesis, we reject the null
hypothesis, and have the knowledge that the error correlation is
nondiagonal with off diagonal values dense with small magnitude.

\section{Conclusion}\label{sec:discuss}

In this paper, for testing cross-sectional independence under
heterogeneous panel data models with serial correlation, we proposed a
max based test based on the maximum of the squared cross-sectional
correlations of residuals to deal with sparse
alternatives. Furthermore, we proposed using Fisher's combined
probability test by combining the p-values of the proposed max based
test and the sum based test, which is applicable regardless
alternatives are sparse or not. In constructing the combined test, the
asymptotic null distribution is established strictly based on the
asymptotic independence of the max based and sum based test
statistics, which is an important contribution of this paper. In
addition, we relaxed the distribution assumption made in the existing
studies for testing independence with serial correlation from Gaussian
to sub-Gaussian. Finally, simulation studies and an empirical application
demonstrate the superiority of the proposed tests in comparison with
some of their competitors.

\newpage
\setcounter{table}{0}\renewcommand{\thetable}{S.\arabic{table}}
\setcounter{figure}{0}\renewcommand{\thefigure}{S.\arabic{figure}}
\setcounter{section}{0}\renewcommand{\thesection}{S.\arabic{section}}
%% Here are the title, author names and addresses
{\Large{Supplementary Material for "Fisher's combined probability test for cross-sectional independence in panel data models with serial correlation"}}

Supplementary Material is organized as follows.
In Section \ref{theorem in section 2}, we provide the proofs of all the theoretical results in Section \ref{sec:statistic}. In Section \ref{proof of lemmas}, we provide the proofs of some lemmas used in Section \ref{theorem in section 2}.

\textsc{Notations.}
For any square matrix $\mathbf{A},$
$(\mathbf{A})_{ij}$ denotes the $(i,j)$-th entry of $\mathbf{A}$,
$\tr(\mathbf{A})$ denotes the trace of $\mathbf{A}$,
$\|\mathbf{A}\|_{\mathrm{F}}$ denotes the Frobenius norm of matrix $\mathbf{A},$ $\|\mathbf{A}\|$ denotes the operator norm of $\mathbf{A}$,
 $\lambda_{max}(\mathbf{A})$ denotes the largest eigenvalue of $\mathbf{A}.$
Let $\mathbf{A}$ and $\mathbf{B}$ be the two matrices, we define
$\mathbf{A}\otimes\mathbf{B}$ as the Kronecker product of $\mathbf{A}$ and $\mathbf{B};$
when $\mathbf{A}$ and $\mathbf{B}$ are two square matrices of the same order, define
$\mathbf{A}\circ\mathbf{B}$ as the Hadamard product of $\mathbf{A}$ and $\mathbf{B}$, that is, $(\mathbf{A}\circ\mathbf{B})_{ij}=(\mathbf{A})_{ij}(\mathbf{B})_{ij}.$
For $1\leq i \leq T,$ let $e_{i}$ be a $T$-dimensional vector, where the $i$-th element is 1 and the rest are 0.
 Let ${\tau}_{T}=(1,1, \ldots, 1)^{\prime}\in \mathbb{R}^{T}.$ The notation $\mathbf{I}_{T}$ denotes the $T\times T$ identity matrix.
For any two real
numbers $x$ and $y$, let $ x \vee y=\max (x, y)$ and $x \wedge y=\min (x, y).$
For any vector $v\in\mathbb{R}^{T},$
$\|v\|$ denotes the Euclidean norm of $v;$
%$v$ follows multivariate distribution with mean vector $\bf{\mu}$
%and covariance matrix $\mathbf{\Sigma},$
%where we omit the subscript $T$ for $\mu$ and $\mathbf{\Sigma}$ throughout entire supplement for convenience.
Throughout the paper,  $C$, $C^{\prime},c,c^{\prime},c_{1},c_{2},\cdots$, denote positive absolute constants.

\section{Proofs of the theorems}\label{theorem in section 2}
Note that we assume that the diagonal elements of $\mathbf{U}$ are all 1, i.e. $\mathbf{U}_{ii}=1,$ for all $1\leq i\leq N.$
Then, under $H_{0},$ we can obtain that $\mathbf{U}=\mathbf{I}_{N}.$
Let $m_{k}=\mE\left(z_{it}^{k}\right),$
where $z_{it}$ is the element at row $i$ and column $t$ of matrix $\mathbf{Z}.$
In order to analyze the moments of the quadratic form, we define some notations:
\begin{align*}
&m_{1}=0, \quad m_{2}=1, \quad m_{3}=\gamma_{1}, \quad m_{4}=\gamma_{2}+3,  \quad m_{5}=\gamma_{3}+10 \gamma_{1}, \quad m_{6}=\gamma_{4}+15 \gamma_{2}+10 \gamma_{1}^{2}+15,\\
&m_{7}=\gamma_{5}+21 \gamma_{3}+35 \gamma_{2} \gamma_{1}+105 \gamma_{1}, \quad m_{8}=\gamma_{6}+28 \gamma_{4}+56 \gamma_{3} \gamma_{1}+35 \gamma_{2}^{2}+210 \gamma_{2}+280 \gamma_{1}^{2}+105.
\end{align*}
Note that by Assumption \ref{assum:E distribution}, we can conclude that $\gamma_{1}, \gamma_{2}, \gamma_{3}, \gamma_{4},\gamma_{5},\gamma_{6}$ are finite.

To facilitate the theoretical derivations, we recall some useful lemmas as follows.

\begin{lemma}\label{lem:laplace approxi}
\cite{lieberman1994,pesaran2012testing} Let $x=(x_{1}\cdots,x_{T})^{\prime}$ be a $T$-dimensional random vector. Let $\mathbf{A}$ and $\mathbf{B}$ be $T \times T$
nonstochastic matrices, where $\mathbf{A}$ is symmetric, $\mathbf{B}$ is semi-positive definite, and
suppose that $x_{1}\cdots,x_{T}$
are independently and identically distributed (iid) observations with zero mean and one variance.
Denote the $l$-th cumulant of $x^{\prime}\mathbf{B}x$ by $\kappa_{l}$ and the
$m+1$ order, $m+k$ degree generalised cumulant
of the product of $(x^{\prime}\mathbf{A}x)^k$ and $x^{\prime}\mathbf{B}x$ by $\kappa_{km}$. If the following three conditions are true:
(i) For $l=1,2, \ldots,$ we have $\kappa_{l}=O(T)$;
(ii) for $k=1,2, \ldots,$ we have $\kappa_{k 0}=\mE\big\{\left(x^{\prime} \mathbf{A} x\right)^{k}\big\}=O\left(T^{k}\right)$;
(iii) for $k, m=1,2, \ldots,$ we have $\kappa_{k m}=O\left(T^{l}\right)$, with $l \leqslant k$;
then, we have
\begin{align}\label{equation:laplace}
\mE\left\{\left(\frac{x^{\prime}\mathbf{A}x}{x^{\prime}\mathbf{B}x}\right)^{k}\right\}=\frac{\mE\big\{
\left(x^{\prime}\mathbf{A}x\right)^k\big\}}{\left\{\mE(x^{\prime}\mathbf{B}x)\right\}^{k}}+b_{k 1}+O\left(T^{-2}\right),
\end{align}
where
\begin{align*}
 b_{k 1}=&\frac{k(k+1)}{2}\left[\frac{\mE\left\{\left(x^{\prime} \mathbf{A} x\right)^{k}\right\} \kappa_{2}}{\left\{\mE\left(x^{\prime} \mathbf{B} x\right)\right\}^{k+2}}\right]-k\left[\frac{\kappa_{k 1}}{\left\{\mE\left(x^{\prime} \mathbf{B} x\right)\right\}^{k+1}}\right]=O(T^{-1}),\\
\kappa_{k 1}=&\mE\left\{\left(x^{\prime} \mathbf{A}  x\right)^{k}\left(x^{\prime} \mathbf{B} x\right)\right\}-\mE\left\{\left(x^{\prime} \mathbf{A}  x\right)^{k}\right\} \mE\left(x^{\prime} \mathbf{B}  x\right).
\end{align*}
\end{lemma}
\begin{remark}\label{laplace}
\cite{lieberman1994} proposed the Laplace approximation of moments of the ratio of quadratic forms, where requires $\mathbf{B}$ to be a positive definite matrix.
\cite{pesaran2012testing} relaxed this condition and allowed $\mathbf{B}$ to be a positive semi-definite matrix.
\end{remark}
\begin{lemma}\label{lem:sigma}
\cite{baltagi2016}
Recall that $\mathbf{M}_{i}=\mathbf{P}_{i}\mathbf{\Sigma } \mathbf{P}_{i}$ for all $1\leq i\leq N.$
For any fixed positive number $k$, we have
(1) $\frac{1}{T} \operatorname{tr}\left(\mathbf{\Sigma}^{k}\right)=O(1);$
(2) $\frac{1}{T} \operatorname{tr}\left(\mathbf{M}_{i}^{k}\right)=O(1);$
(3) $\frac{1}{T} \operatorname{tr}\left(\mathbf{M}_{i_{1}} \mathbf{M}_{i_{2}} \cdots \mathbf{M}_{i_{k}}\right)=O(1),$ for $1\leq i_{1}, i_{2} , \cdots , i_{k}\leq N.$
\end{lemma}
\begin{lemma}\label{lem:matrix inequality}
\cite{abadir2005matrix} Let $\mathbf{A}$ and $\mathbf{B}$ be the same order square matrices.
\begin{enumerate}[(1)]
\item $\tr(\mathbf{A}\circ\mathbf{B})\leq \sqrt{\tr(\mathbf{A}^2)\tr(\mathbf{B}^2)},$ $\tr(\mathbf{A}\circ\mathbf{A})\leq \tr(\mathbf{A}^{2}),$ when $\mathbf{A}$, $\mathbf{ B}$ are symmetric;
\item $\tr(\mathbf{A}\circ\mathbf{B})\leq C\tr(\mathbf{B})$ when $\mathbf{A},$ $\mathbf{B}$ are non-negative definite matrices and $\lambda_{max}(\mathbf{A})\leq C$ for some constant $C\geq 0;$
\item $\mathbf{A} \circ\mathbf{ B}$ is positive semidefinite when $\mathbf{A}$, $\mathbf{ B}$ are positive semidefinite  (See Theorem 7.5.3 in \cite{bernstein2009matrix});
\item ${\tau}_{T}^{\prime}\mathbf{A}\circ\mathbf{B}{\tau}_{T}
= \tr(\mathbf{A}\mathbf{B}^{\prime})$ and $\tr^2(\mathbf{A}\mathbf{B}^{\prime})\leq \tr(\mathbf{A}^{\prime}\mathbf{A})\tr(\mathbf{B}^{\prime}\mathbf{B}) $ for any the same order square matrices $\mathbf{A}$ and $\mathbf{B}$ (See Exercise 12.5 and 12.32 in \cite{abadir2005matrix}).
 \end{enumerate}
\end{lemma}
\begin{lemma}\label{le:moment of quadratic form}\cite{bao2010expectation} Suppose that $\boldsymbol{\xi}=\left(\xi_{1}, \xi_{2}, \ldots, \xi_{T}\right)^{\prime}$, and $\xi_{1}, \xi_{2}, \ldots, \xi_{T}$ are iid observations with zero mean, one variance, $\gamma_{1}=\mE\left(\xi_{t}^{3}\right), \gamma_{2}=\mE\left(\xi_{t}^{4}\right)-3, \gamma_{3}=\mE\left(\xi_{t}^{5}\right)-10 \gamma_{1}, \gamma_{4}=\mE\left(\xi_{t}^{6}\right)-15 \gamma_{2}-10 \gamma_{1}^{2}-15,
\gamma_{5}=\mE\left(\xi_{t}^{7}\right)-21\gamma_{3}-35\gamma_{2}\gamma_{1}-105\gamma_{1}$ and $\gamma_{6}=\mE\left(\xi_{t}^{8}\right)-28 \gamma_{4}-56 \gamma_{3} \gamma_{1}-35 \gamma_{2}^{2}-210 \gamma_{2}-280 \gamma_{1}^{2}-105,$ for all $t=1,2, \ldots, T.$ Suppose that $\mathbf{A}_{j}$, $j=1,2,3,4$ are $T \times T$ real symmetric matrices, and ${\tau}_{T}=(1,1,\cdots,1)^{\prime}$ is a $T$-dimensional vector. Then
\begin{align*}
&\mE\left(\boldsymbol{\xi}^{\prime} \mathbf{A}_{1} \boldsymbol{\xi}\right)=\tr\left(\mathbf{A}_{1}\right),\\
&\mE\left[\left(\boldsymbol{\xi}^{\prime} \mathbf{A}_{1} \boldsymbol{\xi}\right)\left(\boldsymbol{\xi}^{\prime} \mathbf{A}_{2} \boldsymbol{\xi}\right)\right]=\gamma_{2} \tr\left[\left(\mathbf{A}_{1} \circ \mathbf{A}_{2}\right)\right]+\tr\left(\mathbf{A}_{1}\right) \tr\left(\mathbf{A}_{2}\right)+2 \tr\left(\mathbf{A}_{1} \mathbf{A}_{2}\right),\\
&\mE\left[\left(\boldsymbol{\xi}^{\prime} \mathbf{A}_{1} \boldsymbol{\xi}\right)\left(\boldsymbol{\xi}^{\prime} \mathbf{A}_{2} \boldsymbol{\xi}\right)\left(\boldsymbol{\xi}^{\prime} \mathbf{A}_{3} \boldsymbol{\xi}\right)\right]=\gamma_{4} \tr\left(\mathbf{A}_{1} \circ \mathbf{A}_{2} \circ \mathbf{A}_{3}\right)+\gamma_{2} \tr\left(\mathbf{A}_{1}\right) \tr\left(\mathbf{A}_{2} \circ \mathbf{A}_{3}\right) \\
&+\gamma_{2} \tr\left(\mathbf{A}_{2}\right) \tr\left(\mathbf{A}_{1} \circ \mathbf{A}_{3}\right)+\gamma_{2} \tr\left(\mathbf{A}_{3}\right) \tr\left(\mathbf{A}_{1} \circ \mathbf{A}_{2}\right)+4 \gamma_{2} \tr\left[\mathbf{A}_{1} \circ\left(\mathbf{A}_{2} \mathbf{A}_{3}\right)\right] \\
&+4 \gamma_{2} \tr\left[\mathbf{A}_{2} \circ\left(\mathbf{A}_{1} \mathbf{A}_{3}\right)\right]+4 \gamma_{2} \tr\left[\mathbf{A}_{3} \circ\left(\mathbf{A}_{1} \mathbf{A}_{2}\right)\right]+2 \gamma_{1}^{2}\left[{\tau}_{T}^{\prime}\left(\mathbf{I}_{T} \circ \mathbf{A}_{1}\right) \mathbf{A}_{2}\left(\mathbf{I}_{T} \circ \mathbf{A}_{3}\right) {\tau}_{T}\right] \\
&+2 \gamma_{1}^{2}\left[{\tau}_{T}^{\prime}\left(\mathbf{I}_{T} \circ \mathbf{A}_{1}\right) \mathbf{A}_{3}\left(\mathbf{I}_{T} \circ \mathbf{A}_{2}\right) {\tau}_{T}\right]+2 \gamma_{1}^{2}\left[{\tau}_{T}^{\prime}\left(\mathbf{I}_{T} \circ \mathbf{A}_{2}\right) \mathbf{A}_{1}\left(\mathbf{I}_{T} \circ \mathbf{A}_{3}\right) {\tau}_{T}\right] \\
&+4 \gamma_{1}^{2}\left[{\tau}_{T}^{\prime}\left(\mathbf{A}_{1} \circ \mathbf{A}_{2} \circ \mathbf{A}_{3}\right) {\tau}_{T}\right]+\tr\left(\mathbf{A}_{1}\right) \tr\left(\mathbf{A}_{2}\right) \tr\left(\mathbf{A}_{3}\right)+2 \tr\left(\mathbf{A}_{1}\right) \tr\left(\mathbf{A}_{2} \mathbf{A}_{3}\right) \\
&+2 \tr\left(\mathbf{A}_{2}\right) \tr\left(\mathbf{A}_{1} \mathbf{A}_{3}\right)+2 \tr\left(\mathbf{A}_{3}\right) \tr\left(\mathbf{A}_{1} \mathbf{A}_{2}\right)+8 \tr\left(\mathbf{A}_{1} \mathbf{A}_{2} \mathbf{A}_{3}\right),\\
&\mE\left[\left(\boldsymbol{\xi}^{\prime} \mathbf{A}_{1} \boldsymbol{\xi}\right)\left(\boldsymbol{\xi}^{\prime} \mathbf{A}_{2} \boldsymbol{\xi}\right)\left(\boldsymbol{\xi}^{\prime} \mathbf{A}_{3} \boldsymbol{\xi}\right)\left(\boldsymbol{\xi}^{\prime} \mathbf{A}_{4} \boldsymbol{\xi}\right)\right]=\tr\left(\mathbf{A}_{1}\right) \tr\left(\mathbf{A}_{2}\right) \tr\left(\mathbf{A}_{3}\right) \tr\left(\mathbf{A}_{4}\right) \\
&+2\left[\tr\left(\mathbf{A}_{1}\right) \tr\left(\mathbf{A}_{2}\right) \tr\left(\mathbf{A}_{3} \mathbf{A}_{4}\right)+\tr\left(\mathbf{A}_{1}\right) \tr\left(\mathbf{A}_{3}\right) \tr\left(\mathbf{A}_{2} \mathbf{A}_{4}\right)\right.\\
&+\tr\left(\mathbf{A}_{1}\right) \tr\left(\mathbf{A}_{4}\right) \tr\left(\mathbf{A}_{2} \mathbf{A}_{3}\right)+\tr\left(\mathbf{A}_{2}\right) \tr\left(\mathbf{A}_{3}\right) \tr\left(\mathbf{A}_{1} \mathbf{A}_{4}\right) \\
&\left.+\tr\left(\mathbf{A}_{2}\right) \tr\left(\mathbf{A}_{4}\right) \tr\left(\mathbf{A}_{1} \mathbf{A}_{3}\right)+\tr\left(\mathbf{A}_{3}\right) \tr\left(\mathbf{A}_{4}\right)\tr\left(\mathbf{A}_{1} \mathbf{A}_{2}\right)\right] \\
&+4\left[\tr\left(\mathbf{A}_{1} \mathbf{A}_{2}\right) \tr\left(\mathbf{A}_{3} \mathbf{A}_{4}\right)+\tr\left(\mathbf{A}_{1} \mathbf{A}_{3}\right) \tr\left(\mathbf{A}_{2} \mathbf{A}_{4}\right)+\tr\left(\mathbf{A}_{1} \mathbf{A}_{4}\right) \tr\left(\mathbf{A}_{2} \mathbf{A}_{3}\right)\right] \\
&+8\left[\tr\left(\mathbf{A}_{1}\right) \tr\left(\mathbf{A}_{2} \mathbf{A}_{3} \mathbf{A}_{4}\right)+\tr\left(\mathbf{A}_{2}\right) \tr\left(\mathbf{A}_{1} \mathbf{A}_{3} \mathbf{A}_{4}\right)+\tr\left(\mathbf{A}_{3}\right) \tr\left(\mathbf{A}_{1} \mathbf{A}_{2} \mathbf{A}_{4}\right)\right.\\
&\left.+\tr\left(\mathbf{A}_{4}\right) \tr\left(\mathbf{A}_{1} \mathbf{A}_{2} \mathbf{A}_{3}\right)\right]+16\left[\tr\left(\mathbf{A}_{1} \mathbf{A}_{3} \mathbf{A}_{4} \mathbf{A}_{2}\right)+\tr\left(\mathbf{A}_{1} \mathbf{A}_{4} \mathbf{A}_{2} \mathbf{A}_{3}\right)\right. \\
&\left.+\tr\left(\mathbf{A}_{1} \mathbf{A}_{4} \mathbf{A}_{3} \mathbf{A}_{2}\right)\right]
+\gamma_{2} f_{\gamma_{2}}+\gamma_{4} f_{\gamma_{4}}+\gamma_{6} f_{\gamma_{6}}+\gamma_{1}^{2} f_{\gamma_{1}^{2}}+\gamma_{2}^{2} f_{\gamma_{2}^{2}}+\gamma_{1} \gamma_{3} f_{\gamma_{1} \gamma_{3}},
\end{align*}
where expressions for $f_{\gamma_{2}}, f_{\gamma_{4}}, f_{\gamma_{6}}, f_{\gamma_{1}^{2}}, f_{\gamma_{2}^{2}}$ and $f_{\gamma_{1} \gamma_{3}}$  are provided in \cite{bao2010expectation}:
\begin{align*}
f_{\gamma_{2}}=& \tr\left(\mathbf{A}_{1}\right) \tr\left(\mathbf{A}_{2}\right) \tr\left(\mathbf{A}_{3} \circ \mathbf{A}_{4}\right)+\tr\left(\mathbf{A}_{1}\right) \tr\left(\mathbf{A}_{3}\right) \tr\left(\mathbf{A}_{2} \circ\mathbf{A}_{4}\right)\\
&+\tr\left(\mathbf{A}_{1}\right) \tr\left(\mathbf{A}_{4}\right)\tr\left(\mathbf{A}_{2} \circ\mathbf{A}_{3}\right)+\tr\left(\mathbf{A}_{2}\right) \tr\left(\mathbf{A}_{3}\right) \tr\left(\mathbf{A}_{1} \circ \mathbf{A}_{4}\right)\\
&+\tr\left(\mathbf{A}_{2}\right) \tr\left(\mathbf{A}_{4}\right) \tr\left(\mathbf{A}_{1} \circ\mathbf{A}_{3}\right)
+\tr\left(\mathbf{A}_{3}\right) \tr\left(\mathbf{A}_{4}\right) \tr\left(\mathbf{A}_{1}\circ\mathbf{A}_{2}\right) \\
&+2\left[\tau_{T}^{\prime}\left(\mathbf{A}_{1} \circ \mathbf{A}_{2}\right) \tau_{T} \tr\left(\mathbf{A}_{3} \circ \mathbf{A}_{4}\right)+\tau_{T}^{\prime}\left(\mathbf{A}_{1} \circ \mathbf{A}_{3}\right) \tau_{T} \tr\left(\mathbf{A}_{2} \circ \mathbf{A}_{4}\right)\right.\\
&+\tau_{T}^{\prime}\left(\mathbf{A}_{1} \circ \mathbf{A}_{4}\right) \tau_{T} \tr\left(\mathbf{A}_{2} \circ \mathbf{A}_{3}\right)
+\tau_{T}^{\prime}\left(\mathbf{A}_{2} \circ \mathbf{A}_{3}\right)\tau_{T} \tr\left(\mathbf{A}_{1} \circ \mathbf{A}_{4}\right)
\\
&\left.+\tau_{T}^{\prime}\left(\mathbf{A}_{2} \circ \mathbf{A}_{4}\right) \tau_{T} \tr\left(\mathbf{A}_{1} \circ \mathbf{A}_{3}\right)+\tau_{T}^{\prime}\left(\mathbf{A}_{3} \circ \mathbf{A}_{4}\right) \tau_{T} \tr\left(\mathbf{A}_{1} \circ \mathbf{A}_{2}\right)\right]\\
&+4\left[\tr\left(\mathbf{A}_{1}\right) \tr\left\{\mathbf{A}_{2} \circ\left(\mathbf{A}_{3} \mathbf{A}_{4}\right)\right\}+\tr\left(\mathbf{A}_{1}\right) \tr\left\{\mathbf{A}_{3} \circ\left(\mathbf{A}_{2} \mathbf{A}_{4}\right)\right\}+\tr\left(\mathbf{A}_{1}\right) \tr\left\{\mathbf{A}_{4} \circ\left(\mathbf{A}_{2} \mathbf{A}_{3}\right)\right\}\right.\\
&+\tr\left(\mathbf{A}_{2}\right) \tr\left\{\mathbf{A}_{1} \circ\left(\mathbf{A}_{3} \mathbf{A}_{4}\right)\right\}+\tr\left(\mathbf{A}_{2}\right) \tr\left\{\mathbf{A}_{3} \circ\left(\mathbf{A}_{1} \mathbf{A}_{4}\right)\right\}+\tr\left(\mathbf{A}_{2}\right) \tr\left\{\mathbf{A}_{4} \circ\left(\mathbf{A}_{1} \mathbf{A}_{3}\right)\right\}\\
&+\tr\left(\mathbf{A}_{3}\right) \tr\left\{\mathbf{A}_{1} \circ\left(\mathbf{A}_{2} \mathbf{A}_{4}\right)\right\}+\tr\left(\mathbf{A}_{3}\right) \tr\left\{\mathbf{A}_{2} \circ\left(\mathbf{A}_{1} \mathbf{A}_{4}\right)\right\}+\tr\left(\mathbf{A}_{3}\right) \tr\left\}\mathbf{A}_{4} \circ\left(\mathbf{A}_{1} \mathbf{A}_{2}\right)\right\}\\
&\left.+\tr\left(\mathbf{A}_{4}\right)\tr\left\{\mathbf{A}_{1} \circ\left(\mathbf{A}_{2} \mathbf{A}_{3}\right)\right\}+\tr\left(\mathbf{A}_{4}\right) \tr\left\{\mathbf{A}_{2} \circ\left(\mathbf{A}_{1} \mathbf{A}_{3}\right)\right\}+\tr\left(\mathbf{A}_{4}\right) \tr\left\{\mathbf{A}_{3} \circ\left(\mathbf{A}_{1} \mathbf{A}_{2}\right)\right\}\right]
 \\
&+8\left[\tr\left(\left(\mathbf{I}_{T} \circ \mathbf{A}_{1}\right) \mathbf{A}_{2}\mathbf{A}_{3}\mathbf{A}_{4}\right)+\tr\left(\left(\mathbf{I}_{T} \circ \mathbf{A}_{1}\right)\mathbf{A}_{2}\mathbf{A}_{4}\mathbf{A}_{3}\right)+\tr\left(\left(\mathbf{I}_{T} \circ \mathbf{A}_{1}\right) \mathbf{A}_{3}\mathbf{A}_{2} \mathbf{A}_{4}\right)\right.\\
&+\tr\left(\left(\mathbf{I}_{T} \circ \mathbf{A}_{2}\right) \mathbf{A}_{1} \mathbf{A}_{3} \mathbf{A}_{4}\right)+\tr\left(\left(I \circ \mathbf{A}_{2}\right) \mathbf{A}_{1}\mathbf{A}_{4} \mathbf{A}_{3}\right)+\tr\left(\left(\mathbf{I}_{T} \circ \mathbf{A}_{2}\right) \mathbf{A}_{3}\mathbf{A}_{1} \mathbf{A}_{4}\right)\\
&+\tr\left(\left(\mathbf{I}_{T} \circ \mathbf{A}_{3}\right) \mathbf{A}_{1}\mathbf{A}_{2} \mathbf{A}_{4}\right)+\tr\left(\left(\mathbf{I}_{T} \circ \mathbf{A}_{3}\right) \mathbf{A}_{1} \mathbf{A}_{4} \mathbf{A}_{2}\right)+\tr\left(\left(\mathbf{I}_{T} \circ \mathbf{A}_{3}\right) \mathbf{A}_{2}\mathbf{A}_{1} \mathbf{A}_{4}\right)\\
&\left.+\tr\left(\left(\mathbf{I}_{T} \circ \mathbf{A}_{4}\right) \mathbf{A}_{1}\mathbf{A}_{2} \mathbf{A}_{3}\right)+\tr\left(\left(\mathbf{I}_{T} \circ \mathbf{A}_{4}\right) \mathbf{A}_{1} \mathbf{A}_{3} \mathbf{A}_{2}\right)+\tr\left(\left(\mathbf{I}_{T} \circ\mathbf{A}_{4}\right) \mathbf{A}_{2} \mathbf{A}_{1}\mathbf{A}_{3}\right)\right],
\\
&+16\left[\tau_{T}^{\prime}\left\{\mathbf{I}_{T}\circ\left(\mathbf{A}_{1} \mathbf{A}_{2}\right)\right\}\left\{\mathbf{I}_{T} \circ\left(\mathbf{A}_{3} \mathbf{A}_{4}\right)\right\} \tau_{T}+\tau_{T}^{\prime}\left\{\mathbf{I}_{T} \circ\left(\mathbf{A}_{1} \mathbf{A}_{3}\right)\right\}\left\{\mathbf{I}_{T} \circ\left(\mathbf{A}_{2} \mathbf{A}_{4}\right)\right\} \tau_{T}\right. \\
&\left.+\tau_{T}^{\prime}\left\{\mathbf{I}_{T} \circ\left(\mathbf{A}_{1} \mathbf{A}_{4}\right)\right\}\left\{\mathbf{I}_{T} \circ\left(\mathbf{A}_{2} \mathbf{A}_{3}\right)\right\} \tau_{T}\right],\\
f_{\gamma_{4}}=& \tr\left(\mathbf{A}_{1}\right) \tr\left(\mathbf{A}_{2} \circ \mathbf{A}_{3} \circ \mathbf{A}_{4}\right)+\tr\left(\mathbf{A}_{2}\right)\tr\left(\mathbf{A}_{1} \circ \mathbf{A}_{3} \circ \mathbf{A}_{4}\right)+\tr\left(\mathbf{A}_{3}\right) \tr\left(\mathbf{A}_{1} \circ \mathbf{A}_{2} \circ \mathbf{A}_{4}\right)\\
&+\tr\left(\mathbf{A}_{4}\right) \tr\left(\mathbf{A}_{1} \circ \mathbf{A}_{2} \circ \mathbf{A}_{3}\right)+4\left[\tr\left\{\mathbf{A}_{1} \circ \mathbf{A}_{2} \circ\left(\mathbf{A}_{3} \mathbf{A}_{4}\right)\right\}+\tr\left\{\mathbf{A}_{1} \circ \mathbf{A}_{3} \circ\left(\mathbf{A}_{2} \mathbf{A}_{4}\right)\right\}\right.\\
&+\tr\left\{\mathbf{A}_{1} \circ \mathbf{A}_{4} \circ\left(\mathbf{A}_{2} \mathbf{A}_{3}\right)\right\}+\tr\left\{\mathbf{A}_{2} \circ\mathbf{A}_{3} \circ\left(\mathbf{A}_{1} \mathbf{A}_{4}\right)\right\}+\tr\left\{\mathbf{A}_{2} \circ \mathbf{A}_{4} \circ\left(\mathbf{A}_{1} \mathbf{A}_{3}\right)\right\}\\
&\left.+\tr\left\{\mathbf{A}_{3} \circ\mathbf{A}_{4} \circ\left(\mathbf{A}_{1} \mathbf{A}_{2}\right)\right\}\right],\\
f_{\gamma_{6}}=&\tr\left(\mathbf{A}_{1}\circ\mathbf{A}_{2}\circ
\mathbf{A}_{3}\circ\mathbf{A}_{4}\right),\\
f_{\gamma_{1}^{2}}=& 2\left[\tau_{T}^{\prime}\left(\mathbf{I}_{T} \circ \mathbf{A}_{2}\right)  \mathbf{A}_{3}\left(\mathbf{I}_{T} \circ \mathbf{A}_{4}\right) \tau_{T} \tr\left(\mathbf{A}_{1}\right)+\tau_{T}^{\prime}\left(\mathbf{I}_{T} \circ \mathbf{A}_{2}\right) \mathbf{A}_{4}\left(\mathbf{I}_{T} \circ \mathbf{A}_{3}\right) \tau_{T} \tr\left(\mathbf{A}_{1}\right)\right.\\
&+\tau_{T}^{\prime}\left(\mathbf{I}_{T} \circ \mathbf{A}_{3}\right) \mathbf{A}_{2}\left(\mathbf{I}_{T} \circ \mathbf{A}_{4}\right) \tau_{T} \tr\left(\mathbf{A}_{1}\right)
+\tau_{T}^{\prime}\left(\mathbf{I}_{T} \circ \mathbf{A}_{1}\right) \mathbf{A}_{3}\left(\mathbf{I}_{T} \circ \mathbf{A}_{4}\right) \tau_{T} \tr\left(\mathbf{A}_{2}\right)\\
&+\tau_{T}^{\prime}\left(\mathbf{I}_{T}  \circ \mathbf{A}_{1}\right) \mathbf{A}_{4}\left(\mathbf{I}_{T} \circ \mathbf{A}_{3}\right) \tau_{T} \tr\left(\mathbf{A}_{2}\right)+\tau_{T}^{\prime}\left(\mathbf{I}_{T} \circ \mathbf{A}_{3}\right) \mathbf{A}_{1}\left(\mathbf{I}_{T} \circ \mathbf{A}_{4}\right) \tau_{T} \tr\left(\mathbf{A}_{2}\right) \\
&+\tau_{T}^{\prime}\left(\mathbf{I}_{T} \circ \mathbf{A}_{1}\right) \mathbf{A}_{2}\left(\mathbf{I}_{T} \circ \mathbf{A}_{4}\right) \tau_{T} \tr\left(\mathbf{A}_{3}\right)+\tau_{T}^{\prime}\left(\mathbf{I}_{T} \circ \mathbf{A}_{1}\right) \mathbf{A}_{4}\left(\mathbf{I}_{T} \circ \mathbf{A}_{2}\right) \tr\left(\mathbf{A}_{3}\right)\\
&+\tau_{T}^{\prime}\left(\mathbf{I}_{T} \circ \mathbf{A}_{2}\right) \mathbf{A}_{1}\left(\mathbf{I}_{T} \circ \mathbf{A}_{4}\right) \tau_{T} \tr\left(\mathbf{A}_{3}\right) +\tau_{T}^{\prime}\left(\mathbf{I}_{T} \circ \mathbf{A}_{1}\right) \mathbf{A}_{2}\left(\mathbf{I}_{T}\circ \mathbf{A}_{3}\right) \tau_{T} \tr\left(\mathbf{A}_{4}\right)
\\
&\left.+\tau_{T}^{\prime}\left(\mathbf{I}_{T} \circ \mathbf{A}_{1}\right) \mathbf{A}_{3}\left(\mathbf{I}_{T} \circ \mathbf{A}_{2}\right) \tr\left(\mathbf{A}_{4}\right)+\tau_{T}^{\prime}\left(\mathbf{I}_{T} \circ\mathbf{A}_{2}\right) \mathbf{A}_{1}\left(\mathbf{I}_{T} \circ \mathbf{A}_{3}\right) \tau_{T} \right]\\
&+4\left[\tau_{T}^{\prime}\left(\mathbf{I}_{T} \circ \mathbf{A}_{1}\right) \mathbf{A}_{2} \mathbf{A}_{3}\left(\mathbf{I}_{T} \circ \mathbf{A}_{4}\right) \tau_{T}+\tau_{T}^{\prime}\left(\mathbf{I}_{T} \circ \mathbf{A}_{1}\right) \mathbf{A}_{2} \mathbf{A}_{4}\left(\mathbf{I}_{T} \circ \mathbf{A}_{3}\right) \tau_{T}\right.\\
&+\tau_{T}^{\prime}\left(\mathbf{I}_{T} \circ \mathbf{A}_{1}\right) \mathbf{A}_{3} \mathbf{A}_{4}\left(\mathbf{I}_{T} \circ \mathbf{A}_{2}\right) \tau_{T} +\tau_{T}^{\prime}\left(\mathbf{I}_{T} \circ \mathbf{A}_{2}\right) \mathbf{A}_{1} \mathbf{A}_{3}\left(\mathbf{I}_{T} \circ \mathbf{A}_{4}\right) \tau_{T}\\
&+\tau_{T}^{\prime}\left(\mathbf{I}_{T} \circ \mathbf{A}_{2}\right) \mathbf{A}_{1} \mathbf{A}_{4}\left(\mathbf{I}_{T} \mathbf{A}_{3}\right) \tau_{T}+\tau_{T}^{\prime}\left(\mathbf{I}_{T} \circ \mathbf{A}_{2}\right) \mathbf{A}_{3} \mathbf{A}_{4}\left(\mathbf{I}_{T} \circ \mathbf{A}_{1}\right) \tau_{T} \\
&+\tau_{T}^{\prime}\left(\mathbf{I}_{T} \circ \mathbf{A}_{3}\right) \mathbf{A}_{1} \mathbf{A}_{2}\left(\mathbf{I}_{T} \circ \mathbf{A}_{4}\right) \tau_{T}+\tau_{T}^{\prime}\left(\mathbf{I}_{T} \circ \mathbf{A}_{3}\right) \mathbf{A}_{1} \mathbf{A}_{4}\left(\mathbf{I}_{T} \circ \mathbf{A}_{2}\right) \tau_{T}\\
&+\tau_{T}^{\prime}\left(\mathbf{I}_{T} \circ \mathbf{A}_{3}\right) \mathbf{A}_{2} \mathbf{A}_{4}\left(\mathbf{I}_{T} \circ \mathbf{A}_{1}\right) \tau_{T}
+\tau_{T}^{\prime}\left(\mathbf{I}_{T} \circ \mathbf{A}_{4}\right) \mathbf{A}_{1} \mathbf{A}_{2}\left(\mathbf{I}_{T} \circ \mathbf{A}_{3}\right) \tau_{T}\\
&\left.+\tau_{T}^{\prime}\left(\mathbf{I}_{T} \circ \mathbf{A}_{4}\right) \mathbf{A}_{1} \mathbf{A}_{3}\left(\mathbf{I}_{T} \circ \mathbf{A}_{2}\right) \tau_{T}+\tau_{T}^{\prime}\left(\mathbf{I}_{T} \circ \mathbf{A}_{4}\right) \mathbf{A}_{2} \mathbf{A}_{3}\left(\mathbf{I}_{T} \circ \mathbf{A}_{1}\right) \tau_{T}\right]\\
&+4\left[\tau_{T}^{\prime}\left(\mathbf{A}_{2} \circ \mathbf{A}_{3} \circ \mathbf{A}_{4}\right) \tau_{T} \tr\left(\mathbf{A}_{1}\right)+\tau_{T}^{\prime}\left(\mathbf{A}_{1}  \circ \mathbf{A}_{3} \circ \mathbf{A}_{4}\right) \tau_{T} \tr\left(\mathbf{A}_{2}\right)
\right.\\
&\left.+\tau_{T}^{\prime}\left(\mathbf{A}_{1} \circ\mathbf{A}_{2}\circ \mathbf{A}_{4}\right) \tau_{T} \tr\left(\mathbf{A}_{3}\right)+\tau_{T}^{\prime}\left(\mathbf{A}_{1} \circ \mathbf{A}_{2} \circ\mathbf{A}_{3}\right) \tau_{T} \tr\left(\mathbf{A}_{4}\right)\right]\\
&+8\left[\tau_{T}^{\prime}\left(\mathbf{A}_{1} \circ \mathbf{A}_{2}\right) \mathbf{A}_{3}\left(\mathbf{I}_{T} \circ \mathbf{A}_{4}\right) \tau_{T}+\tau_{T}^{\prime}\left(\mathbf{A}_{1} \circ \mathbf{A}_{2}\right) \mathbf{A}_{4}\left(\mathbf{I}_{T} \circ \mathbf{A}_{3}\right) \tau_{T}\right.\\
&+\tau_{T}^{\prime}\left(\mathbf{A}_{1} \circ \mathbf{A}_{3}\right) \mathbf{A}_{2}\left(\mathbf{I}_{T} \circ \mathbf{A}_{4}\right) \tau_{T}
+\tau_{T}^{\prime}\left(\mathbf{A}_{1} \circ \mathbf{A}_{3}\right) \mathbf{A}_{4}\left(\mathbf{I}_{T} \circ \mathbf{A}_{2}\right) \tau_{T}\\
&+\tau_{T}^{\prime}\left(\mathbf{A}_{1} \circ \mathbf{A}_{4}\right) \mathbf{A}_{2}\left(\mathbf{I}_{T} \circ \mathbf{A}_{3}\right) \tau_{T}+
\tau_{T}^{\prime}\left(\mathbf{A}_{1} \circ \mathbf{A}_{4}\right) \mathbf{A}_{3}\left(\mathbf{I}_{T} \circ \mathbf{A}_{2}\right) \tau_{T}\\
&+\tau_{T}^{\prime}\left(\mathbf{A}_{2} \circ \mathbf{A}_{3}\right) \mathbf{A}_{1}\left(\mathbf{I}_{T} \circ \mathbf{A}_{4}\right) \tau_{T}
+\tau_{T}^{\prime}\left(\mathbf{A}_{2} \circ \mathbf{A}_{3}\right) \mathbf{A}_{4}\left(\mathbf{I}_{T} \circ \mathbf{A}_{1}\right) \tau_{T}\\
&+\tau_{T}^{\prime}\left(\mathbf{A}_{2} \circ \mathbf{A}_{4}\right) \mathbf{A}_{1}\left(\mathbf{I}_{T} \circ \mathbf{A}_{3}\right) \tau_{T}
+\tau_{T}^{\prime}\left(\mathbf{A}_{2} \circ \mathbf{A}_{4}\right) \mathbf{A}_{3}\left(\mathbf{I}_{T} \circ \mathbf{A}_{1}\right) \tau_{T}\\
&\left.+\tau_{T}^{\prime}\left(\mathbf{A}_{3} \circ \mathbf{A}_{4}\right) \mathbf{A}_{1}\left(\mathbf{I}_{T} \circ \mathbf{A}_{2}\right) \tau_{T}+\tau_{T}^{\prime}\left(\mathbf{A}_{3} \circ \mathbf{A}_{4}\right) \mathbf{A}_{2}\left(\mathbf{I}_{T} \circ \mathbf{A}_{1}\right) \tau_{T}\right]\\
&+16\left[\tr\left(\mathbf{A}_{1}\left(\mathbf{A}_{2} \circ \mathbf{A}_{3}\right) \mathbf{A}_{4}\right)+\tr\left(\mathbf{A}_{1}\left(\mathbf{A}_{2} \circ \mathbf{A}_{4}\right) \mathbf{A}_{3}\right)+\tr\left(\mathbf{A}_{1}\left(\mathbf{A}_{3} \circ \mathbf{A}_{4}\right) \mathbf{A}_{2}\right)\right. \\
&\left.+\tr\left(\mathbf{A}_{2}\left(\mathbf{A}_{1} \circ \mathbf{A}_{3}\right) \mathbf{A}_{4}\right)+\tr\left(\mathbf{A}_{2}\left(\mathbf{A}_{1} \circ \mathbf{A}_{4}\right) \mathbf{A}_{3}\right)+\tr\left(\mathbf{A}_{3}\left(\mathbf{A}_{1} \circ \mathbf{A}_{2}\right) \mathbf{A}_{4}\right)\right],\\
f_{\gamma_{2}^{2}}=& \tr\left(\mathbf{A}_{1} \circ \mathbf{A}_{2}\right) \tr\left(\mathbf{A}_{3} \circ \mathbf{A}_{4}\right)+\tr\left(\mathbf{A}_{1} \circ \mathbf{A}_{3}\right) \tr\left(\mathbf{A}_{2} \circ \mathbf{A}_{4}\right)+\tr\left(\mathbf{A}_{1} \circ \mathbf{A}_{4}\right) \tr\left(\mathbf{A}_{2} \circ \mathbf{A}_{3}\right) \\
&+4\left[\tau_{T}^{\prime}\left(\mathbf{I}_{T} \circ \mathbf{A}_{1}\right)\left(\mathbf{A}_{2} \circ \mathbf{A}_{3}\right)\left(\mathbf{I}_{T} \circ \mathbf{A}_{4}\right)\tau_{T}+
\tau_{T}^{\prime}\left(\mathbf{I}_{T} \circ \mathbf{A}_{1}\right)\left(\mathbf{A}_{2} \circ \mathbf{A}_{4}\right)\left(\mathbf{I}_{T} \circ \mathbf{A}_{3}\right)\tau_{T}\right.\\
&+\tau_{T}^{\prime}\left(\mathbf{I}_{T} \circ \mathbf{A}_{1}\right)\left(\mathbf{A}_{3} \circ \mathbf{A}_{4}\right)\left(\mathbf{I}_{T} \circ \mathbf{A}_{2}\right)\tau_{T}
+\tau_{T}^{\prime}\left(\mathbf{I}_{T} \circ \mathbf{A}_{2}\right)\left(\mathbf{A}_{1} \circ \mathbf{A}_{3}\right)\left(\mathbf{I}_{T} \circ \mathbf{A}_{4}\right)\tau_{T}\\
&\left.+\tau_{T}^{\prime}\left(\mathbf{I}_{T} \circ \mathbf{A}_{2}\right)\left(\mathbf{A}_{1} \circ \mathbf{A}_{4}\right)\left(\mathbf{I}_{T} \circ \mathbf{A}_{3}\right)\tau_{T}+
\tau_{T}^{\prime}\left(\mathbf{I}_{T} \circ \mathbf{A}_{3}\right)\left(\mathbf{A}_{1} \circ \mathbf{A}_{2}\right)\left(\mathbf{I}_{T} \circ \mathbf{A}_{4}\right)\tau_{T}\right] \\
&+8 \tau_{T}^{\prime}\left(\mathbf{A}_{1} \circ \mathbf{A}_{2} \circ \mathbf{A}_{3} \circ \mathbf{A}_{4}\right) \tau_{T},\\
f_{\gamma_{1} \gamma_{3}} &=2\left[\tau_{T}^{\prime}\left(\mathbf{I}_{T} \circ \mathbf{A}_{1}\right) \mathbf{A}_{2}\left(\mathbf{I}_{T} \circ \mathbf{A}_{3} \circ \mathbf{A}_{4}\right)+\tau_{T}^{\prime}\left(\mathbf{I}_{T} \circ
 \mathbf{A}_{1}\right) \mathbf{A}_{3}\left(\mathbf{I}_{T} \circ \mathbf{A}_{2} \circ \mathbf{A}_{4}\right) \tau_{T}\right.\\
&+\tau_{T}^{\prime}\left(\mathbf{I}_{T} \circ \mathbf{A}_{1}\right) \mathbf{A}_{4}\left(\mathbf{I}_{T} \circ\mathbf{A}_{2}\circ \mathbf{A}_{3}\right) \tau_{T}+\tau_{T}^{\prime}\left(\mathbf{I}_{T} \circ \mathbf{A}_{2}\right) \mathbf{A}_{1}\left(\mathbf{I}_{T} \circ \mathbf{A}_{3} \circ \mathbf{A}_{4}\right) \tau_{T}\\
&+\tau_{T}^{\prime}\left(\mathbf{I}_{T} \circ \mathbf{A}_{2}\right) \mathbf{A}_{3}\left(\mathbf{I}_{T} \circ \mathbf{A}_{1} \circ \mathbf{A}_{4}\right) \tau_{T}
+\tau_{T}^{\prime}\left(\mathbf{I}_{T} \circ \mathbf{A}_{2}\right) \mathbf{A}_{4}\left(\mathbf{I}_{T} \circ\mathbf{A}_{1} \circ \mathbf{A}_{3}\right) \tau_{T}
\\&+\tau_{T}^{\prime}\left(\mathbf{I}_{T} \circ \mathbf{A}_{3}\right) \mathbf{A}_{1}\left(\mathbf{I}_{T} \circ \mathbf{A}_{2} \circ \mathbf{A}_{4}\right) \tau_{T}
+\tau_{T}^{\prime}\left(\mathbf{I}_{T} \circ \mathbf{A}_{3}\right) \mathbf{A}_{2}\left(\mathbf{I}_{T} \circ \mathbf{A}_{1} \circ \mathbf{A}_{4}\right) \tau_{T} \\
&+\tau_{T}^{\prime}\left(\mathbf{I}_{T} \circ \mathbf{A}_{3}\right) \mathbf{A}_{4}\left(\mathbf{I}_{T}\circ\mathbf{A}_{1} \circ \mathbf{A}_{2}\right) \tau_{T}
+\tau_{T}^{\prime}\left(\mathbf{I}_{T} \circ \mathbf{A}_{4}\right) \mathbf{A}_{1}\left(\mathbf{I}_{T} \circ \mathbf{A}_{2} \circ \mathbf{A}_{3}\right) \tau_{T}\\
&\left.+\tau_{T}^{\prime}\left(\mathbf{I}_{T} \circ \mathbf{A}_{4}\right) \mathbf{A}_{2}\left(\mathbf{I}_{T} \circ \mathbf{A}_{1} \circ \mathbf{A}_{3}\right) \tau_{T}
+\tau_{T}^{\prime}\left(\mathbf{I}_{T} \circ \mathbf{A}_{4}\right) \mathbf{A}_{3}\left(\mathbf{I}_{T} \circ\mathbf{A}_{1}\circ \mathbf{A}_{2}\right) \tau_{T}\right]\\
&+8\left[\tau_{T}^{\prime}\left(\mathbf{I}_{T} \circ \mathbf{A}_{1}\right)\left(\mathbf{A}_{2} \circ \mathbf{A}_{3} \circ \mathbf{A}_{4}\right) \tau_{T}+\tau_{T}^{\prime}\left(\mathbf{I}_{T} \circ \mathbf{A}_{2}\right)\left(\mathbf{A}_{1} \circ \mathbf{A}_{3} \circ \mathbf{A}_{4}\right) \tau_{T}\right.\\
&\left.+\tau_{T}^{\prime}\left(\mathbf{I}_{T} \circ \mathbf{A}_{3}\right)\left(\mathbf{A}_{1} \circ\mathbf{A}_{2}\circ \mathbf{A}_{4}\right) \tau_{T}+\tau_{T}^{\prime}\left(\mathbf{I}_{T} \circ \mathbf{A}_{4}\right)\left(\mathbf{A}_{1} \circ \mathbf{A}_{2} \circ \mathbf{A}_{3}\right) \tau_{T}\right].
\end{align*}
\end{lemma}

Now, we are ready to present the proofs of the theorems in Section \ref{sec:statistic}.
\subsection{Proof of Theorem \ref{th:max null}}
Let
$$
\tilde{T}_{i j}\doteq\frac{\epsilon_{i\cdot}^{\prime}\epsilon_{j\cdot}}{\left\|\mathbf{\Sigma}\right\|_{\mathrm{F}}}
=\frac{z_{i}^{\prime}\mathbf{\Sigma}z_{j}}{\left\|\mathbf{\Sigma}\right\|_{\mathrm{F}}},\,\,\,\,\,\,\,\,\,
\rho_{ij}=\frac{\epsilon_{i\cdot}^{\prime}\epsilon_{j\cdot}}{\|\epsilon_{i\cdot}\|\|\epsilon_{j\cdot}\|}.
$$
To simplify notation, we define $z_{i}\doteq(\mathbf{Z})_{i\cdot},$ where $(\mathbf{Z})_{i\cdot}$ represents the i-th row vector of $\mathbf{Z},$ for $1\leq i\leq N.$ Hence, we have
$\hat{\epsilon}_{i\cdot}=\mathbf{P}_{i}\epsilon_{i\cdot}=\mathbf{P}_{i}\mathbf{\Sigma}^{1/2}z_{i}.$
\begin{lemma}\label{14Bi}\cite{feng2020}
Set $\mathbf{B}_{i}=\mathbf{x}_{i}\left(\mathbf{x}_{i}^{\prime} \mathbf{x}_{i}\right)^{-1} \mathbf{x}_{i}^{\prime},$ for $1 \leq i \leq N.$
Then,
$$
\max _{1 \leq i<j \leq N}\left|\hat{\rho}_{i j}-{\rho}_{i j}\right| \leq 14 \cdot\left(\max _{1 \leq i \leq j \leq N} \frac{{\epsilon}_{j\cdot}^{\prime} \mathbf{B}_{i} {\epsilon}_{j\cdot}}{{\epsilon}_{j\cdot}^{\prime}{\epsilon}_{j\cdot}}\right).
$$
\end{lemma}

We now are ready to derive the asymptotic null distribution of $L_{N}.$
First, we will show that for $y\in\mathbb{R}$, under $H_{0}$,
\begin{align}\label{maxrhoijnull}
\mathbb{P}\left(\max _{1 \leq i<j \leq N} \tilde{T}_{i j}^{2}-4\log N+\log \log  N\leq y\right) \rightarrow \exp \left\{-\frac{1}{\sqrt{8 \pi}} \exp \left(-\frac{y}{2}\right)\right\}.
\end{align}
 By Theorem 1 in \cite{arratia1989}, we have
$$
\left|\mathbb{P}\left(\max _{1 \leq i<j \leq N} \tilde{T}_{i j}^{2} \leq t_{N}\right)-e^{-\tau_{N}}\right| \leq b_{1 N}+b_{2N}+b_{3N}
$$
where
$t_{N}=4 \log N-\log \log N+y, \tau_{N}=\underset{1\leq i<j\leq N}{\sum\sum} \mathbb{P}\left(\tilde{T}_{12}^{2}>t_{N}\right),$
\begin{align*}
&b_{1 N} =\underset{1\leq i<j\leq N}{\sum\sum}\sum_{(k,l)\in B_{ij}}\mathbb{P}\left(\tilde{T}_{ij}^{2}>t_{N}\right)\mathbb{P}\left(\tilde{T}_{kl}^{2}>t_{N}\right) \leq N^{3}\left[\mathbb{P}\left(\tilde{T}_{12}^{2}>t_{N}\right)\right]^{2}, \\
&b_{2 N} =\underset{1\leq i<j\leq N}{\sum\sum}\sum_{(k,l)\in B_{ij}\backslash\{(i,j)\}}\mathbb{P}\left(\tilde{T}_{ij}^{2}>t_{N},\tilde{T}_{kl}^{2}>t_{N}\right)\leq N^{3} \mathbb{P}\left(\tilde{T}_{12}^{2}>t_{N}, \tilde{T}_{13}^{2}>t_{N}\right),\\
&b_{3 N} =\underset{1\leq i<j\leq N}{\sum\sum}\mE\left|\mathbb{P}\left\{\tilde{T}_{ij}^{2}>t_{N}|\sigma(\tilde{T}_{kl}^{2}:(k,l)\notin B_{ij})\right\}-\mathbb{P}\left(\tilde{T}_{12}^{2}>t_{N}\right)\right|=0,
\end{align*}
for all $1\leq i<j\leq N$, $B_{ij}\doteq\{(k,l):1\leq k<l\leq N,\{k,l\}\cap\{i,j\}\neq\varnothing\},$
and the third term $b_{3N}$ on the right side of the inequality is equal to zero because for four different indices $i, j, k, l,$ $\tilde{T}_{i j}$ and $\tilde{T}_{k l}$ are independent.
Note that according to Theorem 1.1 in \cite{rudelson2013}, we have for any large $M> 0$, there exists some $C_{1} > 0$ such
that
$$
\mathbb{P}\left(\frac{z_{2}^{\prime}\mathbf{\Sigma}^2z_{2}}{\tr(\mathbf{\Sigma}^{2})}>1+\varepsilon_{1}\right)
\leq2N^{-M},
$$
where $\varepsilon_{1}=C_{1}\sqrt{\log N/\tr(\mathbf{\Sigma}^2)}.$
By Corollary 3.1 in \cite{saulis1991limit} and $\tr(\mathbf{\Sigma}^{2})=\left\|\mathbf{\Sigma}\right\|_{\mathrm{F}}^{2}$, we have
\begin{align}\label{T12}
\mathbb{P}\left(\tilde{T}_{12}^{2}>t_{N}\right)=&
\mathbb{P}\left(\left|\tilde{T}_{12}\right|> \sqrt{t_{N}}\right)\n\\
=&\mathbb{P}\left\{\frac{|z_{1}^{\prime}\mathbf{\Sigma}z_{2}|}{\sqrt{\tr(\mathbf{\Sigma}^{2})}}
>\sqrt{t_{N}}\right\}\n\\
\leq&\mathbb{P}\left\{\frac{|z_{1}^{\prime}\mathbf{\Sigma}z_{2}|}{\sqrt{\tr(\mathbf{\Sigma}^{2})}}
>\sqrt{t_{N}},\frac{z_{2}^{\prime}\mathbf{\Sigma}^2z_{2}}{\tr(\mathbf{\Sigma}^{2})}\leq1+\varepsilon_{1}\right\}
+\mathbb{P}\left(\frac{z_{2}^{\prime}\mathbf{\Sigma}^2z_{2}}{\tr(\mathbf{\Sigma}^{2})}>1+\varepsilon_{1}\right)\n\\
\leq&\mathbb{P}\left\{\frac{|z_{1}^{\prime}\mathbf{\Sigma}z_{2}|}{\sqrt{z_{2}^{\prime}\mathbf{\Sigma}^2z_{2}}}
>\frac{\sqrt{t_{N}}}{1+\varepsilon_{1}}\right\}+2N^{-M}\n\\
\leq&\mE\left[\mE\left\{I\left(\frac{|z_{1}^{\prime}\mathbf{\Sigma}z_{2}|}{\sqrt{z_{2}^{\prime}\mathbf{\Sigma}^2z_{2}}}
>\frac{\sqrt{t_{N}}}{1+\varepsilon_{1}}\right)\Big|z_{2}\right\}\right]+2N^{-M}\n\\
\leq&\mE\left\{\mathbb{P}\left(\frac{|z_{1}^{\prime}\mathbf{\Sigma}z_{2}|}{\sqrt{z_{2}^{\prime}\mathbf{\Sigma}^2z_{2}}}
>\frac{\sqrt{t_{N}}}{1+\varepsilon_{1}}\Big|z_{2}\right)\right\}+2N^{-M}\n\\
=&\{1+o(1)\} \frac{2}{\sqrt{2 \pi t_{N}}} \exp^{-t_{N} / 2}+2N^{-M}\n\\
=&O(N^{-2}),
\end{align}
where last equality holds due to Proposition 2.1.2 in \cite{vershynin2018},
$I(\cdot)$ denotes indicative function, and $M$ is sufficiently large. Similarly, we have
\begin{align*}
\mathbb{P}\left(\tilde{T}_{12}^{2}\leq t_{N}\right)=&
\mathbb{P}\left(\left|\tilde{T}_{12}\right| \leq \sqrt{t_{N}}\right)\n\\
=&\mathbb{P}\left\{\frac{|z_{1}^{\prime}\mathbf{\Sigma}z_{2}|}{\sqrt{\tr(\mathbf{\Sigma}^{2})}}
\leq\sqrt{t_{N}}\right\}\n\\
\leq&\mathbb{P}\left\{\frac{|z_{1}^{\prime}\mathbf{\Sigma}z_{2}|}{\sqrt{\tr(\mathbf{\Sigma}^{2})}}
\leq\sqrt{t_{N}},\frac{z_{2}^{\prime}\mathbf{\Sigma}^2z_{2}}{\tr(\mathbf{\Sigma}^{2})}\geq1-\varepsilon_{1}\right\}
+\mathbb{P}\left(\frac{z_{2}^{\prime}\mathbf{\Sigma}^2z_{2}}{\tr(\mathbf{\Sigma}^{2})}<1-\varepsilon_{1}\right)\n\\
\leq&\mathbb{P}\left\{\frac{|z_{1}^{\prime}\mathbf{\Sigma}z_{2}|}{\sqrt{z_{2}^{\prime}\mathbf{\Sigma}^2z_{2}}}
\leq\frac{\sqrt{t_{N}}}{1-\varepsilon_{1}}\right\}+2N^{-M},
\end{align*}
then $1-\mathbb{P}\left(\tilde{T}_{12}^{2}<t_{N}\right)\geq
1-\mathbb{P}\left\{\frac{|z_{1}^{\prime}\mathbf{\Sigma}z_{2}|}{\sqrt{z_{2}^{\prime}\mathbf{\Sigma}^2z_{2}}}
<\frac{\sqrt{t_{N}}}{1-\varepsilon_{1}}\right\}-2N^{-M}.$
Hence, we have
\begin{align*}
&\mathbb{P}\left(\tilde{T}_{12}^{2}>t_{N}\right)\n\\
\geq&\mathbb{P}\left\{\frac{|z_{1}^{\prime}\mathbf{\Sigma}z_{2}|}{\sqrt{z_{2}^{\prime}\mathbf{\Sigma}^2z_{2}}}
>\frac{\sqrt{t_{N}}}{1-\varepsilon_{1}}\right\}-2N^{-M}\\
=&\{1-o(1)\} \frac{2}{\sqrt{2 \pi t_{N}}} \exp^{-t_{N} / 2}-2N^{-M}\\
=&O(N^{-2}).
\end{align*}
 This shows that $\tau_{N} \sim \frac{1}{\sqrt{8 \pi}} e^{-y / 2}$ and $b_{1 N} \leq C N^{-1}$. For $b_{2 N}$, we have
$$
\mathbb{P}\left(\tilde{T}_{12}^{2}>t_{N}, \tilde{T}_{13}^{2}>t_{N}\right) \leq \mathbb{P}\left(\left|\tilde{T}_{12}-\tilde{T}_{13}\right| \geq 2 \sqrt{t_{N}}\right)+\mathbb{P}\left(\left|\tilde{T}_{12}+\tilde{T}_{13}\right| \geq 2 \sqrt{t_{N}}\right).
$$
Again, by Theorem 1.1 in \cite{rudelson2013}, for any large $M> 0$ , there exists some $C_{2} > 0$ such
that
$$
\mathbb{P}\left(\frac{(z_{2}-z_{3})^{\prime}\mathbf{\Sigma}^2(z_{2}-z_{3})}{\tr(\mathbf{\Sigma}^{2})}>2(1+\varepsilon_{2})\right)
\leq2N^{-M},
$$
where $\varepsilon_{2}=C_{2}\sqrt{\log N/\tr(\mathbf{\Sigma}^2)}.$
By Corollary 3.1 (Cram\'er type moderate deviation results) in \cite{saulis1991limit}, we have
\begin{align*}
&\mathbb{P}\left(\left|\tilde{T}_{12}-\tilde{T}_{13}\right| \geq 2 \sqrt{t_{N}}\right)\\
=&\mathbb{P}\left\{\frac{\left|z_{1}^{\prime}\mathbf{\Sigma}\left(z_{2}-z_{3}\right)\right|}
{\sqrt{\tr(\mathbf{\Sigma}^2)}} \geq 2 \sqrt{t_{N}}\right\}\\
\leq&\mathbb{P}\left\{\frac{\left|z_{1}^{\prime}\mathbf{\Sigma}\left(z_{2}-z_{3}\right)\right|}
{\sqrt{\tr(\mathbf{\Sigma}^2)}} \geq 2 \sqrt{t_{N}},\frac{(z_{2}-z_{3})^{\prime}\mathbf{\Sigma}^2(z_{2}-z_{3})}{\tr(\mathbf{\Sigma}^2)}<2(1+\varepsilon_{2})\right\}\\
&+
\mathbb{P}\left\{\frac{(z_{2}-z_{3})^{\prime}\mathbf{\Sigma}^2(z_{2}-z_{3})}{\tr(\mathbf{\Sigma}^2)}>2(1+\varepsilon_{2})\right\}\\
\leq&\mathbb{P}\left\{\frac{\left|z_{1}^{\prime}\mathbf{\Sigma}\left(z_{2}-z_{3}\right)\right|}
{\sqrt{(z_{2}-z_{3})^{\prime}\mathbf{\Sigma}^2(z_{2}-z_{3})}} \geq \frac{ \sqrt{2t_{N}}}{1+\varepsilon_{2}}\right\}
+2N^{-M}\\
\leq&\{1+o(1)\} \frac{\sqrt{\log N}}{2 \sqrt{\pi}} N^{-4} e^{-y} ,
\end{align*}
where $M$ is sufficiently large.
Similarly, $\mathbb{P}\left(\left|\tilde{T}_{12}+\tilde{T}_{13}\right| \geq 2 \sqrt{t_{n}}\right)=\{1+o(1)\} \frac{\sqrt{\log N}}{2 \sqrt{\pi}} N^{-4} e^{-y}$. Combining these inequalities, we have $b_{2 N} \leq C N^{-1} \sqrt{\log N}$ and
\begin{align}\label{eiej}
\mathbb{P}\left(\max _{1 \leq i<j \leq N} \frac{\left(\epsilon_{i\cdot}^{\prime}\epsilon_{j\cdot}\right)^2}{\left\|\mathbf{\Sigma}\right\|_{\mathrm{F}}^{2}}-4\log N+\log \log  N\leq y\right) \rightarrow G(y)=\exp \left\{-\frac{1}{\sqrt{8 \pi}} \exp \left(-\frac{y}{2}\right)\right\}.
\end{align}
Similarly, due to Theorem 1.1 in \cite{rudelson2013}, we have
\begin{align}\label{ineq-max}
\mathbb{P}\left(\max _{1 \leq i \leq N}\left|\epsilon_{i\cdot}^{\prime}\epsilon_{i\cdot}-\tr\left(\mathbf{\Sigma}\right) \right| \geq C^{\prime} \sqrt{T\log N}\right)=O\left(N^{-M}\right),
\end{align}
where $M>0$ is sufficiently large, $C^{\prime}$ is a constant that depends on $M.$
Due to (\ref{eiej}), we have
\begin{align}\label{dis-rho}
\mathbb{P}\left(\frac{\tr^2(\mathbf{\Sigma})}{T^2\left\|\mathbf{\Sigma}\right\|_{\mathrm{F}}^{2}}\max _{1 \leq i<j \leq N} \frac{\left(\epsilon_{i\cdot}^{\prime}\epsilon_{j\cdot}\right)^2}{\frac{\tr^2(\mathbf{\Sigma})}{T^2}}-4\log N+\log \log  N\leq y\right) \rightarrow \exp \left\{-\frac{1}{\sqrt{8 \pi}} \exp \left(-\frac{y}{2}\right)\right\}.
\end{align}
Set
$$\Omega=\left\{\max_{1\leq i<j\leq N}\left|\frac{\epsilon_{i\cdot}^{\prime}\epsilon_{i\cdot}\tr\left(\mathbf{\Sigma}\right)}{T^2}
+\frac{\epsilon_{j\cdot}^{\prime}\epsilon_{j\cdot}\tr\left(\mathbf{\Sigma}\right)}{T^2}
-\frac{2\tr^2(\mathbf{\Sigma})}{T^2}\right|\leq2C_{1}\sqrt{\frac{\log N}{T}}\right\}.$$
Obviously, due to (\ref{ineq-max}),
we have \begin{align*}
\mathbb{P}\left(\Omega^{c}\right)\leq&\mathbb{P}\left\{\max_{1\leq i<j\leq N}\left|\frac{\epsilon_{i\cdot}^{\prime}\epsilon_{i\cdot}}{T}
+\frac{\epsilon_{j\cdot}^{\prime}\epsilon_{j\cdot}}{T}
-\frac{2\tr\left(\mathbf{\Sigma}\right)}{T}\right|>\frac{2C_{1}T}{\tr\left(\mathbf{\Sigma}\right)}\sqrt{\frac{\log N}{T}}\right\}\\
\leq&\mathbb{P}\left\{\max_{1\leq i<j\leq N}\left|\frac{\epsilon_{i\cdot}^{\prime}\epsilon_{i\cdot}}{T}
+\frac{\epsilon_{j\cdot}^{\prime}\epsilon_{j\cdot}}{T}
-\frac{2\tr\left(\mathbf{\Sigma}\right)}{T}\right|>\frac{2C_{1}}{\lambda_{max}(\mathbf{\Sigma})}\sqrt{\frac{\log N}{T}}\right\}\\
\leq&\mathbb{P}\left\{\max_{1\leq i\leq N}\left|\frac{\epsilon_{i\cdot}^{\prime}\epsilon_{i\cdot}}{T}-\frac{\tr\left(\mathbf{\Sigma}\right)}{T}\right|
>\frac{C_{1}}{\lambda_{max}(\mathbf{\Sigma})}\sqrt{\frac{\log N}{T}}\right\}\\
&+
\mathbb{P}\left\{\max_{1\leq j\leq N}\left|\frac{\epsilon_{j\cdot}^{\prime}\epsilon_{j\cdot}}{T}-\frac{\tr\left(\mathbf{\Sigma}\right)}{T}\right|
>\frac{C_{1}}{\lambda_{max}(\mathbf{\Sigma})}\sqrt{\frac{\log N}{T}}\right\}=O\left(N^{-M}\right).
\end{align*}
We claim that for some constant $C_{3}>0$ such that
$$\mathbb{P}\left(\max_{1\leq i<j\leq N}\left|\frac{\epsilon_{i\cdot}^{\prime}\epsilon_{i\cdot}\epsilon_{j\cdot}^{\prime}\epsilon_{j\cdot}}{T^2}-
\frac{\tr^2(\mathbf{\Sigma})}{T^2}\right|>C_{3}\sqrt{\frac{\log N}{T}}\right)\rightarrow 0$$
as $N,T$ are sufficiently large.
Notice that
\begin{align*}
&\mathbb{P}\left(\max_{1\leq i<j\leq N}\left|\frac{\epsilon_{i\cdot}^{\prime}\epsilon_{i\cdot}\epsilon_{j\cdot}^{\prime}\epsilon_{j\cdot}}{T^2}-
\frac{\tr^2(\mathbf{\Sigma})}{T^2}\right|>C_{3}\sqrt{\frac{\log N}{T}}\right)\\
\leq&
\mathbb{P}\left(\max_{1\leq i<j\leq N}\left|\frac{\epsilon_{i\cdot}^{\prime}\epsilon_{i\cdot}\epsilon_{j\cdot}^{\prime}\epsilon_{j\cdot}}{T^2}-
\frac{\tr^2(\mathbf{\Sigma})}{T^2}\right|>C_{3}\sqrt{\frac{\log N}{T}},\Omega\right)+\mathbb{P}\left(\Omega^{c}\right),
\end{align*}
we just need to prove that
$$
\mathbb{P}\left(\max_{1\leq i<j\leq N}\left|\frac{\epsilon_{i\cdot}^{\prime}\epsilon_{i\cdot}\epsilon_{j\cdot}^{\prime}\epsilon_{j\cdot}}{T^2}-
\frac{\tr^2(\mathbf{\Sigma})}{T^2}\right|>C_{3}\sqrt{\frac{\log N}{T}},\Omega\right)\rightarrow 0.$$
In fact, due to (\ref{ineq-max}), for sufficiently large $C_{3}$ satisfying
$C_{3}-2C_{1}>C_{3}\sqrt{\frac{1}{2}}
$ and $\sqrt{\frac{C_{3}}{2}}>C_{1},$
we have
\begin{align*}
&\mathbb{P}\left(\max_{1\leq i<j\leq N}\left|\frac{\epsilon_{i\cdot}^{\prime}\epsilon_{i\cdot}\epsilon_{j\cdot}^{\prime}\epsilon_{j\cdot}}{T^2}-
\frac{\tr^2(\mathbf{\Sigma})}{T^2}\right|>C_{3}\sqrt{\frac{\log N}{T}},\Omega\right)\\
\leq&\mathbb{P}\left(\max_{1\leq i<j\leq N}\left|\frac{\epsilon_{i\cdot}^{\prime}\epsilon_{i\cdot}\epsilon_{j\cdot}^{\prime}\epsilon_{j\cdot}
-\tr^2(\mathbf{\Sigma})-\epsilon_{i\cdot}^{\prime}\epsilon_{i\cdot}\tr\left(\mathbf{\Sigma}\right)
-\epsilon_{j\cdot}^{\prime}\epsilon_{j\cdot}\tr\left(\mathbf{\Sigma}\right)+2\tr^2(\mathbf{\Sigma})}{T^2}\right|
>C_{3}\sqrt{\frac{\log N}{2T}}\right)\\
\leq&\mathbb{P}\left\{\max_{1\leq i<j\leq N}\left|\left(\frac{\epsilon_{i\cdot}^{\prime}\epsilon_{i\cdot}}{T}-
\frac{\tr\left(\mathbf{\Sigma}\right)}{T}\right)\left(\frac{\epsilon_{j\cdot}^{\prime}\epsilon_{j\cdot}}{T}-
\frac{\tr\left(\mathbf{\Sigma}\right)}{T}\right)\right|>C_{3}\sqrt{\frac{\log N}{2T}}\right\}\\
\leq&\mathbb{P}\left\{\max_{1\leq i\leq N}\left|\frac{\epsilon_{i\cdot}^{\prime}\epsilon_{i\cdot}}{T}-
\frac{\tr\left(\mathbf{\Sigma}\right)}{T}\right|>\sqrt{C_{3}}\left(\frac{\log N}{2T}\right)^{1/4}\right\}\\
&+\mathbb{P}\left\{\max_{1\leq j\leq N}\left|\frac{\epsilon_{j\cdot}^{\prime}\epsilon_{j\cdot}}{T}-
\frac{\tr\left(\mathbf{\Sigma}\right)}{T}\right|>\sqrt{C_{3}}\left(\frac{\log N}{2T}\right)^{1/4}\right\}\\
\leq&\mathbb{P}\left\{\max_{1\leq i\leq N}\left|\frac{\epsilon_{i\cdot}^{\prime}\epsilon_{i\cdot}}{T}-
\frac{\tr\left(\mathbf{\Sigma}\right)}{T}\right|>\sqrt{\frac{C_{3}\log N}{2T}}\right\}+\mathbb{P}\left\{\max_{1\leq j\leq N}\left|\frac{\epsilon_{j\cdot}^{\prime}\epsilon_{j\cdot}}{T}-
\frac{\tr\left(\mathbf{\Sigma}\right)}{T}\right|>\sqrt{\frac{C_{3}\log N}{2T}}\right\}\\
=&O\left(N^{-M}\right),
\end{align*}
where the last inequality holds due to (\ref{ineq-max}).
Then, we can conclude that
$$\mathbb{P}\left(\max_{1\leq i<j\leq N}\left|\frac{\epsilon_{i\cdot}^{\prime}\epsilon_{i\cdot}\epsilon_{j\cdot}^{\prime}\epsilon_{j\cdot}}{T^2}-
\frac{\tr^2(\mathbf{\Sigma})}{T^2}\right|>C_{3}\sqrt{\frac{\log N}{T}}\right)
=O\left(N^{-M}\right)
$$
and
$$\mathbb{P}\left(\max_{1\leq i<j\leq N}\left|\frac{\epsilon_{i\cdot}^{\prime}\epsilon_{i\cdot}\epsilon_{j\cdot}^{\prime}\epsilon_{j\cdot}/T^2}{\tr^2(\mathbf{\Sigma})/T^2}-
1\right|<\frac{C_{3}}{\lambda_{min}^2(\mathbf{\Sigma})}\sqrt{\frac{\log N}{T}}\right)\rightarrow 1.
$$
Thus, with probability tending to one, we have
$$
\begin{aligned}
\max _{1 \leq i <j\leq N}\left|\frac{\tr^2(\mathbf{\Sigma})/T^2}{\epsilon_{i\cdot}^{\prime}\epsilon_{i\cdot}\epsilon_{j\cdot}^{\prime}\epsilon_{j\cdot}/T^2}-1\right| & \leq \max _{1 \leq i <j\leq N} \frac{T^2}{\epsilon_{i\cdot}^{\prime}\epsilon_{i\cdot}\epsilon_{j\cdot}^{\prime}\epsilon_{j\cdot}} \max _{1 \leq i  <j \leq N}\left|\frac{\tr^2(\mathbf{\Sigma})}{T^2}-\frac{\epsilon_{i\cdot}^{\prime}\epsilon_{i\cdot}\epsilon_{j\cdot}^{\prime}\epsilon_{j\cdot}}{T^2}\right| \\
& \leq 2 \max _{1 \leq i <j\leq N} \frac{T^2}{\tr^2(\mathbf{\Sigma})} \max _{1 \leq i <j\leq  N}\left|\frac{\tr^2(\mathbf{\Sigma})}{T^2}-\frac{\epsilon_{i\cdot}^{\prime}\epsilon_{i\cdot}\epsilon_{j\cdot}^{\prime}\epsilon_{j\cdot}}{T^2}\right|
=O_{p}\left\{\sqrt{\frac{\log N}{T}}\right\}.
\end{aligned}
$$
This, together with (\ref{dis-rho}), we have
\begin{align}\label{dis-rho/}
&\mathbb{P}\left(\frac{\tr^2(\mathbf{\Sigma})}
{\left\|\mathbf{\Sigma}\right\|_{\mathrm{F}}^{2}}\max _{1 \leq i<j \leq N} \frac{\left(\epsilon_{i\cdot}^{\prime}\epsilon_{j\cdot}\right)^2}
{\epsilon_{i\cdot}^{\prime}\epsilon_{i\cdot}\epsilon_{j\cdot}^{\prime}\epsilon_{j\cdot}
}-4 \log N+\log \log  N\leq y\right)\n\\
=&\mathbb{P}\left(\frac{\tr^2(\mathbf{\Sigma})}
{\left\|\mathbf{\Sigma}\right\|_{\mathrm{F}}^{2}}\max _{1 \leq i<j \leq N} \rho^2_{ij}-4 \log N+\log \log  N\leq y\right)
\rightarrow \exp \left\{-\frac{1}{\sqrt{8 \pi}} \exp \left(-\frac{y}{2}\right)\right\},
\end{align}
for any $y\in \mathbb{R}.$

Then, we want to prove that
$$\mathbb{P}\left(\frac{\tr^2(\mathbf{\Sigma})}
{\left\|\mathbf{\Sigma}\right\|_{\mathrm{F}}^{2}}\max _{1 \leq i<j \leq N} \hat{\rho}^2_{ij}-4 \log N+\log \log  N\leq y\right)
 \rightarrow \exp \left\{-\frac{1}{\sqrt{8 \pi}} \exp \left(-\frac{y}{2}\right)\right\}.$$
So, we just need to show that for any $\epsilon>0$
$$\mathbb{P}\left(\max_{1 \leq i<j \leq N}\left|\hat{\rho}_{i j}-{\rho}_{i j}\right|>\frac{\epsilon\sqrt{\tr\left(\mathbf{\Sigma}^2\right)}}{\sqrt{\log N}\tr\left(\mathbf{\Sigma}\right)}\right)\rightarrow 0.$$
This is because if it holds, we have
$$
\sqrt{\frac{\tr^2(\mathbf{\Sigma})}
{\left\|\mathbf{\Sigma}\right\|_{\mathrm{F}}^2} \log N} \cdot\left(\max_{1 \leq i<j \leq N}\left|\hat{\rho}_{i j}\right|-\max_{1 \leq i<j \leq N}\left|{\rho}_{i j}\right|\right) \rightarrow 0
$$
in probability. Set $\Delta=\max_{1 \leq i<j \leq N}\left|\hat{\rho}_{i j}\right|-\max_{1 \leq i<j \leq N}\left|{\rho}_{i j}\right| .$ Then
\begin{align*}
&\frac{\tr^2(\mathbf{\Sigma})}
{\left\|\mathbf{\Sigma}\right\|_{\mathrm{F}}^2}  \max_{1 \leq i<j \leq N}\left|\hat{\rho}_{i j}\right|^{2}\\
=&\frac{\tr^2(\mathbf{\Sigma})}
{\left\|\mathbf{\Sigma}\right\|_{\mathrm{F}}^2}\left( \max_{1 \leq i<j \leq N}\left|{\rho}_{i j}\right|+\Delta\right)^{2}\\
=&\frac{\tr^2(\mathbf{\Sigma})}
{\left\|\mathbf{\Sigma}\right\|_{\mathrm{F}}^2}  \max_{1 \leq i<j \leq N}\left|{\rho}_{i j}\right|^{2}+2 \frac{\tr^2(\mathbf{\Sigma})}
{\left\|\mathbf{\Sigma}\right\|_{\mathrm{F}}^2}  \max_{1 \leq i<j \leq N}\left|{\rho}_{i j}\right| \Delta+\frac{\tr^2(\mathbf{\Sigma})}
{\left\|\mathbf{\Sigma}\right\|_{\mathrm{F}}^2}  \Delta^{2}.
\end{align*}
The Slutsky lemma and (\ref{dis-rho/}) say that $(\frac{\tr^2(\mathbf{\Sigma})}
{\left\|\mathbf{\Sigma}\right\|_{\mathrm{F}}^2} / \log N)^{1 / 2} \max_{1 \leq i<j \leq N}\left|{\rho}_{i j}\right| \rightarrow 2$ in probability. Consequently,
$$
\begin{aligned}
&\frac{\tr^2(\mathbf{\Sigma})}
{\left\|\mathbf{\Sigma}\right\|_{\mathrm{F}}^2} \max_{1 \leq i<j \leq N}\left|{\rho}_{i j}\right| \Delta=\left(\frac{\frac{\tr^2(\mathbf{\Sigma})}
{\left\|\mathbf{\Sigma}\right\|_{\mathrm{F}}^2}}{\log N}\right)^{1 / 2} \max_{1 \leq i<j \leq N}\left|{\rho}_{i j}\right| \cdot\Big(\sqrt{\frac{\tr^2(\mathbf{\Sigma})}
{\left\|\mathbf{\Sigma}\right\|_{\mathrm{F}}^2}  \log N} \Delta\Big) \rightarrow 0 \\
&\frac{\tr^2(\mathbf{\Sigma})}
{\left\|\mathbf{\Sigma}\right\|_{\mathrm{F}}^2}  \Delta^{2}=\Big[\sqrt{\frac{\tr^2(\mathbf{\Sigma})}
{\left\|\mathbf{\Sigma}\right\|_{\mathrm{F}}^2}  \log N} \Delta\Big]^{2} \cdot \frac{1}{\log N} \rightarrow 0
\end{aligned}
$$
in probability. Then, we can obtain that
$$
\frac{\tr^2(\mathbf{\Sigma})}
{\left\|\mathbf{\Sigma}\right\|_{\mathrm{F}}^2} \max_{1 \leq i<j \leq N}\left|\hat{\rho}_{i j}\right|^{2}=\frac{\tr^2(\mathbf{\Sigma})}
{\left\|\mathbf{\Sigma}\right\|_{\mathrm{F}}^2} \max_{1 \leq i<j \leq N}\left|{\rho}_{i j}\right|^{2}+o_{p}(1).
$$
So, we will prove
\begin{align}\label{pijhat-pij}
\mathbb{P}\left(\max_{1 \leq i<j \leq N}\left|\hat{\rho}_{i j}-{\rho}_{i j}\right|>\frac{\epsilon\sqrt{\tr\left(\mathbf{\Sigma}^2\right)}}{\sqrt{\log N}\tr\left(\mathbf{\Sigma}\right)}\right)\rightarrow 0.
\end{align}
By Lemma \ref{14Bi},
\begin{align*}
&\mathbb{P}\left(\max_{1 \leq i<j \leq N}\left|\hat{\rho}_{i j}-{\rho}_{i j}\right|>\frac{\epsilon\sqrt{\tr\left(\mathbf{\Sigma}^2\right)}}{\sqrt{\log N}\tr\left(\mathbf{\Sigma}\right)}\right)\\
\leq&N^2\max_{1 \leq i<j \leq N}\mathbb{P}\left( \frac{{\epsilon}_{j\cdot}^{\prime} \mathbf{B}_{i} {\epsilon}_{j\cdot}}{{\epsilon}_{j\cdot}^{\prime}{\epsilon}_{j\cdot}}>\frac{\epsilon\sqrt{\tr\left(\mathbf{\Sigma}^2\right)}}
{14\sqrt{\log N}\tr\left(\mathbf{\Sigma}\right)}\right)\\
\leq&N^2\max_{1 \leq i \leq N}\mathbb{P}\left( \frac{{\epsilon}_{1\cdot}^{\prime} \mathbf{B}_{i} {\epsilon}_{1\cdot}}{{\epsilon}_{1\cdot}^{\prime}{\epsilon}_{1\cdot}}>\frac{\epsilon\sqrt{\tr\left(\mathbf{\Sigma}^2\right)}}
{14\sqrt{\log N}\tr\left(\mathbf{\Sigma}\right)}\right)\\
\leq&N^2\max_{1 \leq i \leq N}\left[\mathbb{P}\left( \frac{{\epsilon}_{1\cdot}^{\prime} \mathbf{B}_{i} {\epsilon}_{1\cdot}}{{\epsilon}_{1\cdot}^{\prime}{\epsilon}_{1\cdot}}>\frac{\epsilon\sqrt{\tr\left(\mathbf{\Sigma}^2\right)}}
{14\sqrt{\log N}\tr\left(\mathbf{\Sigma}\right)},{\epsilon}_{1\cdot}^{\prime}{\epsilon}_{1\cdot}>\frac{\tr(\mathbf{\Sigma})}{2}\right)+
\mathbb{P}\left({\epsilon}_{1\cdot}^{\prime}{\epsilon}_{1\cdot}<\frac{\tr(\mathbf{\Sigma})}{2}\right)\right]\\
\leq&N^2\max_{1 \leq i \leq N}\left[\mathbb{P}\left( {\epsilon}_{1\cdot}^{\prime} \mathbf{B}_{i} {\epsilon}_{1\cdot}>\frac{\epsilon\sqrt{\tr\left(\mathbf{\Sigma}^2\right)}}
{28\sqrt{\log N}}\right)+\mathbb{P}\left(z_{1}^{\prime}\mathbf{\Sigma}z_{1}<\frac{\tr(\mathbf{\Sigma})}{2}\right)\right]\\
\leq&N^2\max_{1 \leq i \leq N}\left[\mathbb{P}\left( {\epsilon}_{1\cdot}^{\prime} \mathbf{B}_{i} {\epsilon}_{1\cdot}>\frac{\epsilon\sqrt{\tr\left(\mathbf{\Sigma}^2\right)}}
{28\sqrt{\log N}}\right)+2\exp\left(-\eta_{0} T\right)\right].
\end{align*}
Here, by Theorem 1.1 in \cite{rudelson2013}, there exists a constant $\eta_{0},\,C>0$ such
that
$$\mathbb{P}\left(z_{1}^{\prime}\mathbf{\Sigma}z_{1}<\frac{\tr(\mathbf{\Sigma})}{2}\right)\leq C\exp\left(-\eta_{0} T\right).$$
Note that $\tr(\mathbf{\Sigma}^{1/2}\mathbf{B}_{i}\mathbf{\Sigma}^{1/2})=\tr(\mathbf{B}_{i}\mathbf{\Sigma})\leq p\lambda_{max}(\mathbf{\Sigma})$
and $\tr(\mathbf{\Sigma}^{1/2}\mathbf{B}_{i}\mathbf{\Sigma}\mathbf{B}_{i}\mathbf{\Sigma}^{1/2})\leq\tr(\mathbf{B}_{i}\mathbf{\Sigma}^{2})\leq p\lambda_{max}(\mathbf{\Sigma})^{2}.$
Again, by Theorem 1.1 in \cite{rudelson2013} and Assumption \ref{assum:matrix}, for some constant $C>0,$ we have
\begin{align*}
&\mathbb{P}\left({\epsilon}_{1\cdot}^{\prime} \mathbf{B}_{i} {\epsilon}_{1\cdot}>\frac{\epsilon\sqrt{\tr\left(\mathbf{\Sigma}^2\right)}}
{28\sqrt{\log N}}\right)\\
\leq&\mathbb{P}\left({\epsilon}_{1\cdot}^{\prime} \mathbf{B}_{i} {\epsilon}_{1\cdot}>\tr(\mathbf{\Sigma}^{1/2}\mathbf{B}_{i}\mathbf{\Sigma}^{1/2})+\frac{\epsilon\sqrt{\tr\left(\mathbf{\Sigma}^2\right)}}
{56\sqrt{\log N}}\right)\\
\leq&2\exp\left\{-C\sqrt{T/\log N}\right\},
\end{align*}
where the first inequality holds because that for sufficiently large $N,T$, $\tr(\mathbf{\Sigma}^{1/2}\mathbf{B}_{i}\mathbf{\Sigma}^{1/2})\leq p\lambda_{max}(\mathbf{\Sigma})
\leq\epsilon\sqrt{\tr\left(\mathbf{\Sigma}^2\right)}/
\{56\sqrt{\log N}\},$ for all $i=1,\cdots,N.$
As $p> 0$ is fixed, we have
\begin{align*}
&\mathbb{P}\left(\max_{1 \leq i<j \leq N}\left|\hat{\rho}_{i j}-{\rho}_{i j}\right|>\frac{\epsilon\sqrt{\tr\left(\mathbf{\Sigma}^2\right)}}{\sqrt{\log N}\tr\left(\mathbf{\Sigma}\right)}\right)\\
\leq&2N^2\exp\left(-C\sqrt{T/\log N}\right)+2N^2\exp\left(-\eta_{0} T\right)\rightarrow0.
\end{align*}
By the slutsky lemma again, we can obtain that
\begin{align}\label{tilde Tmax}
\mathbb{P}\left(\frac{\tr^2(\mathbf{\Sigma})}
{\left\|\mathbf{\Sigma}\right\|_{\mathrm{F}}^{2}}\max _{1 \leq i<j \leq N} \hat{\rho}^2_{ij}-4 \log N+\log \log  N\leq y\right)
 \rightarrow \exp \left\{-\frac{1}{\sqrt{8 \pi}} \exp \left(-\frac{y}{2}\right)\right\}.
\end{align}
Then, we complete the proof. \hfill$\Box$
\subsection{Proof of Theorem \ref{th:max cons}}
\begin{lemma}\label{lemax1}
Under the same assumptions as in Theorem \ref{th:max cons}.
For any $\varepsilon\in(0,1)$ and sufficiently large $T,$
$$
\mathbb{P}\bigg(\left|\hat{\sigma}_{i j}-\sigma_{i j}\right| \geq  x\sqrt{\frac{\sigma_{i j}^{2}+(1+|\gamma_{2}|)\sigma_{i i} \sigma_{j j}}{N}}\bigg) \leq C \exp \left(-\frac{x^2}{2}(1-\varepsilon)\right),
$$
uniformly for $x\in(0,N^{\frac{1}{2}\wedge(\frac{1}{\tau}-\frac{1}{2})}),$ where $C$ does not depend on $i, j.$
\end{lemma}
\begin{lemma}\label{lemax2}
Under the same assumptions as in Theorem \ref{th:max cons}. Let $\hat{\mathbf{\Gamma}}=\big(\frac{\hat{\epsilon}_{i\cdot}^{\prime}\hat{\epsilon}_{j\cdot}}{T}\big)_{1 \leq i, j \leq N}, \hat{\gamma}_{N}=\|\hat{\mathbf{\Gamma}}\|_{\mathrm{F}}^{2}-\frac{1}{T}\{\tr(\hat{\mathbf{\Gamma}})\}^{2}$ and $\gamma_{N}=\left(\frac{\tr(\mathbf{\Sigma})}{T}\right)^{2}\|\mathbf{U}\|_{\mathrm{F}}^{2} .$ We have
\begin{align}\label{gamman}
\frac{\hat{\gamma}_{N}}{N}=\frac{\gamma_{N}}{N} a_{N}+b_{N},
\end{align}
where $\left\{a_{N}\right\}$ are real numbers satisfying $1-c_{1}/T \leq a_{N} \leq 1+c_{2}N / T$ for some constant $c_{1},c_{2}>0,\left\{b_{N}\right\}$ are random variables satisfying
$$
\mathbb{E}\left(b_{N}^{2}\right)=O\left(\frac{1}{NT}\right).$$
\end{lemma}
With preparations earlier, we are now ready to prove Theorem \ref{th:max cons}.
According to (\ref{tilde Tmax}), we have $$\tr^2(\mathbf{\Sigma})
/\left\|\mathbf{\Sigma}\right\|_{\mathrm{F}}^{2}\max _{1 \leq i<j \leq N} \hat{\rho}^2_{ij}=O_{p}(\log N).$$
Notice that
\begin{align*}
& \Big|\frac{\tr^2(\tilde{\mathbf{\Sigma}})}
{\|\tilde{\mathbf{\Sigma}}\|_{\mathrm{F}}^{2}}\max _{1 \leq i<j \leq N} \hat{\rho}^2_{ij}-\frac{\tr^2(\mathbf{\Sigma})}
{\left\|\mathbf{\Sigma}\right\|_{\mathrm{F}}^{2}}\max _{1 \leq i<j \leq N} \hat{\rho}^2_{ij} \Big| \\
\leq & \frac{\tr^2(\mathbf{\Sigma})}
{\left\|\mathbf{\Sigma}\right\|_{\mathrm{F}}^{2}}\max _{1 \leq i<j \leq N} \hat{\rho}^2_{ij} \Big|\frac{\tr^2(\tilde{\mathbf{\Sigma}})}{\|\tilde{\mathbf{\Sigma}}\|^2_{\mathrm{F}}}
\frac{\left\|{\mathbf{\Sigma}}\right\|^2_{\mathrm{F}}}{\tr^2({\mathbf{\Sigma}})}
-1\Big|.
\end{align*}
To obtain the limit null distribution of $L_{N},$ we only need to show that
$$\frac{\tr^2(\tilde{\mathbf{\Sigma}})}{\|\tilde{\mathbf{\Sigma}}\|^2_{\mathrm{F}}}
\frac{\left\|{\mathbf{\Sigma}}\right\|^2_{\mathrm{F}}}{\tr^2({\mathbf{\Sigma}})}
-1=o_{p}\{\log^{-1}N\}.$$
%Note that the lemma \ref{lemax1} and \ref{lemax2} are very similar to the lemma B.1 and B.2 in the Supplement to \cite{chen2018}.
So,
similar to the proof of Theorem 2.1 in \cite{chen2018}, we can obtain that
\begin{align}\label{max:cons}
\frac{\tr^2(\tilde{\mathbf{\Sigma}})}{\|\tilde{\mathbf{\Sigma}}\|^2_{\mathrm{F}}}
\frac{\left\|{\mathbf{\Sigma}}\right\|^2_{\mathrm{F}}}{\tr^2({\mathbf{\Sigma}})}
=1+O_{P}\bigg\{\Big(\sqrt{\frac{\log T}{N}}\Big)^{\min (1,2-\tau)}\bigg\}
=1+o_{p}\{\log^{-1}N\},
\end{align}
due to Lemmas \ref{lemax1} and \ref{lemax2}. Then, we complete the proof and obtain that
$$\mathbb{P}\left(\frac{\tr^2(\tilde{\mathbf{\Sigma}})}
{\|\tilde{\mathbf{\Sigma}}\|^2_{\mathrm{F}}}L_{N}-4\log N+\log \log  N\leq y\right) \rightarrow \exp \left\{-\frac{1}{\sqrt{8 \pi}} \exp \left(-\frac{y}{2}\right)\right\}.$$
\hfill$\Box$
%\begin{remark}
%It is worth noting that our proof idea is similar to that of \cite{chen2018}, but since we prove the limit distribution of max type statistic under sub-Gaussian distributions, the proof techniques used are not completely similar, such as Hanson-Wright inequality in \cite{rudelson2013}, and we also need to prove that the error caused by the residuals is negligible.
%\end{remark}
\subsection{Proof of Theorem \ref{th:maxpower}}
Under the assumptions in Theorem \ref{th:maxpower}, we establish two similar lemmas.
\begin{lemma}\label{H1:lemax1}
Under the same assumptions as in Theorem \ref{th:maxpower}. For any $\varepsilon\in(0,1)$ and sufficiently large $T,$
$$
\mathbb{P}\bigg(\left|\hat{\sigma}_{i j}-\sigma_{i j}\right| \geq  x\sqrt{\frac{\tr(\mathbf{U}^2)(\sigma_{i i} \sigma_{j j}+\sigma_{i j}^{2})}{N^2}}\bigg) \leq C \exp \left(-C^{\prime}x^2(1-\varepsilon)\right),
$$
uniformly for $x\in(0,N^{\frac{1}{2}\wedge(\frac{1}{\tau}-\frac{1}{2})}),$ where $C$ and $C^{\prime} $do not depend on $i, j.$
\end{lemma}
\begin{lemma}\label{H1:lemax2}
Under the same assumptions as in Theorem \ref{th:maxpower}. Let $\hat{\mathbf{\Gamma}}=\big(\frac{\hat{\epsilon}_{i\cdot}^{\prime}\hat{\epsilon}_{j\cdot}}{T}\big)_{1 \leq i, j \leq N}, \hat{\gamma}_{N}=\|\hat{\mathbf{\Gamma}}\|_{\mathrm{F}}^{2}-\frac{1}{T}\{\tr(\hat{\mathbf{\Gamma}})\}^{2}$ and $\gamma_{N}=\left(\frac{\tr(\mathbf{\Sigma})}{T}\right)^{2}\|\mathbf{U}\|_{\mathrm{F}}^{2} .$ We have
\begin{align}\label{H1:gamman}
\frac{\hat{\gamma}_{N}}{N}=\frac{\gamma_{N}}{N} a_{N}+b_{N},
\end{align}
where $\left\{a_{N}\right\}$ are real numbers satisfying $1-c_{1}/T\leq a_{N} \leq 1+c_{2}N / T$ for some constant $c_{1},c_{2}>0,\left\{b_{N}\right\}$ are random variables satisfying
$$
\mathbb{E}\left(b_{N}^{2}\right)=O\left(\frac{1}{T^{\frac{1}{2}\vee \left(\frac{1}{\tau}-\frac{1}{2}\right)}}\right).$$
\end{lemma}
We are now ready to prove Theorem \ref{th:maxpower}.
Because for any $1\leq i\leq j\leq N,$ ${\epsilon}_{i\cdot}^{\prime}{\epsilon}_{j\cdot}={Z}^{\prime}\left[\{u_{i}^{1/2}(u_{j}^{1/2})^{\prime}\}\otimes \mathbf{\Sigma}\right]{Z}$
and
${\epsilon}_{j\cdot}^{\prime}\mathbf{B}_{i}{\epsilon}_{j\cdot}={Z}^{\prime}\left[\{u_{j}^{1/2}(u_{j}^{1/2})^{\prime}\}\otimes (\mathbf{\Sigma}^{1/2}\mathbf{B}_{i}\mathbf{\Sigma}^{1/2})\right]{Z}.$
Similar to (\ref{pijhat-pij}), we can use the similar approach to prove that
$$\mathbb{P}\left(\max_{1\leq i<j\leq N}\left|\hat{\rho}_{ij}-{\rho}_{ij}\right|>\frac{\epsilon\sqrt{\tr\left(\mathbf{\Sigma}^2\right)}}{\sqrt{\log N}\tr\left(\mathbf{\Sigma}\right)}\right)\rightarrow 0,$$
and
$$\mathbb{P}\left(\max_{1\leq i<j\leq N}\left|\frac{\epsilon_{i\cdot}^{\prime}\epsilon_{i\cdot}\epsilon_{j\cdot}^{\prime}\epsilon_{j\cdot}}{T^2}-
\frac{\tr^2(\mathbf{\Sigma})}{T^2}\right|>C_{3}\sqrt{\frac{\log N}{T}}\right)
=O\left(N^{-M}\right).
$$
In addition, similar to the proof of Theorem 2.1 in \cite{chen2018}, we can also prove that
\begin{align}\label{max:cons}
\frac{\|\tilde{\mathbf{\Sigma}}\|^2_{\mathrm{F}}}{\tr^2(\tilde{\mathbf{\Sigma}})}
\frac{\tr^2({\mathbf{\Sigma}})}{\left\|{\mathbf{\Sigma}}\right\|^2_{\mathrm{F}}}
=1+O_{P}\bigg\{\Big(\sqrt{\frac{\log T}{N}}\Big)^{\min (1,2-\tau)}\bigg\},
\end{align}
due to Lemmas \ref{H1:lemax1} and \ref{H1:lemax2}.
Note that $$\sqrt{\var\left(\epsilon_{i\cdot}^{\prime}\epsilon_{j\cdot}\right)}= \sqrt{\|\mathbf{\Sigma}\|_{\mathrm{F}}^{2}\left(u_{i i} u_{j j}+u_{i j}^{2}\right)+\gamma_{2}\tr\left[\{u_{i}^{1/2}(u_{j}^{1/2})^{\prime}\}\circ
\{u_{i}^{1/2}(u_{j}^{1/2})^{\prime}\}\right]\tr(\mathbf{\Sigma}\circ
\mathbf{\Sigma})},$$
where
$u_{l}^{1/2}\in \mathbb{R}^{N}$ denotes the $l$-th row vector of matrix $\mathbf{U}^{1/2}.$ Note that we assume that $(\mathbf{U})_{ii}=1,$ for all $1\leq i\leq N,$
so, we have $$\tr\left[\{u_{i}^{1/2}(u_{j}^{1/2})^{\prime}\}\circ
\{u_{i}^{1/2}(u_{j}^{1/2})^{\prime}\}\right]\tr(\mathbf{\Sigma}\circ
\mathbf{\Sigma})\leq \tr(\mathbf{\Sigma}^{2}).$$
Without loss of generality, we can assume that $u_{12} \geq \delta \sqrt{\frac{\|\mathbf{\Sigma}\|_{\mathrm{F}}^{2}}{\tr^{2}(\mathbf{\Sigma})}\log N }$ for some constant $\delta>2$.
Note that for some constant $C>0,$
by Theorem 1.1 in \cite{rudelson2013} and Assumption \ref{assum:matrix},
we have
\begin{align*}
\mathbb{P}\left(\sqrt{\frac{\log N}{T}}\frac{\epsilon_{1\cdot}^{\prime}\epsilon_{2\cdot}-u_{12}\tr(\mathbf{\Sigma})}
{\sqrt{\|\mathbf{\Sigma}\|_{\mathrm{F}}^{2}}}>\varepsilon\right)
\leq2\exp\left\{-C\varepsilon\log N\right\}\rightarrow 0,
\end{align*}
for all $\varepsilon>0.$
Then, there exist a constant $M>0$ satisfying
\begin{align*}
&\mathbb{P}\left(\sqrt{\frac{\tr^2(\tilde{\mathbf{\Sigma}})}
{\|\tilde{\mathbf{\Sigma}}\|_{\mathrm{F}}^{2}}}\max_{1\leq i< j\leq N}\hat{\rho}_{ij} > \sqrt{4 \log N-\log \log N+w_{\alpha}}\right)\\
\geq&
\mathbb{P}\left(\sqrt{\frac{\tr^2(\tilde{\mathbf{\Sigma}})}
{\|\tilde{\mathbf{\Sigma}}\|_{\mathrm{F}}^{2}}}\hat{\rho}_{12} > \sqrt{4 \log N-\log \log N+w_{\alpha}}\right)\\
\geq&\mathbb{P}\left(\sqrt{\frac{\tr^{2}(\mathbf{\Sigma})}{\|\mathbf{\Sigma}\|_{\mathrm{F}}^{2}} }\hat{\rho}_{12} > \sqrt{4 \log N-\log \log N+w_{\alpha}}\Big(1+\frac{M\sqrt{\log N}}{\sqrt{N}}\Big)\right)+o(1)
\\
\geq&\mathbb{P}\left(\sqrt{\frac{\tr^{2}(\mathbf{\Sigma})}{\|\mathbf{\Sigma}\|_{\mathrm{F}}^{2}} }{\rho}_{12}>\sqrt{4 \log N-\log \log N+w_{\alpha}}\Big(1+\frac{M\sqrt{\log N}}{\sqrt{N}}\Big)+o(1)\right)+o(1)\\
\geq&\mathbb{P}\left(\sqrt{\frac{\tr^{2}(\mathbf{\Sigma})}{\|\mathbf{\Sigma}\|_{\mathrm{F}}^{2}} }{\rho}_{12} > \sqrt{4 \log N-\log \log N+w_{\alpha}}+o(1)\right)+o(1)\\
\geq&
\mathbb{P}\left(\sqrt{\frac{\tr^{2}(\mathbf{\Sigma})}{\|\mathbf{\Sigma}\|_{\mathrm{F}}^{2}} } \frac{\epsilon_{1\cdot}^{\prime}\epsilon_{2\cdot}-\operatorname{tr}(\mathbf{\Sigma}) u_{12}}{\sqrt{\epsilon_{1\cdot}^{\prime}\epsilon_{1\cdot}\epsilon_{2\cdot}^{\prime}\epsilon_{2\cdot}}}
>-(\delta-2) \sqrt{\log N} / 2\right)+o(1),\\
\geq&
\mathbb{P}\left(\frac{\epsilon_{1\cdot}^{\prime}\epsilon_{2\cdot}-\operatorname{tr}(\mathbf{\Sigma}) u_{12}}{\sqrt{\|\mathbf{\Sigma}\|_{\mathrm{F}}^{2}}}
>-(\delta-2) \sqrt{\log N} / 2\right)+o(1)\\
=&1-\mathbb{P}\left(\frac{\epsilon_{1\cdot}^{\prime}\epsilon_{2\cdot}-\operatorname{tr}(\mathbf{\Sigma}) u_{12}}{\sqrt{\|\mathbf{\Sigma}\|_{\mathrm{F}}^{2}}}
\leq-(\delta-2) \sqrt{\log N} / 2\right)+o(1)\\
\geq&1-\mathbb{P}\left(\Big|\frac{\epsilon_{1\cdot}^{\prime}\epsilon_{2\cdot}-\operatorname{tr}(\mathbf{\Sigma}) u_{12}}{\sqrt{\|\mathbf{\Sigma}\|_{\mathrm{F}}^{2}}}\Big|
\geq(\delta-2) \sqrt{\log N} / 2\right)+o(1)\\
=&1-O(N^{-\varepsilon_{1}})+o(1)\rightarrow 1,
\end{align*}
where $\varepsilon_{1}=C\{(\delta-2)/2\}^{2}>0,$ for some constant $C>0.$
Consequently, we complete the proof of this theorem.\hfill$\Box$

\subsection{Proof of Theorem \ref{th:sum-max}}
Define
\begin{align}\label{signdef}
&\tilde{T}_{max}=
\max_{1\leq i<j\leq N}\frac{(\epsilon_{i\cdot}^{\prime}\epsilon_{j\cdot})^2}
{\left\|\mathbf{\Sigma}\right\|_{\mathrm{F}}^{2}}\n\\
&\Lambda_{N}=\{(i, j) ; 1 \leq i<j \leq N\} \nonumber\n\\
&A_{N}=\left\{S_{N}/\sigma_{S_{N}} \leq x\right\} \quad \text { and } \quad
 B_{I}=\left\{\left|\epsilon_{i\cdot}^{\prime}\epsilon_{j\cdot}\right|\geq l_{N}\right\}
\end{align}
for any $I=(i, j) \in \Lambda_{N},$ where $a_{N}=4\log N-\log \log  N+y$
and
$$l_{N}=\sqrt{\left\|\mathbf{\Sigma}\right\|_{\mathrm{F}}^{2}\left[4\log N-\log \log  N+y\right]}
=\sqrt{\left\|\mathbf{\Sigma}\right\|_{\mathrm{F}}^{2}a_{N}}.$$
To make a clear presentation, we impose a trivial ordering for elements in $\Lambda_{N} .$ For any $I_{1}=\left(i_{1}, j_{1}\right) \in \Lambda_{N}$ and $I_{2}=\left(i_{2}, j_{2}\right) \in$ $\Lambda_{N},$ we say $I_{1}<I_{2}$ if $i_{1}<i_{2}$ or $i_{1}=i_{2}$ but $j_{1}<j_{2}.$

Recall that under certain assumptions, \cite{baltagi2016} established the asymptotic property of $S_{N}$ that under the null hypothesis,
$S_{N}/{\hat{\sigma}}_{S_{N}}\to \mathcal{N}(0,1)$ in distribution when $\min(N,T)\rightarrow \infty.$
Since the assumptions in this paper are slightly different from those in \cite{baltagi2016},
we reconsider the asymptotic properties of $S_{N}$ under our
Assumptions \ref{assum:E distribution}-\ref{assum:matrix}. Similar to Theorems 2-3 in \cite{baltagi2016}, Lemma \ref{th:sum null} presents the asymptotic null distribution of $S_{N}$,
and Lemma \ref{th:sum cons} presents that $\hat{\sigma}_{S_{N}}$ is a ratio-consistent estimator of the variance of $S_{N}$.

\begin{lemma}\label{th:sum null}
 Under Assumptions \ref{assum:E distribution}-\ref{assum:matrix} and the null hypothesis,
we have $S_{N}/{\sigma}_{S_{N}}\to \mathcal{N}(0,1)$ in distribution, as $\min(N,T)\rightarrow \infty$ with $\underset{\min(N,T) \rightarrow \infty}{\lim}N/T=\gamma\in (0,+\infty).$
Here,
$\mathbf{M}_{i}=\mathbf{P}_{i}\mathbf{\Sigma}\mathbf{P}_{i},$ for any $1\leq i\leq N,$
and $$\sigma^2_{S_{N}}=\frac{2}{N(N-1)}\underset{1\leq i<j\leq N}{\sum\sum}\tr\left(\mathbf{M}_{i}\mathbf{M}_{j}\right)
/\{\tr\left(\mathbf{M}_{i}\right)\tr\left(\mathbf{M}_{j}\right)\}.$$
\end{lemma}
%{\color{red}Since $\mathbf{\Sigma}$ is unobservable, $\hat{\sigma}^2_{S_{N}}$ can be estimated by}
Recall that
\begin{align}\label{sigmasumhat}
\hat{\sigma}^2_{S_{N}}=\frac{2}{N(N-1)}\underset{1\leq i<j\leq N}{\sum\sum} v_{j}^{\prime}\left(v_{i}-\bar{v}_{ij}\right) v_{i}^{\prime}\left(v_{j}-\bar{v}_{i j}\right),
\end{align}
$\bar{v}_{ij}=\sum_{1<k \neq i, j<N} v_{k}/(N-2)$ and $v_{k}=\hat{\epsilon}_{k\cdot}/\|\hat{\epsilon}_{k\cdot}\|$ for all $1\leq k\leq N.$
\begin{lemma}\label{th:sum cons}
 Under Assumptions \ref{assum:E distribution}-\ref{assum:matrix} and the null hypothesis, as $\min(N,T)\rightarrow \infty$ with $\underset{\min(N,T) \rightarrow \infty}{\lim}N/T=\gamma\in (0,+\infty)$, we have $\hat{\sigma}^2_{S_{N}}/{\sigma}^2_{S_{N}}\to 1$ in probability.
\end{lemma}

Next, the following lemmas are provided for establishing the asymptotic independence between the two test statistics $S_{N}$
 and $L_{N}.$
\begin{lemma}\label{indep hat}
Assume that Assumptions \ref{assum:E distribution}-\ref{assum:matrix} hold, under $H_{0},$ if $\tilde{T}_{max}-4 \log N+\log \log  N$ and $S_{N}/\sigma_{S_{N}}$ are asymptotically independent,
then $L_{N}{\tr^2(\tilde{\mathbf{\Sigma}})}/{\|\tilde{\mathbf{\Sigma}}\|_{\mathrm{F}}^{2}}-4 \log N+\log \log  N$ and $S_{N}/\hat{\sigma}_{S_{N}}$ are also asymptotically
independent.
\end{lemma}
\begin{lemma}\label{linear}
Under $H_0$ and same assumptions in Lemma \ref{indep hat}, let
$$
H(N, k)=\sum_{I_{1}<I_{2}<\cdots<I_{k} \in \Lambda_{N}} \mathbb{P}\left(B_{I_{1}} B_{I_{2}} \cdots B_{I_{k}}\right).
$$
Then $\lim _{k \rightarrow \infty} \limsup _{\min(N,T) \rightarrow \infty} H(N, k)=0.$
\end{lemma}
\begin{lemma}\label{linear2}
Under the assumptions of Lemma \ref{indep hat},
$$
\sum_{I_{1}<I_{2}<\cdots<I_{k} \in \Lambda_{N}}\left[\mathbb{P}\left(A_{N} B_{I_{1}} B_{I_{2}} \cdots B_{I_{k}}\right)-\mathbb{P}\left(A_{N}\right) \cdot \mathbb{P}\left(B_{I_{1}} B_{I_{2}} \cdots B_{I_{k}}\right)\right] \rightarrow 0
$$
as $\min(N,T) \rightarrow \infty$ for each $k \geq 1.$
\end{lemma}

Next, we are ready to prove asymptotic independence stated in Theorem \ref{th:sum-max}.
By (\ref{eiej}) and Lemma \ref{th:sum null}, the following hold,
\begin{align}
 &\max _{1 \leq i<j \leq N} \frac{\left(\epsilon_{i\cdot}^{\prime}\epsilon_{j\cdot}\right)^2}{\left\|\mathbf{\Sigma}\right\|_{\mathrm{F}}^{2}}-4\log N+\log \log  N\rightarrow G(y)\,\text { in distribution; }\label{demaxQ}\\
&\frac{S_{N}}{\sigma_{S_{N}}} \rightarrow N(0,1)\text { in distribution, }\label{desumQ}
\end{align}
where $G(y)=\exp \left\{- \exp \left(-y/2\right)/\sqrt{8\pi}\right\}.$
To show asymptotic independence, according to Lemma \ref{indep hat}, it is enough to show the limit of
\begin{align}
\lim _{\min(N,T) \rightarrow \infty} \mathbb{P}\left(\frac{S_{N}}{\sigma_{S_{N}}}\leq x, \max _{1 \leq i<j \leq N} \frac{\left(\epsilon_{i\cdot}^{\prime}\epsilon_{j\cdot}\right)^2}{\left\|\mathbf{\Sigma}\right\|_{\mathrm{F}}^{2}}\leq a_{N}\right)=\Phi(x) \cdot G(y),
\end{align}
for any $x \in \mathbb{R}$ and $y \in \mathbb{R}$, where $\Phi(x)=(2 \pi)^{-1 / 2} \int_{-\infty}^{x} e^{-t^{2} / 2} d t$ and
$$a_{N}=4\log N-\log\log N+y,$$
which makes sense for large $N$. Because of (\ref{demaxQ}) and (\ref{desumQ}), the above is equivalent to that
\begin{align}\label{indedengjia1Q}
\lim _{\min(N,T) \rightarrow \infty} \mathbb{P}\left(\frac{S_{N}}{\sigma_{S_{N}}}\leq x, \max _{1 \leq i<j \leq N} \frac{\left(\epsilon_{i\cdot}^{\prime}\epsilon_{j\cdot}\right)^2}{\left\|\mathbf{\Sigma}\right\|_{\mathrm{F}}^{2}}> a_{N}\right)=\Phi(x) \cdot\left\{1-G(y)\right\},
\end{align}
for any $x \in \mathbb{R}$ and $y \in \mathbb{R}$. Review notation $\Lambda_{N}, A_{N}$ and $B_{I}$ for any $I=$ $(i, j) \in \Lambda_{N}$ in (\ref{signdef}). Write
\begin{align}\label{unionQ}
\mathbb{P}\left(\frac{S_{N}}{\sigma_{S_{N}}}\leq x, \max _{1 \leq i<j \leq N} \frac{\left(\epsilon_{i\cdot}^{\prime}\epsilon_{j\cdot}\right)^2}{\left\|\mathbf{\Sigma}\right\|_{\mathrm{F}}^{2}}> a_{N}\right)=\mathbb{P}\left(\bigcup_{I \in \Lambda_{N}} A_{N} B_{I}\right).
\end{align}
Here the notation ${A}_{N} B_{I}$ stands for ${A}_{N} \cap B_{I}$. From the inclusion-exclusion principle,
\begin{align}\label{UAB2k+1Q}
\mathbb{P}\left(\bigcup_{I \in \Lambda_{N}} A_{N} B_{I}\right) \leq & \sum_{I_{1} \in \Lambda_{N}} \mathbb{P}\left(A_{N} B_{I_{1}}\right)-\sum_{I_{1}<I_{2} \in \Lambda_{N}} \mathbb{P}\left(A_{N} B_{I_{1}} B_{I_{2}}\right)+\cdots+ \n\\
&\sum_{I_{1}<I_{2}<\cdots<I_{2 l+1} \in \Lambda_{N}}\mathbb{P}\left(A_{N} B_{I_{1}} B_{I_{2}} \cdots B_{I_{2 l+1}}\right)
\end{align}
and
\begin{align}\label{UAB2kQ}
\mathbb{P}\left(\bigcup_{I \in \Lambda_{N}} A_{N} B_{I}\right) \geq& \sum_{I_{1} \in \Lambda_{N}} \mathbb{P}\left(A_{N} B_{I_{1}}\right)-\sum_{I_{1}<I_{2} \in \Lambda_{N}} \mathbb{P}\left(A_{N} B_{I_{1}} B_{I_{2}}\right)+\cdots-\\
&\sum_{I_{1}<I_{2}<\cdots<I_{2 l} \in \Lambda_{N}} \mathbb{P}\left(A_{N} B_{I_{1}} B_{I_{2}} \cdots B_{I_{2 l}}\right)
\end{align}
for any integer $l \geq 1$. Reviewing the definition
$$
H(N, k)=\sum_{I_{1}<I_{2}<\cdots<I_{k} \in \Lambda_{N}} \mathbb{P}\left(B_{I_{1}} B_{I_{2}} \cdots B_{I_{k}}\right),
$$
for $k \geq 1$ in Lemma $\ref{linear},$ we have from the lemma that
\begin{align}\label{lHpkQ}
\lim _{k \rightarrow \infty} \limsup _{\min(N,T) \rightarrow \infty} H(N, k)=0.
\end{align}
Set
$$\zeta(p,k)
=\sum_{I_{1}<I_{2}<\cdots<I_{k} \in \Lambda_{N}}\left\{\mathbb{P}\left(A_{N} B_{I_{1}} B_{I_{2}} \cdots B_{I_{k}}\right)-\mathbb{P}\left(A_{N}\right) \cdot \mathbb{P}\left(B_{I_{1}} B_{I_{2}} \cdots B_{I_{k}}\right)\right\},
$$
for $T \geq 1 .$ By Lemma \ref{linear2},
\begin{align}\label{zetaQ}
\lim _{\min(N,T) \rightarrow \infty} \zeta(N, k)=0,
\end{align}
for each $k \geq 1$. The assertion (\ref{UAB2k+1Q}) implies that
\begin{align}\label{UzetaH2k+1Q}
& \mathbb{P}\left(\bigcup_{I \in \Lambda_{N}} A_{N} B_{I}\right) \n\\
\leq & \mathbb{P}\left(A_{N}\right)\left\{\sum_{I_{1} \in \Lambda_{N}} \mathbb{P}\left(B_{I_{1}}\right)-\sum_{I_{1}<I_{2} \in \Lambda_{N}} \mathbb{P}\left(B_{I_{1}} B_{I_{2}}\right)+\cdots-\right.\n\\
&\left.\sum_{I_{1}<I_{2}<\cdots<I_{2 l} \in \Lambda_{N}} \mathbb{P}\left(B_{I_{1}} B_{I_{2}} \cdots B_{I_{2 l}}\right)\right\}+\left\{\sum_{k=1}^{2 l} \zeta(N, k)\right\}+H(N, 2 l+1) \n\\
\leq P &\left(A_{N}\right) \cdot \mathbb{P}\left(\bigcup_{I \in \Lambda_{N}} B_{I}\right)+\left\{\sum_{k=1}^{2 l} \zeta(N, k)\right\}+H(N, 2 l+1),
\end{align}
where the inclusion-exclusion formula is used again in the last inequality, that is
$$
\begin{array}{r}
\mathbb{P}\left(\bigcup_{I \in \Lambda_{N}} B_{I}\right) \geq\left\{\sum_{I_{1} \in \Lambda_{N}} \mathbb{P}\left(B_{I_{1}}\right)-\sum_{I_{1}<I_{2} \in \Lambda_{N}} \mathbb{P}\left(B_{I_{1}} B_{I_{2}}\right)+\cdots-\right. \\
\left.\sum_{I_{1}<I_{2}<\cdots<I_{2 l} \in \Lambda_{N}} \mathbb{P}\left(B_{I_{1}} B_{I_{2}} \cdots B_{I_{2 l}}\right)\right\},
\end{array}
$$
for all $l \geq 1$. By the definition of $a_{N}$ and (\ref{demaxQ}),
\begin{align*}
\mathbb{P}\left(\bigcup_{I \in \Lambda_{N}} B_{I}\right)&=\mathbb{P}\left(\max _{1 \leq i<j \leq N} \frac{\left(\epsilon_{i\cdot}^{\prime}\epsilon_{j\cdot}\right)^2}{\left\|\mathbf{\Sigma}\right\|_{\mathrm{F}}^{2}}>a_{N}\right)\n\\
&=\mathbb{P}\left\{\max _{1 \leq i<j \leq N} \frac{\left(\epsilon_{i\cdot}^{\prime}\epsilon_{j\cdot}\right)^2}{\left\|\mathbf{\Sigma}\right\|_{\mathrm{F}}^{2}}
-4 \log N+\log \log  N>y\right\} \n\\
&\rightarrow 1-G(y),
\end{align*}
as $\min(N,T) \rightarrow \infty$. By (\ref{desumQ}), $\mathbb{P}\left(A_{N}\right) \rightarrow \Phi(x)$ as $\min(N,T) \rightarrow \infty$. From (\ref{unionQ}), by fixing $l$ first and sending $\min(N,T) \rightarrow \infty$, we get from (\ref{zetaQ}) that
$$
\begin{aligned}
& \limsup _{\min(N,T) \rightarrow \infty} \mathbb{P}\left(\frac{S_{N}}{\sigma_{S_{N}}}\leq x, \max _{1 \leq i<j \leq N} \frac{\left(\epsilon_{i\cdot}^{\prime}\epsilon_{j\cdot}\right)^2}{\left\|\mathbf{\Sigma}\right\|_{\mathrm{F}}^{2}}>a_{N}\right) \\
\leq & \Phi(x) \cdot\left\{1-G(y)\right\}+\limsup _{\min(N,T) \rightarrow \infty} H(N, 2 l+1).
\end{aligned}
$$
Now, let $l\rightarrow \infty$ and use (\ref{lHpkQ}) to see
\begin{align}\label{limitrightQ}
\limsup _{\min(N,T) \rightarrow \infty} \mathbb{P}\left(\frac{S_{N}}{\sigma_{S_{N}}}\leq x, \max _{1 \leq i<j \leq N} \frac{\left(\epsilon_{i\cdot}^{\prime}\epsilon_{j\cdot}\right)^2}{\left\|\mathbf{\Sigma}\right\|_{\mathrm{F}}^{2}}>a_{N}\right) \leq \Phi(x) \cdot\left\{1-G(y)\right\}.
\end{align}
By applying the same argument to (\ref{UAB2kQ}), we see that the counterpart of
(\ref{UzetaH2k+1Q}) becomes
\begin{align*}
\mathbb{P}\left(\bigcup_{I \in \Lambda_{N}} A_{N} B_{I}\right)
\geq & \mathbb{P}\left(A_{N}\right)\left\{\sum_{I_{1} \in \Lambda_{N}} \mathbb{P}\left(B_{I_{1}}\right)-\sum_{I_{1}<I_{2} \in \Lambda_{N}} \mathbb{P}\left(B_{I_{1}} B_{I_{2}}\right)+\cdots+\right.\n\\
&\left.\sum_{I_{1}<I_{2}<\cdots<I_{2 l-1} \in \Lambda_{N}} \mathbb{P}\left(B_{I_{1}} B_{I_{2}} \cdots B_{I_{2 l-1}}\right)\right\}+\n\\
&\left\{\sum_{k=1}^{2 l-1} \zeta(N, k)\right\}-H(N, 2 l) \n\\
\geq & \mathbb{P}\left(A_{N}\right) \cdot \mathbb{P}\left(\bigcup_{I \in \Lambda_{N}} B_{I}\right)+\left\{\sum_{k=1}^{2 l-1} \zeta(N, k)\right\}-H(N, 2 l),
\end{align*}
where in the last step we use the inclusion-exclusion principle such that
\begin{align*}
\mathbb{P}\left(\bigcup_{I \in \Lambda_{N}} B_{I}\right) \leq\left\{\sum_{I_{1} \in \Lambda_{N}} \mathbb{P}\left(B_{I_{1}}\right)-\sum_{I_{1}<I_{2} \in \Lambda_{N}} \mathbb{P}\left(B_{I_{1}} B_{I_{2}}\right)+\cdots+\right. \\
\left.\sum_{I_{1}<I_{2}<\cdots<I_{2 l-1} \in \Lambda_{N}} \mathbb{P}\left(B_{I_{1}} B_{I_{2}} \cdots B_{I_{2 l-1}}\right)\right\},
\end{align*}
for all $l \geq 1$. Review (\ref{unionQ}) and repeat the earlier procedure to see
$$
\limsup _{\min(N,T) \rightarrow \infty} \mathbb{P}\left(\frac{S_{N}}{\sigma_{S_{N}}}\leq x, \max _{1 \leq i<j \leq N} \frac{\left(\epsilon_{i\cdot}^{\prime}\epsilon_{j\cdot}\right)^2}{\left\|\mathbf{\Sigma}\right\|_{\mathrm{F}}^{2}}>a_{N}\right) \geq \Phi(x) \cdot\left\{1-G(y)\right\},
$$
by sending $\min(N,T) \rightarrow \infty$ and then sending $l \rightarrow \infty$. This and (\ref{limitrightQ}) yield (\ref{indedengjia1Q}). The proof is completed due to Lemma \ref{indep hat}.
\hfill$\Box$
\section{Proofs of the lemmas}\label{proof of lemmas}
In this section, we will prove the lemmas used in the previous section.
\subsection{Proof of Lemma \ref{lemax1}}
Recall that $\mathbf{M}_{i}=\mathbf{P}_{i}\mathbf{\Sigma}\mathbf{P}_{i},$
$\mathbf{B}_{i}= \mathbf{x}_{i}\left(\mathbf{x}_{i}^{\prime} \mathbf{x}_{i}\right)^{-1} \mathbf{x}_{i}^{\prime}.$
$(\mathbf{A})_{ij}$ is the element at row $i$ and column $j$ of matrix
$\mathbf{A},$ for any square matrix $\mathbf{A}.$ Hence, we define the element at row $i$ and column $j$ of matrix $\mathbf{M}_{l}=\mathbf{P}_{l}\mathbf{\Sigma}\mathbf{P}_{l}$ as $\sigma_{ij,l},$ $1\leq i \leq j \leq N.$ Let $\omega_{ij,l}=\tr\{(\mathbf{\Sigma}^{1/2}\mathbf{P}_{l}e_{i}e_{j}^{\prime}\mathbf{P}_{l}\mathbf{\Sigma}^{1/2}
)\circ(\mathbf{\Sigma}^{1/2}\mathbf{P}_{l}e_{i}e_{j}^{\prime}\mathbf{P}_{l}\mathbf{\Sigma}^{1/2})\}\leq \sigma_{ii,l}\sigma_{jj,l},$
for $1\leq i <j\leq T;1\leq l\leq N.$

Then, we claim that
\begin{align}\label{sigmaij,l-sigmaij}
\Big|\Big(\mathbf{B}_{l}\mathbf{\Sigma}\Big)_{i j}\Big|=O(T^{-1 \vee-\frac{1}{\tau}}),\,\,\,\,\Big|\Big(\mathbf{\Sigma}\mathbf{B}_{l}\Big)_{i j}\Big|=O(T^{-1 \vee-\frac{1}{\tau}}),\,\,\,\,\,\Big|\Big(\mathbf{B}_{l}\mathbf{\Sigma}\mathbf{B}_{s}\Big)_{i j}\Big|=O(T^{-1\vee-\frac{1}{\tau}}),
\end{align}
for any $1\leq l,s\leq N,\,1\leq i,j\leq T.$
To simplify the expression, we let $b_{ij,l}\doteq\Big(\mathbf{B}_{l}\Big)_{i j} ,$
 for any $1\leq l\leq N,\,1\leq i,j\leq T.$ Under Assumption \ref{assum:x_it}, we have $\max_{1\leq j\leq N}|b_{ij,l}|=O(T^{-1}).$
\begin{align*}
\Big|\Big(\mathbf{B}_{l}\mathbf{\Sigma}\Big)_{i j}\Big|=
\Big|\sum_{k=1}^{T}b_{ik,l}\sigma_{kj}\Big|
&\leq \sum_{k=1}^{T}|b_{ik,l}||\sigma_{kj}|\\
&\leq O(T^{-1})\sum_{k=1}^{T}|\sigma_{kj}|.
\end{align*}
Due to Assumption \ref{assum:matrix}, we have $\sum_{k=1}^{T}\left|\sigma_{j k}\right|^{\tau} \leq C$ for some $0<\tau<2$ and $1\leq j \leq T.$ So, if $0<\tau<1,$ we have $\sum_{k=1}^{T}|\sigma_{kj}|\leq\sum_{k=1}^{T}|\sigma_{kj}|^{\tau}|\sigma_{kj}|^{1-\tau}=O(1).$
If $1<\tau<2,$ from a convex inequality we have $$\left\{\big(\sum_{k=1}^{T}|\sigma_{kj}|\big)^{\tau}\right\}^{1/\tau}\leq \left\{T^{\tau-1}\sum_{k=1}^{T}|\sigma_{kj}|^{\tau}\right\}^{1/\tau}\leq O(T^{1-\frac{1}{\tau}}).$$
So, $
\Big|\Big(\mathbf{B}_{l}\mathbf{\Sigma}\Big)_{i j}\Big|=O(T^{-1\vee-\frac{1}{\tau}}).$
Similarly, we have $\Big|\Big(\mathbf{\Sigma}\mathbf{B}_{l}\Big)_{i j}\Big|=O(T^{-1\vee-\frac{1}{\tau}})$ and $\Big|\Big(\mathbf{B}_{l}\mathbf{\Sigma}\mathbf{B}_{s}\Big)_{i j}\Big|=O(T^{-1\vee-\frac{1}{\tau}}).$
Then, due to $\mathbf{P}_{l}\mathbf{\Sigma}\mathbf{P}_{l}=\mathbf{\Sigma}-
\mathbf{B}_{l}\mathbf{\Sigma}-\mathbf{\Sigma}\mathbf{B}_{l}
+\mathbf{B}_{l}\mathbf{\Sigma}\mathbf{B}_{l},$
we can conclude that
$$\left|\sum_{l=1}^{N}\frac{\sigma_{i j,l}-\sigma_{ij}}{N}\right|=O(T^{-1\vee-\frac{1}{\tau}})$$
and
$$\left|\sum_{l=1}^{N}\frac{\sigma_{i i,l} \sigma_{j j,l}+\sigma_{i j,l}^{2}-\sigma_{ii}\sigma_{jj}-\sigma^2_{ij}}{N}\right|=O(T^{-1\vee-\frac{1}{\tau}}).$$

Recall that $\bar{\hat{\epsilon}}_{\cdot j}=\sum_{l=1}^{N}\hat{\epsilon}_{lj}/N,$ $1\leq j\leq N,$
we have $\hat{\sigma}_{i j}=\frac{1}{N-1} \sum_{l=1}^{N} \hat{\epsilon}_{l i} \hat{\epsilon}_{l j} -\frac{N}{N-1} \bar{\hat{\epsilon}}_{\cdot i} \bar{\hat{\epsilon}}_{ \cdot j}$ Since $\cov\left(\hat{\epsilon}_{l i}, \hat{\epsilon}_{l j}\right)=\sigma_{i j,l}$, we obtain that $\var\left(\hat{\epsilon}_{l i} \hat{\epsilon}_{l j}\right)=\sigma_{i j,l}^{2}+\sigma_{i i,l} \sigma_{j j,l}+\gamma_{2}\omega_{ij,l}.$
 By classical Cram\'er type large deviation results for independent random variables (see Corollary 3.1 in \cite{saulis1991limit}), we have for any $\varepsilon>0$,
\begin{align}\label{large deviation}
&\mathbb{P}\Big(\Big|\frac{\sum_{l=1}^{N}\left(\hat{\epsilon}_{l i} \hat{\epsilon}_{l j}-\sigma_{i j,l}\right)}{\sqrt{\sum_{l=1}^{N}\left(\sigma_{i j,l}^{2}+(1+|\gamma_{2}|)\sigma_{i i,l} \sigma_{j j,l}\right)}}\Big| \geq x\Big)\\
\leq&
\mathbb{P}\Big(\Big|\frac{\sum_{l=1}^{N}\left(\hat{\epsilon}_{l i} \hat{\epsilon}_{l j}-\sigma_{i j,l}\right)}{\sqrt{\sum_{l=1}^{N}\left(\sigma_{i i,l} \sigma_{j j,l}+\sigma_{i j,l}^{2}+\gamma_{2}\omega_{ij,l}\right)}}\Big| \geq x\Big) \\
\leq& C \exp \Big\{-\frac{x^{2}}{2}(1-\varepsilon)\Big\}
\end{align}
uniformly in $x \in\big[0, o(\sqrt{N})\big) .$ For $\bar{\hat{\epsilon}}_{\cdot j}$, we have $\var\left(\bar{\hat{\epsilon}}_{\cdot j}\right)=\sum_{l=1}^{ N} \sigma_{jj,l}/N^{2} .$ By (\ref{sigmaij,l-sigmaij}), $\var\left(\bar{\hat{\epsilon}}_{\cdot j}\right)\leq C N^{-1},$ uniformly in $1 \leq j \leq T .$ Again, by classical Cram\'er type large deviation results for independent random variables, we have for any $\varepsilon>0$,
$$
\mathbb{P}\left(\left|\bar{\hat{\epsilon}}_{\cdot j}\right| \geq x \sqrt{\var\left(\bar{\hat{\epsilon}}_{\cdot j}\right)}\right) \leq C \exp \Big\{-\frac{x^{2}}{2}(1-\varepsilon)\Big\}
$$
uniformly in $x \in\big[0, o(\sqrt{N})\big) .$ So
$$
\mathbb{P}\left(\left|\bar{\hat{\epsilon}}_{\cdot i}\bar{\hat{\epsilon}}_{\cdot j} \right| \geq x^{2} \sqrt{\var\left(\bar{\hat{\epsilon}}_{\cdot i}\right) \var\left(\bar{\hat{\epsilon}}_{\cdot j}\right)}\right) \leq 2C \exp \Big\{-\frac{x^{2}}{2}(1-\varepsilon)\Big\}
$$
uniformly in $x \in\big[0, o(\sqrt{N})\big) .$ We have, uniformly for $x \in\big[0, o(\sqrt{N})\big), x^{2} \sqrt{\var\left(\bar{\hat{\epsilon}}_{\cdot i}\right) \var\left(\bar{\hat{\epsilon}}_{\cdot i}\right)}=o(x / \sqrt{N})$. So for any $\delta>0$ and large $N$
\begin{align}\label{tail normal}
\mathbb{P}\left(\left|\bar{\hat{\epsilon}}_{\cdot i}\bar{\hat{\epsilon}}_{\cdot j}\right| \geq \delta \frac{x}{\sqrt{N}}\right) \leq 2C \exp \Big\{-\frac{x^{2}}{2}(1-\varepsilon)\Big\}
\end{align}
uniformly for $x \in\big[0, o(\sqrt{N})\big).$ Hence,
the lemma follows from (\ref{large deviation}) and (\ref{tail normal}).
%, for any $\varepsilon\in(0,1),$ we have
%\begin{align}\label{sigmaij,l}
%\mathbb{P}\bigg\{\Big|\hat{\sigma}_{ij}-\sum_{l=1}^{N}\frac{\sigma_{i j,l}}{N-1}\Big| \geq \frac{x}{N-1}\sqrt{\sum_{l=1}^{N}\sigma_{i j,l}^{2}+(1+|\gamma_{2}|)\sigma_{i i,l} \sigma_{j j,l}}\Big\} \leq C \exp \Big\{-\frac{x^{2}}{2}(1-\varepsilon)\bigg\}.
%\end{align}
%This
%proves the lemma.
\hfill$\Box$
\subsection{Proof of Lemma \ref{lemax2}}

So, according to the definitions in Lemma \ref{lemax2},
we have
\begin{align*}
\|\hat{\mathbf{\Gamma}}\|_{\mathrm{F}}^{2}&=\frac{1}{T^2}\sum_{i=1}^{N} \sum_{j=1}^{N}( \hat{\epsilon}_{i\cdot}^{\prime}\hat{\epsilon}_{j\cdot})^{2},\\
\{\tr(\hat{\mathbf{\Gamma}})\}^{2}&=\frac{1}{T^2}\left(\sum_{i=1}^{N}
\hat{\epsilon}_{i\cdot}^{\prime}\hat{\epsilon}_{i\cdot}\right)^{2},\\
T^2a_{N}&=\frac{1}{\gamma_{N}}\left[\sum_{i=1}^{N} \sum_{j=1}^{N} \mathbb{E} (\hat{\epsilon}_{i\cdot}^{\prime}\hat{\epsilon}_{j\cdot})^{2}-\frac{1}{T} \mathbb{E}\left(\sum_{i=1}^{N} \hat{\epsilon}_{i\cdot}^{\prime}\hat{\epsilon}_{i\cdot}\right)^{2}\right],\\
T^2b_{N}&=\frac{1}{N}\left[\sum_{i=1} \sum_{j=1} (\hat{\epsilon}_{i\cdot}^{\prime}\hat{\epsilon}_{j\cdot})^{2}-\frac{1}{T}\left(\sum_{i=1} \hat{\epsilon}_{i\cdot}^{\prime}\hat{\epsilon}_{i\cdot}\right)^{2}-\sum_{i=1} \sum_{j=1} \mathbb{E} (\hat{\epsilon}_{i\cdot}^{\prime}\hat{\epsilon}_{j\cdot})^{2}+\frac{1}{T} \mathbb{E}\left(\sum_{i=1} \hat{\epsilon}_{i\cdot}^{\prime}\hat{\epsilon}_{i\cdot}\right)^{2}\right].
\end{align*}
It is easy to verify that $a_{N}$ and $b_{N}$ will make the equation (\ref{gamman}) true. In the following, we will prove that $a_{N}, b_{N}$ satisfy the properties in the lemma.

We first deal with the term $a_{N} .$ Recall that $\hat{\epsilon}_{i\cdot}^{\prime}\hat{\epsilon}_{j\cdot}$ and $\mathbf{P}_{i}\mathbf{\Sigma}\mathbf{P}_{i}=\mathbf{M}_{i},$ we have
\begin{equation*}
\mathbb{E} (\hat{\epsilon}_{i\cdot}^{\prime}\hat{\epsilon}_{j\cdot})^{2} =\left\{
\begin{aligned}
&\tr(\mathbf{M}_{i}\mathbf{M}_{j})  , & i\neq j, \\
&2\tr(\mathbf{M}^2_{i})+\tr^2(\mathbf{M}_{i})+\gamma_{2}\tr\{(\mathbf{\Sigma}^{1/2}
\mathbf{P}_{i}\mathbf{\Sigma}^{1/2})\circ(\mathbf{\Sigma}^{1/2}
\mathbf{P}_{i}\mathbf{\Sigma}^{1/2})\} , & i=j.
\end{aligned}
\right.
\end{equation*}
Therefore,
 $$\sum_{i=1}^{N}\sum_{j=1}^{N}\mathbb{E} (\hat{\epsilon}_{i\cdot}^{\prime}\hat{\epsilon}_{j\cdot})^{2} =\underset{i\neq j}{\sum^{N}\sum^{N}}\tr(\mathbf{M}_{i}\mathbf{M}_{j}) +\sum_{i=1}^{N}2\tr(\mathbf{M}^2_{i})+\tr^2(\mathbf{M}_{i})+\gamma_{2}\tr\{(\mathbf{\Sigma}^{1/2}
\mathbf{P}_{i}\mathbf{\Sigma}^{1/2})\circ(\mathbf{\Sigma}^{1/2}
\mathbf{P}_{i}\mathbf{\Sigma}^{1/2})\}.$$ Moreover, by Lemma \ref{le:moment of quadratic form},
\begin{align*}
\mathbb{E}\left(\sum_{i=1}^{N} \hat{\epsilon}_{i\cdot}^{\prime}\hat{\epsilon}_{i\cdot}\right)^{2} =&\sum_{i=1}^{N} \sum_{j=1}^{N} \mathbb{E}( \hat{\epsilon}_{i\cdot}^{\prime}\hat{\epsilon}_{i\cdot} \hat{\epsilon}_{j\cdot}^{\prime}\hat{\epsilon}_{j\cdot})\\
=&\sum_{i=1}^{N}\mathbb{E}( \hat{\epsilon}_{i\cdot}^{\prime}\hat{\epsilon}_{i\cdot} \hat{\epsilon}_{i\cdot}^{\prime}\hat{\epsilon}_{i\cdot})+\underset{i\neq j}{\sum^{N}\sum^{N}}\mathbb{E}( \hat{\epsilon}_{i\cdot}^{\prime}\hat{\epsilon}_{i\cdot} \hat{\epsilon}_{j\cdot}^{\prime}\hat{\epsilon}_{j\cdot})\\
=&\sum_{i=1}^{N}2\tr(\mathbf{M}^2_{i})+\tr^2(\mathbf{M}_{i})+\gamma_{2}\tr\{(\mathbf{\Sigma}^{1/2}
\mathbf{P}_{i}\mathbf{\Sigma}^{1/2})\circ(\mathbf{\Sigma}^{1/2}
\mathbf{P}_{i}\mathbf{\Sigma}^{1/2})\}\\
&+\underset{i\neq j}{\sum^{N}\sum^{N}}\tr(\mathbf{M}_{i})\tr(\mathbf{M}_{j}).
\end{align*}
So we have
\begin{align*}
{T^2} \gamma_{N}a_{N}&=\left[\sum_{i=1}^{N} \sum_{j=1}^{N} \mathbb{E} (\hat{\epsilon}_{i\cdot}^{\prime}\hat{\epsilon}_{j\cdot})^{2}-\frac{1}{T} \mathbb{E}\left(\sum_{i=1}^{N} \hat{\epsilon}_{i\cdot}^{\prime}\hat{\epsilon}_{i\cdot}\right)^{2}\right]\\
=&\underset{i\neq j}{\sum^{N}\sum^{N}}\tr(\mathbf{M}_{i}\mathbf{M}_{j}) +\sum_{i=1}^{N}2\tr(\mathbf{M}^2_{i})+\tr^2(\mathbf{M}_{i})+\gamma_{2}\tr\{(\mathbf{\Sigma}^{1/2}
\mathbf{P}_{i}\mathbf{\Sigma}^{1/2})\circ(\mathbf{\Sigma}^{1/2}
\mathbf{P}_{i}\mathbf{\Sigma}^{1/2})\}\\
&-\frac{1}{T}\Big[\sum_{i=1}^{N}2\tr(\mathbf{M}^2_{i})+\tr^2(\mathbf{M}_{i})+
\gamma_{2}\tr\{(\mathbf{\Sigma}^{1/2}
\mathbf{P}_{i}\mathbf{\Sigma}^{1/2})\circ(\mathbf{\Sigma}^{1/2}
\mathbf{P}_{i}\mathbf{\Sigma}^{1/2})\}\\
&\quad\quad\quad+\underset{i\neq j}{\sum^{N}\sum^{N}}\tr(\mathbf{M}_{i})\tr(\mathbf{M}_{j})\Big]\\
\geq&\sum_{i=1}^{N}\tr^2(\mathbf{M}_{i})\left(1-\frac{1}{T}\right)+\underset{i\neq j}{\sum^{N}\sum^{N}}\left\{\tr(\mathbf{M}_{i}\mathbf{M}_{j})-\frac{1}{T}\tr(\mathbf{M}_{i})\tr(\mathbf{M}_{j})\right\}\\
&-(1-\frac{1}{T})|\gamma_{2}|\sum_{i=1}^{N}\tr\{(\mathbf{\Sigma}^{1/2}
\mathbf{P}_{i}\mathbf{\Sigma}^{1/2})\circ(\mathbf{\Sigma}^{1/2}
\mathbf{P}_{i}\mathbf{\Sigma}^{1/2})\}\\
\geq&\left(1-\frac{1}{T}\right)\sum_{i=1}^{N}\left\{\tr^2(\mathbf{\Sigma})-2\tr(\mathbf{\Sigma})
\tr(\mathbf{B}_{i}\mathbf{\Sigma})\right\}+\underset{i\neq j}{\sum^{N}\sum^{N}}
\left\{\tr(\mathbf{\Sigma}^2)-2\tr(\mathbf{\Sigma}^2\mathbf{B}_{j})\right.\\
&-\left.2\tr(\mathbf{\Sigma}^2\mathbf{B}_{i})-2\tr(\mathbf{\Sigma}\mathbf{B}_{i}
\mathbf{B}_{j}\mathbf{\Sigma}\mathbf{B}_{j})-2\tr(\mathbf{\Sigma}\mathbf{B}_{j}
\mathbf{B}_{i}\mathbf{\Sigma}\mathbf{B}_{i})+2\tr(\mathbf{B}_{i}\mathbf{\Sigma}^{2}\mathbf{B}_{j})\right\}\\
&-\frac{1}{T}\underset{i\neq j}{\sum^{N}\sum^{N}}\left\{\tr^2(\mathbf{\Sigma})
+\tr(\mathbf{B}_{i}\mathbf{\Sigma})\tr(\mathbf{B}_{j}\mathbf{\Sigma})\right\}
-|\gamma_{2}|\sum_{i=1}^{N}\{\tr(\mathbf{\Sigma}^2)+\tr(\mathbf{B}_{i}\mathbf{\Sigma}\mathbf{B}_{i}\mathbf{\Sigma})\}\\
\geq &N\tr^2(\mathbf{\Sigma})-2pN\tr(\mathbf{\Sigma})C-\frac{N\tr^2(\mathbf{\Sigma})}{T}
+\underset{i\neq j}{\sum^{N}\sum^{N}}\left\{\tr(\mathbf{\Sigma}^2)-\frac{\tr^2(\mathbf{\Sigma})}{T}\right\}\\
&-10pC^2N(N-1)-\frac{p^2C^2N(N-1)}{T}-|\gamma_{2}|C^2NT-|\gamma_{2}|NpC^2\\
\geq&T^2\gamma_{N}\left(1-\frac{c_{1}}{T}\right),
\end{align*}
for some constant $c_{1}>0.$
Moreover, we have
\begin{align*}
{T^2} \gamma_{N}a_{N}&=\left[\sum_{i=1}^{N} \sum_{j=1}^{N} \mathbb{E} (\hat{\epsilon}_{i\cdot}^{\prime}\hat{\epsilon}_{j\cdot})^{2}-\frac{1}{T} \mathbb{E}\left(\sum_{i=1}^{N} \hat{\epsilon}_{i\cdot}^{\prime}\hat{\epsilon}_{i\cdot}\right)^{2}\right]\\
=&\underset{i\neq j}{\sum^{N}\sum^{N}}\tr(\mathbf{M}_{i}\mathbf{M}_{j}) +\sum_{i=1}^{N}2\tr(\mathbf{M}^2_{i})+\tr^2(\mathbf{M}_{i})
+\gamma_{2}\tr\{(\mathbf{\Sigma}^{1/2}
\mathbf{P}_{i}\mathbf{\Sigma}^{1/2})\circ(\mathbf{\Sigma}^{1/2}
\mathbf{P}_{i}\mathbf{\Sigma}^{1/2})\}\\
&-\frac{1}{T}\Big[\sum_{i=1}^{N}2\tr(\mathbf{M}^2_{i})
+\tr^2(\mathbf{M}_{i})+\gamma_{2}\tr\{(\mathbf{\Sigma}^{1/2}
\mathbf{P}_{i}\mathbf{\Sigma}^{1/2})\circ(\mathbf{\Sigma}^{1/2}
\mathbf{P}_{i}\mathbf{\Sigma}^{1/2})\}\\
&\quad\quad\quad+\underset{i\neq j}{\sum^{N}\sum^{N}}\tr(\mathbf{M}_{i})\tr(\mathbf{M}_{j})\Big]\\
\leq&\underset{i\neq j}{\sum^{N}\sum^{N}}\left\{\tr(\mathbf{\Sigma}^2)+2\tr(\mathbf{\Sigma}\mathbf{B}_{i}
\mathbf{\Sigma}\mathbf{B}_{j})+2\tr(\mathbf{B}_{i}\mathbf{\Sigma}^2\mathbf{B}_{j})
+\tr(\mathbf{\Sigma}\mathbf{B}_{j}\mathbf{\Sigma}\mathbf{B}_{j})
+\tr(\mathbf{\Sigma}\mathbf{B}_{i}\mathbf{\Sigma}\mathbf{B}_{i})\right.\\
&\left.+\tr(\mathbf{B}_{i}
\mathbf{\Sigma}\mathbf{B}_{i}\mathbf{B}_{j}
\mathbf{\Sigma}\mathbf{B}_{j})-2\tr(\mathbf{\Sigma}\mathbf{B}_{i}
\mathbf{B}_{j}\mathbf{\Sigma}\mathbf{B}_{j})-2\tr(\mathbf{\Sigma}\mathbf{B}_{j}
\mathbf{B}_{i}\mathbf{\Sigma}\mathbf{B}_{i})\right\}+2C^2NT\\
&+\sum_{i=1}^{N}\left\{
\tr^2(\mathbf{\Sigma})+\tr^2(\mathbf{\Sigma}\mathbf{B}_{i})\right\}+(1-\frac{1}{T})|\gamma_{2}|\sum_{i=1}^{N}\{\tr(\mathbf{\Sigma}^{2})
+\tr(\mathbf{B}_{i}\mathbf{\Sigma}\mathbf{B}_{i}\mathbf{\Sigma})\}\\
\leq&C^2N(N-1)T+11pC^2N(N-1)+2C^2NT+N\tr^2(\mathbf{\Sigma})+p^2C^2N\\
&+|\gamma_{2}|C^2NT+|\gamma_{2}|NpC^2\\
\leq&T^2\gamma_{N}\left(1+\frac{c_{2}N}{T}\right),
\end{align*}
for some constant $c_{2}>0.$ This proves that $a_{N}$ satisfies the inequality in the lemma.

It remains to calculate $b_{N}.$ We have
\begin{align*}
\var\bigg\{\sum_{i=1}^{N} \sum_{j=1}^{N} (\hat{\epsilon}_{i\cdot}^{\prime}\hat{\epsilon}_{j\cdot})^{2}\bigg\}
&=\mE\Bigg[\bigg\{\sum_{i=1}^{N} \sum_{j=1}^{N} (\hat{\epsilon}_{i\cdot}^{\prime}\hat{\epsilon}_{j\cdot})^{2}\bigg\}^{2}\Bigg]
-\mE^2\bigg\{\sum_{i=1}^{N} \sum_{j=1}^{N} (\hat{\epsilon}_{i\cdot}^{\prime}\hat{\epsilon}_{j\cdot})^{2}\bigg\}.
\end{align*}
Note that $i\neq j\neq k$ means that $i,j,k$ are not equal, and
  $i\neq j\neq k\neq l$ means that $i,j,k,l$ are not equal.
First, by Lemma \ref{le:moment of quadratic form}, we have
\begin{align*}
&\mE\Bigg[\bigg\{\sum_{i=1}^{N} \sum_{j=1}^{N} (\hat{\epsilon}_{i\cdot}^{\prime}\hat{\epsilon}_{j\cdot})^{2}\bigg\}^{2}\Bigg]\\
=&\sum_{i=1}^{N}\mE\left\{(\hat{\epsilon}_{i\cdot}^{\prime}\hat{\epsilon}_{i\cdot})^{4}\right\}+2\underset{i\neq j}{\sum^{N}\sum^{N}}\mE\left\{(\hat{\epsilon}_{i\cdot}^{\prime}\hat{\epsilon}_{j\cdot})^{4}\right\}+\underset{i\neq j}{\sum^{N}\sum^{N}}\mE\left\{(\hat{\epsilon}_{i\cdot}^{\prime}\hat{\epsilon}_{i\cdot})^{2}(\hat{\epsilon}_{j\cdot}^{\prime}\hat{\epsilon}_{j\cdot})^{2}\right\}\\
&+4\underset{i\neq j}{\sum^{N}\sum^{N}}\mE\left\{(\hat{\epsilon}_{i\cdot}^{\prime}\hat{\epsilon}_{j\cdot})^{2}
(\hat{\epsilon}_{i\cdot}^{\prime}\hat{\epsilon}_{i\cdot})^{2}\right\}+4\underset{i\neq j\neq k}{\sum^{N}\sum^{N}\sum^{N}}
\mE\left\{(\hat{\epsilon}_{i\cdot}^{\prime}\hat{\epsilon}_{j\cdot})^{2}
(\hat{\epsilon}_{i\cdot}^{\prime}\hat{\epsilon}_{k\cdot})^{2}\right\}\\
&+2\underset{i\neq j\neq k}{\sum^{N}\sum^{N}\sum^{N}}
\mE\left\{(\hat{\epsilon}_{i\cdot}^{\prime}\hat{\epsilon}_{i\cdot})^{2}
(\hat{\epsilon}_{j\cdot}^{\prime}\hat{\epsilon}_{k\cdot})^{2}\right\}+\underset{i\neq j\neq k\neq l}{\sum^{N}\sum^{N}\sum^{N}\sum^{N}}\mE\left\{(\hat{\epsilon}_{i\cdot}^{\prime}\hat{\epsilon}_{j\cdot})^{2}
(\hat{\epsilon}_{k\cdot}^{\prime}\hat{\epsilon}_{l\cdot})^{2}\right\}\\
=&\sum_{i=1}^{N}\Big\{48\tr(\mathbf{M}^4_{i})+32\tr(\mathbf{M}^3_{i})\tr(\mathbf{M}_{i})+12\tr^2(\mathbf{M}^2_{i})
+12\tr(\mathbf{M}^2_{i})\tr^2(\mathbf{M}_{i})+\tr^4(\mathbf{M}_{i})\\
&+\gamma_{2} f_{\gamma_{2}}+\gamma_{4} f_{\gamma_{4}}+\gamma_{6} f_{\gamma_{6}}+\gamma_{1}^{2} f_{\gamma_{1}^{2}}+\gamma_{2}^{2} f_{\gamma_{2}^{2}}+\gamma_{1} \gamma_{3} f_{\gamma_{1} \gamma_{3}}\Big\}\\
&+2\underset{i\neq j}{\sum^{N}\sum^{N}}\Big[6\tr(\mathbf{M}_{i}\mathbf{M}_{j})^2+3\tr^2(\mathbf{M}_{i}\mathbf{M}_{j})+
3\gamma_{2}\tr\{(\mathbf{\Sigma}^{1/2}\mathbf{P}_{j}\mathbf{M}_{i}\mathbf{P}_{j}\mathbf{\Sigma}^{1/2}
)\circ(\mathbf{\Sigma}^{1/2}\mathbf{P}_{j}\mathbf{M}_{i}\mathbf{P}_{j}\mathbf{\Sigma}^{1/2})\}\\
&+\gamma_{2}
\mE\tr\{(\mathbf{\Sigma}^{1/2}\mathbf{P}_{i}\mathbf{P}_{j}\epsilon_{j\cdot}\epsilon_{j\cdot}^{\prime}\mathbf{P}_{j}\mathbf{P}_{i}\mathbf{\Sigma}^{1/2}
)\circ(\mathbf{\Sigma}^{1/2}\mathbf{P}_{i}\mathbf{P}_{j}\epsilon_{j\cdot}\epsilon_{j\cdot}^{\prime}\mathbf{P}_{j}\mathbf{P}_{i}\mathbf{\Sigma}^{1/2})\}
\Big]\\
&+\underset{i\neq j}{\sum^{N}\sum^{N}}\Big(\left[2\tr(\mathbf{M}^2_{i})+\tr^2(\mathbf{M}_{i})
+\gamma_{2}\tr\{(\mathbf{\Sigma}^{1/2}\mathbf{P}_{i}\mathbf{\Sigma}^{1/2}
)\circ(\mathbf{\Sigma}^{1/2}\mathbf{P}_{i}\mathbf{\Sigma}^{1/2})\}\right]\\
&\,\,\,\,\,\,\,\,\,\,\,\,\,\,\,\,\,\,\,\,\,\,\,\,\,\,\,\times
\left[2\tr(\mathbf{M}^2_{j})+\tr^2(\mathbf{M}_{j})+\gamma_{2}\tr\{(\mathbf{\Sigma}^{1/2}\mathbf{P}_{j}\mathbf{\Sigma}^{1/2}
)\circ(\mathbf{\Sigma}^{1/2}\mathbf{P}_{j}\mathbf{\Sigma}^{1/2})\}\right]\Big)\\
&+4\underset{i\neq j}{\sum^{N}\sum^{N}}\Big[\gamma_{4} \tr\left\{(\mathbf{\Sigma}^{1/2}\mathbf{P}_{i}\mathbf{\Sigma}^{1/2}) \circ (\mathbf{\Sigma}^{1/2}\mathbf{P}_{i}\mathbf{\Sigma}^{1/2}) \circ (\mathbf{\Sigma}^{1/2}\mathbf{P}_{i}\mathbf{M}_{j}\mathbf{P}_{i}\mathbf{\Sigma}^{1/2})\right\}\\
&+2\gamma_{2} \tr\left(\mathbf{\Sigma}^{1/2}\mathbf{P}_{i}\mathbf{\Sigma}^{1/2}\right) \tr\left\{(\mathbf{\Sigma}^{1/2}\mathbf{P}_{i}\mathbf{\Sigma}^{1/2}) \circ( \mathbf{\Sigma}^{1/2}\mathbf{P}_{i}\mathbf{M}_{j}\mathbf{P}_{i}\mathbf{\Sigma}^{1/2})\right\} \\
&+\gamma_{2} \tr\left(\mathbf{\Sigma}^{1/2}\mathbf{P}_{i}\mathbf{M}_{j}\mathbf{P}_{i}\mathbf{\Sigma}^{1/2}\right) \tr\left\{(\mathbf{\Sigma}^{1/2}\mathbf{P}_{i}\mathbf{\Sigma}^{1/2}) \circ (\mathbf{\Sigma}^{1/2}\mathbf{P}_{i}\mathbf{\Sigma}^{1/2})\right\}\\
&+8 \gamma_{2} \tr\left[(\mathbf{\Sigma}^{1/2}\mathbf{P}_{i}\mathbf{\Sigma}^{1/2} ) \circ\left(\mathbf{\Sigma}^{1/2}\mathbf{P}_{i}\mathbf{\Sigma}^{1/2} \mathbf{\Sigma}^{1/2}\mathbf{P}_{i}\mathbf{M}_{j}\mathbf{P}_{i}\mathbf{\Sigma}^{1/2}\right)\right] \\
&+4 \gamma_{2} \tr\left[(\mathbf{\Sigma}^{1/2}\mathbf{P}_{i}\mathbf{M}_{j}\mathbf{P}_{i}\mathbf{\Sigma}^{1/2} )\circ\left(\mathbf{\Sigma}^{1/2}\mathbf{P}_{i}\mathbf{\Sigma}^{1/2} \mathbf{\Sigma}^{1/2}\mathbf{P}_{i}\mathbf{\Sigma}^{1/2}\right)\right]\\
&+4 \gamma_{1}^{2}\left[{\tau}_{T}^{\prime}\left\{\mathbf{I}_{T} \circ (\mathbf{\Sigma}^{1/2}\mathbf{P}_{i}\mathbf{\Sigma}^{1/2})\right\} \mathbf{\Sigma}^{1/2}\mathbf{P}_{i}\mathbf{\Sigma}^{1/2}\left\{\mathbf{I}_{T} \circ (\mathbf{\Sigma}^{1/2}\mathbf{P}_{i}\mathbf{M}_{j}\mathbf{P}_{i}\mathbf{\Sigma}^{1/2})\right\} {\tau}_{T}\right] \\
&+2 \gamma_{1}^{2}\left[{\tau}_{T}^{\prime}\left\{\mathbf{I}_{T} \circ (\mathbf{\Sigma}^{1/2}\mathbf{P}_{i}\mathbf{\Sigma}^{1/2})\right\} \mathbf{\Sigma}^{1/2}\mathbf{P}_{i}\mathbf{M}_{j}\mathbf{P}_{i}\mathbf{\Sigma}^{1/2}\left\{\mathbf{I}_{T} \circ (\mathbf{\Sigma}^{1/2}\mathbf{P}_{i}\mathbf{\Sigma}^{1/2})\right\} {\tau}_{T}\right] \\
&+4 \gamma_{1}^{2}\left[{\tau}_{T}^{\prime}\left\{(\mathbf{\Sigma}^{1/2}\mathbf{P}_{i}\mathbf{\Sigma}^{1/2} ) \circ (\mathbf{\Sigma}^{1/2}\mathbf{P}_{i}\mathbf{\Sigma}^{1/2}) \circ ( \mathbf{\Sigma}^{1/2}\mathbf{P}_{i}\mathbf{M}_{j}\mathbf{P}_{i}\mathbf{\Sigma}^{1/2})\right\} {\tau}_{T}\right]\\
&+\tr\left(\mathbf{\Sigma}^{1/2}\mathbf{P}_{i}\mathbf{\Sigma}^{1/2}\right) \tr\left(\mathbf{\Sigma}^{1/2}\mathbf{P}_{i}\mathbf{\Sigma}^{1/2}\right) \tr\left(\mathbf{\Sigma}^{1/2}\mathbf{P}_{i}\mathbf{M}_{j}\mathbf{P}_{i}\mathbf{\Sigma}^{1/2}\right)\\
&+4 \tr\left(\mathbf{\Sigma}^{1/2}\mathbf{P}_{i}\mathbf{\Sigma}^{1/2}\right) \tr\left(\mathbf{\Sigma}^{1/2}\mathbf{P}_{i}\mathbf{\Sigma}^{1/2} \mathbf{\Sigma}^{1/2}\mathbf{P}_{i}\mathbf{M}_{j}\mathbf{P}_{i}\mathbf{\Sigma}^{1/2}\right) \\
&+2 \tr\left(\mathbf{\Sigma}^{1/2}\mathbf{P}_{i}\mathbf{M}_{j}\mathbf{P}_{i}\mathbf{\Sigma}^{1/2}\right) \tr\left(\mathbf{\Sigma}^{1/2}\mathbf{P}_{i}\mathbf{\Sigma}^{1/2} \mathbf{\Sigma}^{1/2}\mathbf{P}_{i}\mathbf{\Sigma}^{1/2}\right)\\
&+8 \tr\left(\mathbf{\Sigma}^{1/2}\mathbf{P}_{i}\mathbf{\Sigma}^{1/2} \mathbf{\Sigma}^{1/2}\mathbf{P}_{i}\mathbf{\Sigma}^{1/2} \mathbf{\Sigma}^{1/2}\mathbf{P}_{i}\mathbf{M}_{j}\mathbf{P}_{i}\mathbf{\Sigma}^{1/2}\right)\Big]\\
&+4\underset{i\neq j\neq k}{\sum^{N}\sum^{N}\sum^{N}}
\Big[2\tr(\mathbf{M}_{j}\mathbf{M}_{i}\mathbf{M}_{k}\mathbf{M}_{i})+\tr(\mathbf{M}_{j}\mathbf{M}_{i})\tr(\mathbf{M}_{k}\mathbf{M}_{i})\\
&+\gamma_{2}\tr\{(\mathbf{\Sigma}^{1/2}\mathbf{P}_{i}\mathbf{M}_{j}\mathbf{P}_{i}\mathbf{\Sigma}^{1/2})
\circ(\mathbf{\Sigma}^{1/2}\mathbf{P}_{i}\mathbf{M}_{k}\mathbf{P}_{i}\mathbf{\Sigma}^{1/2})\}\Big]\\
&+2\underset{i\neq j\neq k}{\sum^{N}\sum^{N}\sum^{N}}
\mE\left[2\tr(\mathbf{M}^2_{i})+\tr^2(\mathbf{M}_{i})+\gamma_{2}\tr\{(\mathbf{\Sigma}^{1/2}\mathbf{P}_{i}\mathbf{\Sigma}^{1/2}
)\circ(\mathbf{\Sigma}^{1/2}\mathbf{P}_{i}\mathbf{\Sigma}^{1/2})\}\right]\tr(\mathbf{M}_{j}\mathbf{M}_{k})\\
&+\underset{i\neq j\neq k\neq l}{\sum^{N}\sum^{N}\sum^{N}\sum^{N}}\left\{\tr(\mathbf{M}_{i}\mathbf{M}_{j})\tr(\mathbf{M}_{k}\mathbf{M}_{l})\right\}.
\end{align*}
Here, because the diagonal elements of $(\mathbf{\Sigma}^{1/2}\mathbf{P}_{i}\mathbf{\Sigma}^{1/2})_{ii}$ are greater than 0 and less than $C,$ we can obtain that
\begin{align*}
f_{\gamma_{2}}=&6 \tr\left(\mathbf{\Sigma}^{1/2}\mathbf{P}_{i}\mathbf{\Sigma}^{1/2}\right) \tr\left(\mathbf{\Sigma}^{1/2}\mathbf{P}_{i}\mathbf{\Sigma}^{1/2}\right) \tr\left\{(\mathbf{\Sigma}^{1/2}\mathbf{P}_{i}\mathbf{\Sigma}^{1/2} )\circ (\mathbf{\Sigma}^{1/2}\mathbf{P}_{i}\mathbf{\Sigma}^{1/2})\right\}\\
&+12{\tau}_{T}^{\prime}\left\{(\mathbf{\Sigma}^{1/2}\mathbf{P}_{i}\mathbf{\Sigma}^{1/2} )\circ ( \mathbf{\Sigma}^{1/2}\mathbf{P}_{i}\mathbf{\Sigma}^{1/2})\right\}{\tau}_{T}\tr\left\{(\mathbf{\Sigma}^{1/2}\mathbf{P}_{i}\mathbf{\Sigma}^{1/2} )\circ (\mathbf{\Sigma}^{1/2}\mathbf{P}_{i}\mathbf{\Sigma}^{1/2})\right\} \\
&+48\tr\left(\mathbf{\Sigma}^{1/2}\mathbf{P}_{i}\mathbf{\Sigma}^{1/2}\right) \tr\left\{(\mathbf{\Sigma}^{1/2}\mathbf{P}_{i}\mathbf{\Sigma}^{1/2} ) \circ\left(\mathbf{\Sigma}^{1/2}\mathbf{P}_{i}\mathbf{\Sigma}^{1/2} \mathbf{\Sigma}^{1/2}\mathbf{P}_{i}\mathbf{\Sigma}^{1/2}\right)\right\}\\
&+96\tr\left[\left\{\mathbf{I}_{T} \circ (\mathbf{\Sigma}^{1/2}\mathbf{P}_{i}\mathbf{\Sigma}^{1/2})\right\} \mathbf{\Sigma}^{1/2}\mathbf{P}_{i}\mathbf{\Sigma}^{1/2} \mathbf{\Sigma}^{1/2}\mathbf{P}_{i}\mathbf{\Sigma}^{1/2} \mathbf{\Sigma}^{1/2}\mathbf{P}_{i}\mathbf{\Sigma}^{1/2}\right]\\
&+48{\tau}_{T}^{\prime}\left\{\mathbf{I}_{T} \circ\left(\mathbf{\Sigma}^{1/2}\mathbf{P}_{i}\mathbf{\Sigma}^{1/2} \mathbf{\Sigma}^{1/2}\mathbf{P}_{i}\mathbf{\Sigma}^{1/2}\right)\right\}\left\{\mathbf{I}_{T}  \circ\left(\mathbf{\Sigma}^{1/2}\mathbf{P}_{i}\mathbf{\Sigma}^{1/2} \mathbf{\Sigma}^{1/2}\mathbf{P}_{i}\mathbf{\Sigma}^{1/2}\right)\right\} {\tau}_{T}\\
\leq&6 \tr\left(\mathbf{\Sigma}^{1/2}\mathbf{P}_{i}\mathbf{\Sigma}^{1/2}\right) \tr\left(\mathbf{\Sigma}^{1/2}\mathbf{P}_{i}\mathbf{\Sigma}^{1/2}\right) \tr\left\{\left(\mathbf{\Sigma}^{1/2}\mathbf{P}_{i}\mathbf{\Sigma}^{1/2} \right)^{2}\right\}+12\tr^{2}\left\{\left(\mathbf{\Sigma}^{1/2}\mathbf{P}_{i}\mathbf{\Sigma}^{1/2}\right)^{2}\right\}\\
&+48\tr\left(\mathbf{\Sigma}^{1/2}\mathbf{P}_{i}\mathbf{\Sigma}^{1/2}\right)
\sqrt{\tr\left\{\left(\mathbf{\Sigma}^{1/2}\mathbf{P}_{i}\mathbf{\Sigma}^{1/2}\right)^{2}\right\}
\tr\left\{\left(\mathbf{\Sigma}^{1/2}\mathbf{P}_{i}\mathbf{\Sigma}^{1/2}\right)^{4}\right\}}\\
&+96C\tr\left\{\left(\mathbf{\Sigma}^{1/2}\mathbf{P}_{i}\mathbf{\Sigma}^{1/2}\right)^{3}\right\}
+48\tr^{2}\left(\mathbf{\Sigma}^{1/2}\mathbf{P}_{i}\mathbf{\Sigma}^{1/2} \mathbf{\Sigma}^{1/2}\mathbf{P}_{i}\mathbf{\Sigma}^{1/2}\right)\\
=&O(T^3),\\
f_{\gamma_{4}}=& 4\tr\left(\mathbf{\Sigma}^{1/2}\mathbf{P}_{i}\mathbf{\Sigma}^{1/2}\right) \tr\left\{(\mathbf{\Sigma}^{1/2}\mathbf{P}_{i}\mathbf{\Sigma}^{1/2}) \circ (\mathbf{\Sigma}^{1/2}\mathbf{P}_{i}\mathbf{\Sigma}^{1/2}) \circ (\mathbf{\Sigma}^{1/2}\mathbf{P}_{i}\mathbf{\Sigma}^{1/2})\right\} \\
&+24\tr\left\{(\mathbf{\Sigma}^{1/2}\mathbf{P}_{i}\mathbf{\Sigma}^{1/2}) \circ (\mathbf{\Sigma}^{1/2}\mathbf{P}_{i}\mathbf{\Sigma}^{1/2} ) \circ\left(\mathbf{\Sigma}^{1/2}\mathbf{P}_{i}\mathbf{\Sigma}^{1/2} \mathbf{\Sigma}^{1/2}\mathbf{P}_{i}\mathbf{\Sigma}^{1/2}\right)\right\}=O(T^2),\\
f_{\gamma_{6}}=& \tr\left\{(\mathbf{\Sigma}^{1/2}\mathbf{P}_{i}\mathbf{\Sigma}^{1/2}) \circ (\mathbf{\Sigma}^{1/2}\mathbf{P}_{i}\mathbf{\Sigma}^{1/2}) \circ (\mathbf{\Sigma}^{1/2}\mathbf{P}_{i}\mathbf{\Sigma}^{1/2}) \circ (\mathbf{\Sigma}^{1/2}\mathbf{P}_{i}\mathbf{\Sigma}^{1/2})\right\}=O(T),\\
f_{\gamma_{1}^{2}}=& 24{\tau}_{T}^{\prime}\left\{\mathbf{I}_{T} \circ (\mathbf{\Sigma}^{1/2}\mathbf{P}_{i}\mathbf{\Sigma}^{1/2})\right\} \mathbf{\Sigma}^{1/2}\mathbf{P}_{i}\mathbf{\Sigma}^{1/2}\left\{\mathbf{I}_{T} \circ (\mathbf{\Sigma}^{1/2}\mathbf{P}_{i}\mathbf{\Sigma}^{1/2})\right\}{\tau}_{T} \tr\left(\mathbf{\Sigma}^{1/2}\mathbf{P}_{i}\mathbf{\Sigma}^{1/2}\right)\\
&+48{\tau}_{T}^{\prime}\left\{\mathbf{I}_{T} \circ (\mathbf{\Sigma}^{1/2}\mathbf{P}_{i}\mathbf{\Sigma}^{1/2})\right\} \mathbf{\Sigma}^{1/2}\mathbf{P}_{i}\mathbf{\Sigma}^{1/2} \mathbf{\Sigma}^{1/2}\mathbf{P}_{i}\mathbf{\Sigma}^{1/2}\left\{\mathbf{I}_{T} \circ (\mathbf{\Sigma}^{1/2}\mathbf{P}_{i}\mathbf{\Sigma}^{1/2})\right\}\tau_{T}\\
&+16{\tau}_{T}^{\prime}\left\{(\mathbf{\Sigma}^{1/2}\mathbf{P}_{i}\mathbf{\Sigma}^{1/2} )\circ (\mathbf{\Sigma}^{1/2}\mathbf{P}_{i}\mathbf{\Sigma}^{1/2}) \circ (\mathbf{\Sigma}^{1/2}\mathbf{P}_{i}\mathbf{\Sigma}^{1/2})\right\}{\tau}_{T} \tr\left(\mathbf{\Sigma}^{1/2}\mathbf{P}_{i}\mathbf{\Sigma}^{1/2}\right)\\
&+96{\tau}_{T}^{\prime}\left\{(\mathbf{\Sigma}^{1/2}\mathbf{P}_{i}\mathbf{\Sigma}^{1/2} )\circ (\mathbf{\Sigma}^{1/2}\mathbf{P}_{i}\mathbf{\Sigma}^{1/2})\right\} \mathbf{\Sigma}^{1/2}\mathbf{P}_{i}\mathbf{\Sigma}^{1/2}\left\{\mathbf{I}_{T} \circ (\mathbf{\Sigma}^{1/2}\mathbf{P}_{i}\mathbf{\Sigma}^{1/2})\right\} \tau_{T}\\
&+96\tr\left[\mathbf{\Sigma}^{1/2}\mathbf{P}_{i}\mathbf{\Sigma}^{1/2}\left\{(\mathbf{\Sigma}^{1/2}\mathbf{P}_{i}\mathbf{\Sigma}^{1/2} )\circ (\mathbf{\Sigma}^{1/2}\mathbf{P}_{i}\mathbf{\Sigma}^{1/2})\right\} \mathbf{\Sigma}^{1/2}\mathbf{P}_{i}\mathbf{\Sigma}^{1/2}\right]\\
\leq&24C^2\sum_{s=1}^{T}\sum_{t=1}^{T}|\left(\mathbf{\Sigma}^{1/2}
\mathbf{P}_{i}\mathbf{\Sigma}^{1/2}\right)_{st}|
\tr\left(\mathbf{\Sigma}^{1/2}\mathbf{P}_{i}\mathbf{\Sigma}^{1/2}\right)\\
&+48C^{2}\sum_{s=1}^{T}\sum_{t=1}^{T}|(\mathbf{\Sigma}^{1/2}\mathbf{P}_{i}\mathbf{\Sigma}^{1/2} \mathbf{\Sigma}^{1/2}\mathbf{P}_{i}\mathbf{\Sigma}^{1/2})|_{st}\\
&+16\sum_{s=1}^{T}\sum_{t=1}^{T}|(\mathbf{\Sigma}^{1/2}\mathbf{P}_{i}\mathbf{\Sigma}^{1/2} \mathbf{\Sigma}^{1/2}\mathbf{P}_{i}\mathbf{\Sigma}^{1/2})|_{st}^{3} \tr\left(\mathbf{\Sigma}^{1/2}\mathbf{P}_{i}\mathbf{\Sigma}^{1/2}\right)\\
&+96\sum_{s=1}^{T}\sum_{t=1}^{T}\sum_{k=1}^{T}\left(\mathbf{\Sigma}^{1/2}
\mathbf{P}_{i}\mathbf{\Sigma}^{1/2}\right)^{2}_{sk}\left(\mathbf{\Sigma}^{1/2}
\mathbf{P}_{i}\mathbf{\Sigma}^{1/2}\right)_{kt}\left(\mathbf{\Sigma}^{1/2}
\mathbf{P}_{i}\mathbf{\Sigma}^{1/2}\right)_{tt}\\
&+96\sum_{s=1}^{T}\sum_{t=1}^{T}
\left(\mathbf{\Sigma}^{1/2}\mathbf{P}_{i}\mathbf{\Sigma}^{1/2} \right)^{2}_{st}\left(\mathbf{\Sigma}^{1/2}\mathbf{P}_{i}\mathbf{\Sigma}^{1/2}\mathbf{\Sigma}^{1/2}\mathbf{P}_{i}\mathbf{\Sigma}^{1/2}  \right)_{ts}=O(T^3),\\
f_{\gamma_{2}^{2}}=& 3\tr\left\{(\mathbf{\Sigma}^{1/2}\mathbf{P}_{i}\mathbf{\Sigma}^{1/2} )\circ (\mathbf{\Sigma}^{1/2}\mathbf{P}_{i}\mathbf{\Sigma}^{1/2})\right\} \tr\left\{(\mathbf{\Sigma}^{1/2}\mathbf{P}_{i}\mathbf{\Sigma}^{1/2}) \circ (\mathbf{\Sigma}^{1/2}\mathbf{P}_{i}\mathbf{\Sigma}^{1/2})\right\}\\
&+24{\tau}_{T}^{\prime}\left\{\mathbf{I}_{T} \circ (\mathbf{\Sigma}^{1/2}\mathbf{P}_{i}\mathbf{\Sigma}^{1/2})\right\}\left\{(\mathbf{\Sigma}^{1/2}\mathbf{P}_{i}\mathbf{\Sigma}^{1/2} )\circ (\mathbf{\Sigma}^{1/2}\mathbf{P}_{i}\mathbf{\Sigma}^{1/2})\right\}\left\{\mathbf{I}_{T} \circ (\mathbf{\Sigma}^{1/2}\mathbf{P}_{i}\mathbf{\Sigma}^{1/2})\right\}{\tau}_{T}\\
&+8{\tau}_{T}^{\prime}\left\{(\mathbf{\Sigma}^{1/2}\mathbf{P}_{i}\mathbf{\Sigma}^{1/2} )\circ (\mathbf{\Sigma}^{1/2}\mathbf{P}_{i}\mathbf{\Sigma}^{1/2}) \circ (\mathbf{\Sigma}^{1/2}\mathbf{P}_{i}\mathbf{\Sigma}^{1/2}) \circ (\mathbf{\Sigma}^{1/2}\mathbf{P}_{i}\mathbf{\Sigma}^{1/2})\right\} {\tau}_{T}\\
\leq&3\tr^2\left(\mathbf{\Sigma}^{1/2}\mathbf{P}_{i}\mathbf{\Sigma}^{1/2} \mathbf{\Sigma}^{1/2}\mathbf{P}_{i}\mathbf{\Sigma}^{1/2}\right)\\
&+24\sum_{s=1}^{T}\sum_{t=1}^{T}\left(\mathbf{\Sigma}^{1/2}\mathbf{P}_{i}
\mathbf{\Sigma}^{1/2}\right)_{st}^{2}\left(\mathbf{\Sigma}^{1/2}\mathbf{P}_{i}
\mathbf{\Sigma}^{1/2}\right)_{ss}\left(\mathbf{\Sigma}^{1/2}\mathbf{P}_{i}
\mathbf{\Sigma}^{1/2}\right)_{tt}\\
&+8\sum_{s=1}^{T}\sum_{t=1}^{T}\left(\mathbf{\Sigma}^{1/2}\mathbf{P}_{i}
\mathbf{\Sigma}^{1/2}\right)_{st}^{4}\\
=&O(T^2),\\
f_{\gamma_{1} \gamma_{3}}=& 24{\tau}_{T}^{\prime}\left\{\mathbf{I}_{T} \circ (\mathbf{\Sigma}^{1/2}\mathbf{P}_{i}\mathbf{\Sigma}^{1/2})\right\} \mathbf{\Sigma}^{1/2}\mathbf{P}_{i}\mathbf{\Sigma}^{1/2}\left\{\mathbf{I}_{T} \circ (\mathbf{\Sigma}^{1/2}\mathbf{P}_{i}\mathbf{\Sigma}^{1/2}) \circ (\mathbf{\Sigma}^{1/2}\mathbf{P}_{i}\mathbf{\Sigma}^{1/2})\right\}{\tau}_{T}\\
&+32{\tau}_{T}^{\prime}\left\{\mathbf{I}_{T} \circ (\mathbf{\Sigma}^{1/2}\mathbf{P}_{i}\mathbf{\Sigma}^{1/2})\right\}\left\{(\mathbf{\Sigma}^{1/2}\mathbf{P}_{i}\mathbf{\Sigma}^{1/2} )\circ (\mathbf{\Sigma}^{1/2}\mathbf{P}_{i}\mathbf{\Sigma}^{1/2}) \circ ( \mathbf{\Sigma}^{1/2}\mathbf{P}_{i}\mathbf{\Sigma}^{1/2})\right\}{\tau}_{T}\\
\leq&24\sum_{s=1}^{T}\sum_{t=1}^{T}
\left(\mathbf{\Sigma}^{1/2}\mathbf{P}_{i}\mathbf{\Sigma}^{1/2}\right)_{st}
\left(\mathbf{\Sigma}^{1/2}\mathbf{P}_{i}\mathbf{\Sigma}^{1/2}\right)_{ss}
\left(\mathbf{\Sigma}^{1/2}\mathbf{P}_{i}\mathbf{\Sigma}^{1/2}\right)_{tt}^{2}\\
&+32\sum_{s=1}^{T}\sum_{t=1}^{T}\left(\mathbf{\Sigma}^{1/2}\mathbf{P}_{i}\mathbf{\Sigma}^{1/2}\right)_{st}^{3}
\left(\mathbf{\Sigma}^{1/2}\mathbf{P}_{i}\mathbf{\Sigma}^{1/2}\right)_{ss}\\
=&O(T^2),
\end{align*}
Then, by Lemma \ref{le:moment of quadratic form},
\begin{align*}
&\mE^2\Big\{\sum_{i=1}^{N} \sum_{j=1}^{N} (\hat{\epsilon}_{i\cdot}^{\prime}\hat{\epsilon}_{j\cdot})^{2}\Big\}\\
=&
\left[\sum_{1\leq i\neq j\leq N}\tr(\mathbf{M}_{i}\mathbf{M}_{j}) +\sum_{i=1}^{N}2\tr(\mathbf{M}^2_{i})+\tr^2(\mathbf{M}_{i})+\gamma_{2}\tr\{(\mathbf{\Sigma}^{1/2}\mathbf{P}_{i}\mathbf{\Sigma}^{1/2}
)\circ(\mathbf{\Sigma}^{1/2}\mathbf{P}_{i}\mathbf{\Sigma}^{1/2})\}\right]^2\\
=&\sum_{i=1}^{N}\Big[4\tr^2(\mathbf{M}^2_{i})+\tr^4(\mathbf{M}_{i})+4\tr(\mathbf{M}^2_{i})\tr^2(\mathbf{M}_{i})
+\gamma_{2}^2\tr^2\{(\mathbf{\Sigma}^{1/2}\mathbf{P}_{i}\mathbf{\Sigma}^{1/2}
)\circ(\mathbf{\Sigma}^{1/2}\mathbf{P}_{i}\mathbf{\Sigma}^{1/2})\}\\
&+4\gamma_{2}\tr(\mathbf{M}^2_{i})\tr\{(\mathbf{\Sigma}^{1/2}\mathbf{P}_{i}\mathbf{\Sigma}^{1/2}
)\circ(\mathbf{\Sigma}^{1/2}\mathbf{P}_{i}\mathbf{\Sigma}^{1/2})\}+2\gamma_{2}\tr^2(\mathbf{M}_{i})\tr\{(\mathbf{\Sigma}^{1/2}\mathbf{P}_{i}\mathbf{\Sigma}^{1/2}
)\circ(\mathbf{\Sigma}^{1/2}\mathbf{P}_{i}\mathbf{\Sigma}^{1/2})\}\Big]\\
+&\underset{i\neq j}{\sum^{N}\sum^{N}}\Big[4\tr(\mathbf{M}^2_{i})\tr(\mathbf{M}^2_{j})+2\tr(\mathbf{M}^2_{i})\tr^2(\mathbf{M}_{j})
+2\tr(\mathbf{M}^2_{j})\tr^2(\mathbf{M}_{i})+\tr^2(\mathbf{M}_{i})\tr^2(\mathbf{M}_{j})\\
&+\gamma_{2}^2\tr\{(\mathbf{\Sigma}^{1/2}\mathbf{P}_{i}\mathbf{\Sigma}^{1/2}
)\circ(\mathbf{\Sigma}^{1/2}\mathbf{P}_{i}\mathbf{\Sigma}^{1/2})\}\tr\{(\mathbf{\Sigma}^{1/2}\mathbf{P}_{j}\mathbf{\Sigma}^{1/2})
\circ(\mathbf{\Sigma}^{1/2}\mathbf{P}_{j}\mathbf{\Sigma}^{1/2})\}\\
&+2\gamma_{2}\tr(\mathbf{M}^2_{i})\tr\{(\mathbf{\Sigma}^{1/2}\mathbf{P}_{j}\mathbf{\Sigma}^{1/2})
\circ(\mathbf{\Sigma}^{1/2}\mathbf{P}_{j}\mathbf{\Sigma}^{1/2})\}
+2\gamma_{2}\tr\{(\mathbf{M}^2_{j})\tr(\mathbf{\Sigma}^{1/2}\mathbf{P}_{i}\mathbf{\Sigma}^{1/2})
\circ(\mathbf{\Sigma}^{1/2}\mathbf{P}_{i}\mathbf{\Sigma}^{1/2})\}\\
&+\gamma_{2}\tr^2\{(\mathbf{M}_{i})\tr(\mathbf{\Sigma}^{1/2}\mathbf{P}_{j}\mathbf{\Sigma}^{1/2}
)\circ(\mathbf{\Sigma}^{1/2}\mathbf{P}_{j}\mathbf{\Sigma}^{1/2})\}+\gamma_{2}\tr^2(\mathbf{M}_{j})\tr\{(\mathbf{\Sigma}^{1/2}\mathbf{P}_{i}\mathbf{\Sigma}^{1/2})
\circ(\mathbf{\Sigma}^{1/2}\mathbf{P}_{i}\mathbf{\Sigma}^{1/2})\}\Big]\\
&+\underset{i\neq j}{\sum^{N}\sum^{N}}\Big[2\tr^2(\mathbf{M}_{i}\mathbf{M}_{j})+8\tr(\mathbf{M}^2_{i})\tr(\mathbf{M}_{i}\mathbf{M}_{j})
+4\tr(\mathbf{M}_{i}\mathbf{M}_{j})\tr^2(\mathbf{M}_{i})\\
&+4\gamma_{2}\tr(\mathbf{M}_{i}\mathbf{M}_{j})\tr\{(\mathbf{\Sigma}^{1/2}\mathbf{P}_{i}\mathbf{\Sigma}^{1/2}
)\circ(\mathbf{\Sigma}^{1/2}\mathbf{P}_{i}\mathbf{\Sigma}^{1/2})\}\Big]\\
+&\underset{i\neq j\neq k}{\sum^{N}\sum^{N}\sum^{N}}\Big[4\tr(\mathbf{M}_{i}\mathbf{M}_{j})\tr(\mathbf{M}_{i}\mathbf{M}_{k})
+4\tr(\mathbf{M}_{i}\mathbf{M}_{j})\tr(\mathbf{M}^2_{k})+2\tr(\mathbf{M}_{i}\mathbf{M}_{j})\tr^2(\mathbf{M}_{k})\\
&+2\gamma_{2}\tr(\mathbf{M}_{i}\mathbf{M}_{j})\tr\{(\mathbf{\Sigma}^{1/2}\mathbf{P}_{k}\mathbf{\Sigma}^{1/2}
)\circ(\mathbf{\Sigma}^{1/2}\mathbf{P}_{k}\mathbf{\Sigma}^{1/2})\}\Big]\\
&+\underset{i\neq j\neq k\neq l}{\sum^{N}\sum^{N}\sum^{N}\sum^{N}}\left\{\tr(\mathbf{M}_{i}\mathbf{M}_{j})\tr(\mathbf{M}_{k}\mathbf{M}_{l})\right\}.
\end{align*}
So, we have
\begin{align*}
\var\left\{\sum_{i=1}^{N} \sum_{j=1}^{N} (\hat{\epsilon}_{i\cdot}^{\prime}\hat{\epsilon}_{j\cdot})^{2}\right\}
=&O\left(NT^3+N^2T^2+N^2T+N^3T\right).
\end{align*}
Similarly, we can obtain that
\begin{align*}
&\mE\Big(\sum_{i=1}^{N}\hat{\epsilon}_{i\cdot}^{\prime}\hat{\epsilon}_{i\cdot}\Big)^{4}\\
=&\mE\Big(\sum_{i=1}^{N}\sum_{j=1}^{N}\sum_{k=1}^{N}\sum_{l=1}^{N}\hat{\epsilon}_{i\cdot}^{\prime}\hat{\epsilon}_{i\cdot}
\hat{\epsilon}_{j\cdot}^{\prime}\hat{\epsilon}_{j\cdot}\hat{\epsilon}_{k\cdot}^{\prime}\hat{\epsilon}_{k\cdot}
\hat{\epsilon}_{l\cdot}^{\prime}\hat{\epsilon}_{l\cdot}\Big)\\
=&\sum_{i=1}^{N}\mE\left(\hat{\epsilon}_{i\cdot}^{\prime}\hat{\epsilon}_{i\cdot}\right)^{4}
+3\underset{i \neq j}{\sum^{N}\sum^{N}}\mE(\hat{\epsilon}_{i\cdot}^{\prime}\hat{\epsilon}_{i\cdot})^2(\hat{\epsilon}_{j\cdot}^{\prime}\hat{\epsilon}_{j\cdot})^2
+4\underset{i \neq j}{\sum^{N}\sum^{N}}\mE\{(\hat{\epsilon}_{i\cdot}^{\prime}\hat{\epsilon}_{i\cdot})^3\hat{\epsilon}_{j\cdot}^{\prime}\hat{\epsilon}_{j\cdot}\}\\
&+6\underset{i \neq j\neq k}{\sum^{N}\sum^{N}\sum^{N}}\mE\{(\hat{\epsilon}_{i\cdot}^{\prime}\hat{\epsilon}_{i\cdot})^2
\hat{\epsilon}_{j\cdot}^{\prime}\hat{\epsilon}_{j\cdot}\hat{\epsilon}_{k\cdot}^{\prime}\hat{\epsilon}_{k\cdot}\}
+\underset{i \neq j\neq k\neq l}{\sum^{N}\sum^{N}\sum^{N}\sum^{N}}
\mE(\hat{\epsilon}_{i\cdot}^{\prime}\hat{\epsilon}_{i\cdot}\hat{\epsilon}_{j\cdot}^{\prime}\hat{\epsilon}_{j\cdot}
\hat{\epsilon}_{k\cdot}^{\prime}\hat{\epsilon}_{k\cdot}\hat{\epsilon}_{l\cdot}^{\prime}\hat{\epsilon}_{l\cdot})\\
=&\sum_{i=1}^{N}\Big\{48\tr(\mathbf{M}^4_{i})+32\tr(\mathbf{M}^3_{i})\tr(\mathbf{M}_{i})+12\tr^2(\mathbf{M}^2_{i})
+12\tr(\mathbf{M}^2_{i})\tr^2(\mathbf{M}_{i})
+\tr^4(\mathbf{M}_{i})\\
&+\gamma_{2} f_{\gamma_{2}}+\gamma_{4} f_{\gamma_{4}}+\gamma_{6} f_{\gamma_{6}}+\gamma_{1}^{2} f_{\gamma_{1}^{2}}+\gamma_{2}^{2} f_{\gamma_{2}^{2}}+\gamma_{1} \gamma_{3} f_{\gamma_{1} \gamma_{3}}\Big\}\\
&+3\underset{i \neq j}{\sum^{N}\sum^{N}}\Big(\left[2\tr(\mathbf{M}^2_{i})+\tr^2(\mathbf{M}_{i})
+\gamma_{2}\tr\{(\mathbf{\Sigma}^{1/2}\mathbf{P}_{i}\mathbf{\Sigma}^{1/2})\circ(
\mathbf{\Sigma}^{1/2}\mathbf{P}_{i}\mathbf{\Sigma}^{1/2})\}\right]\\
&\,\,\,\,\,\,\,\,\,\,\,\,\times
\left[2\tr(\mathbf{M}^2_{j})+\tr^2(\mathbf{M}_{j})
+\gamma_{2}\tr\{(\mathbf{\Sigma}^{1/2}\mathbf{P}_{j}\mathbf{\Sigma}^{1/2})\circ
(\mathbf{\Sigma}^{1/2}\mathbf{P}_{j}\mathbf{\Sigma}^{1/2})\}\right]\Big)
\\&+4\underset{i\neq j}{\sum^{N}\sum^{N}}\Big(
\gamma_{4} \tr\left\{(\mathbf{\Sigma}^{1/2}\mathbf{P}_{i}\mathbf{\Sigma}^{1/2} )\circ( \mathbf{\Sigma}^{1/2}\mathbf{P}_{i}\mathbf{\Sigma}^{1/2} )\circ (\mathbf{\Sigma}^{1/2}\mathbf{P}_{i}\mathbf{\Sigma}^{1/2})\right\}\\
&+3\gamma_{2} \tr\left(\mathbf{\Sigma}^{1/2}\mathbf{P}_{i}\mathbf{\Sigma}^{1/2}\right) \tr\left\{(\mathbf{\Sigma}^{1/2}\mathbf{P}_{i}\mathbf{\Sigma}^{1/2}) \circ( \mathbf{\Sigma}^{1/2}\mathbf{P}_{i}\mathbf{\Sigma}^{1/2})\right\} \\
&+12\gamma_{2} \tr\left\{(\mathbf{\Sigma}^{1/2}\mathbf{P}_{i}\mathbf{\Sigma}^{1/2} )\circ\left(\mathbf{\Sigma}^{1/2}\mathbf{P}_{i}\mathbf{\Sigma}^{1/2} \mathbf{\Sigma}^{1/2}\mathbf{P}_{i}\mathbf{\Sigma}^{1/2}\right)\right\} \\
&+6 \gamma_{1}^{2}\left[{\tau}_{T}^{\prime}\left\{\mathbf{I}_{T} \circ (\mathbf{\Sigma}^{1/2}\mathbf{P}_{i}\mathbf{\Sigma}^{1/2})\right\} \mathbf{\Sigma}^{1/2}\mathbf{P}_{i}\mathbf{\Sigma}^{1/2}\left\{(\mathbf{I}_{T} )\circ (\mathbf{\Sigma}^{1/2}\mathbf{P}_{i}\mathbf{\Sigma}^{1/2})\right\} {\tau}_{T}\right]\\
& +4 \gamma_{1}^{2}\left[{\tau}_{T}^{\prime}\left\{(\mathbf{\Sigma}^{1/2}\mathbf{P}_{i}\mathbf{\Sigma}^{1/2} )\circ( \mathbf{\Sigma}^{1/2}\mathbf{P}_{i}\mathbf{\Sigma}^{1/2}) \circ ( \mathbf{\Sigma}^{1/2}\mathbf{P}_{i}\mathbf{\Sigma}^{1/2})\right\} {\tau}_{T}\right]\\
&+\tr^3\left(\mathbf{\Sigma}^{1/2}\mathbf{P}_{i}\mathbf{\Sigma}^{1/2}\right)+6 \tr\left(\mathbf{\Sigma}^{1/2}\mathbf{P}_{i}\mathbf{\Sigma}^{1/2}\right) \tr\left(\mathbf{\Sigma}^{1/2}\mathbf{P}_{i}\mathbf{\Sigma}^{1/2} \mathbf{\Sigma}^{1/2}\mathbf{P}_{i}\mathbf{\Sigma}^{1/2}\right)\\
&+8 \tr\left(\mathbf{\Sigma}^{1/2}\mathbf{P}_{i}\mathbf{\Sigma}^{1/2} \mathbf{\Sigma}^{1/2}\mathbf{P}_{i}\mathbf{\Sigma}^{1/2} \mathbf{\Sigma}^{1/2}\mathbf{P}_{i}\mathbf{\Sigma}^{1/2}\right)\Big)\tr(\mathbf{M}_{j})\\
&+6\underset{i \neq j\neq k}{\sum^{N}\sum^{N}\sum^{N}}[2\tr(\mathbf{M}^2_{i})+\tr^2(\mathbf{M}_{i})
+\gamma_{2}\tr\{(\mathbf{\Sigma}^{1/2}\mathbf{P}_{i}\mathbf{\Sigma}^{1/2})\circ
(\mathbf{\Sigma}^{1/2}\mathbf{P}_{i}\mathbf{\Sigma}^{1/2})\}]
\tr(\mathbf{M}_{j})\tr(\mathbf{M}_{k})\\&+\underset{i \neq j\neq k\neq l}{\sum^{N}\sum^{N}\sum^{N}\sum^{N}}\tr(\mathbf{M}_{i})\tr(\mathbf{M}_{j})
\tr(\mathbf{M}_{k})\tr(\mathbf{M}_{l}),
\end{align*}
where $f_{\gamma_{2}},\,f_{\gamma_{4}},\, f_{\gamma_{6}},\, f_{\gamma_{1}^{2}},\, f_{\gamma_{2}^{2}},\,f_{\gamma_{1} \gamma_{3}}$ are the same as above,
and
\begin{align*}
&\Big\{\mE\Big(\sum_{i=1}^{N}\hat{\epsilon}_{i\cdot}^{\prime}\hat{\epsilon}_{i\cdot}\Big)^{2}\Big\}^2\\
=&\Big[\sum_{i=1}^{N}2\tr(\mathbf{M}^2_{i})+\tr^2(\mathbf{M}_{i})+\gamma_{2}\tr\{(\mathbf{\Sigma}^{1/2}\mathbf{P}_{i}\mathbf{\Sigma}^{1/2}
)\circ(\mathbf{\Sigma}^{1/2}\mathbf{P}_{i}\mathbf{\Sigma}^{1/2})\}+\underset{i\neq j}{\sum^{N}\sum^{N}}\tr(\mathbf{M}_{i})\tr(\mathbf{M}_{j})\Big]^2\\
=&\sum_{i=1}^{N}\Big[4\tr^2(\mathbf{M}^2_{i})+\tr^4(\mathbf{M}_{i})+4\tr(\mathbf{M}^2_{i})\tr^2(\mathbf{M}_{i})
+\gamma_{2}^2\tr^2\{(\mathbf{\Sigma}^{1/2}\mathbf{P}_{i}\mathbf{\Sigma}^{1/2}
)\circ(\mathbf{\Sigma}^{1/2}\mathbf{P}_{i}\mathbf{\Sigma}^{1/2})\}\\
&+4\gamma_{2}\tr(\mathbf{M}^2_{i})\tr\{(\mathbf{\Sigma}^{1/2}\mathbf{P}_{i}\mathbf{\Sigma}^{1/2}
)\circ(\mathbf{\Sigma}^{1/2}\mathbf{P}_{i}\mathbf{\Sigma}^{1/2})\}+2\gamma_{2}\tr^2(\mathbf{M}_{i})
\tr\{(\mathbf{\Sigma}^{1/2}\mathbf{P}_{i}\mathbf{\Sigma}^{1/2})
\circ(\mathbf{\Sigma}^{1/2}\mathbf{P}_{i}\mathbf{\Sigma}^{1/2})\}\Big]\\
+&\underset{i\neq j}{\sum^{N}\sum^{N}}\Big[4\tr(\mathbf{M}^2_{i})\tr(\mathbf{M}^2_{j})+2\tr(\mathbf{M}^2_{i})\tr^2(\mathbf{M}_{j})
+2\tr(\mathbf{M}^2_{j})\tr^2(\mathbf{M}_{i})+\tr^2(\mathbf{M}_{i})\tr^2(\mathbf{M}_{j})\\
&+\gamma_{2}^2\tr\{(\mathbf{\Sigma}^{1/2}\mathbf{P}_{i}\mathbf{\Sigma}^{1/2}
)\circ(\mathbf{\Sigma}^{1/2}\mathbf{P}_{i}\mathbf{\Sigma}^{1/2})\}
\tr\{(\mathbf{\Sigma}^{1/2}\mathbf{P}_{j}\mathbf{\Sigma}^{1/2})
\circ(\mathbf{\Sigma}^{1/2}\mathbf{P}_{j}\mathbf{\Sigma}^{1/2})\}\\
&+2\gamma_{2}\tr(\mathbf{M}^2_{i})
\tr\{(\mathbf{\Sigma}^{1/2}\mathbf{P}_{j}\mathbf{\Sigma}^{1/2})
\circ(\mathbf{\Sigma}^{1/2}\mathbf{P}_{j}\mathbf{\Sigma}^{1/2})\}
+2\gamma_{2}\tr(\mathbf{M}^2_{j})\tr\{(\mathbf{\Sigma}^{1/2}\mathbf{P}_{i}\mathbf{\Sigma}^{1/2}
)\circ(\mathbf{\Sigma}^{1/2}\mathbf{P}_{i}\mathbf{\Sigma}^{1/2})\}\\
&+\gamma_{2}\tr^2\{(\mathbf{M}_{i})\tr(\mathbf{\Sigma}^{1/2}\mathbf{P}_{j}\mathbf{\Sigma}^{1/2}
)\circ(\mathbf{\Sigma}^{1/2}\mathbf{P}_{j}\mathbf{\Sigma}^{1/2})\}+\gamma_{2}\tr^2(\mathbf{M}_{j})\tr\{(\mathbf{\Sigma}^{1/2}\mathbf{P}_{i}\mathbf{\Sigma}^{1/2}
)\circ(\mathbf{\Sigma}^{1/2}\mathbf{P}_{i}\mathbf{\Sigma}^{1/2})\}\Big]\\
&+\underset{i\neq j}{\sum^{N}\sum^{N}}\Big[2\tr^2(\mathbf{M}_{i})\tr^2(\mathbf{M}_{j})+8\tr(\mathbf{M}^2_{i})\tr(\mathbf{M}_{i})\tr(\mathbf{M}_{j})
+4\tr^3(\mathbf{M}_{i})\tr(\mathbf{M}_{j})\\
&+4\gamma_{2}\tr(\mathbf{M}_{i})\tr(\mathbf{M}_{j})\tr\{(\mathbf{\Sigma}^{1/2}\mathbf{P}_{i}\mathbf{\Sigma}^{1/2}
)\circ(\mathbf{\Sigma}^{1/2}\mathbf{P}_{i}\mathbf{\Sigma}^{1/2})\}\Big]\\
+&\underset{i\neq j\neq k}{\sum^{N}\sum^{N}\sum^{N}}\Big[4\tr^2(\mathbf{M}_{i})\tr(\mathbf{M}_{j})\tr(\mathbf{M}_{k})
+4\tr(\mathbf{M}_{i})\tr(\mathbf{M}_{j})\tr(\mathbf{M}^2_{k})+2\tr(\mathbf{M}_{i})\tr(\mathbf{M}_{j})\tr^2(\mathbf{M}_{k})\\
&+2\gamma_{2}\tr(\mathbf{M}_{i})\tr(\mathbf{M}_{j})\tr\{(\mathbf{\Sigma}^{1/2}\mathbf{P}_{k}\mathbf{\Sigma}^{1/2}
)\circ(\mathbf{\Sigma}^{1/2}\mathbf{P}_{k}\mathbf{\Sigma}^{1/2})\}\Big]\\
&+\underset{i\neq j\neq k\neq l}{\sum^{N}\sum^{N}\sum^{N}\sum^{N}}\left\{\tr(\mathbf{M}_{i})\tr(\mathbf{M}_{j})\tr(\mathbf{M}_{k})\tr(\mathbf{M}_{l})\right\}.
\end{align*}
So, we have
\begin{align*}
\var\bigg\{\Big(\sum_{i=1}^{N}\hat{\epsilon}_{i\cdot}^{\prime}\hat{\epsilon}_{i\cdot}\Big)^{2}\bigg\}=O(NT^3+N^2T^3+N^3T^3).
\end{align*}
Finally, we have
\begin{align*}
\mE(b^2_{N})\leq&\frac{1}{N^2T^4}\Bigg[2\var\bigg\{\sum_{i=1}^{N} \sum_{j=1}^{N} (\hat{\epsilon}_{i\cdot}^{\prime}\hat{\epsilon}_{j\cdot})^{2}\bigg\}+\frac{2}{T^2}
\var\bigg\{\Big(\sum_{i=1}^{N}\hat{\epsilon}_{i\cdot}^{\prime}\hat{\epsilon}_{i\cdot}\Big)^{2}\bigg\}\Bigg]\\
=&O\left(\frac{1}{NT}\right).
\end{align*}
\hfill$\Box$
\subsection{Proof of Lemma \ref{H1:lemax1}}
Recall that $\mathbf{E}=\mathbf{U}^{1/2}\mathbf{Z}\mathbf{\Sigma}^{1/2},$
then we have $$\hat{\epsilon}_{li}\hat{\epsilon}_{lj}={Z}^{\prime}\left[\{u_{l}^{1/2}(u_{l}^{1/2})^{\prime}\}\otimes \{m_{i}^{l}(m_{j}^{l})^{\prime}\}\right]{Z},$$ where
$Z=(z_{11},\cdots,z_{1T},z_{21},\cdots,z_{2T},\cdots,z_{N1},\cdots,z_{NT})^{\prime}\in
\mathbb{R}^{NT},$
$u_{l}^{1/2}\in \mathbb{R}^{N}$ denotes the $l$-th row vector of matrix $\mathbf{U}^{1/2},$ and $m_{i}^{l} \in \mathbb{R}^{T}$ represents the $i$-th row vector of matrix $\mathbf{P}_{l}\mathbf{\Sigma}^{1/2}.$
Hence, $\sum_{l=1}^{N}\hat{\epsilon}_{li}\hat{\epsilon}_{lj}={Z}^{\prime}\left[\sum_{l=1}^{N}\{u_{l}^{1/2}(u_{l}^{1/2})^{\prime}\}\otimes \{m_{i}^{l}(m_{j}^{l})^{\prime}\}\right]{Z}.$
To simplify notation, we write $\mathbf{W}_{ij}$ as matrix $\sum_{l=1}^{N}\{u_{l}^{1/2}(u_{l}^{1/2})^{\prime}\}\otimes \{m_{i}^{l}(m_{j}^{l})^{\prime}\}.$
 Recall that $||\mathbf{A}||$ denotes the operator norm of matrix $\mathbf{A}.$
So, by the Hanson-Wright inequality in \cite{rudelson2013},
we can obtain that for every $t>0$,
$$
\mathbb{P}\left\{\left|{Z}^{\prime}\mathbf{W}_{ij}{Z}-\mathbb{E} \big({Z}^{\prime}\mathbf{W}_{ij}{Z}\big)\right|>t\right\}
\leq 2 \exp \left[-c \min \left(\frac{t^{2}}{K^{4}\|\mathbf{W}_{ij}\|_{\mathrm{F}}^{2}}, \frac{t}{K^{2}\|\mathbf{W}_{ij}\|}\right)\right].
$$
Obviously,
$\mathbb{E} \big({Z}^{\prime}\mathbf{W}_{ij}{Z}\big)=\tr(\mathbf{W}_{ij})
=\sum_{l=1}^{N}u_{ll}\sigma_{ij,l},$ where $u_{ll}=(\mathbf{U})_{ll},$ for $1\leq l\leq N.$
Recall that we assume that $u_{ll}=1,$
for $1\leq l\leq N.$
According to the properties of the Kronecker product,
we have
$$\tr\{\mathbf{W}_{ij}(\mathbf{W}_{ij})^{\prime}\}
=\underset{l,s=1}{\sum^{N}\sum^{N}}u_{ls}^2 (\mathbf{P}_{l}\mathbf{\Sigma}\mathbf{P}_{s})_{ii}
(\mathbf{P}_{l}\mathbf{\Sigma}\mathbf{P}_{s})_{jj}.$$
Due to $(\mathbf{P}_{l}\mathbf{\Sigma}\mathbf{P}_{s})_{jj}
=\sigma_{jj}-(\mathbf{B}_{l}\mathbf{\Sigma})_{jj}-(\mathbf{B}_{s}\mathbf{\Sigma})_{jj}-(\mathbf{B}_{l}
\mathbf{\Sigma}\mathbf{B}_{s})_{jj}$ and (\ref{sigmaij,l-sigmaij}),
we have $\tr\{\mathbf{W}_{ij}(\mathbf{W}_{ij})^{\prime}\}=\tr(\mathbf{U}^2)\sigma_{ii}\sigma_{jj}\{1+o(1)\}$
 and $\tr\{\mathbf{W}_{ij}\mathbf{W}_{ij}\}=\tr(\mathbf{U}^2)\sigma_{ij}^{2}\{1+o(1)\}.$ Then, we focus on $\|\mathbf{W}_{ij}\|.$
According to the triangle inequality of norms, we have
\begin{align*}
\|\mathbf{W}_{ij}\|&\leq \|\sum_{l=1}^{N}u_{l}^{1/2}(u_{l}^{1/2})^{\prime}\otimes \sigma_{i}^{1/2}(\sigma_{j}^{1/2})^{\prime}\|+\|\sum_{l=1}^{N}u_{l}^{1/2}(u_{l}^{1/2})^{\prime}\otimes \big\{m_{i}^{l}(m_{j}^{l})^{\prime}-\sigma_{i}^{1/2}(\sigma_{j}^{1/2})^{\prime}\big\}\|\\\n
&\leq \sqrt{\lambda_{max}(\mathbf{U}^{2})\sigma_{ii}\sigma_{jj}}+\|\sum_{l=1}^{N}u_{l}^{1/2}(u_{l}^{1/2})^{\prime}\otimes \big\{m_{i}^{l}(m_{j}^{l})^{\prime}-\sigma_{i}^{1/2}(\sigma_{j}^{1/2})^{\prime}\big\}\|,
\end{align*}
where $\sigma_{i}^{1/2}\in \mathbb{R}^{N}$ denotes the $i$-th row vector of matrix $\mathbf{\Sigma}^{1/2}.$
Then, using the definition of spectral norm, we can obtain that for any vector $x \in \mathbb{R}^{NT}$ with norm $\|x\|=1,$
\begin{align*}
&\|\sum_{l=1}^{N}u_{l}^{1/2}(u_{l}^{1/2})^{\prime}\otimes \big\{m_{i}^{l}(m_{j}^{l})^{\prime}-\sigma_{i}^{1/2}(\sigma_{j}^{1/2})^{\prime}\big\}\|^2\\\n
%=&\max_{\|x\|=1}x^{\prime}\Big[\sum_{s=1}^{N}u_{s}^{1/2}(u_{s}^{1/2})^{\prime}\otimes \big\{m_{j}^{s}(m_{i}^{s})^{\prime}-\sigma_{j}^{1/2}(\sigma_{i}^{1/2})^{\prime}\big\}\Big]
%\Big[\sum_{l=1}^{N}u_{l}^{1/2}(u_{l}^{1/2})^{\prime}\otimes \big\{m_{i}^{l}(m_{j}^{l})^{\prime}-\sigma_{i}^{1/2}(\sigma_{j}^{1/2})^{\prime}\big\}\Big]x\\\n
=&\max_{\|x\|=1}x^{\prime}\Big[\sum_{s=1}^{N}\sum_{l=1}^{N}u_{sl}u_{s}^{1/2}(u_{l}^{1/2})^{\prime}\otimes \big\{m_{j}^{s}(m_{i}^{s})^{\prime}-\sigma_{j}^{1/2}(\sigma_{i}^{1/2})^{\prime}\big\}
\big\{m_{i}^{l}(m_{j}^{l})^{\prime}-\sigma_{i}^{1/2}(\sigma_{j}^{1/2})^{\prime}\big\}\Big]x\\\n
\leq&\max_{\|x\|=1}x^{\prime}\Big[\sum_{l=1}^{N}u_{ll}u_{l}^{1/2}(u_{l}^{1/2})^{\prime}\otimes\big\{m_{j}^{l}(m_{i}^{l})^{\prime}-\sigma_{j}^{1/2}(\sigma_{i}^{1/2})^{\prime}\big\}
\big\{m_{i}^{l}(m_{j}^{l})^{\prime}-\sigma_{i}^{1/2}(\sigma_{j}^{1/2})^{\prime}\big\}\Big]x\\\n
&+\max_{\|x\|=1}x^{\prime}\Big[\underset{l\neq s}{\sum^{N}\sum^{N}}u_{ls}u_{l}^{1/2}(u_{s}^{1/2})^{\prime}\otimes\big\{m_{j}^{l}(m_{i}^{l})^{\prime}-\sigma_{j}^{1/2}(\sigma_{i}^{1/2})^{\prime}\big\}
\big\{m_{i}^{s}(m_{j}^{s})^{\prime}-\sigma_{i}^{1/2}(\sigma_{j}^{1/2})^{\prime}\big\}\Big]x\\\n
=&\Delta_{1}+\Delta_{2}.
\end{align*}
Next, we first calculate $\Delta_{1},$
\begin{align*}
\Delta_{1}=&\max_{\|x\|=1}x^{\prime}\Big[\sum_{l=1}^{N}u_{ll}u_{l}^{1/2}(u_{l}^{1/2})^{\prime}\otimes\big\{m_{j}^{l}(m_{i}^{l})^{\prime}-\sigma_{j}^{1/2}(\sigma_{i}^{1/2})^{\prime}\big\}
\big\{m_{i}^{l}(m_{j}^{l})^{\prime}-\sigma_{i}^{1/2}(\sigma_{j}^{1/2})^{\prime}\big\}\Big]x\\\n
\leq & \tr\Big[\sum_{l=1}^{N}u_{ll}u_{l}^{1/2}(u_{l}^{1/2})^{\prime}\otimes\big\{m_{j}^{l}(m_{i}^{l})^{\prime}-\sigma_{j}^{1/2}(\sigma_{i}^{1/2})^{\prime}\big\}
\big\{m_{i}^{l}(m_{j}^{l})^{\prime}-\sigma_{i}^{1/2}(\sigma_{j}^{1/2})^{\prime}\big\}\Big]\\
\leq &\sum_{l=1}^{N}\tr\Big[u_{ll}u_{l}^{1/2}(u_{l}^{1/2})^{\prime}\otimes\big\{m_{j}^{l}(m_{i}^{l})^{\prime}-\sigma_{j}^{1/2}(\sigma_{i}^{1/2})^{\prime}\big\}
\big\{m_{i}^{l}(m_{j}^{l})^{\prime}-\sigma_{i}^{1/2}(\sigma_{j}^{1/2})^{\prime}\big\}\Big]\\
\leq&\sum_{l=1}^{N}u_{ll}^2\tr\Big[\big\{m_{j}^{l}(m_{i}^{l})^{\prime}-\sigma_{j}^{1/2}(\sigma_{i}^{1/2})^{\prime}\big\}
\big\{m_{i}^{l}(m_{j}^{l})^{\prime}-\sigma_{i}^{1/2}(\sigma_{j}^{1/2})^{\prime}\big\}\Big]\\
\leq &\sum_{l=1}^{N}u_{ll}^{2}\Big\{\sigma_{ii,l}\sigma_{jj,l}-2
(\mathbf{P}_{l}\mathbf{\Sigma})_{ii}(\mathbf{P}_{l}\mathbf{\Sigma})_{jj}+\sigma_{ii}\sigma_{jj}\Big\}\\
=&O(T^{0\vee(1-\frac{1}{\tau})})
\end{align*}
where the last inequality holds due to (\ref{sigmaij,l-sigmaij}). Similarly, we have
\begin{align*}
\Delta_{2}=&\max_{\|x\|=1}x^{\prime}\Big[\underset{l\neq s}{\sum^{N}\sum^{N}}u_{ls}u_{l}^{1/2}(u_{s}^{1/2})^{\prime}\otimes\big\{m_{j}^{l}(m_{i}^{l})^{\prime}-\sigma_{j}^{1/2}(\sigma_{i}^{1/2})^{\prime}\big\}
\big\{m_{i}^{s}(m_{j}^{s})^{\prime}-\sigma_{i}^{1/2}(\sigma_{j}^{1/2})^{\prime}\big\}\Big]x\\
\leq &\tr\Big[\underset{l\neq s}{\sum^{N}\sum^{N}}u_{ls}u_{l}^{1/2}(u_{s}^{1/2})^{\prime}\otimes\big\{m_{j}^{l}(m_{i}^{l})^{\prime}-\sigma_{j}^{1/2}(\sigma_{i}^{1/2})^{\prime}\big\}
\big\{m_{i}^{s}(m_{j}^{s})^{\prime}-\sigma_{i}^{1/2}(\sigma_{j}^{1/2})^{\prime}\big\}\Big]\\
\leq&\underset{l\neq s}{\sum^{N}\sum^{N}}\tr\Big[u_{ls}u_{l}^{1/2}(u_{s}^{1/2})^{\prime}\otimes\big\{m_{j}^{l}(m_{i}^{l})^{\prime}-\sigma_{j}^{1/2}(\sigma_{i}^{1/2})^{\prime}\big\}
\big\{m_{i}^{s}(m_{j}^{s})^{\prime}-\sigma_{i}^{1/2}(\sigma_{j}^{1/2})^{\prime}\big\}\Big]\\
\leq&\underset{l\neq s}{\sum^{N}\sum^{N}}u_{ls}^2\tr\Big[\big\{m_{j}^{l}(m_{i}^{l})^{\prime}-\sigma_{j}^{1/2}(\sigma_{i}^{1/2})^{\prime}\big\}
\big\{m_{i}^{s}(m_{j}^{s})^{\prime}-\sigma_{i}^{1/2}(\sigma_{j}^{1/2})^{\prime}\big\}\Big]\\
\leq&\underset{l\neq s}{\sum^{N}\sum^{N}}u_{ls}^2\Big\{(\mathbf{P}_{l}\mathbf{\Sigma}\mathbf{P}_{s})_{ii}
(\mathbf{P}_{l}\mathbf{\Sigma}\mathbf{P}_{s})_{jj}-(\mathbf{P}_{l}\mathbf{\Sigma})_{ii}
(\mathbf{P}_{l}\mathbf{\Sigma})_{jj}-(\mathbf{P}_{s}\mathbf{\Sigma})_{jj}
(\mathbf{P}_{s}\mathbf{\Sigma})_{jj}+\sigma_{ii}\sigma_{jj}\Big\}\\
=&O(T^{0\vee(1-\frac{1}{\tau})}).
\end{align*}
So, we can conclude that $\|\mathbf{W}_{ij}\|=O(T^{0\vee(\frac{1}{2}-\frac{1}{2\tau})}).$
Then, according to Theorem 1.1 in \cite{rudelson2013},
for every $t\in (0,o(N^{\frac{1}{2}\wedge \frac{1}{2\tau}})),$
there exist a constant $c>0,$
\begin{align}\label{H1:ehatlilj}
&\mathbb{P}\left(\left|\sum_{l=1}^{N}\hat{\epsilon}_{li}\hat{\epsilon}_{lj
}-\sum_{l=1}^{N}u_{ll}\sigma_{ij,l}\right|>t\sqrt{\tr(\mathbf{U}^{2})
(\sigma_{ii}\sigma_{jj}+\sigma_{ij}^{2})}\right)\\\n
\leq&2\exp\Big(-c\frac{t^2\tr(\mathbf{U}^{2})(\sigma_{ii}\sigma_{jj}+\sigma_{ij}^{2})}
{K^{4}\tr\{\mathbf{W}_{ij}(\mathbf{W}_{ij})^{\prime}\}}\Big)\\\n
\leq&2\exp\Big(-c^{\prime}\frac{t^2}
{K^{4}}\Big)
\end{align}
where $\tr\{\mathbf{W}_{ij}(\mathbf{W}_{ij})^{\prime}\}=\tr(\mathbf{U}^2)\sigma_{ii}\sigma_{jj}\{1+o(1)\},$
and $c^{\prime}>0$ is a positive constant independent of $N$ and $ T.$

In fact, $\bar{\hat{\epsilon}}_{\cdot j}=\sum_{l=1}^{N}\hat{\epsilon}_{li}/N$ can be seen as a linear combination of $z_{ks},$ $1\leq k,s\leq N,$ where $z_{ks}$ is the element at row $k$ and column $s$ of random matrix $\mathbf{Z}.$ Then, by classical Cram\'{e}r type large deviation results for independent random variables(see Corollary 3.1 in \cite{saulis1991limit}), we have for any $\varepsilon > 0,$
$$\mathbb{P}\left(|\bar{\hat{\epsilon}}_{\cdot j}|>x\sqrt{\var(\bar{\hat{\epsilon}}_{\cdot j})}\right)\leq C \exp\left\{-\frac{x^2}{2}(1-\varepsilon)\right\},$$
uniformly in $x \in[0, o(\sqrt{N})).$
For $\bar{\hat{\epsilon}}_{\cdot j},$ we have
$$\var(\bar{\hat{\epsilon}}_{\cdot j})=\frac{1}{N^2}\sum_{l=1}^{N}u_{ll}\sigma_{ii,l}
+\frac{1}{N^2}\underset{l\neq s}{\sum^{N}\sum^{N}}u_{ls}(\mathbf{P}_{l}\mathbf{\Sigma}\mathbf{P}_{s})_{ii}=O(N^{-1\vee-\frac{1}{\tau}}),$$
uniformly in $1 \leq j \leq N. $
 So, for any $\varepsilon > 0,$
$$
\mathbb{P}\left(\left|\bar{\hat{\epsilon}}_{\cdot j}\bar{\hat{\epsilon}}_{\cdot i}\right| \geq x^{2} \sqrt{\var\left(\bar{\hat{\epsilon}}_{\cdot j}\right) \var\left(\bar{\hat{\epsilon}}_{\cdot i}\right)}\right) \leq 2C \exp\left\{-\frac{x^2}{2}(1-\varepsilon)\right\}
$$
uniformly in $x\in[0, o(\sqrt{N}))$. We have, uniformly for $x \in[0, o(N^{\frac{1}{2} \wedge\left(\frac{1}{\tau}-\frac{1}{2}\right)})),$ $x^{2} \sqrt{\var\left(\bar{\hat{\epsilon}}_{\cdot i}\right) \var\left(\bar{\hat{\epsilon}}_{\cdot j}\right)}=$ $o(x / \sqrt{N})$. So for any
$\varepsilon>0,$ large $N$, and any $\delta>0,$
\begin{align}\label{H1:ehatbarij}
\mathbb{P}\left(\left|\bar{\hat{\epsilon}}_{\cdot i} \bar{\hat{\epsilon}}_{\cdot j}\right| \geq \delta \frac{x}{\sqrt{N}}\right) \leq 2C \exp\left\{-\frac{x^2}{2}(1-\varepsilon)\right\},
\end{align}
uniformly  and $x\in[0, o(N^{\frac{1}{2} \wedge\left(\frac{1}{\tau}-\frac{1}{2}\right)})).$
Then, the lemma follows from (\ref{H1:ehatlilj}) and (\ref{H1:ehatbarij}).\hfill$\Box$
\subsection{Proof of Lemma \ref{H1:lemax2}}
Recall that without loss of generality, we assume that $u_{ll}=1,$
for $1\leq l\leq N.$
Similarly, we have
\begin{align*}
\|\hat{\mathbf{\Gamma}}\|_{\mathrm{F}}^{2}&=\frac{1}{T^2}\sum_{i=1}^{N} \sum_{j=1}^{N} \hat{\epsilon}_{i\cdot}^{\prime}\hat{\epsilon}_{j\cdot}^{2}\\
\{\tr(\hat{\mathbf{\Gamma}})\}^{2}&=\frac{1}{T^2}\left(\sum_{i=1}^{N}\hat{\epsilon}_{i\cdot}^{\prime}\hat{\epsilon}_{i\cdot}\right)^{2}\\
T^2a_{N}&=\frac{1}{\gamma_{N}}\left[\sum_{i=1}^{N} \sum_{j=1}^{N} \mathbb{E} (\hat{\epsilon}_{i\cdot}^{\prime}\hat{\epsilon}_{j\cdot})^{2}-\frac{1}{T} \mathbb{E}\left(\sum_{i=1}^{N} \hat{\epsilon}_{i\cdot}^{\prime}\hat{\epsilon}_{i\cdot}\right)^{2}\right]\\
T^2b_{N}&=\frac{1}{N}\left[\sum_{i=1} \sum_{j=1} (\hat{\epsilon}_{i\cdot}^{\prime}\hat{\epsilon}_{j\cdot})^{2}-\frac{1}{T}\left(\sum_{i=1} \hat{\epsilon}_{i\cdot}^{\prime}\hat{\epsilon}_{i\cdot}\right)^{2}-\sum_{i=1} \sum_{j=1} \mathbb{E} (\hat{\epsilon}_{i\cdot}^{\prime}\hat{\epsilon}_{j\cdot})^{2}+\frac{1}{T} \mathbb{E}\left(\sum_{i=1} \hat{\epsilon}_{i\cdot}^{\prime}\hat{\epsilon}_{i\cdot}\right)^{2}\right]
\end{align*}
it is easy to verify that $a_{N}$ and $b_{N}$ will make the equation (\ref{H1:gamman}) true.

 In the following, we prove that $a_{N}, b_{N}$ satisfy the properties in the lemma.
We first deal with the term $a_{N} .$
Recall that $Z=(z_{11},\cdots,z_{1T},z_{21},\cdots,z_{2T},\cdots,z_{N1},\cdots,z_{NT})^{\prime}$
and $u_{l}^{1/2}\in \mathbb{R}^{N}$ denotes the $l$-th row vector of matrix $\mathbf{U}^{1/2},$
we have
\begin{equation*}
 (\hat{\epsilon}_{i\cdot}^{\prime}\hat{\epsilon}_{j\cdot})={Z}^{\prime}
 \left[\{u_{i}^{1/2}(u_{j}^{1/2})^{\prime}\}\otimes(\mathbf{\Sigma}^{1/2}
 \mathbf{P}_{i}\mathbf{P}_{j}\mathbf{\Sigma}^{1/2})\right]Z.
\end{equation*}
Therefore, due to Lemma \ref{le:moment of quadratic form},
\begin{align*}
\sum_{1 \leq i,j \leq N} \mathbb{E} (\hat{\epsilon}_{i\cdot}^{\prime}\hat{\epsilon}_{j\cdot})^{2} =&\sum_{1\leq i\neq j\leq N}u_{ij}^2\tr\{(\mathbf{\Sigma}^{1/2}
 \mathbf{P}_{i}\mathbf{P}_{j}\mathbf{\Sigma}^{1/2})^2\}+u_{ij}^2\tr^2(\mathbf{\Sigma}^{1/2}
 \mathbf{P}_{i}\mathbf{P}_{j}\mathbf{\Sigma}^{1/2})\\
&+ u_{ii}u_{jj}\tr\{(\mathbf{\Sigma}^{1/2}
 \mathbf{P}_{i}\mathbf{P}_{j}\mathbf{\Sigma}^{1/2})(\mathbf{\Sigma}^{1/2}
 \mathbf{P}_{i}\mathbf{P}_{j}\mathbf{\Sigma}^{1/2})^{\prime}\}\\
 &+\gamma_{2}\tr\left[\{u_{i}^{1/2}(u_{j}^{1/2})^{\prime}\}\circ
 \{u_{i}^{1/2}(u_{j}^{1/2})^{\prime}\}\right]\tr\left\{(\mathbf{\Sigma}^{1/2}\mathbf{P}_{i}
 \mathbf{P}_{j}\mathbf{\Sigma}^{1/2})\circ(\mathbf{\Sigma}^{1/2}\mathbf{P}_{i}
 \mathbf{P}_{j}\mathbf{\Sigma}^{1/2})\right\}\\
 &+\sum_{i=1}^{N}2u_{ii}^{2}\tr\{(\mathbf{\Sigma}^{1/2}
 \mathbf{P}_{i}\mathbf{\Sigma}^{1/2})^{2}\}+u_{ii}^{2}\tr^{2}(\mathbf{\Sigma}^{1/2}
 \mathbf{P}_{i}\mathbf{\Sigma}^{1/2})\\
 &+\gamma_{2}\tr\left[\{u_{i}^{1/2}(u_{i}^{1/2})^{\prime}\}\circ
 \{u_{i}^{1/2}(u_{i}^{1/2})^{\prime}\}\right]\tr\left\{(\mathbf{\Sigma}^{1/2}\mathbf{P}_{i}
\mathbf{\Sigma}^{1/2})\circ(\mathbf{\Sigma}^{1/2}\mathbf{P}_{i}
\mathbf{\Sigma}^{1/2})\right\}.
\end{align*} Moreover,
\begin{align*}
\mathbb{E}\left(\sum_{i=1}^{N} \hat{\epsilon}_{i\cdot}^{\prime}\hat{\epsilon}_{i\cdot}\right)^{2} =&\sum_{i=1}^{N} \sum_{j=1}^{N} \mathbb{E}( \hat{\epsilon}_{i\cdot}^{\prime}\hat{\epsilon}_{i\cdot} \hat{\epsilon}_{j\cdot}^{\prime}\hat{\epsilon}_{j\cdot})\\
=&\sum_{i=1}^{N}\mathbb{E}( \hat{\epsilon}_{i\cdot}^{\prime}\hat{\epsilon}_{i\cdot} \hat{\epsilon}_{i\cdot}^{\prime}\hat{\epsilon}_{i\cdot})+\underset{i\neq j}{\sum^{N}\sum^{N}}\mathbb{E}( \hat{\epsilon}_{i\cdot}^{\prime}\hat{\epsilon}_{i\cdot} \hat{\epsilon}_{j\cdot}^{\prime}\hat{\epsilon}_{j\cdot})\\
=&\sum_{i=1}^{N}2u_{ii}^{2}\tr\{(\mathbf{\Sigma}^{1/2}
 \mathbf{P}_{i}\mathbf{\Sigma}^{1/2})^2\}+u_{ii}^{2}\tr^2(\mathbf{\Sigma}^{1/2}
 \mathbf{P}_{i}\mathbf{\Sigma}^{1/2})\\
 &+\gamma_{2}\tr\left[\{u_{i}^{1/2}(u_{i}^{1/2})^{\prime}\}\circ
 \{u_{i}^{1/2}(u_{i}^{1/2})^{\prime}\}\right]\tr\left\{(\mathbf{\Sigma}^{1/2}\mathbf{P}_{i}
\mathbf{\Sigma}^{1/2})\circ(\mathbf{\Sigma}^{1/2}\mathbf{P}_{i}
\mathbf{\Sigma}^{1/2})\right\}\\
&+\underset{i\neq j}{\sum^{N}\sum^{N}}2u_{ij}^{2}\tr(\mathbf{\Sigma}^{1/2}
 \mathbf{P}_{i}\mathbf{\Sigma}^{1/2}\mathbf{\Sigma}^{1/2}
 \mathbf{P}_{j}\mathbf{\Sigma}^{1/2})+u_{ii}u_{jj}\tr(\mathbf{\Sigma}^{1/2}
 \mathbf{P}_{i}\mathbf{\Sigma}^{1/2})\tr(\mathbf{\Sigma}^{1/2}
 \mathbf{P}_{j}\mathbf{\Sigma}^{1/2})\\
 &+\gamma_{2}\tr\left[\{u_{i}^{1/2}(u_{i}^{1/2})^{\prime}\}\circ
 \{u_{j}^{1/2}(u_{j}^{1/2})^{\prime}\}\right]\tr\left\{(\mathbf{\Sigma}^{1/2}
 \mathbf{P}_{i}\mathbf{\Sigma}^{1/2})\circ(\mathbf{\Sigma}^{1/2}
 \mathbf{P}_{j}\mathbf{\Sigma}^{1/2})\right\}.
\end{align*}
Similarly, due to $\sum_{1\leq i\neq j\leq N}\tr\left[\{u_{i}^{1/2}(u_{j}^{1/2})^{\prime}\}\circ
 \{u_{i}^{1/2}(u_{j}^{1/2})^{\prime}\}\right]=O(N),$ we have
\begin{align*}
{T^2} \gamma_{N}a_{N}&=\left[\sum_{i=1}^{N} \sum_{j=1}^{N} \mathbb{E} (\hat{\epsilon}_{i\cdot}^{\prime}\hat{\epsilon}_{j\cdot})^{2}-\frac{1}{T} \mathbb{E}\left(\sum_{i=1}^{N} \hat{\epsilon}_{i\cdot}^{\prime}\hat{\epsilon}_{i\cdot}\right)^{2}\right]\\
=&\sum_{1\leq i\neq j\leq N}u_{ij}^2\tr\{(\mathbf{\Sigma}^{1/2}
 \mathbf{P}_{i}\mathbf{P}_{j}\mathbf{\Sigma}^{1/2})^2\}+u_{ij}^2\tr^2(\mathbf{\Sigma}^{1/2}
 \mathbf{P}_{i}\mathbf{P}_{j}\mathbf{\Sigma}^{1/2})\\
&+ u_{ii}u_{jj}\tr\{(\mathbf{\Sigma}^{1/2}
 \mathbf{P}_{i}\mathbf{P}_{j}\mathbf{\Sigma}^{1/2})(\mathbf{\Sigma}^{1/2}
 \mathbf{P}_{i}\mathbf{P}_{j}\mathbf{\Sigma}^{1/2})^{\prime}\}\\
 &+\gamma_{2}\tr\left[\{u_{i}^{1/2}(u_{j}^{1/2})^{\prime}\}\circ
 \{u_{i}^{1/2}(u_{j}^{1/2})^{\prime}\}\right]\tr\left\{(\mathbf{\Sigma}^{1/2}\mathbf{P}_{i}
 \mathbf{P}_{j}\mathbf{\Sigma}^{1/2})\circ(\mathbf{\Sigma}^{1/2}\mathbf{P}_{i}
 \mathbf{P}_{j}\mathbf{\Sigma}^{1/2})\right\}\\
 &+\sum_{i=1}^{N}2u_{ii}^{2}\tr\{(\mathbf{\Sigma}^{1/2}
 \mathbf{P}_{i}\mathbf{\Sigma}^{1/2})^{2}\}+u_{ii}^{2}\tr^{2}(\mathbf{\Sigma}^{1/2}
 \mathbf{P}_{i}\mathbf{\Sigma}^{1/2})\\
 &+\gamma_{2}\tr\left[\{u_{i}^{1/2}(u_{i}^{1/2})^{\prime}\}\circ
 \{u_{i}^{1/2}(u_{i}^{1/2})^{\prime}\}\right]\tr\left\{(\mathbf{\Sigma}^{1/2}\mathbf{P}_{i}
\mathbf{\Sigma}^{1/2})\circ(\mathbf{\Sigma}^{1/2}\mathbf{P}_{i}
\mathbf{\Sigma}^{1/2})\right\}\\
&-\frac{1}{T}\Big[\sum_{i=1}^{N}2u_{ii}^{2}\tr\{(\mathbf{\Sigma}^{1/2}
 \mathbf{P}_{i}\mathbf{\Sigma}^{1/2})^2\}+u_{ii}^{2}\tr^2(\mathbf{\Sigma}^{1/2}
 \mathbf{P}_{i}\mathbf{\Sigma}^{1/2})\\
 &+\gamma_{2}\tr\left[\{u_{i}^{1/2}(u_{i}^{1/2})^{\prime}\}\circ
 \{u_{i}^{1/2}(u_{i}^{1/2})^{\prime}\}\right]\tr\left\{(\mathbf{\Sigma}^{1/2}\mathbf{P}_{i}
\mathbf{\Sigma}^{1/2})\circ(\mathbf{\Sigma}^{1/2}\mathbf{P}_{i}
\mathbf{\Sigma}^{1/2})\right\}\\
&+\underset{i\neq j}{\sum^{N}\sum^{N}}2u_{ij}^{2}\tr(\mathbf{\Sigma}^{1/2}
 \mathbf{P}_{i}\mathbf{\Sigma}^{1/2}\mathbf{\Sigma}^{1/2}
 \mathbf{P}_{j}\mathbf{\Sigma}^{1/2})+u_{ii}u_{jj}\tr(\mathbf{\Sigma}^{1/2}
 \mathbf{P}_{i}\mathbf{\Sigma}^{1/2})\tr(\mathbf{\Sigma}^{1/2}
 \mathbf{P}_{j}\mathbf{\Sigma}^{1/2})\\
 &+\gamma_{2}\tr\left[\{u_{i}^{1/2}(u_{i}^{1/2})^{\prime}\}\circ
 \{u_{j}^{1/2}(u_{j}^{1/2})^{\prime}\}\right]\tr\left\{(\mathbf{\Sigma}^{1/2}
 \mathbf{P}_{i}\mathbf{\Sigma}^{1/2})\circ(\mathbf{\Sigma}^{1/2}
 \mathbf{P}_{j}\mathbf{\Sigma}^{1/2})\right\}\Big]\\
\geq&\tr(\mathbf{U}^2)\tr^2(\mathbf{\Sigma})\{1-O(T^{-1})\}.
\end{align*}
Moreover, we have
\begin{align*}
{T^2} \gamma_{N}a_{N}&=\left[\sum_{i=1}^{N} \sum_{j=1}^{N} \mathbb{E} (\hat{\epsilon}_{i\cdot}^{\prime}\hat{\epsilon}_{j\cdot})^{2}-\frac{1}{T} \mathbb{E}\left(\sum_{i=1}^{N} \hat{\epsilon}_{i\cdot}^{\prime}\hat{\epsilon}_{i\cdot}\right)^{2}\right]\\
=&\sum_{1\leq i\neq j\leq N}u_{ij}^2\tr\{(\mathbf{\Sigma}^{1/2}
 \mathbf{P}_{i}\mathbf{P}_{j}\mathbf{\Sigma}^{1/2})^2\}+u_{ij}^2\tr^2(\mathbf{\Sigma}^{1/2}
 \mathbf{P}_{i}\mathbf{P}_{j}\mathbf{\Sigma}^{1/2})\\
&+ u_{ii}u_{jj}\tr\{(\mathbf{\Sigma}^{1/2}
 \mathbf{P}_{i}\mathbf{P}_{j}\mathbf{\Sigma}^{1/2})(\mathbf{\Sigma}^{1/2}
 \mathbf{P}_{i}\mathbf{P}_{j}\mathbf{\Sigma}^{1/2})^{\prime}\}\\
 &+\gamma_{2}\tr\left[\{u_{i}^{1/2}(u_{j}^{1/2})^{\prime}\}\circ\{
 u_{i}^{1/2}(u_{j}^{1/2})^{\prime}\}\right]\tr\left\{(\mathbf{\Sigma}^{1/2}\mathbf{P}_{i}
 \mathbf{P}_{j}\mathbf{\Sigma}^{1/2})\circ(\mathbf{\Sigma}^{1/2}\mathbf{P}_{i}
 \mathbf{P}_{j}\mathbf{\Sigma}^{1/2})\right\}\\
 &+\sum_{i=1}^{N}2u_{ii}^{2}\tr\{(\mathbf{\Sigma}^{1/2}
 \mathbf{P}_{i}\mathbf{\Sigma}^{1/2})^{2}\}+u_{ii}^{2}\tr^{2}(\mathbf{\Sigma}^{1/2}
 \mathbf{P}_{i}\mathbf{\Sigma}^{1/2})\\
 &+\gamma_{2}\tr\left[\{u_{i}^{1/2}(u_{i}^{1/2})^{\prime}\}\circ
 \{u_{i}^{1/2}(u_{i}^{1/2})^{\prime}\}\right]\tr\left\{(\mathbf{\Sigma}^{1/2}\mathbf{P}_{i}
\mathbf{\Sigma}^{1/2})\circ(\mathbf{\Sigma}^{1/2}\mathbf{P}_{i}
\mathbf{\Sigma}^{1/2})\right\}\\
&-\frac{1}{T}\Big[\sum_{i=1}^{N}2u_{ii}^{2}\tr\{(\mathbf{\Sigma}^{1/2}
 \mathbf{P}_{i}\mathbf{\Sigma}^{1/2})^2\}+u_{ii}^{2}\tr^2(\mathbf{\Sigma}^{1/2}
 \mathbf{P}_{i}\mathbf{\Sigma}^{1/2})\\
 &+\gamma_{2}\tr\left[\{u_{i}^{1/2}(u_{i}^{1/2})^{\prime}\}\circ
 \{u_{i}^{1/2}(u_{i}^{1/2})^{\prime}\}\right]\tr\left\{(\mathbf{\Sigma}^{1/2}\mathbf{P}_{i}
\mathbf{\Sigma}^{1/2})\circ(\mathbf{\Sigma}^{1/2}\mathbf{P}_{i}
\mathbf{\Sigma}^{1/2})\right\}\\
&+\underset{i\neq j}{\sum^{N}\sum^{N}}2u_{ij}^{2}\tr(\mathbf{\Sigma}^{1/2}
 \mathbf{P}_{i}\mathbf{\Sigma}^{1/2}\mathbf{\Sigma}^{1/2}
 \mathbf{P}_{j}\mathbf{\Sigma}^{1/2})+u_{ii}u_{jj}\tr(\mathbf{\Sigma}^{1/2}
 \mathbf{P}_{i}\mathbf{\Sigma}^{1/2})\tr(\mathbf{\Sigma}^{1/2}
 \mathbf{P}_{j}\mathbf{\Sigma}^{1/2})\\
 &+\gamma_{2}\tr\left[\{u_{i}^{1/2}(u_{i}^{1/2})^{\prime}\}\circ\{
 u_{j}^{1/2}(u_{j}^{1/2})^{\prime}\}\right]\tr\left\{(\mathbf{\Sigma}^{1/2}
 \mathbf{P}_{i}\mathbf{\Sigma}^{1/2})\circ(\mathbf{\Sigma}^{1/2}
 \mathbf{P}_{j}\mathbf{\Sigma}^{1/2})\right\}\Big]\\
\leq&\tr(\mathbf{U}^2)\tr^2(\mathbf{\Sigma})\{1+O(NT^{-1})\}.
\end{align*}
This proves that $a_{N}$ satisfies the inequality in the lemma.
It remains to calculate $b_{N}.$ We have
\begin{align*}
\var\left\{\sum_{i=1}^{N} \sum_{j=1}^{N} (\hat{\epsilon}_{i\cdot}^{\prime}\hat{\epsilon}_{j\cdot})^{2}\right\}
&=\mE\left[\left\{\sum_{i=1}^{N} \sum_{j=1}^{N} (\hat{\epsilon}_{i\cdot}^{\prime}\hat{\epsilon}_{j\cdot})^{2}\right\}^{2}\right]
-\mE^2\left\{\sum_{i=1}^{N} \sum_{j=1}^{N} (\hat{\epsilon}_{i\cdot}^{\prime}\hat{\epsilon}_{j\cdot})^{2}\right\}.
\end{align*}
First,
\begin{align*}
&\var\left[\left\{\sum_{i=1}^{N} \sum_{j=1}^{N} (\hat{\epsilon}_{i\cdot}^{\prime}\hat{\epsilon}_{j\cdot})^{2}\right\}^{2}\right]\\
=&\sum_{i=1}^{N}\mE\left\{(\hat{\epsilon}_{i\cdot}^{\prime}\hat{\epsilon}_{i\cdot})^{4}\right\}
-\mE^2\{(\hat{\epsilon}_{i\cdot}^{\prime}\hat{\epsilon}_{i\cdot})^{2}\}+2\underset{i\neq j}{\sum^{N}\sum^{N}}\mE\left\{(\hat{\epsilon}_{i\cdot}^{\prime}\hat{\epsilon}_{j\cdot})^{4}\right\}
-\mE^2\{(\hat{\epsilon}_{i\cdot}^{\prime}\hat{\epsilon}_{j\cdot})^{2}\}\\
&+\underset{i\neq j}{\sum^{N}\sum^{N}}\mE\left\{(\hat{\epsilon}_{i\cdot}^{\prime}\hat{\epsilon}_{i\cdot})^{2}
(\hat{\epsilon}_{j\cdot}^{\prime}\hat{\epsilon}_{j\cdot})^{2}\right\}
-\mE\{(\hat{\epsilon}_{i\cdot}^{\prime}\hat{\epsilon}_{i\cdot})^{2}\}
\mE\{(\hat{\epsilon}_{j\cdot}^{\prime}\hat{\epsilon}_{j\cdot})^{2}\}\\
&+4\underset{i\neq j}{\sum^{N}\sum^{N}}\mE\left\{(\hat{\epsilon}_{i\cdot}^{\prime}\hat{\epsilon}_{j\cdot})^{2}
(\hat{\epsilon}_{i\cdot}^{\prime}\hat{\epsilon}_{i\cdot})^{2}\right\}
-\mE\{(\hat{\epsilon}_{i\cdot}^{\prime}\hat{\epsilon}_{i\cdot})^{2}\}
\mE\{(\hat{\epsilon}_{i\cdot}^{\prime}\hat{\epsilon}_{j\cdot})^{2}\}\\
&+2\underset{i\neq j\neq k}{\sum^{N}\sum^{N}\sum^{N}}
\mE\left\{(\hat{\epsilon}_{i\cdot}^{\prime}\hat{\epsilon}_{i\cdot})^{2}
(\hat{\epsilon}_{j\cdot}^{\prime}\hat{\epsilon}_{k\cdot})^{2}\right\}
-\mE\{(\hat{\epsilon}_{i\cdot}^{\prime}\hat{\epsilon}_{i\cdot})^{2}\}
\mE\{(\hat{\epsilon}_{j\cdot}^{\prime}\hat{\epsilon}_{k\cdot})^{2}\}\\
&+4\underset{i\neq j\neq k}{\sum^{N}\sum^{N}\sum^{N}}
\mE\left\{(\hat{\epsilon}_{i\cdot}^{\prime}\hat{\epsilon}_{j\cdot})^{2}
(\hat{\epsilon}_{i\cdot}^{\prime}\hat{\epsilon}_{k\cdot})^{2}\right\}
-\mE\{(\hat{\epsilon}_{i\cdot}^{\prime}\hat{\epsilon}_{j\cdot})^{2}\}
\mE\{(\hat{\epsilon}_{i\cdot}^{\prime}\hat{\epsilon}_{k\cdot})^{2}\}\\
&+\underset{i\neq j\neq k\neq l}{\sum^{N}\sum^{N}\sum^{N}\sum^{N}}\mE\left\{(\hat{\epsilon}_{i\cdot}^{\prime}\hat{\epsilon}_{j\cdot})^{2}
(\hat{\epsilon}_{k\cdot}^{\prime}\hat{\epsilon}_{l\cdot})^{2}\right\}
-\mE\{(\hat{\epsilon}_{i\cdot}^{\prime}\hat{\epsilon}_{j\cdot})^{2}\}
\mE\{(\hat{\epsilon}_{k\cdot}^{\prime}\hat{\epsilon}_{l\cdot})^{2}\}\\
=&G_{1}+G_{2}+G_{3}+G_{4}+G_{5}+G_{6}+G_{7}.
\end{align*}
Recall that $\mathbf{Q}_{ij}=\mathbf{\Sigma}^{1/2}\mathbf{P}_{i}\mathbf{P}_{j}
\mathbf{\Sigma}^{1/2}.$ Let $\mathbf{\Omega}_{ij}\doteq u_{i}^{1/2}(u_{j}^{1/2})^{\prime}$ for $1\leq i,j\leq N.$
First, we focus on $G_{1}.$
According to Lemma \ref{le:moment of quadratic form}, we have
$$
G_{1}=G_{1_{1}}+\gamma_{2}G_{1_{2}}+\gamma_{4}G_{1_{3}}+\gamma_{6}G_{1_{4}}
+\gamma_{1}^{2}G_{1_{5}}+\gamma_{2}^{2}G_{1_{6}}+\gamma_{1}\gamma_{3}G_{1_{7}},$$
where
\begin{align*}
G_{1_{1}}=&\sum_{i=1}^{N}8u_{ii}^{4}
\tr\left(\mathbf{Q}^{2}_{ii}\right)\tr^{2}\left(\mathbf{Q}_{ii}\right)
+8u_{ii}^{4}\tr^2\left(\mathbf{Q}_{ii}^{2}\right)
+32u_{ii}^{4}\tr\left(\mathbf{Q}_{ii}\right)\tr\left(\mathbf{Q}^{3}_{ii}\right)\\
&+48u_{ii}^{4}\tr\left(\mathbf{Q}_{ii}^{4}\right),\\
G_{1_{2}}=&\sum_{i=1}^{N}
4u_{ii}^{2}\tr^2\left(\mathbf{Q}_{ii}\right)
\tr\left\{\mathbf{\Omega}_{ii}\circ \mathbf{\Omega}_{ii}\right\}
\tr\left(\mathbf{Q}_{ii}\circ\mathbf{Q}_{ii}\right)\\
&+8u_{ii}^{2}
\tr\left(\mathbf{Q}_{ii}^{2}\right)\tr\left\{\mathbf{\Omega}_{ii}\circ \mathbf{\Omega}_{ii}\right\}
\tr\left(\mathbf{Q}_{ii}\circ\mathbf{Q}_{ii}\right)\\
&+48u_{ii}
\tr\left(\mathbf{Q}_{ii}\right)\tr\left\{\mathbf{\Omega}_{ii}\circ u_{ii}\mathbf{\Omega}_{ii}\right\}\tr\left(\mathbf{Q}_{ii}\circ \mathbf{Q}_{ii}^{2}\right)\\
&+96\tr\left\{\left(\mathbf{I}_{N}\circ \mathbf{\Omega}_{ii}\right)u_{ii}^{2}\mathbf{\Omega}_{ii}\right\}
\tr\left\{\left(\mathbf{I}_{T}\circ \mathbf{Q}_{ii}\right)\mathbf{Q}_{ii}^{3}\right\}\\
&+48\tau_{N}^{\prime}\left\{\left(\mathbf{I}_{N}\circ u_{ii}\mathbf{\Omega}_{ii}\right)\left(\mathbf{I}_{N}\circ u_{ii}\mathbf{\Omega}_{ii}\right)\right\}\tau_{N}
\tau_{T}^{\prime}\left\{\left(\mathbf{I}_{T}\circ \mathbf{Q}_{ii}^{2}\right)\left(\mathbf{I}_{T}\circ \mathbf{Q}_{ii}^{2}\right)\right\}\tau_{T},\\
G_{1_{3}}=&\sum_{i=1}^{N}
4u_{ii}\tr\left(\mathbf{Q}_{ii}\right)\tr\left\{\mathbf{\Omega}_{ii}\circ \mathbf{\Omega}_{ii}\circ \mathbf{\Omega}_{ii}\right\}\tr\left(\mathbf{Q}_{ii}\circ\mathbf{Q}_{ii}\circ\mathbf{Q}_{ii}\right)\\
&+24\tr\left\{\mathbf{\Omega}_{ii}\circ
\mathbf{\Omega}_{ii}\circ
u_{ii}\mathbf{\Omega}_{ii}\right\}\tr(\mathbf{Q}_{ii}\circ
\mathbf{Q}_{ii}\circ\mathbf{Q}_{ii}^{2}),\\
G_{1_{4}}=&\sum_{i=1}^{N}
\tr\left\{\mathbf{\Omega}_{ii}\circ
\mathbf{\Omega}_{ii}\circ
\mathbf{\Omega}_{ii}\circ
\mathbf{\Omega}_{ii}\right\}\tr\left(\mathbf{Q}_{ii}\circ
\mathbf{Q}_{ii}\circ\mathbf{Q}_{ii}\circ\mathbf{Q}_{ii}\right),\\
G_{1_{5}}=&\sum_{i=1}^{N}
24\tau_{N}^{\prime}\left[\left\{\mathbf{I}_{N}\circ \mathbf{\Omega}_{ii}\right\}\mathbf{\Omega}_{ii}
\left\{\mathbf{I}_{N}\circ \mathbf{\Omega}_{ii}\right\}\right]\tau_{N}
\tau_{T}^{\prime}\left[\left\{\mathbf{I}_{T}\circ \mathbf{Q}_{ii}\right\}\mathbf{Q}_{ii}
\left\{\mathbf{I}_{T}\circ \mathbf{Q}_{ii}\right\}\right]\tau_{T}u_{ii}\tr(\mathbf{Q}_{ii})\\
&+48\tau_{N}^{\prime}\left\{\left(\mathbf{I}_{N}\circ\mathbf{\Omega}_{ii}\right)
u_{ii}\mathbf{\Omega}_{ii}\left\{\mathbf{I}_{N}\circ\mathbf{\Omega}_{ii}\right\}\right\}\tau_{N}
\tau_{T}^{\prime}\left\{\left(\mathbf{I}_{N}\circ\mathbf{Q}_{ii}\right)
\mathbf{Q}^{2}_{ii}\left(\mathbf{I}_{N}\circ\mathbf{Q}_{ii}\right)\right\}\tau_{T}\\
&+16\tau_{N}^{\prime}\left(\mathbf{\Omega}_{ii}\circ\mathbf{\Omega}_{ii}
\circ\mathbf{\Omega}_{ii}\right)\tau_{N}
\tau_{T}^{\prime}\left(\mathbf{Q}_{ii}\circ\mathbf{Q}_{ii}
\circ\mathbf{Q}_{ii}\right)\tau_{T}u_{ii}\tr\left(\mathbf{Q}_{ii}\right)\\
&+96\tau_{N}^{\prime}\left(\mathbf{\Omega}_{ii}\circ\mathbf{\Omega}_{ii}\right)
\mathbf{\Omega}_{ii}\left(\mathbf{I}_{N}\circ\mathbf{\Omega}_{ii}\right)\tau_{N}
\tau_{T}\prime\left(\mathbf{Q}_{ii}\circ\mathbf{Q}_{ii}\right)
\mathbf{Q}_{ii}\left(\mathbf{I}_{T}\circ\mathbf{Q}_{ii}\right)\tau_{T}\\
&+96\tr\left\{\mathbf{\Omega}_{ii}\left(\mathbf{\Omega}_{ii}\circ
\mathbf{\Omega}_{ii}\right)\mathbf{\Omega}_{ii}\right\}
\tr\left\{\mathbf{Q}_{ii}\left(\mathbf{Q}_{ii}\circ
\mathbf{Q}_{ii}\right)\mathbf{Q}_{ii}\right\},\\
G_{1_{6}}=&\sum_{i=1}^{N}
2\tr^{2}\left(\mathbf{\Omega}_{ii}\circ
\mathbf{\Omega}_{ii}\right)\tr^{2}\left(\mathbf{Q}_{ii}\circ
\mathbf{Q}_{ii}\right)\\
&+24\tau_{N}^{\prime}
\left(\mathbf{I}_{N}\circ\mathbf{\Omega}_{ii}\right)
\left(\mathbf{\Omega}_{ii}\circ\mathbf{\Omega}_{ii}\right)
\left(\mathbf{I}_{N}\circ\mathbf{\Omega}_{ii}\right)\tau_{N}
\tau_{T}^{\prime}
\left(\mathbf{I}_{T}\circ\mathbf{Q}_{ii}\right)
\left(\mathbf{Q}_{ii}\circ\mathbf{Q}_{ii}\right)
\left(\mathbf{I}_{T}\circ\mathbf{Q}_{ii}\right)\tau_{T}\\
&+8\tau_{N}^{\prime}\left(\mathbf{\Omega}_{ii}\circ\mathbf{\Omega}_{ii}
\circ\mathbf{\Omega}_{ii}\circ\mathbf{\Omega}_{ii}\right)\tau_{N}
\tau_{T}^{\prime}\left(\mathbf{Q}_{ii}\circ\mathbf{Q}_{ii}
\circ\mathbf{Q}_{ii}\circ\mathbf{Q}_{ii}\right)\tau_{T},\\
G_{1_{7}}=&\sum_{i=1}^{N}24\tau_{N}^{\prime}
\left(\mathbf{I}_{N}\circ\mathbf{\Omega}_{ii}\right)\mathbf{\Omega}_{ii}
\left(\mathbf{I}_{N}\circ\mathbf{\Omega}_{ii}\circ\mathbf{\Omega}_{ii}\right)\tau_{N}
\tau_{T}^{\prime}
\left(\mathbf{I}_{T}\circ\mathbf{Q}_{ii}\right)\mathbf{Q}_{ii}
\left(\mathbf{I}_{T}\circ\mathbf{Q}_{ii}\circ\mathbf{Q}_{ii}\right)\tau_{T}\\
&+32\tau_{N}^{\prime}
\left(\mathbf{I}_{N}\circ\mathbf{\Omega}_{ii}\right)
\left(\mathbf{\Omega}_{ii}\circ\mathbf{\Omega}_{ii}\circ\mathbf{\Omega}_{ii}\right)\tau_{N}
\tau_{T}^{\prime}
\left(\mathbf{I}_{T}\circ\mathbf{Q}_{ii}\right)
\left(\mathbf{Q}_{ii}\circ\mathbf{Q}_{ii}\circ\mathbf{Q}_{ii}\right)\tau_{T}.
\end{align*}

Obviously, $G_{1_{1}}=O(NT^{3}).$
Due to Lemma \ref{lem:matrix inequality} (1),
 we have
\begin{align*}
G_{1_{2}}=&\sum_{i=1}^{N}
4u_{ii}^{2}\tr^2\left(\mathbf{Q}_{ii}\right)
\tr\left\{\mathbf{\Omega}_{ii}\circ \mathbf{\Omega}_{ii}\right\}
\tr\left(\mathbf{Q}_{ii}\circ\mathbf{Q}_{ii}\right)\\
&+8u_{ii}^{2}
\tr\left(\mathbf{Q}_{ii}^{2}\right)\tr\left\{\mathbf{\Omega}_{ii}\circ \mathbf{\Omega}_{ii}\right\}
\tr\left(\mathbf{Q}_{ii}\circ\mathbf{Q}_{ii}\right)\\
&+48u_{ii}
\tr\left(\mathbf{Q}_{ii}\right)\tr\left\{\mathbf{\Omega}_{ii}\circ u_{ii}\mathbf{\Omega}_{ii}\right\}\tr\left(\mathbf{Q}_{ii}\circ \mathbf{Q}_{ii}^{2}\right)\\
&+96\tr\left\{\left(\mathbf{I}_{N}\circ \mathbf{\Omega}_{ii}\right)u_{ii}^{2}\mathbf{\Omega}_{ii}\right\}
\tr\left\{\left(\mathbf{I}_{T}\circ \mathbf{Q}_{ii}\right)\mathbf{Q}_{ii}^{3}\right\}\\
&+48\tau_{N}^{\prime}\left\{\left(\mathbf{I}_{N}\circ u_{ii}\mathbf{\Omega}_{ii}\right)\left(\mathbf{I}_{N}\circ u_{ii}\mathbf{\Omega}_{ii}\right)\right\}\tau_{N}
\tau_{T}^{\prime}\left\{\left(\mathbf{I}_{T}\circ \mathbf{Q}_{ii}^{2}\right)\left(\mathbf{I}_{T}\circ \mathbf{Q}_{ii}^{2}\right)\right\}\tau_{T}\\
=&O(NT^{3})+O(NT^{2})+\sum_{i=1}^{N}48u_{ii}
\tr\left(\mathbf{Q}_{ii}\right)\tr\left\{\mathbf{\Omega}_{ii}\circ u_{ii}\mathbf{\Omega}_{ii}\right\}\tr\left(\mathbf{Q}_{ii}\circ \mathbf{Q}_{ii}^{2}\right)\\
&+96\tr\left\{\left(\mathbf{I}_{N}\circ \mathbf{\Omega}_{ii}\right)u_{ii}^{2}\mathbf{\Omega}_{ii}\right\}
\tr\left\{\left(\mathbf{I}_{T}\circ \mathbf{Q}_{ii}\right)\mathbf{Q}_{ii}^{3}\right\}\\
&+48\tau_{N}^{\prime}\left\{\left(\mathbf{I}_{N}\circ u_{ii}\mathbf{\Omega}_{ii}\right)\left(\mathbf{I}_{N}\circ u_{ii}\mathbf{\Omega}_{ii}\right)\right\}\tau_{N}
\tau_{T}^{\prime}\left\{\left(\mathbf{I}_{T}\circ \mathbf{Q}_{ii}^{2}\right)\left(\mathbf{I}_{T}\circ \mathbf{Q}_{ii}^{2}\right)\right\}\tau_{T}\\
=&O(NT^{3})+O(NT^{2})+\sum_{i=1}^{N}48u_{ii}u_{ii}\tr\left(\mathbf{Q}_{ii}\right)
\sum_{s=1}^{N}(\mathbf{\Omega}_{ii})_{ss}^{2}
\sum_{s=1}^{T}(\mathbf{Q}_{ii}^{2})_{ss}(\mathbf{Q}_{ii})_{ss}\\
&+96u_{ii}^{2}
\sum_{s=1}^{N}(\mathbf{\Omega}_{ii})_{ss}^{2}
\sum_{s=1}^{T}(\mathbf{Q}_{ii}^{3})_{ss}(\mathbf{Q}_{ii})_{ss}
+48u_{ii}^{2}
\sum_{s=1}^{N}(\mathbf{\Omega}_{ii})_{ss}^{2}
\sum_{s=1}^{T}(\mathbf{Q}_{ii}^{2})_{ss}^{2}\\
\leq&O(NT^{3})+O(NT^{2})+\sum_{i=1}^{N}48u_{ii}^{4}
|\tr\left(\mathbf{Q}_{ii}\right)|
|\sum_{s=1}^{T}(\mathbf{Q}_{ii}^{2})_{ss}(\mathbf{Q}_{ii})_{ss}|\\
&+96u_{ii}^{4}|
\sum_{s=1}^{T}(\mathbf{Q}_{ii}^{3})_{ss}(\mathbf{Q}_{ii})_{ss}|
+48u_{ii}^{4}|
\sum_{s=1}^{T}(\mathbf{Q}_{ii}^{2})_{ss}^{2}|\\
=&O(NT^{3})+O(NT^{2})+O(NT).
 \end{align*}
Then,
\begin{align*}
G_{1_{3}}=&\sum_{i=1}^{N}
4u_{ii}\tr\left(\mathbf{Q}_{ii}\right)\tr\left\{\mathbf{\Omega}_{ii}\circ \mathbf{\Omega}_{ii}\circ \mathbf{\Omega}_{ii}\right\}\tr\left(\mathbf{Q}_{ii}\circ\mathbf{Q}_{ii}\circ\mathbf{Q}_{ii}\right)\\
&+24\tr\left\{\mathbf{\Omega}_{ii}\circ
\mathbf{\Omega}_{ii}\circ
u_{ii}\mathbf{\Omega}_{ii}\right\}\tr(\mathbf{Q}_{ii}\circ
\mathbf{Q}_{ii}\circ\mathbf{Q}_{ii}^{2})\\
=&\sum_{i=1}^{N}
4u_{ii}\tr\left(\mathbf{Q}_{ii}\right)
\sum_{s=1}^{N}(\mathbf{\Omega}_{ii})_{ss}^{3}
\sum_{s=1}^{T}(\mathbf{Q}_{ii})_{ss}^{3}+24u_{ii}
\sum_{s=1}^{N}(\mathbf{\Omega}_{ii})_{ss}^{3}
\sum_{s=1}^{T}(\mathbf{Q}_{ii})_{ss}^{2}(\mathbf{Q}_{ii}^{2})_{ss}\\
\leq&\sum_{i=1}^{N}
4u_{ii}^{4}
|\tr\left(\mathbf{Q}_{ii}\right)|
|\sum_{s=1}^{T}(\mathbf{Q}_{ii})_{ss}^{3}|+
24u_{ii}^{4}
|\sum_{s=1}^{T}(\mathbf{Q}_{ii})_{ss}^{2}(\mathbf{Q}_{ii}^{2})_{ss}|\\
=&O(NT^{2})+O(NT).
\end{align*}
Similarly, we can easily obtain that $G_{1_{4}}=O(NT).$
Note that for any square matrix $\mathbf{A}\in \mathbb{R}^{T\times T},$
$$
|\sum_{s=1}^{T}\sum_{t=1}^{T}(\mathbf{A})_{st}|
\leq\sqrt{T^{2}\tr\left(\mathbf{A}\mathbf{A}^{\prime}\right)}.$$
So,
we next have
\begin{align*}
G_{1_{5}}=&\sum_{i=1}^{N}
24\tau_{N}^{\prime}\left[\left\{\mathbf{I}_{N}\circ \mathbf{\Omega}_{ii}\right\}\mathbf{\Omega}_{ii}
\left\{\mathbf{I}_{N}\circ \mathbf{\Omega}_{ii}\right\}\right]\tau_{N}
\tau_{T}^{\prime}\left[\left\{\mathbf{I}_{T}\circ \mathbf{Q}_{ii}\right\}\mathbf{Q}_{ii}
\left\{\mathbf{I}_{T}\circ \mathbf{Q}_{ii}\right\}\right]\tau_{T}u_{ii}\tr(\mathbf{Q}_{ii})\\
&+48\tau_{N}^{\prime}\left\{\left(\mathbf{I}_{N}\circ\mathbf{\Omega}_{ii}\right)
u_{ii}\mathbf{\Omega}_{ii}\left\{\mathbf{I}_{N}\circ\mathbf{\Omega}_{ii}\right\}\right\}\tau_{N}
\tau_{T}^{\prime}\left\{\left(\mathbf{I}_{N}\circ\mathbf{Q}_{ii}\right)
\mathbf{Q}^{2}_{ii}\left(\mathbf{I}_{N}\circ\mathbf{Q}_{ii}\right)\right\}\tau_{T}\\
&+16\tau_{N}^{\prime}\left(\mathbf{\Omega}_{ii}\circ\mathbf{\Omega}_{ii}
\circ\mathbf{\Omega}_{ii}\right)\tau_{N}
\tau_{T}^{\prime}\left(\mathbf{Q}_{ii}\circ\mathbf{Q}_{ii}
\circ\mathbf{Q}_{ii}\right)\tau_{T}u_{ii}\tr\left(\mathbf{Q}_{ii}\right)\\
&+96\tau_{N}^{\prime}\left(\mathbf{\Omega}_{ii}\circ\mathbf{\Omega}_{ii}\right)
\mathbf{\Omega}_{ii}\left(\mathbf{I}_{N}\circ\mathbf{\Omega}_{ii}\right)\tau_{N}
\tau_{T}\prime\left(\mathbf{Q}_{ii}\circ\mathbf{Q}_{ii}\right)
\mathbf{Q}_{ii}\left(\mathbf{I}_{T}\circ\mathbf{Q}_{ii}\right)\tau_{T}\\
&+96\tr\left\{\mathbf{\Omega}_{ii}\left(\mathbf{\Omega}_{ii}\circ
\mathbf{\Omega}_{ii}\right)\mathbf{\Omega}_{ii}\right\}
\tr\left\{\mathbf{Q}_{ii}\left(\mathbf{Q}_{ii}\circ
\mathbf{Q}_{ii}\right)\mathbf{Q}_{ii}\right\}\\
=&\sum_{i=1}^{N}
24\sum_{s=1}^{N}\sum_{t=1}^{N}(\mathbf{\Omega}_{ii})_{ss}
(\mathbf{\Omega}_{ii})_{st}
(\mathbf{\Omega}_{ii})_{tt}
\sum_{s=1}^{T}\sum_{t=1}^{T}(\mathbf{Q}_{ii})_{ss}
(\mathbf{Q}_{ii})_{st}
(\mathbf{Q}_{ii})_{tt}\\
&+48u_{ii}
\sum_{s=1}^{N}\sum_{t=1}^{N}(\mathbf{\Omega}_{ii})_{ss}
(\mathbf{\Omega}_{ii})_{st}
(\mathbf{\Omega}_{ii})_{tt}
\sum_{s=1}^{T}\sum_{t=1}^{T}(\mathbf{Q}_{ii})_{ss}
(\mathbf{Q}_{ii}^{2})_{st}
(\mathbf{Q}_{ii})_{tt}
\\
&+16u_{ii}\tr\left(\mathbf{Q}_{ii}\right)
\sum_{s=1}^{N}\sum_{t=1}^{N}(\mathbf{\Omega}_{ii})_{st}^{3}
\sum_{s=1}^{T}\sum_{t=1}^{T}(\mathbf{Q}_{ii})_{st}^{3}
\\
&+96
\sum_{s=1}^{N}\sum_{t=1}^{N}\sum_{q=1}^{N}(\mathbf{\Omega}_{ii})_{qt}
(\mathbf{\Omega}_{ii})_{tt}(\mathbf{\Omega}_{ii})_{sq}^{2}
\sum_{s=1}^{N}\sum_{t=1}^{N}\sum_{q=1}^{N}(\mathbf{Q}_{ii})_{qt}
(\mathbf{Q}_{ii})_{tt}(\mathbf{Q}_{ii})_{sq}^{2}
\\
&+96u_{ii}\sum_{s=1}^{N}\sum_{t=1}^{N}
(\mathbf{\Omega}_{ii})_{st}^{2}(\mathbf{\Omega}_{ii})_{ts}
\sum_{s=1}^{T}\sum_{t=1}^{T}
(\mathbf{Q}_{ii})_{st}^{2}(\mathbf{Q}_{ii}^{2})_{ts}\\
\leq&\sum_{i=1}^{N}
24u_{ii}^{2}
\sum_{s=1}^{T}\sum_{t=1}^{T}C|
(\mathbf{Q}_{ii})_{st}|+48u_{ii}^{3}
\sum_{s=1}^{T}\sum_{t=1}^{T}C|
(\mathbf{Q}_{ii}^{2})_{st}|
\\
&+16u_{ii}^{3}
\tr\left(\mathbf{Q}_{ii}\right)
\sum_{s=1}^{T}\sum_{t=1}^{T}C(\mathbf{Q}_{ii})_{st}^{2}
+96u_{ii}^{3}
\sum_{t=1}^{N}\sum_{q=1}^{N}|(\mathbf{Q}_{ii})_{qt}|
(\mathbf{Q}_{ii})_{tt}(\mathbf{Q}_{ii}^{2})_{qq}
\\
&+96u_{ii}^{3}
\sum_{s=1}^{T}\sum_{t=1}^{T}C
(\mathbf{Q}_{ii})_{st}^{2}\\=&O(NT^{3/2})+O(NT^{2})+O(NT).
\end{align*}
Similarly, for $G_{1_{6}},$
we have
\begin{align*}
G_{1_{6}}=&\sum_{i=1}^{N}
2\tr^{2}\left(\mathbf{\Omega}_{ii}\circ
\mathbf{\Omega}_{ii}\right)\tr^{2}\left(\mathbf{Q}_{ii}\circ
\mathbf{Q}_{ii}\right)\\
&+24\tau_{N}^{\prime}
\left(\mathbf{I}_{N}\circ\mathbf{\Omega}_{ii}\right)
\left(\mathbf{\Omega}_{ii}\circ\mathbf{\Omega}_{ii}\right)
\left(\mathbf{I}_{N}\circ\mathbf{\Omega}_{ii}\right)\tau_{N}
\tau_{T}^{\prime}
\left(\mathbf{I}_{T}\circ\mathbf{Q}_{ii}\right)
\left(\mathbf{Q}_{ii}\circ\mathbf{Q}_{ii}\right)
\left(\mathbf{I}_{T}\circ\mathbf{Q}_{ii}\right)\tau_{T}\\
&+8\tau_{N}^{\prime}\left(\mathbf{\Omega}_{ii}\circ\mathbf{\Omega}_{ii}
\circ\mathbf{\Omega}_{ii}\circ\mathbf{\Omega}_{ii}\right)\tau_{N}
\tau_{T}^{\prime}\left(\mathbf{Q}_{ii}\circ\mathbf{Q}_{ii}
\circ\mathbf{Q}_{ii}\circ\mathbf{Q}_{ii}\right)\tau_{T}\\
=&\sum_{i=1}^{N}
2\big\{\sum_{s=1}^{N}(\mathbf{\Omega}_{ii})_{ss}^{2}\big\}^{2}
\big\{\sum_{s=1}^{N}(\mathbf{Q}_{ii})_{ss}^{2}\big\}^{2}
+24\sum_{s=1}^{N}\sum_{t=1}^{N}(\mathbf{\Omega}_{ii})_{st}^{2}
(\mathbf{\Omega}_{ii})_{ss}(\mathbf{\Omega}_{ii})_{tt}
\sum_{s=1}^{T}\sum_{t=1}^{T}(\mathbf{Q}_{ii})_{st}^{2}
(\mathbf{Q}_{ii})_{ss}(\mathbf{Q}_{ii})_{tt}\\
&+8\sum_{s=1}^{N}\sum_{t=1}^{N}
(\mathbf{\Omega}_{ii})_{st}^{4}
\sum_{s=1}^{T}\sum_{t=1}^{T}
(\mathbf{Q}_{ii})_{st}^{4}\\
\leq&\sum_{i=1}^{N}
2u_{ii}^{4}\tr^{2}(\mathbf{Q}_{ii}^{2})
+24u_{ii}^{4}
\sum_{s=1}^{T}\sum_{t=1}^{T}C(\mathbf{Q}_{ii})_{st}^{2}+8u_{ii}^{4}
\sum_{s=1}^{T}\sum_{t=1}^{T}C
(\mathbf{Q}_{ii})_{st}^{2}\\
=&O(NT^{2})+O(NT).
\end{align*}
Furthermore, we have
\begin{align*}
G_{1_{7}}=&\sum_{i=1}^{N}24\tau_{N}^{\prime}
\left(\mathbf{I}_{N}\circ\mathbf{\Omega}_{ii}\right)\mathbf{\Omega}_{ii}
\left(\mathbf{I}_{N}\circ\mathbf{\Omega}_{ii}\circ\mathbf{\Omega}_{ii}\right)\tau_{N}
\tau_{T}^{\prime}
\left(\mathbf{I}_{T}\circ\mathbf{Q}_{ii}\right)\mathbf{Q}_{ii}
\left(\mathbf{I}_{T}\circ\mathbf{Q}_{ii}\circ\mathbf{Q}_{ii}\right)\tau_{T}\\
&+32\tau_{N}^{\prime}
\left(\mathbf{I}_{N}\circ\mathbf{\Omega}_{ii}\right)
\left(\mathbf{\Omega}_{ii}\circ\mathbf{\Omega}_{ii}\circ\mathbf{\Omega}_{ii}\right)\tau_{N}
\tau_{T}^{\prime}
\left(\mathbf{I}_{T}\circ\mathbf{Q}_{ii}\right)
\left(\mathbf{Q}_{ii}\circ\mathbf{Q}_{ii}\circ\mathbf{Q}_{ii}\right)\tau_{T}\\
=&\sum_{i=1}^{N}24
\sum_{s=1}^{N}\sum_{t=1}^{N}(\mathbf{\Omega}_{ii})_{st}(\mathbf{\Omega}_{ii})_{ss}
(\mathbf{\Omega}_{ii})_{tt}^{2}
\sum_{s=1}^{T}\sum_{t=1}^{T}(\mathbf{Q}_{ii})_{st}(\mathbf{Q}_{ii})_{ss}
(\mathbf{Q}_{ii})_{tt}^{2}\\
&+32\sum_{s=1}^{N}\sum_{t=1}^{N}
(\mathbf{\Omega}_{ii})_{st}^{3}(\mathbf{\Omega}_{ii})_{ss}
\sum_{s=1}^{T}\sum_{t=1}^{T}
(\mathbf{Q}_{ii})_{st}^{3}(\mathbf{Q}_{ii})_{ss}\\
\leq&\sum_{i=1}^{N}24
u_{ii}^{3}
\sum_{s=1}^{T}\sum_{t=1}^{T}C|(\mathbf{Q}_{ii})_{st}|+32u_{ii}^{3}
\sum_{s=1}^{T}\sum_{t=1}^{T}C
(\mathbf{Q}_{ii})_{st}^{2}\\
=&O(NT^{3/2})+O(NT).
\end{align*}

Combining the above calculations, we can get $G_{1}=O(NT^{3}).$
We next deal with $G_{2}$. Similarly, due to Lemma \ref{le:moment of quadratic form}, we have
$$
G_{2}=G_{2_{1}}+\gamma_{2}G_{2_{2}}+\gamma_{4}G_{2_{3}}+\gamma_{6}G_{2_{4}}
+\gamma_{1}^{2}G_{2_{5}}+\gamma_{2}^{2}G_{2_{6}}+\gamma_{1}\gamma_{3}G_{2_{7}},$$
where
\begin{align*}
G_{2_{1}}=&2\underset{i\neq j}{\sum^{N}\sum^{N}}4u_{ij}^{4}\tr^{2}\left(\mathbf{Q}_{ij}\right)
\tr\left(\mathbf{Q}_{ij}^{2}\right)+4u_{ij}^{2}u_{ii}u_{jj}\tr^{2}\left(\mathbf{Q}_{ij}\right)
\tr\left(\mathbf{Q}_{ij}\mathbf{Q}_{ji}\right)\\
&+2u_{ij}^{4}\tr^{2}\left(\mathbf{Q}_{ij}^{2}\right)
+2u_{ii}^{2}u_{jj}^{2}\tr^{2}\left(\mathbf{Q}_{ij}\mathbf{Q}_{ji}\right)
+4u_{ij}^{2}u_{ii}u_{jj}\tr\left(\mathbf{Q}_{ij}^{2}\right)
\tr\left(\mathbf{Q}_{ij}\mathbf{Q}_{ji}\right)\\
&+8u_{ij}^{4}\tr\left(\mathbf{Q}_{ij}\right)
\tr\left(\mathbf{Q}_{ij}^{3}\right)
+24u_{ii}u_{jj}u_{ij}^{2}\tr\left(\mathbf{Q}_{ij}\right)
\tr\left(\mathbf{Q}_{ij}^{2}\mathbf{Q}_{ji}\right)
+6u_{ij}^{4}\tr\left(\mathbf{Q}_{ij}^{4}\right)\\
&+24u_{ij}^{2}u_{ii}u_{jj}\tr\left(\mathbf{Q}_{ij}^{3}\mathbf{Q}_{ji}\right)
+12u_{ij}^{2}u_{ii}u_{jj}\tr\left(\mathbf{Q}_{ij}^{2}\mathbf{Q}_{ji}^{2}\right)
+6u_{ii}^{2}u_{jj}^{2}\tr\left(\mathbf{Q}_{ij}\mathbf{Q}_{ji}\mathbf{Q}_{ij}\mathbf{Q}_{ji}\right),\\
G_{2_{2}}=&2\underset{i\neq j}{\sum^{N}\sum^{N}}
4u_{ij}^{2}\tr^{2}\left(\mathbf{Q}_{ij}\right)\tr\left\{\mathbf{\Omega}_{ij}\circ \mathbf{\Omega}_{ij}\right\}\tr\left(\mathbf{Q}_{ij}\circ\mathbf{Q}_{ij}\right)\\
&+8\Big\{\frac{1}{2}u_{ij}^{2}\tr\left(\mathbf{Q}_{ij}^{2}\right)
+\frac{1}{2}u_{ii}u_{jj}\tr\left(\mathbf{Q}_{ij}\mathbf{Q}_{ji}\right)\Big\}
\tr\big(\mathbf{\Omega}_{ij}\circ \mathbf{\Omega}_{ij}\big)\tr\left(\mathbf{Q}_{ij}\circ\mathbf{Q}_{ij}\right)\\
&+48u_{ij}\tr\left(\mathbf{Q}_{ij}\right)\Big[\frac{1}{2}u_{ij}
\tr\left(\mathbf{\Omega}_{ij}\circ\mathbf{\Omega}_{ij}\right)
\tr\left(\mathbf{Q}_{ij}\circ\mathbf{Q}_{ij}^{2}\right)\\
&\quad+
\frac{1}{4}u_{jj}
\tr\left(\mathbf{\Omega}_{ij}\circ\mathbf{\Omega}_{ii}\right)
\tr\left\{\mathbf{Q}_{ij}\circ(\mathbf{Q}_{ij}\mathbf{Q}_{ji})\right\}\\
&\quad+
\frac{1}{4}u_{ii}
\tr\left(\mathbf{\Omega}_{ij}\circ\mathbf{\Omega}_{jj}\right)
\tr\left\{\mathbf{Q}_{ij}\circ(\mathbf{Q}_{ji}\mathbf{Q}_{ij})\right\}
\Big]\\
&+24\Big[u_{ii}u_{jj}
\tr\left\{\left(\mathbf{I}_{N}\circ\mathbf{\Omega}_{ij}\right)\mathbf{\Omega}_{ji}\right\}
\tr\left\{\left(\mathbf{I}_{T}\circ\mathbf{Q}_{ij}\right)\mathbf{Q}_{ji}
\mathbf{Q}_{ij}\mathbf{Q}_{ji}\right\}\\
&\quad+u_{ij}^{2}\tr\left\{\left(\mathbf{I}_{N}\circ
\mathbf{\Omega}_{ij}\right)\mathbf{\Omega}_{ij}\right\}
\tr\left\{\left(\mathbf{I}_{T}\circ\mathbf{Q}_{ij}\right)\mathbf{Q}_{ij}^{3}\right\}\\
&\quad+u_{ij}u_{jj}
tr\left\{\left(\mathbf{I}_{N}\circ
\mathbf{\Omega}_{ij}\right)\mathbf{\Omega}_{ii}\right\}
\tr\left\{\left(\mathbf{I}_{T}\circ\mathbf{Q}_{ij}\right)\mathbf{Q}_{ij}
\mathbf{Q}_{ij}\mathbf{Q}_{ji}\right\}\\
&\quad+u_{ij}u_{ii}
tr\left\{\left(\mathbf{I}_{N}\circ
\mathbf{\Omega}_{ij}\right)\mathbf{\Omega}_{jj}\right\}
\tr\left\{\left(\mathbf{I}_{T}\circ\mathbf{Q}_{ij}\right)\mathbf{Q}_{ji}
\mathbf{Q}_{ji}\mathbf{Q}_{ij}\right\}\Big]\\
&+3\Big[4
\tau_{N}^{\prime}\left(\mathbf{I}_{N}\circ u_{ij}\mathbf{\Omega}_{ij}\right)^{2}\tau_{N}
\tau_{T}^{\prime}\left\{\mathbf{I}_{T}\circ (\mathbf{Q}_{ij}\mathbf{Q}_{ij})\right\}^{2}\tau_{T}\\
&\quad+
\tau_{N}^{\prime}\left(\mathbf{I}_{N}\circ u_{jj}\mathbf{\Omega}_{ii}\right)^{2}\tau_{N}
\tau_{T}^{\prime}\left\{\mathbf{I}_{T}\circ (\mathbf{Q}_{ij}\mathbf{Q}_{ji})\right\}^{2}\tau_{T}\\
&\quad+
\tau_{N}^{\prime}\left(\mathbf{I}_{N}\circ u_{ii}\mathbf{\Omega}_{jj}\right)^{2}\tau_{N}
\tau_{T}^{\prime}\left\{\mathbf{I}_{T}\circ (\mathbf{Q}_{ji}\mathbf{Q}_{ij})\right\}^{2}\tau_{T}\\
&\quad+4\tau_{N}^{\prime}\left(\mathbf{I}_{N}\circ u_{ij}\mathbf{\Omega}_{ij}\right)\left(\mathbf{I}_{N}\circ u_{jj}\mathbf{\Omega}_{ii}\right)\tau_{N}
\tau_{T}^{\prime}\left(\mathbf{I}_{T}\circ \mathbf{Q}_{ij}^{2}\right)
\left\{\mathbf{I}_{T}\circ (\mathbf{Q}_{ij}\mathbf{Q}_{ji})\right\}\tau_{T}\\
&\quad+4\tau_{N}^{\prime}\left(\mathbf{I}_{N}\circ u_{ij}\mathbf{\Omega}_{ij}\right)\left(\mathbf{I}_{N}\circ u_{ii}\mathbf{\Omega}_{jj}\right)\tau_{N}
\tau_{T}^{\prime}\left(\mathbf{I}_{T}\circ \mathbf{Q}_{ij}^{2}\right)
\left\{\mathbf{I}_{T}\circ (\mathbf{Q}_{ji}\mathbf{Q}_{ij})\right\}\tau_{T}\\
&\quad+2\tau_{N}^{\prime}\left(\mathbf{I}_{N}\circ u_{jj}\mathbf{\Omega}_{ii}\right)\left(\mathbf{I}_{N}\circ u_{ii}\mathbf{\Omega}_{jj}\right)\tau_{N}
\tau_{T}^{\prime}\left\{\mathbf{I}_{T}\circ (\mathbf{Q}_{ij}\mathbf{Q}_{ji})\right\}
\left\{\mathbf{I}_{T}\circ (\mathbf{Q}_{ji}\mathbf{Q}_{ij})\right\}\tau_{T}\Big],\\
G_{2_{3}}=&2\underset{i\neq j}{\sum^{N}\sum^{N}}
4u_{ij}\tr\left(\mathbf{Q}_{ij}\right)
\tr\left(\mathbf{\Omega}_{ij}\circ\mathbf{\Omega}_{ij}
\circ\mathbf{\Omega}_{ij}\right)
\tr\left(\mathbf{Q}_{ij}\circ\mathbf{Q}_{ij}
\circ\mathbf{Q}_{ij}\right)\\
&+24\Big[\frac{1}{2}u_{ij}\tr\left(\mathbf{\Omega}_{ij}
\circ\mathbf{\Omega}_{ij}\circ\mathbf{\Omega}_{ij}\right)
\tr\left(\mathbf{Q}_{ij}
\circ\mathbf{Q}_{ij}\circ\mathbf{Q}^{2}_{ij}\right)\\
&\quad+\frac{1}{4}u_{jj}\tr\left(\mathbf{\Omega}_{ij}
\circ\mathbf{\Omega}_{ij}\circ\mathbf{\Omega}_{ii}\right)
\tr\left\{\mathbf{Q}_{ij}
\circ\mathbf{Q}_{ij}\circ(\mathbf{Q}_{ij}\mathbf{Q}_{ji})\right\}
\\
&\quad+\frac{1}{4}u_{ii}\tr\left(\mathbf{\Omega}_{ij}
\circ\mathbf{\Omega}_{ij}\circ\mathbf{\Omega}_{jj}\right)
\tr\left\{\mathbf{Q}_{ij}
\circ\mathbf{Q}_{ij}\circ(\mathbf{Q}_{ji}\mathbf{Q}_{ij})\right\}
\Big],\\
G_{2_{4}}=&2\underset{i\neq j}{\sum^{N}\sum^{N}}
\tr\left(\mathbf{\Omega}_{ij}\circ
\mathbf{\Omega}_{ij}\circ
\mathbf{\Omega}_{ij}\circ\mathbf{\Omega}_{ij}\right)
\tr\left(\mathbf{Q}_{ij}\circ
\mathbf{Q}_{ij}\circ
\mathbf{Q}_{ij}\circ\mathbf{Q}_{ij}\right),\\
G_{2_{5}}=&2\underset{i\neq j}{\sum^{N}\sum^{N}}
24u_{ij}\tr\left(\mathbf{Q}_{ij}\right)
\tau_{N}^{\prime}\left(\mathbf{I}_{N}\circ\mathbf{\Omega}_{ij}\right)
\mathbf{\Omega}_{ij}\left(\mathbf{I}_{N}\circ\mathbf{\Omega}_{ij}\right)\tau_{N}
\tau_{T}^{\prime}\left(\mathbf{I}_{T}\circ\mathbf{Q}_{ij}\right)
\mathbf{Q}_{ij}\left(\mathbf{I}_{T}\circ\mathbf{Q}_{ij}\right)\tau_{T}\\
&+12\Big\{2u_{ij}\tau_{N}^{\prime}
\left(\mathbf{I}_{N}\circ\mathbf{\Omega}_{ij}\right)
\mathbf{\Omega}_{ij}\left(\mathbf{I}_{N}\circ\mathbf{\Omega}_{ij}\right)\tau_{N}
\tau_{T}^{\prime}\left(\mathbf{I}_{T}\circ\mathbf{Q}_{ij}\right)
\mathbf{Q}_{ij}^{2}\left(\mathbf{I}_{T}\circ\mathbf{Q}_{ij}\right)\tau_{T}\\
&\quad+u_{jj}\tau_{N}^{\prime}
\left(\mathbf{I}_{N}\circ\mathbf{\Omega}_{ij}\right)
\mathbf{\Omega}_{ii}\left(\mathbf{I}_{N}\circ\mathbf{\Omega}_{ij}\right)\tau_{N}
\tau_{T}^{\prime}\left(\mathbf{I}_{T}\circ\mathbf{Q}_{ij}\right)
\mathbf{Q}_{ij}\mathbf{Q}_{ji}\left(\mathbf{I}_{T}\circ\mathbf{Q}_{ij}\right)\tau_{T}\\
&\quad+
u_{ii}\tau_{N}^{\prime}
\left(\mathbf{I}_{N}\circ\mathbf{\Omega}_{ij}\right)
\mathbf{\Omega}_{jj}\left(\mathbf{I}_{N}\circ\mathbf{\Omega}_{ij}\right)\tau_{N}
\tau_{T}^{\prime}\left(\mathbf{I}_{T}\circ\mathbf{Q}_{ij}\right)
\mathbf{Q}_{ji}\mathbf{Q}_{ij}\left(\mathbf{I}_{T}\circ\mathbf{Q}_{ij}\right)\tau_{T}\Big\}\\
&+16u_{ij}\tr\left(\mathbf{Q}_{ij}\right)\frac{1}{4}
\Big[\tr\left\{\mathbf{\Omega}_{ij}\left(\mathbf{\Omega}_{ij}\circ
\mathbf{\Omega}_{ij}\right)\right\}\tr\left\{\mathbf{Q}_{ij}\left(\mathbf{Q}_{ij}\circ
\mathbf{Q}_{ij}\right)\right\}\\
&\quad+2\tr\left\{\mathbf{\Omega}_{ij}\left(\mathbf{\Omega}_{ij}\circ
\mathbf{\Omega}_{ji}\right)\right\}\tr\left\{\mathbf{Q}_{ij}\left(\mathbf{Q}_{ij}\circ
\mathbf{Q}_{ji}\right)\right\}\\
&\quad+\tr\left\{\mathbf{\Omega}_{ij}\left(\mathbf{\Omega}_{ji}\circ
\mathbf{\Omega}_{ji}\right)\right\}\tr\left\{\mathbf{Q}_{ij}\left(\mathbf{Q}_{ji}\circ
\mathbf{Q}_{ji}\right)\right\}\Big]\\
&+12\Big\{\tau_{N}^{\prime}
\left(\mathbf{\Omega}_{ij}\circ\mathbf{\Omega}_{ij}\right)\mathbf{\Omega}_{ij}
\left(\mathbf{I}_{N}\circ\mathbf{\Omega}_{ij}\right)\tau_{N}
\tau_{T}^{\prime}
\left(\mathbf{Q}_{ij}\circ\mathbf{Q}_{ij}\right)\mathbf{Q}_{ij}
\left(\mathbf{I}_{T}\circ\mathbf{Q}_{ij}\right)\tau_{T}\\
&\quad+\tau_{N}^{\prime}
\left(\mathbf{\Omega}_{ij}\circ\mathbf{\Omega}_{ij}\right)\mathbf{\Omega}_{ji}
\left(\mathbf{I}_{N}\circ\mathbf{\Omega}_{ij}\right)\tau_{N}
\tau_{T}^{\prime}
\left(\mathbf{Q}_{ij}\circ\mathbf{Q}_{ij}\right)\mathbf{Q}_{ji}
\left(\mathbf{I}_{T}\circ\mathbf{Q}_{ij}\right)\tau_{T}\\
&\quad+\tau_{N}^{\prime}
\left(\mathbf{\Omega}_{ij}\circ\mathbf{\Omega}_{ji}\right)\mathbf{\Omega}_{ij}
\left(\mathbf{I}_{N}\circ\mathbf{\Omega}_{ij}\right)\tau_{N}
\tau_{T}^{\prime}
\left(\mathbf{Q}_{ij}\circ\mathbf{Q}_{ji}\right)\mathbf{Q}_{ij}
\left(\mathbf{I}_{T}\circ\mathbf{Q}_{ij}\right)\tau_{T}\\
&\quad+\tau_{N}^{\prime}
\left(\mathbf{\Omega}_{ji}\circ\mathbf{\Omega}_{ij}\right)\mathbf{\Omega}_{ij}
\left(\mathbf{I}_{N}\circ\mathbf{\Omega}_{ij}\right)\tau_{N}
\tau_{T}^{\prime}
\left(\mathbf{Q}_{ji}\circ\mathbf{Q}_{ij}\right)\mathbf{Q}_{ij}
\left(\mathbf{I}_{T}\circ\mathbf{Q}_{ij}\right)\tau_{T}\\
&\quad+\tau_{N}^{\prime}
\left(\mathbf{\Omega}_{ij}\circ\mathbf{\Omega}_{ji}\right)\mathbf{\Omega}_{ji}
\left(\mathbf{I}_{N}\circ\mathbf{\Omega}_{ij}\right)\tau_{N}
\tau_{T}^{\prime}
\left(\mathbf{Q}_{ij}\circ\mathbf{Q}_{ji}\right)\mathbf{Q}_{ji}
\left(\mathbf{I}_{T}\circ\mathbf{Q}_{ij}\right)\tau_{T}
\\
&\quad+\tau_{N}^{\prime}
\left(\mathbf{\Omega}_{ji}\circ\mathbf{\Omega}_{ij}\right)\mathbf{\Omega}_{ji}
\left(\mathbf{I}_{N}\circ\mathbf{\Omega}_{ij}\right)\tau_{N}
\tau_{T}^{\prime}
\left(\mathbf{Q}_{ji}\circ\mathbf{Q}_{ij}\right)\mathbf{Q}_{ji}
\left(\mathbf{I}_{T}\circ\mathbf{Q}_{ij}\right)\tau_{T}
\\
&\quad+\tau_{N}^{\prime}
\left(\mathbf{\Omega}_{ji}\circ\mathbf{\Omega}_{ji}\right)\mathbf{\Omega}_{ij}
\left(\mathbf{I}_{N}\circ\mathbf{\Omega}_{ij}\right)\tau_{N}
\tau_{T}^{\prime}
\left(\mathbf{Q}_{ji}\circ\mathbf{Q}_{ji}\right)\mathbf{Q}_{ij}
\left(\mathbf{I}_{T}\circ\mathbf{Q}_{ij}\right)\tau_{T}
\\
&\quad+\tau_{N}^{\prime}
\left(\mathbf{\Omega}_{ji}\circ\mathbf{\Omega}_{ji}\right)\mathbf{\Omega}_{ji}
\left(\mathbf{I}_{N}\circ\mathbf{\Omega}_{ij}\right)\tau_{N}
\tau_{T}^{\prime}
\left(\mathbf{Q}_{ji}\circ\mathbf{Q}_{ji}\right)\mathbf{Q}_{ji}
\left(\mathbf{I}_{T}\circ\mathbf{Q}_{ij}\right)\tau_{T}
\Big\}\\
&+12\Big[
\tr\left\{\left(\mathbf{\Omega}_{ij}\circ\mathbf{\Omega}_{ij}\right)u_{ij}
\mathbf{\Omega}_{ij}\right\}\tr\left\{\left(\mathbf{Q}_{ij}\circ\mathbf{Q}_{ij}\right)
\mathbf{Q}_{ij}^{2}\right\}\\
&\quad+\tr\left\{\left(\mathbf{\Omega}_{ij}\circ
\mathbf{\Omega}_{ij}\right)u_{ij}
\mathbf{\Omega}_{ji}\right\}\tr\left\{\left(\mathbf{Q}_{ij}\circ
\mathbf{Q}_{ij}\right)
\mathbf{Q}_{ji}^{2}\right\}\\
&\quad+
\tr\left\{\left(\mathbf{\Omega}_{ij}\circ
\mathbf{\Omega}_{ij}\right)u_{jj}
\mathbf{\Omega}_{ii}\right\}\tr\left\{\left(\mathbf{Q}_{ij}\circ
\mathbf{Q}_{ij}\right)\mathbf{Q}_{ij}
\mathbf{Q}_{ji}\right\}\\
&\quad+
\tr\left\{\left(\mathbf{\Omega}_{ij}\circ
\mathbf{\Omega}_{ij}\right)u_{ii}
\mathbf{\Omega}_{jj}\right\}\tr\left\{\left(\mathbf{Q}_{ij}\circ
\mathbf{Q}_{ij}\right)
\mathbf{Q}_{ji}\mathbf{Q}_{ij}
\right\}\\
&\quad+
\tr\left\{\left(\mathbf{\Omega}_{ij}\circ\mathbf{\Omega}_{ji}\right)u_{ij}
\mathbf{\Omega}_{ij}\right\}\tr\left\{\left(\mathbf{Q}_{ij}\circ\mathbf{Q}_{ji}\right)
\mathbf{Q}_{ij}^{2}\right\}
\\
&\quad+\tr\left\{\left(\mathbf{\Omega}_{ij}\circ
\mathbf{\Omega}_{ji}\right)u_{ij}
\mathbf{\Omega}_{ji}\right\}\tr\left\{\left(\mathbf{Q}_{ij}\circ
\mathbf{Q}_{ji}\right)
\mathbf{Q}_{ji}^{2}\right\}\\
&\quad+
\tr\left\{\left(\mathbf{\Omega}_{ij}\circ
\mathbf{\Omega}_{ji}\right)u_{jj}
\mathbf{\Omega}_{ii}\right\}\tr\left\{\left(\mathbf{Q}_{ij}\circ
\mathbf{Q}_{ji}\right)\mathbf{Q}_{ij}
\mathbf{Q}_{ji}\right\}\\
&\quad+
\tr\left\{\left(\mathbf{\Omega}_{ij}\circ
\mathbf{\Omega}_{ji}\right)u_{ii}
\mathbf{\Omega}_{jj}\right\}\tr\left\{\left(\mathbf{Q}_{ij}\circ
\mathbf{Q}_{ji}\right)
\mathbf{Q}_{ji}\mathbf{Q}_{ij}
\right\}\Big],\\
G_{2_{6}}=&2\underset{i\neq j}{\sum^{N}\sum^{N}}
2\tr^{2}\left(\mathbf{\Omega}_{ij}\circ
\mathbf{\Omega}_{ij}\right)\tr^{2}\left(\mathbf{Q}_{ij}\circ
\mathbf{Q}_{ij}\right)\\
&+12\Big\{\tau_{N}^{\prime}
\left(\mathbf{I}_{N}\circ\mathbf{\Omega}_{ij}\right)
\left(\mathbf{\Omega}_{ij}\circ\mathbf{\Omega}_{ij}\right)
\left(\mathbf{I}_{N}\circ\mathbf{\Omega}_{ij}\right)\tau_{N}
\tau_{T}^{\prime}
\left(\mathbf{I}_{T}\circ\mathbf{Q}_{ij}\right)
\left(\mathbf{Q}_{ij}\circ\mathbf{Q}_{ij}\right)
\left(\mathbf{I}_{T}\circ\mathbf{Q}_{ij}\right)\tau_{T}\\
&\quad+
\tau_{N}^{\prime}
\left(\mathbf{I}_{N}\circ\mathbf{\Omega}_{ij}\right)
\left(\mathbf{\Omega}_{ij}\circ\mathbf{\Omega}_{ji}\right)
\left(\mathbf{I}_{N}\circ\mathbf{\Omega}_{ij}\right)\tau_{N}
\tau_{T}^{\prime}
\left(\mathbf{I}_{T}\circ\mathbf{Q}_{ij}\right)
\left(\mathbf{Q}_{ij}\circ\mathbf{Q}_{ji}\right)
\left(\mathbf{I}_{T}\circ\mathbf{Q}_{ij}\right)\tau_{T}
\Big\}\\
&+\frac{1}{2}\big\{
\tau_{N}^{\prime}\left(\mathbf{\Omega}_{ij}\circ
\mathbf{\Omega}_{ij}\circ
\mathbf{\Omega}_{ij}\circ
\mathbf{\Omega}_{ij}\right)\tau_{N}
\tau_{T}^{\prime}\left(\mathbf{Q}_{ij}\circ
\mathbf{Q}_{ij}\circ
\mathbf{Q}_{ij}\circ
\mathbf{Q}_{ij}\right)\tau_{T}\\
&\quad+
\tau_{N}^{\prime}\left(\mathbf{\Omega}_{ji}\circ
\mathbf{\Omega}_{ji}\circ
\mathbf{\Omega}_{ji}\circ
\mathbf{\Omega}_{ji}\right)\tau_{N}
\tau_{T}^{\prime}\left(\mathbf{Q}_{ji}\circ
\mathbf{Q}_{ji}\circ
\mathbf{Q}_{ji}\circ
\mathbf{Q}_{ji}\right)\tau_{T}\\
&\quad+4
\tau_{N}^{\prime}\left(\mathbf{\Omega}_{ij}\circ
\mathbf{\Omega}_{ij}\circ
\mathbf{\Omega}_{ij}\circ
\mathbf{\Omega}_{ji}\right)\tau_{N}
\tau_{T}^{\prime}\left(\mathbf{Q}_{ij}\circ
\mathbf{Q}_{ij}\circ
\mathbf{Q}_{ij}\circ
\mathbf{Q}_{ji}\right)\tau_{T}
\\
&\quad+4
\tau_{N}^{\prime}\left(\mathbf{\Omega}_{ij}\circ
\mathbf{\Omega}_{ji}\circ
\mathbf{\Omega}_{ji}\circ
\mathbf{\Omega}_{ji}\right)\tau_{N}
\tau_{T}^{\prime}\left(\mathbf{Q}_{ij}\circ
\mathbf{Q}_{ji}\circ
\mathbf{Q}_{ji}\circ
\mathbf{Q}_{ji}\right)\tau_{T}\\
&\quad+6
\tau_{N}^{\prime}\left(\mathbf{\Omega}_{ij}\circ
\mathbf{\Omega}_{ij}\circ
\mathbf{\Omega}_{ji}\circ
\mathbf{\Omega}_{ji}\right)\tau_{N}
\tau_{T}^{\prime}\left(\mathbf{Q}_{ij}\circ
\mathbf{Q}_{ij}\circ
\mathbf{Q}_{ji}\circ
\mathbf{Q}_{ji}\right)\tau_{T}
\big\},
\\
G_{2_{7}}=&2\underset{i\neq j}{\sum^{N}\sum^{N}}
12\Big\{\tau_{N}^{\prime}
\left(\mathbf{I}_{N}\circ\mathbf{\Omega}_{ij}\right)\mathbf{\Omega}_{ij}
\left(\mathbf{I}_{N}\circ\mathbf{\Omega}_{ij}\circ\mathbf{\Omega}_{ij}\right)\tau_{N}
\tau_{T}^{\prime}
\left(\mathbf{I}_{T}\circ\mathbf{Q}_{ij}\right)\mathbf{Q}_{ij}
\left(\mathbf{I}_{T}\circ\mathbf{Q}_{ij}\circ\mathbf{Q}_{ij}\right)\tau_{T}\\
&\quad+\tau_{N}^{\prime}
\left(\mathbf{I}_{N}\circ\mathbf{\Omega}_{ij}\right)\mathbf{\Omega}_{ji}
\left(\mathbf{I}_{N}\circ\mathbf{\Omega}_{ij}\circ\mathbf{\Omega}_{ij}\right)\tau_{N}
\tau_{T}^{\prime}
\left(\mathbf{I}_{T}\circ\mathbf{Q}_{ij}\right)\mathbf{Q}_{ji}
\left(\mathbf{I}_{T}\circ\mathbf{Q}_{ij}\circ\mathbf{Q}_{ij}\right)\tau_{T}
\Big\}\\
&+4\Big\{\tau_{N}^{\prime}
\left(\mathbf{I}_{N}\circ\mathbf{\Omega}_{ij}\right)\left(\mathbf{\Omega}_{ji}\circ
\mathbf{\Omega}_{ij}\circ\mathbf{\Omega}_{ji}\right)\tau_{N}
\tau_{T}^{\prime}
\left(\mathbf{I}_{T}\circ\mathbf{Q}_{ij}\right)\left(\mathbf{Q}_{ji}\circ
\mathbf{Q}_{ij}\circ\mathbf{Q}_{ji}\right)\tau_{T}\\
&\quad+\tau_{N}^{\prime}
\left(\mathbf{I}_{N}\circ\mathbf{\Omega}_{ij}\right)\left(\mathbf{\Omega}_{ij}\circ
\mathbf{\Omega}_{ij}\circ\mathbf{\Omega}_{ij}\right)\tau_{N}
\tau_{T}^{\prime}
\left(\mathbf{I}_{T}\circ\mathbf{Q}_{ij}\right)\left(\mathbf{Q}_{ij}\circ
\mathbf{Q}_{ij}\circ\mathbf{Q}_{ij}\right)\tau_{T}
\\
&\quad+\tau_{N}^{\prime}
\left(\mathbf{I}_{N}\circ\mathbf{\Omega}_{ij}\right)\left(\mathbf{\Omega}_{ij}\circ
\mathbf{\Omega}_{ij}\circ\mathbf{\Omega}_{ji}\right)\tau_{N}
\tau_{T}^{\prime}
\left(\mathbf{I}_{T}\circ\mathbf{Q}_{ij}\right)\left(\mathbf{Q}_{ij}\circ
\mathbf{Q}_{ij}\circ\mathbf{Q}_{ji}\right)\tau_{T}
\\
&\quad+\tau_{N}^{\prime}
\left(\mathbf{I}_{N}\circ\mathbf{\Omega}_{ij}\right)\left(\mathbf{\Omega}_{ji}\circ
\mathbf{\Omega}_{ji}\circ\mathbf{\Omega}_{ij}\right)\tau_{N}
\tau_{T}^{\prime}
\left(\mathbf{I}_{T}\circ\mathbf{Q}_{ij}\right)\left(\mathbf{Q}_{ji}\circ
\mathbf{Q}_{ji}\circ\mathbf{Q}_{ij}\right)\tau_{T}
\\
&\quad+\tau_{N}^{\prime}
\left(\mathbf{I}_{N}\circ\mathbf{\Omega}_{ij}\right)\left(\mathbf{\Omega}_{ji}\circ
\mathbf{\Omega}_{ji}\circ\mathbf{\Omega}_{ji}\right)\tau_{N}
\tau_{T}^{\prime}
\left(\mathbf{I}_{T}\circ\mathbf{Q}_{ij}\right)\left(\mathbf{Q}_{ji}\circ
\mathbf{Q}_{ji}\circ\mathbf{Q}_{ji}\right)\tau_{T}
\\
&\quad+\tau_{N}^{\prime}
\left(\mathbf{I}_{N}\circ\mathbf{\Omega}_{ij}\right)\left(\mathbf{\Omega}_{ij}\circ
\mathbf{\Omega}_{ji}\circ\mathbf{\Omega}_{ij}\right)\tau_{N}
\tau_{T}^{\prime}
\left(\mathbf{I}_{T}\circ\mathbf{Q}_{ij}\right)\left(\mathbf{Q}_{ij}\circ
\mathbf{Q}_{ji}\circ\mathbf{Q}_{ij}\right)\tau_{T}
\\
&\quad+\tau_{N}^{\prime}
\left(\mathbf{I}_{N}\circ\mathbf{\Omega}_{ij}\right)\left(\mathbf{\Omega}_{ij}\circ
\mathbf{\Omega}_{ji}\circ\mathbf{\Omega}_{ji}\right)\tau_{N}
\tau_{T}^{\prime}
\left(\mathbf{I}_{T}\circ\mathbf{Q}_{ij}\right)\left(\mathbf{Q}_{ij}\circ
\mathbf{Q}_{ji}\circ\mathbf{Q}_{ji}\right)\tau_{T}
\\
&\quad+\tau_{N}^{\prime}
\left(\mathbf{I}_{N}\circ\mathbf{\Omega}_{ij}\right)\left(\mathbf{\Omega}_{ji}\circ
\mathbf{\Omega}_{ij}\circ\mathbf{\Omega}_{ij}\right)\tau_{N}
\tau_{T}^{\prime}
\left(\mathbf{I}_{T}\circ\mathbf{Q}_{ij}\right)\left(\mathbf{Q}_{ji}\circ
\mathbf{Q}_{ij}\circ\mathbf{Q}_{ij}\right)\tau_{T}
\Big\}.
\end{align*}
Because $\underset{i\neq j}{\sum\sum}u_{ij}^{4}<\underset{i\neq j}{\sum\sum}u_{ij}^{2}<\tr\left(\mathbf{U}^{2}\right)=O(N),$
we have $G_{2_{1}}=O(NT^{3})+O(N^{2}T^{2}).$
Similarly,
due to $\underset{i\neq j}{\sum\sum}|u_{ij}|=O(N^{1\vee(2-1/\tau)})$
and Lemma \ref{lem:sigma},
we have $G_{2_{2}}=O(NT^{3})+O(N^{2}T^{2}),$
$G_{2_{3}}=O(N^{3/2}T^{2})+O(N^{2}T),$
$G_{2_{4}}=O(N^{2}T),$
$G_{2_{5}}=O(N^{2}T^{2}),$
$G_{2_{6}}=O(N^{2}T^{2})$
and $G_{2_{7}}=O(N^{2}T^{3/2}).$
Combining the above calculations, we can get $G_{2}= O(NT^{3}+N^{2}T^{2}$. We further deal with $G_{3}$.
Similarly, due to Lemma \ref{le:moment of quadratic form}, we have
$$
G_{3}=G_{3_{1}}+\gamma_{2}G_{3_{2}}+\gamma_{4}G_{3_{3}}+\gamma_{6}G_{3_{4}}
+\gamma_{1}^{2}G_{3_{5}}+\gamma_{2}^{2}G_{3_{6}}+\gamma_{1}\gamma_{3}G_{3_{7}},$$
where
\begin{align*}
G_{3_{1}}=&\underset{i\neq j}{\sum^{N}\sum^{N}}8 u_{ii}u_{jj}u_{ij}^{2}\tr\left(\mathbf{Q}_{ii}\right)
\tr\left(\mathbf{Q}_{jj}\right)\tr\left(\mathbf{Q}_{ii}\mathbf{Q}_{jj}\right)
+8u_{ij}^{4}\tr^{2}\left(\mathbf{Q}_{ii}\mathbf{Q}_{jj}\right)\\
&+16u_{ii}u_{jj}u_{ij}^{2}\tr\left(\mathbf{Q}_{ii}\mathbf{Q}_{jj}^{2}\right)
\tr\left(\mathbf{Q}_{ii}\right)+16u_{jj}u_{ii}u_{ij}^{2}
\tr\left(\mathbf{Q}_{ii}^{2}\mathbf{Q}_{jj}\right)
\tr\left(\mathbf{Q}_{jj}\right)\\
&+32u_{ii}u_{jj}u_{ij}^{2}\tr\left(\mathbf{Q}_{ii}^{2}\mathbf{Q}_{jj}^{2}\right)
+16u_{ij}^{4}\tr\left(\mathbf{Q}_{ii}\mathbf{Q}_{jj}\mathbf{Q}_{ii}
\mathbf{Q}_{jj}\right),\\
G_{3_{2}}=&\underset{i\neq j}{\sum^{N}\sum^{N}}
4
u_{ii}u_{jj}\tr\left(\mathbf{Q}_{ii}\right)\tr\left(\mathbf{Q}_{jj}\right)
\tr\left(\mathbf{\Omega}_{ii}\circ\mathbf{\Omega}_{jj}\right)
\tr\left(\mathbf{Q}_{ii}\circ\mathbf{Q}_{jj}\right)\\
&+8\tau_{N}^{\prime}
\left(\mathbf{\Omega}_{ii}\circ\mathbf{\Omega}_{jj}\right)\tau_{N}
\tau_{T}^{\prime}
\left(\mathbf{Q}_{ii}\circ\mathbf{Q}_{jj}\right)\tau_{T}
\tr\left(\mathbf{\Omega}_{ii}\circ\mathbf{\Omega}_{jj}\right)
\tr\left(\mathbf{Q}_{ii}\circ\mathbf{Q}_{jj}\right)\\
&+4\Big[
2u_{ii}\tr\left(\mathbf{Q}_{ii}\right)\tr\left(\mathbf{\Omega}_{ii}\circ
u_{jj}\mathbf{\Omega}_{jj}\right)
\tr\left(\mathbf{Q}_{ii}\circ\mathbf{Q}_{jj}^{2}\right)\\
&\quad+4u_{ii}\tr\left(\mathbf{Q}_{ii}\right)\tr\left(\mathbf{\Omega}_{jj}\circ
u_{ij}\mathbf{\Omega}_{ij}\right)
\tr\left\{\mathbf{Q}_{jj}\circ(\mathbf{Q}_{ii}\mathbf{Q}_{jj})\right\}\\
&\quad+2u_{jj}\tr\left(\mathbf{Q}_{jj}\right)\tr\left(\mathbf{\Omega}_{jj}\circ
u_{ii}\mathbf{\Omega}_{ii}\right)
\tr\left(\mathbf{Q}_{jj}\circ\mathbf{Q}_{ii}^{2}\right)\\
&\quad+4u_{jj}\tr\left(\mathbf{Q}_{jj}\right)
\tr\left(\mathbf{\Omega}_{ii}\circ
u_{ij}\mathbf{\Omega}_{ij}\right)
\tr\left\{\mathbf{Q}_{ii}\circ(\mathbf{Q}_{ii}\mathbf{Q}_{jj})\right\}\Big]\\
&+8\Big[4\tr\left\{\left(\mathbf{I}_{N}\circ\mathbf{\Omega}_{ii}\right)
u_{ij}u_{jj}\mathbf{\Omega}_{ij}\right\}\tr\left\{\left(\mathbf{I}_{T}
\circ\mathbf{Q}_{ii}\right)
\mathbf{Q}_{ii}\mathbf{Q}_{jj}^{2}\right\}\\
&\quad+2
\tr\left\{\left(\mathbf{I}_{N}\circ\mathbf{\Omega}_{ii}\right)
u_{ij}^{2}\mathbf{\Omega}_{jj}\right\}\tr\left\{\left(\mathbf{I}_{T}
\circ\mathbf{Q}_{ii}\right)
\mathbf{Q}_{jj}\mathbf{Q}_{ii}\mathbf{Q}_{jj}\right\}\\
&\quad+4
\tr\left\{\left(\mathbf{I}_{N}\circ\mathbf{\Omega}_{jj}\right)
u_{ij}u_{ii}\mathbf{\Omega}_{ij}\right\}\tr\left\{\left(\mathbf{I}_{T}
\circ\mathbf{Q}_{jj}\right)
\mathbf{Q}_{ii}^{2}\mathbf{Q}_{jj}\right\}\\
&\quad+2
\tr\left\{\left(\mathbf{I}_{N}\circ\mathbf{\Omega}_{jj}\right)
u_{ij}^{2}\mathbf{\Omega}_{ii}\right\}\tr\left\{\left(\mathbf{I}_{T}
\circ\mathbf{Q}_{jj}\right)
\mathbf{Q}_{ii}\mathbf{Q}_{jj}\mathbf{Q}_{ii}\right\}\Big]\\
&+16\Big[\tau_{N}^{\prime}
\left(\mathbf{I}_{N}\circ u_{ii}\mathbf{\Omega}_{ii}\right)
\left(\mathbf{I}_{N}\circ u_{jj}\mathbf{\Omega}_{jj}\right)\tau_{N}
\tau_{T}^{\prime}
\left(\mathbf{I}_{T}\circ \mathbf{Q}_{ii}^{2}\right)
\left(\mathbf{I}_{T}\circ \mathbf{Q}_{jj}^{2}\right)\tau_{T}\\
&\quad+2\tau_{N}^{\prime}
\left(\mathbf{I}_{N}\circ u_{ij}\mathbf{\Omega}_{ij}\right)
\left(\mathbf{I}_{N}\circ u_{ij}\mathbf{\Omega}_{ij}\right)\tau_{N}
\tau_{T}^{\prime}
\left\{\mathbf{I}_{T}\circ (\mathbf{Q}_{ii}\mathbf{Q}_{jj})\right\}
\left\{\mathbf{I}_{T}\circ (\mathbf{Q}_{ii}\mathbf{Q}_{jj})\right\}\tau_{T}
\Big],\\
G_{3_{3}}=&\underset{i\neq j}{\sum^{N}\sum^{N}}
2u_{ii}\tr\left(\mathbf{Q}_{ii}\right)
\tr\left(\mathbf{\Omega}_{ii}\circ\mathbf{\Omega}_{jj}\circ\mathbf{\Omega}_{jj}
\right)\tr\left(\mathbf{Q}_{ii}\circ\mathbf{Q}_{jj}\circ\mathbf{Q}_{jj}
\right)\\
&+2u_{jj}\tr\left(\mathbf{Q}_{jj}\right)
\tr\left(\mathbf{\Omega}_{ii}\circ\mathbf{\Omega}_{ii}\circ\mathbf{\Omega}_{jj}
\right)\tr\left(\mathbf{Q}_{ii}\circ\mathbf{Q}_{ii}\circ\mathbf{Q}_{jj}
\right)\\
&+4\Big[
\tr\left(\mathbf{\Omega}_{ii}\circ\mathbf{\Omega}_{ii}\circ u_{jj}\mathbf{\Omega}_{jj}\right)
\tr\left(\mathbf{Q}_{ii}\circ\mathbf{Q}_{ii}\circ\mathbf{Q}_{jj}^{2}
\right)\\
&\quad+4\tr\left(\mathbf{\Omega}_{ii}\circ\mathbf{\Omega}_{jj}\circ u_{ij}\mathbf{\Omega}_{ij}\right)
\tr\left\{\mathbf{Q}_{ii}\circ\mathbf{Q}_{jj}\circ(\mathbf{Q}_{ii}\mathbf{Q}_{jj})
\right\}\\
&\quad+\tr\left(\mathbf{\Omega}_{jj}\circ\mathbf{\Omega}_{jj}\circ u_{ii}\mathbf{\Omega}_{ii}\right)
\tr\left(\mathbf{Q}_{jj}\circ\mathbf{Q}_{jj}\circ\mathbf{Q}_{ii}^{2}
\right)\Big],\\
G_{3_{4}}=&\underset{i\neq j}{\sum^{N}\sum^{N}}
\tr\left(\mathbf{\Omega}_{ii}\circ\mathbf{\Omega}_{ii}\circ
\mathbf{\Omega}_{jj}\circ\mathbf{\Omega}_{jj}\right)
\tr\left(\mathbf{Q}_{ii}\circ\mathbf{Q}_{ii}\circ
\mathbf{Q}_{jj}\circ\mathbf{Q}_{jj}\right),\\
G_{3_{5}}=&\underset{i\neq j}{\sum^{N}\sum^{N}}
2\Big\{4u_{ii}\tr\left(\mathbf{Q}_{ii}\right)
\tau_{N}^{\prime}\left(\mathbf{I}_{N}\circ \mathbf{\Omega}_{ii}\right)\mathbf{\Omega}_{jj}
\left(\mathbf{I}_{N}\circ \mathbf{\Omega}_{jj}\right)\tau_{N}
\tau_{T}^{\prime}\left(\mathbf{I}_{T}\circ \mathbf{Q}_{ii}\right)\mathbf{Q}_{jj}
\left(\mathbf{I}_{T}\circ \mathbf{Q}_{jj}\right)\tau_{T}\\
&\quad+2u_{ii}\tr\left(\mathbf{Q}_{ii}\right)
\tau_{N}^{\prime}\left(\mathbf{I}_{N}\circ \mathbf{\Omega}_{jj}\right)\mathbf{\Omega}_{ii}
\left(\mathbf{I}_{N}\circ \mathbf{\Omega}_{jj}\right)\tau_{N}
\tau_{T}^{\prime}\left(\mathbf{I}_{T}\circ \mathbf{Q}_{jj}\right)\mathbf{Q}_{ii}
\left(\mathbf{I}_{T}\circ \mathbf{Q}_{jj}\right)\tau_{T}
\\
&\quad+4u_{jj}\tr\left(\mathbf{Q}_{jj}\right)
\tau_{N}^{\prime}\left(\mathbf{I}_{N}\circ \mathbf{\Omega}_{ii}\right)\mathbf{\Omega}_{ii}
\left(\mathbf{I}_{N}\circ \mathbf{\Omega}_{jj}\right)\tau_{N}
\tau_{T}^{\prime}\left(\mathbf{I}_{T}\circ \mathbf{Q}_{ii}\right)\mathbf{Q}_{ii}
\left(\mathbf{I}_{T}\circ \mathbf{Q}_{jj}\right)\tau_{T}
\\
&\quad+2u_{jj}\tr\left(\mathbf{Q}_{jj}\right)
\tau_{N}^{\prime}\left(\mathbf{I}_{N}\circ \mathbf{\Omega}_{ii}\right)\mathbf{\Omega}_{jj}
\left(\mathbf{I}_{N}\circ \mathbf{\Omega}_{ii}\right)\tau_{N}
\tau_{T}^{\prime}\left(\mathbf{I}_{T}\circ \mathbf{Q}_{ii}\right)\mathbf{Q}_{jj}
\left(\mathbf{I}_{T}\circ \mathbf{Q}_{ii}\right)\tau_{T}\Big\}\\
&+4\Big\{4
\tau_{N}^{\prime}\left\{\left(\mathbf{I}_{N}\circ\mathbf{\Omega}_{ii}\right)
u_{ij}\mathbf{\Omega}_{ij}
\left\{\mathbf{I}_{N}\circ\mathbf{\Omega}_{jj}\right\}\right\}\tau_{N}
\tau_{T}^{\prime}\left\{\left(\mathbf{I}_{N}\circ\mathbf{Q}_{ii}\right)
\mathbf{Q}_{ii}\mathbf{Q}_{jj}
\left(\mathbf{I}_{N}\circ\mathbf{Q}_{jj}\right)\right\}\tau_{T}\\
&\quad+2
\tau_{N}^{\prime}\left\{\left(\mathbf{I}_{N}\circ\mathbf{\Omega}_{ii}\right)
u_{jj}\mathbf{\Omega}_{jj}
\left\{\mathbf{I}_{N}\circ\mathbf{\Omega}_{ii}\right\}\right\}\tau_{N}
\tau_{T}^{\prime}\left\{\left(\mathbf{I}_{N}\circ\mathbf{Q}_{ii}\right)
\mathbf{Q}_{jj}^{2}
\left(\mathbf{I}_{N}\circ\mathbf{Q}_{ii}\right)\right\}\tau_{T}\\
&\quad+2
\tau_{N}^{\prime}\left\{\left(\mathbf{I}_{N}\circ\mathbf{\Omega}_{jj}\right)
u_{ii}\mathbf{\Omega}_{ii}
\left\{\mathbf{I}_{N}\circ\mathbf{\Omega}_{jj}\right\}\right\}\tau_{N}
\tau_{T}^{\prime}\left\{\left(\mathbf{I}_{N}\circ\mathbf{Q}_{jj}\right)
\mathbf{Q}_{ii}^{2}
\left(\mathbf{I}_{N}\circ\mathbf{Q}_{jj}\right)\right\}\tau_{T}\\
&\quad+4
\tau_{N}^{\prime}\left\{\left(\mathbf{I}_{N}\circ\mathbf{\Omega}_{jj}\right)
u_{ij}\mathbf{\Omega}_{ij}
\left\{\mathbf{I}_{N}\circ\mathbf{\Omega}_{ii}\right\}\right\}\tau_{N}
\tau_{T}^{\prime}\left\{\left(\mathbf{I}_{N}\circ\mathbf{Q}_{jj}\right)
\mathbf{Q}_{ii}\mathbf{Q}_{jj}
\left(\mathbf{I}_{N}\circ\mathbf{Q}_{ii}\right)\right\}\tau_{T}\Big\}\\
&+4\Big\{2u_{ii}\tr\left(\mathbf{Q}_{ii}\right)
\tau_{N}^{\prime}\left(\mathbf{\Omega}_{ii}\circ\mathbf{\Omega}_{jj}
\circ\mathbf{\Omega}_{jj}\right)\tau_{N}
\tau_{T}^{\prime}\left(\mathbf{Q}_{ii}\circ\mathbf{Q}_{jj}
\circ\mathbf{Q}_{jj}\right)\tau_{T}\\
&\quad+2u_{jj}\tr\left(\mathbf{Q}_{jj}\right)
\tau_{N}^{\prime}\left(\mathbf{\Omega}_{ii}\circ\mathbf{\Omega}_{ii}
\circ\mathbf{\Omega}_{jj}\right)\tau_{N}
\tau_{T}^{\prime}\left(\mathbf{Q}_{ii}\circ\mathbf{Q}_{ii}
\circ\mathbf{Q}_{jj}\right)\tau_{T}\Big\}\\
&+8\Big\{2
\tau_{N}^{\prime}\left(\mathbf{\Omega}_{ii}\circ\mathbf{\Omega}_{ii}\right)
\mathbf{\Omega}_{jj}\left(\mathbf{I}_{N}\circ\mathbf{\Omega}_{jj}\right)\tau_{N}
\tau_{T}^{\prime}\left(\mathbf{Q}_{ii}\circ\mathbf{Q}_{ii}\right)
\mathbf{Q}_{jj}\left(\mathbf{I}_{T}\circ\mathbf{Q}_{jj}\right)\tau_{T}\\
&\quad+4
\tau_{N}^{\prime}\left(\mathbf{\Omega}_{ii}\circ\mathbf{\Omega}_{jj}\right)
\mathbf{\Omega}_{ii}\left(\mathbf{I}_{N}\circ\mathbf{\Omega}_{jj}\right)\tau_{N}
\tau_{T}^{\prime}\left(\mathbf{Q}_{ii}\circ\mathbf{Q}_{jj}\right)
\mathbf{Q}_{ii}\left(\mathbf{I}_{T}\circ\mathbf{Q}_{jj}\right)\tau_{T}\\
&\quad+4
\tau_{N}^{\prime}\left(\mathbf{\Omega}_{ii}\circ\mathbf{\Omega}_{jj}\right)
\mathbf{\Omega}_{jj}\left(\mathbf{I}_{N}\circ\mathbf{\Omega}_{ii}\right)\tau_{N}
\tau_{T}^{\prime}\left(\mathbf{Q}_{ii}\circ\mathbf{Q}_{jj}\right)
\mathbf{Q}_{jj}\left(\mathbf{I}_{T}\circ\mathbf{Q}_{ii}\right)\tau_{T}\\
&\quad+2
\tau_{N}^{\prime}\left(\mathbf{\Omega}_{jj}\circ\mathbf{\Omega}_{jj}\right)
\mathbf{\Omega}_{ii}\left(\mathbf{I}_{N}\circ\mathbf{\Omega}_{ii}\right)\tau_{N}
\tau_{T}^{\prime}\left(\mathbf{Q}_{jj}\circ\mathbf{Q}_{jj}\right)
\mathbf{Q}_{ii}\left(\mathbf{I}_{T}\circ\mathbf{Q}_{ii}\right)\tau_{T}\Big\}\\
&+16\Big[4\tr\left\{\mathbf{\Omega}_{ii}\left(\mathbf{\Omega}_{ii}\circ
\mathbf{\Omega}_{jj}\right)\mathbf{\Omega}_{jj}\right\}
\tr\left\{\mathbf{Q}_{ii}\left(\mathbf{Q}_{ii}\circ
\mathbf{Q}_{jj}\right)\mathbf{Q}_{jj}\right\}\\
&\quad+\tr\left\{\mathbf{\Omega}_{ii}\left(\mathbf{\Omega}_{jj}\circ
\mathbf{\Omega}_{jj}\right)\mathbf{\Omega}_{ii}\right\}
\tr\left\{\mathbf{Q}_{ii}\left(\mathbf{Q}_{jj}\circ
\mathbf{Q}_{jj}\right)\mathbf{Q}_{ii}\right\}\\
&\quad+\tr\left\{\mathbf{\Omega}_{jj}\left(\mathbf{\Omega}_{ii}\circ
\mathbf{\Omega}_{ii}\right)\mathbf{\Omega}_{jj}\right\}
\tr\left\{\mathbf{Q}_{jj}\left(\mathbf{Q}_{ii}\circ
\mathbf{Q}_{ii}\right)\mathbf{Q}_{jj}\right\}\Big],\\
G_{3_{6}}=&\underset{i\neq j}{\sum^{N}\sum^{N}}
2\tr^{2}\left(\mathbf{\Omega}_{ii}\circ
\mathbf{\Omega}_{jj}\right)\tr^{2}\left(\mathbf{Q}_{ii}\circ
\mathbf{Q}_{jj}\right)\\
&+4\Big\{4\tau_{N}^{\prime}
\left(\mathbf{I}_{N}\circ\mathbf{\Omega}_{ii}\right)
\left(\mathbf{\Omega}_{ii}\circ\mathbf{\Omega}_{jj}\right)
\left(\mathbf{I}_{N}\circ\mathbf{\Omega}_{jj}\right)\tau_{N}
\tau_{T}^{\prime}
\left(\mathbf{I}_{T}\circ\mathbf{Q}_{ii}\right)
\left(\mathbf{Q}_{ii}\circ\mathbf{Q}_{jj}\right)
\left(\mathbf{I}_{T}\circ\mathbf{Q}_{jj}\right)\tau_{T}\\
&\quad+\tau_{N}^{\prime}
\left(\mathbf{I}_{N}\circ\mathbf{\Omega}_{ii}\right)
\left(\mathbf{\Omega}_{jj}\circ\mathbf{\Omega}_{jj}\right)
\left(\mathbf{I}_{N}\circ\mathbf{\Omega}_{ii}\right)\tau_{N}
\tau_{T}^{\prime}
\left(\mathbf{I}_{T}\circ\mathbf{Q}_{ii}\right)
\left(\mathbf{Q}_{jj}\circ\mathbf{Q}_{jj}\right)
\left(\mathbf{I}_{T}\circ\mathbf{Q}_{ii}\right)\tau_{T}\\
&\quad+\tau_{N}^{\prime}
\left(\mathbf{I}_{N}\circ\mathbf{\Omega}_{jj}\right)
\left(\mathbf{\Omega}_{ii}\circ\mathbf{\Omega}_{ii}\right)
\left(\mathbf{I}_{N}\circ\mathbf{\Omega}_{jj}\right)\tau_{N}
\tau_{T}^{\prime}
\left(\mathbf{I}_{T}\circ\mathbf{Q}_{jj}\right)
\left(\mathbf{Q}_{ii}\circ\mathbf{Q}_{ii}\right)
\left(\mathbf{I}_{T}\circ\mathbf{Q}_{jj}\right)\tau_{T}\Big\}\\
&+8\tau_{N}^{\prime}
\left(\mathbf{\Omega}_{ii}\circ\mathbf{\Omega}_{ii}
\circ\mathbf{\Omega}_{jj}\circ\mathbf{\Omega}_{jj}\right)\tau_{N}
\tau_{T}^{\prime}\left(\mathbf{Q}_{ii}\circ\mathbf{Q}_{ii}
\circ\mathbf{Q}_{jj}\circ\mathbf{Q}_{jj}\right)\tau_{T},\\
G_{3_{7}}=&\underset{i\neq j}{\sum^{N}\sum^{N}}
2\Big\{2
\tau_{N}^{\prime}
\left(\mathbf{I}_{N}\circ\mathbf{\Omega}_{ii}\right)\mathbf{\Omega}_{ii}
\left(\mathbf{I}_{N}\circ\mathbf{\Omega}_{jj}\circ\mathbf{\Omega}_{jj}\right)\tau_{N}
\tau_{T}^{\prime}
\left(\mathbf{I}_{T}\circ\mathbf{Q}_{ii}\right)\mathbf{Q}_{ii}
\left(\mathbf{I}_{T}\circ\mathbf{Q}_{jj}\circ\mathbf{Q}_{jj}\right)\tau_{T}\\
&\quad+4\tau_{N}^{\prime}
\left(\mathbf{I}_{N}\circ\mathbf{\Omega}_{ii}\right)\mathbf{\Omega}_{jj}
\left(\mathbf{I}_{N}\circ\mathbf{\Omega}_{ii}\circ\mathbf{\Omega}_{jj}\right)\tau_{N}
\tau_{T}^{\prime}
\left(\mathbf{I}_{T}\circ\mathbf{Q}_{ii}\right)\mathbf{Q}_{jj}
\left(\mathbf{I}_{T}\circ\mathbf{Q}_{ii}\circ\mathbf{Q}_{jj}\right)\tau_{T}\\
&\quad+4\tau_{N}^{\prime}
\left(\mathbf{I}_{N}\circ\mathbf{\Omega}_{jj}\right)\mathbf{\Omega}_{ii}
\left(\mathbf{I}_{N}\circ\mathbf{\Omega}_{ii}\circ\mathbf{\Omega}_{jj}\right)\tau_{N}
\tau_{T}^{\prime}
\left(\mathbf{I}_{T}\circ\mathbf{Q}_{jj}\right)\mathbf{Q}_{ii}
\left(\mathbf{I}_{T}\circ\mathbf{Q}_{ii}\circ\mathbf{Q}_{jj}\right)\tau_{T}\\
&\quad+2\tau_{N}^{\prime}
\left(\mathbf{I}_{N}\circ\mathbf{\Omega}_{jj}\right)\mathbf{\Omega}_{jj}
\left(\mathbf{I}_{N}\circ\mathbf{\Omega}_{ii}\circ\mathbf{\Omega}_{ii}\right)\tau_{N}
\tau_{T}^{\prime}
\left(\mathbf{I}_{T}\circ\mathbf{Q}_{jj}\right)\mathbf{Q}_{jj}
\left(\mathbf{I}_{T}\circ\mathbf{Q}_{ii}\circ\mathbf{Q}_{ii}\right)\tau_{T}\Big\}\\
&+8\Big\{2\tau_{N}^{\prime}
\left(\mathbf{I}_{N}\circ\mathbf{\Omega}_{ii}\right)
\left(\mathbf{\Omega}_{ii}\circ\mathbf{\Omega}_{jj}\circ\mathbf{\Omega}_{jj}\right)\tau_{N}
\tau_{T}^{\prime}
\left(\mathbf{I}_{T}\circ\mathbf{Q}_{ii}\right)
\left(\mathbf{Q}_{ii}\circ\mathbf{Q}_{jj}\circ\mathbf{Q}_{jj}\right)\tau_{T}\\
&\quad+2\tau_{N}^{\prime}
\left(\mathbf{I}_{N}\circ\mathbf{\Omega}_{jj}\right)
\left(\mathbf{\Omega}_{ii}\circ\mathbf{\Omega}_{ii}\circ\mathbf{\Omega}_{jj}\right)\tau_{N}
\tau_{T}^{\prime}
\left(\mathbf{I}_{T}\circ\mathbf{Q}_{jj}\right)
\left(\mathbf{Q}_{ii}\circ\mathbf{Q}_{ii}\circ\mathbf{Q}_{jj}\right)\tau_{T}\Big\}.
\end{align*}
Because $\underset{i\neq j}{\sum\sum}u_{ij}^{4}<\underset{i\neq j}{\sum\sum}u_{ij}^{2}<\tr\left(\mathbf{U}^{2}\right)=O(N),$
we have $G_{3_{1}}=O(NT^{3})+O(NT^{2}).$
Similarly,
due to $\underset{i\neq j}{\sum\sum}|u_{ij}|=O(N^{1\vee(2-1/\tau)})$
and Lemma \ref{lem:sigma},
we have $G_{3_{2}}=O(N^{2}T^{3})+O(N^{2}T^{2}),$
$G_{3_{3}}=O(N^{2}T),$
$G_{3_{4}}=O(N^{2}T),$
$G_{3_{5}}=O(N^{2}T^{5/2}),$
$G_{3_{6}}=O(N^{2}T^{2})$
and $G_{3_{7}}=O(N^{2}T^{3/2}).$
Combining the above calculations, we can get $G_{3}= O(N^{2}T^{3})$. We next deal with $G_{4}$.
Similarly,
$$
G_{4}=G_{4_{1}}+\gamma_{2}G_{4_{2}}+\gamma_{4}G_{4_{3}}+\gamma_{6}G_{4_{4}}
+\gamma_{1}^{2}G_{4_{5}}+\gamma_{2}^{2}G_{4_{6}}+\gamma_{1}\gamma_{3}G_{4_{7}},$$
where

Because $\underset{i\neq j}{\sum\sum}u_{ij}^{4}<\underset{i\neq j}{\sum\sum}u_{ij}^{2}<\tr\left(\mathbf{U}^{2}\right)=O(N),$
we have $G_{4_{1}}=O(NT^{3})+O(N^{2}T^{2}).$
Similarly,
due to $\underset{i\neq j}{\sum\sum}|u_{ij}|=O(N^{1\vee(2-1/\tau)})$
and Lemma \ref{lem:sigma},
we have $G_{4_{2}}=O(N^{1\vee(2-1/\tau)}T^{3})+O(N^{2}T^{2}),$
$G_{4_{3}}=O(N^{2}T^{2}),$
$G_{4_{4}}=O(N^{2}T),$
$G_{4_{5}}=O(N^{2}T^{5/2}),$
$G_{4_{6}}=O(N^{2}T^{2})$
and $G_{4_{7}}=O(N^{2}T^{2}).$
Combining the above calculations, we can get $G_{4}= O(N^{2}T^{5/2}+N^{1\vee(2-1/\tau)}T^{3})$. We next deal with $G_{5}$.
Similarly, due to Lemma \ref{le:moment of quadratic form}, we have
$$
G_{5}=G_{5_{1}}+\gamma_{2}G_{5_{2}}+\gamma_{4}G_{5_{3}}+\gamma_{6}G_{5_{4}}
+\gamma_{1}^{2}G_{5_{5}}+\gamma_{2}^{2}G_{5_{6}}+\gamma_{1}\gamma_{3}G_{5_{7}},$$
where

Because $\underset{i\neq j\neq k}{\sum\sum\sum}|u_{ij}u_{jk}|=O(N^{1\vee(3-2/\tau)})$
and
$\underset{i\neq j\neq k}{\sum\sum\sum}u_{ij}^{2}=O(N^{2}),$
we have $G_{5_{1}}=O(N^{1\vee(3-2/\tau)}T^{3})+O(N^{2}T^{2}).$
Similarly,
due to $\underset{i\neq j\neq k}{\sum\sum}|u_{ij}|=O(N^{2\vee (3-1/\tau)}),$
and Lemma \ref{lem:sigma},
we have $G_{5_{2}}=O(N^{2\vee (3-1/\tau)}T^{3})+O(N^{3}T^{2}),$
$G_{5_{3}}=O(N^{2\vee (3-1/\tau)}T^{2})+O(N^{3}T),$
$G_{5_{4}}=O(N^{3}T),$
$G_{5_{5}}=O(N^{3}T^{5/2}),$
$G_{5_{6}}=O(N^{3}T^{2})$
and $G_{5_{7}}=O(N^{3}T^{2}).$
Combining the above calculations, we can get $G_{5}= O(N^{1\vee(3-2/\tau)}T^{3})+ O(N^{3}T^{5/2})$. We next deal with $G_{6}$.
Similarly, due to Lemma \ref{le:moment of quadratic form}, we have
$$
G_{6}=G_{6_{1}}+\gamma_{2}G_{6_{2}}+\gamma_{4}G_{6_{3}}+\gamma_{6}G_{6_{4}}
+\gamma_{1}^{2}G_{6_{5}}+\gamma_{2}^{2}G_{6_{6}}+\gamma_{1}\gamma_{3}G_{6_{7}},$$
where

 Due to $$\underset{i\neq j\neq k}{\sum^{N}\sum^{N}\sum^{N}}u_{ij}^{2}u_{ik}^{2}\leq\underset{i\neq j\neq k}{\sum^{N}\sum^{N}\sum^{N}}u_{ij}^{2}=O(N^{2})$$
and
$$\underset{i\neq j\neq k}{\sum^{N}\sum^{N}\sum^{N}}|u_{ij}u_{kj}u_{ik}|\leq\underset{i\neq j\neq k}{\sum^{N}\sum^{N}\sum^{N}}|u_{ij}u_{kj}|=O(N^{1\vee(3-2/\tau)}),$$
we have $G_{6_{1}}=O(N^{2}T^{3}+N^{1\vee(3-2/\tau)}T^{3}).$
Similarly,
due to $\underset{i\neq j\neq k}{\sum\sum}|u_{ij}|=O(N^{2\vee (3-1/\tau)})$
and Lemma \ref{lem:sigma},
we have $G_{6_{2}}=O(N^{1\vee(3-2/\tau)}T^{3})+O(N^{2\vee(3-1/\tau)}T^{2})+O(N^{3}T),$
$G_{6_{3}}=O(N^{2\vee(3-1/\tau)}T^{2})+O(N^{3}T),$
$G_{6_{4}}=O(N^{3}T),$
$G_{6_{5}}=O(N^{2\vee(3-1/\tau)}T^{5/2})+O(N^{3}T^{2}),$
$G_{6_{6}}=O(N^{3}T^{2})$
and $G_{6_{7}}=O(N^{3}T^{2}).$
Combining the above calculations, we can get $G_{6}= O(N^{1\vee(3-2/\tau)}T^{3})+ O(N^{3}T^{2}+O(N^{2\vee(3-1/\tau)}T^{5/2})$. We next deal with $G_{7}$.
Similarly, due to Lemma \ref{le:moment of quadratic form}, we have
$$
G_{7}=G_{7_{1}}+\gamma_{2}G_{7_{2}}+\gamma_{4}G_{7_{3}}+\gamma_{6}G_{7_{4}}
+\gamma_{1}^{2}G_{7_{5}}+\gamma_{2}^{2}G_{7_{6}}+\gamma_{1}\gamma_{3}G_{7_{7}},$$
where

 Due to $$\underset{i\neq j\neq k\neq l}{\sum^{N}\sum^{N}\sum^{N}\sum^{N}}u_{kj}^{2}u_{il}^{2}=O(N^{2}),$$
 $$\underset{i\neq j\neq k\neq l}{\sum^{N}\sum^{N}\sum^{N}\sum^{N}}|u_{ij}u_{jl}u_{kk}u_{li}|\leq
\sum_{i=1}^{N}\sum_{k=1}^{N}\sum_{j=1}^{N}|u_{ij}|\sum_{l=1}^{N}|u_{li}|
=O(N^{2\vee(4-2/\tau)})$$
and
$$\underset{i\neq j\neq k\neq l}{\sum^{N}\sum^{N}\sum^{N}\sum^{N}}|u_{ij}u_{jk}u_{kl}u_{li}|\leq\underset{i\neq j}{\sum^{N}\sum^{N}}|u_{ij}|\underset{k\neq l}{\sum^{N}\sum^{N}}|u_{kl}|=O(N^{2\vee(4-2/\tau)}),$$
we have $G_{7_{1}}=O(N^{2\vee(4-2/\tau)}T^{3}).$
Similarly,
due to $$\underset{i\neq j\neq k\neq l}{\sum^{N}\sum^{N}\sum^{N}\sum^{N}}|u_{ij}|=O(N^{3\vee(4-1/\tau)}),$$
and Lemma \ref{lem:sigma},
we have $G_{7_{2}}=O(N^{2\vee(4-2/\tau)}T^{3})+O(N^{3\vee(4-1/\tau)}T^{2}),$
$G_{7_{3}}=O(N^{3\vee(4-1/\tau)}T^{2})+O(N^{4}T),$
$G_{7_{4}}=O(N^{4}T),$
$G_{7_{5}}=O(N^{3\vee(4-1/\tau)}T^{5/2})+O(N^{4}T^{3/2}),$
and $G_{7_{7}}=O(N^{4}T^{3/2}).$
Here, we focus on calculating the first term of $G_{7_{6}}.$
Because
\begin{align*}
&\underset{i\neq j\neq k \neq l}{\sum^{N}\sum^{N}\sum^{N}\sum^{N}}2\tr^{2}\left(
\mathbf{\Omega}_{ij}\circ\mathbf{\Omega}_{kl}\right)
\tr^{2}\left(
\mathbf{Q}_{ij}\circ\mathbf{Q}_{kl}\right)\\
=&O(T^{2})\underset{i\neq j\neq k \neq l}{\sum^{N}\sum^{N}\sum^{N}\sum^{N}}\tr^{2}\left(
\mathbf{\Omega}_{ij}\circ\mathbf{\Omega}_{kl}\right)\\
=&O(T^{2})\underset{i\neq j\neq k \neq l}{\sum^{N}\sum^{N}\sum^{N}\sum^{N}}
\Big(\sum_{s=1}^{N}(\mathbf{\Omega}_{ij})_{ss}(\mathbf{\Omega}_{kl})_{ss}
\Big)^{2}\\
\leq&O(T^{2})\underset{i\neq j\neq k \neq l}{\sum^{N}\sum^{N}\sum^{N}\sum^{N}}
\sum_{s=1}^{N}(\mathbf{\Omega}_{ij})_{ss}^{2}(\mathbf{\Omega}_{kl})_{ss}^{2}\\
&+O(T^{2})\underset{i\neq j\neq k \neq l}{\sum^{N}\sum^{N}\sum^{N}\sum^{N}}
\underset{s_{1}\neq s_{2}}{\sum^{N}\sum^{N}}|(\mathbf{\Omega}_{ij})_{s_{1}s_{1}}(\mathbf{\Omega}_{kl})_{s_{1}s_{1}}
(\mathbf{\Omega}_{ij})_{s_{2}s_{2}}(\mathbf{\Omega}_{kl})_{s_{2}s_{2}}|\\
\leq&O(NT^{2}+N^{2}T^{2}),
\end{align*}
where the last inequality holds due to $\mathbf{\Omega}_{ij}=u_{i}^{1/2}(u_{j}^{1/2})^{\prime}$
and $(u_{i}^{1/2})^{\prime}u_{i}^{1/2}=1,$ for all $1\leq i,j\leq N.$
Hence, we have $G_{7_{6}}=O(N^{2}T^{2})+O(N^{4}T^{3/2}).$
and $G_{7}=O(N^{2\vee(4-2/\tau)}T^{3})+O(N^{3\vee(4-1/\tau)}T^{2})+
O(N^{3\vee(4-1/\tau)}T^{5/2})+O(N^{4}T^{3/2})$.
Combining $G_{1},\cdots,G_{7},$
we can conclude that
$$\var\Big[\big\{\sum_{i=1}^{N} \sum_{j=1}^{N} (\hat{\epsilon}_{i\cdot}^{\prime}\hat{\epsilon}_{j\cdot})^{2}\big\}^{2}\Big]
=O(N^{4}T^{3/2}+N^{2}T^{3}+N^{3}T^{5/2}+N^{2\vee(4-2/\tau)}T^{3}+N^{3\vee(4-1/\tau)}T^{5/2}).$$
Similarly, for $$\sum_{i=1}^{N}\hat{\epsilon}_{i\cdot}^{\prime}
\hat{\epsilon}_{i\cdot}=\sum_{i=1}^{N}Z^{\prime}\left(\mathbf{\Omega}_{ii}\otimes
\mathbf{Q}_{ii}\right)Z=Z^{\prime}\left\{\sum_{i=1}^{N}\left(\mathbf{\Omega}_{ii}\otimes
\mathbf{Q}_{ii}\right)\right\}Z\doteq Z^{\prime}\mathbf{J}Z.$$
So, we have
$\var\bigg\{\Big(\sum_{i=1}^{N}\hat{\epsilon}_{i\cdot}^{\prime}
\hat{\epsilon}_{i\cdot}\Big)^{2}\bigg\}=\var\bigg\{\Big( Z^{\prime}\mathbf{J}Z\Big)^{2}\bigg\}.$
Again using Lemma \ref{le:moment of quadratic form},
we have
\begin{align*}
&\var\bigg\{\Big( Z^{\prime}\mathbf{J}Z\Big)^{2}\bigg\}\\
=&\mE\Big( Z^{\prime}\mathbf{J}Z\Big)^{4}-\mE^{2}\Big( Z^{\prime}\mathbf{J}Z\Big)^{2}\\
=&K_{1}+\gamma_{2}K_{2}+\gamma_{4}K_{3}+\gamma_{6}K_{4}+\gamma_{1}^{2}K_{5}
+\gamma_{2}^{2}K_{6}+\gamma_{1}\gamma_{3}K_{7},
\end{align*}
where
\begin{align*}
K_{1}=&8\tr^{2}\left(\mathbf{J}\right)\tr\left(\mathbf{J}^{2}\right)
+8\tr^{2}\left(\mathbf{J}^{2}\right)+32\tr\left(\mathbf{J}\right)\tr\left(\mathbf{J}^{3}\right)
+48\tr\left(\mathbf{J}^{4}\right),\\
K_{2}=&4\tr\left(\mathbf{J}\right) \tr\left(\mathbf{J}\right) \tr\left(\mathbf{J} \circ \mathbf{J}\right)+8{\tau}_{NT}^{\prime}\left(\mathbf{J} \circ \mathbf{J}\right){\tau}_{NT}\tr\left(\mathbf{J} \circ \mathbf{J}\right) +48\tr\left(\mathbf{J}\right) \tr\left(\mathbf{J} \circ\mathbf{J}^{2}\right)\\
&+96\tr\left\{\left(\mathbf{I}_{NT} \circ \mathbf{J}\right) \mathbf{J}^{3}\right\}+48{\tau}_{NT}^{\prime}\left(\mathbf{I}_{NT} \circ\mathbf{J}^{2}\right)\left(\mathbf{I}_{NT}  \circ\mathbf{J}^{2}\right) {\tau}_{NT},\\
K_{3}=& 4\tr\left(\mathbf{J}\right) \tr\left(\mathbf{J} \circ \mathbf{J} \circ \mathbf{J}\right) +24\tr\left(\mathbf{J} \circ \mathbf{J} \circ\mathbf{J}^{2}\right),\\
K_{4}=& \tr\left(\mathbf{J} \circ \mathbf{J} \circ \mathbf{J} \circ \mathbf{J}\right),\\
K_{5}=& 24{\tau}_{NT}^{\prime}\left(\mathbf{I}_{NT} \circ \mathbf{J}\right) \mathbf{J}\left(\mathbf{I}_{NT} \circ \mathbf{J}\right){\tau}_{NT} \tr\left(\mathbf{J}\right)+48{\tau}_{NT}^{\prime}\left(\mathbf{I}_{NT} \circ \mathbf{J}\right) \mathbf{J} ^{2}\left(\mathbf{I}_{NT} \circ \mathbf{J}\right)\tau_{NT}\\
&+16{\tau}_{NT}^{\prime}\left(\mathbf{J} \circ \mathbf{J} \circ \mathbf{J}\right){\tau}_{NT} \tr\left(\mathbf{J}\right)+96{\tau}_{NT}^{\prime}\left(\mathbf{J} \circ \mathbf{J}\right) \mathbf{J}\left(\mathbf{I}_{NT} \circ \mathbf{J}\right) \tau_{NT}\\
&+96\tr\left(\mathbf{J}\left(\mathbf{J} \circ \mathbf{J}\right) \mathbf{J}\right),\\
K_{6}=& 3\tr\left(\mathbf{J} \circ \mathbf{J}\right) \tr\left(\mathbf{J} \circ \mathbf{J}\right)+24{\tau}_{NT}^{\prime}\left(\mathbf{I}_{NT} \circ \mathbf{J}\right)\left(\mathbf{J} \circ \mathbf{J}\right)\left(\mathbf{I}_{NT} \circ \mathbf{J}\right){\tau}_{NT}\\
&+8{\tau}_{NT}^{\prime}\left(\mathbf{J} \circ \mathbf{J} \circ \mathbf{J} \circ \mathbf{J}\right) {\tau}_{NT},\\
K_{7}=&24{\tau}_{NT}^{\prime}\left(\mathbf{I}_{NT} \circ \mathbf{J}\right) \mathbf{J}\left(\mathbf{I}_{NT} \circ \mathbf{J} \circ \mathbf{J}\right){\tau}_{NT}+32{\tau}_{NT}^{\prime}\left(\mathbf{I}_{NT} \circ \mathbf{J}\right)\left(\mathbf{J} \circ \mathbf{J} \circ \mathbf{J}\right){\tau}_{NT}.
\end{align*}
First, we have
\begin{align*}
\tr\left(\mathbf{J}\right)=&\sum_{i=1}^{N}u_{ii}\tr(\mathbf{Q}_{ii})=O(NT),\\
\tr\left(\mathbf{J}^{2}\right)=&\sum_{i=1}^{N}\sum_{j=1}^{N}u_{ij}^{2}
\tr(\mathbf{Q}_{ii}\mathbf{Q}_{jj})=O(NT).
\end{align*}
Then, due to Assumption \ref{assum:matrix} and
$$\sum_{i=1}^{N}\sum_{j=1}^{N}\sum_{k=1}^{N}|u_{ij}u_{jk}u_{ik}|
=O(N^{1\vee(3-2/\tau)}),$$
we have
\begin{align*}
\tr\left(\mathbf{J}^{3}\right)=&\sum_{i=1}^{N}\sum_{j=1}^{N}\sum_{k=1}^{N}
u_{ij}u_{jk}u_{ik}
\tr(\mathbf{Q}_{ii}\mathbf{Q}_{jj}\mathbf{Q}_{kk})\\
=&\tr\left(\mathbf{U}^{3}\right)
\tr\left(\mathbf{\Sigma}^{3}\right)+\sum_{i=1}^{N}\sum_{j=1}^{N}\sum_{k=1}^{N}u_{ij}u_{jk}u_{ik}  \left\{\tr(\mathbf{Q}_{ii}\mathbf{Q}_{jj}\mathbf{Q}_{kk})-\tr\left(\mathbf{\Sigma}^{3}\right)\right\}\\
\leq&\tr\left(\mathbf{U}^{3}\right)
\tr\left(\mathbf{\Sigma}^{3}\right)+O(N^{1\vee(3-2/\tau)})=O(NT).
\end{align*}
Next, due to Assumption \ref{assum:matrix} and
\begin{align*}
&\sum_{i=1}^{N}\sum_{j=1}^{N}\sum_{k=1}^{N}\sum_{l=1}^{N}|u_{ij}u_{jk}u_{kl}u_{li}|\\
\leq&\sum_{i=1}^{N}\sum_{k=1}^{N}\left(\sum_{j=1}^{N}|u_{ij}u_{jk}|
\sum_{l=1}^{N}|u_{kl}u_{li}|\right)\\
\leq&\sum_{i=1}^{N}\sum_{k=1}^{N}\left(\sum_{j=1}^{N}u_{ij}^{2}\sum_{j=1}^{N}u_{jk}^{2}\right)\\
\leq&\sum_{i=1}^{N}\sum_{k=1}^{N}C
=O(N^{2}),
\end{align*}
we can get
\begin{align*}
\tr\left(\mathbf{J}^{4}\right)=&\sum_{i=1}^{N}\sum_{j=1}^{N}\sum_{k=1}^{N}
\sum_{l=1}^{N}
u_{ij}u_{jk}u_{kl}u_{li}
\tr(\mathbf{Q}_{ii}\mathbf{Q}_{jj}\mathbf{Q}_{kk}\mathbf{Q}_{ll})\\
=&\tr\left(\mathbf{U}^{4}\right)
\tr\left(\mathbf{\Sigma}^{4}\right)+\sum_{i=1}^{N}\sum_{j=1}^{N}
\sum_{k=1}^{N}\sum_{l=1}^{N}u_{ij}u_{jk}u_{kl}u_{li} \left\{\tr(\mathbf{Q}_{ii}\mathbf{Q}_{jj}\mathbf{Q}_{kk}\mathbf{Q}_{ll})
-\tr\left(\mathbf{\Sigma}^{4}\right)\right\}\\
\leq&\tr\left(\mathbf{U}^{4}\right)
\tr\left(\mathbf{\Sigma}^{4}\right)+O(N^{2})=O(NT).
\end{align*}
Similarly, according to the above formulas, we have
$K_{1}=O(N^{3}T^{3})=K_{2},$
$K_{3}=O(N^{2}T^{2}),$ $K_{4}=O(NT),$
$K_{5}=O(N^{3}T^{3}),$
$K_{6}=O(N^{2}T^{2})=K_{7}.$
Then, we can conclude that
$$\var\bigg\{\Big(\sum_{i=1}^{N}\hat{\epsilon}_{i\cdot}^{\prime}
\hat{\epsilon}_{i\cdot}\Big)^{2}\bigg\}=O(N^{3}T^{3}).$$
Finally, by Assumption \ref{assum:matrix}, we can obtain that
\begin{align*}
\mE(b^2_{N})\leq&\frac{1}{N^2T^4}\Bigg[2\var\bigg\{\sum_{i=1}^{N} \sum_{j=1}^{N} (\hat{\epsilon}_{i\cdot}^{\prime}\hat{\epsilon}_{j\cdot})^{2}\bigg\}+\frac{2}{T^2}
\var\bigg\{\Big(\sum_{i=1}^{N}\hat{\epsilon}_{i\cdot}^{\prime}\hat{\epsilon}_{i\cdot}\Big)^{2}\bigg\}\Bigg]\\
=&O\left(\frac{1}{T^{\frac{1}{2}\vee \left(\frac{1}{\tau}-\frac{1}{2}\right)}}\right).
\end{align*}
\hfill$\Box$
\subsection{Proof of Lemma \ref{th:sum null}}
Recall that $\mathbf{M}_{i}=\mathbf{P}_{i}\mathbf{\Sigma}\mathbf{P}_{i}.$
Let
$\hat{\epsilon}_{i\cdot}=\mathbf{P}_{i}\epsilon_{i\cdot}=\mathbf{P}_{i}\mathbf{\Sigma}^{1/2}z_{i},$
where $z_{i}=(z_{i1},\cdots,z_{iT})^{\prime}$ is the $i$-th row vector of $\mathbf{Z}.$ Define $\mathbf{Q}_{ij}\doteq\mathbf{\Sigma}^{1/2}\mathbf{P}_{i}\mathbf{P}_{j}\mathbf{\Sigma}^{1/2}
,$ then, we have $\mathbf{Q}_{ii}=\mathbf{\Sigma}^{1/2}\mathbf{P}_{i}\mathbf{\Sigma}^{1/2}.$
Obviously, by Assumption \ref{assum:E distribution}, we have $\mE\left(S_{N}\right)=0.$ Then, we have
\begin{align*}
\mE\left(S^2_{N}\right)&=\frac{2}{N(N-1)}\underset{i<j}{\sum\sum}\underset{s<t}{\sum\sum}
\mE\left(\frac{\epsilon_{i\cdot}^{\prime} \mathbf{P}_{i} \mathbf{P}_{j} \epsilon_{j\cdot}}{\left\|\mathbf{P}_{i} \epsilon_{i\cdot}\right\| \cdot\left\|\mathbf{P}_{j} \epsilon_{j\cdot}\right\|}\frac{\epsilon_{s}^{\prime} \mathbf{P}_{s} \mathbf{P}_{t} \epsilon_{t}}{\left\|\mathbf{P}_{s} \epsilon_{s}\right\| \cdot\left\|\mathbf{P}_{t} \epsilon_{t}\right\|}\right)\\
&=\frac{2}{N(N-1)}\underset{i<j}{\sum\sum}
\mE\left\{\frac{\left(\epsilon_{i\cdot}^{\prime} \mathbf{P}_{i} \mathbf{P}_{j} \epsilon_{j\cdot}\right)^2}{\left\|\mathbf{P}_{i} \epsilon_{i\cdot}\right\|^2 \cdot\left\|\mathbf{P}_{j} \epsilon_{j\cdot}\right\|^2}\right\}\\
&=\frac{2}{N(N-1)}\underset{i<j}{\sum\sum}\mE\left[\frac{\epsilon_{j\cdot}^{\prime} \mathbf{P}_{j} \mathbf{M}_{i}\mathbf{P}_{j} \epsilon_{j\cdot}}{\tr(\mathbf{M}_{i})\left\|\mathbf{P}_{j} \epsilon_{j\cdot}\right\|^2}+\frac{(\epsilon_{j\cdot}^{\prime} \mathbf{P}_{j} \mathbf{M}_{i}\mathbf{P}_{j} \epsilon_{j\cdot})\cdot\{2\tr(\mathbf{M}^2_{i})+\gamma_{
2}\tr(\mathbf{Q}_{ii}\circ\mathbf{Q}_{ii})\}}{\tr^3(\mathbf{M}_{i})\left\|\mathbf{P}_{j} \epsilon_{j\cdot}\right\|^2}\right.\\
&\left.\,\,\,\,-\frac{[2(\epsilon_{j\cdot}^{\prime} \mathbf{P}_{j} \mathbf{M}^2_{i}\mathbf{P}_{j} \epsilon_{j\cdot})+\gamma_{2}\tr\{(\mathbf{\Sigma}^{1/2}\mathbf{P}_{i}\mathbf{P}_{j}
\epsilon_{j\cdot}\epsilon_{j\cdot}^{\prime}\mathbf{P}_{j}\mathbf{P}_{i}\mathbf{\Sigma}^{1/2})
\circ \mathbf{Q}_{ii}\}]}
{\tr^2(\mathbf{M}_{i})\left\|\mathbf{P}_{j} \epsilon_{j\cdot}\right\|^2}+O(T^{-2})\right]\\
&=\frac{2}{N(N-1)}\underset{i<j}{\sum\sum}\frac{\tr\left(\mathbf{M}_{i}
\mathbf{M}_{j}\right)}
{\tr\left(\mathbf{M}_{i}\right)\tr\left(\mathbf{M}_{j}\right)}
+O(T^{-2})\\
&=\sigma^2_{S_{N}}\left\{1+o(1)\right\},
\end{align*}
where the third equality holds due to Lemmas \ref{lem:laplace approxi} and \ref{le:moment of quadratic form}, and the last equality above holds because of Lemma \ref{lem:sigma}, and the last second equality above holds because of Lemmas \ref{lem:laplace approxi}, \ref{lem:sigma} and \ref{le:moment of quadratic form},
$\tr(\mathbf{Q}_{ii}\circ\mathbf{Q}_{ii})\leq \tr(\mathbf{Q}^2_{ii})=\tr(\mathbf{M}^2_{i})=O(T)$
and $$0\leq \tr\{(\mathbf{\Sigma}^{1/2}\mathbf{P}_{i}\mathbf{P}_{j}
\epsilon_{j\cdot}\epsilon_{j\cdot}^{\prime}\mathbf{P}_{j}\mathbf{P}_{i}\mathbf{\Sigma}^{1/2})
\circ \mathbf{Q}_{ii}\}/\left\|\mathbf{P}_{j} \epsilon_{j\cdot}\right\|\leq C \tr(\mathbf{\Sigma}^{1/2}\mathbf{P}_{i}\mathbf{P}_{j}
\epsilon_{j\cdot}\epsilon_{j\cdot}^{\prime}\mathbf{P}_{j}\mathbf{P}_{i}\mathbf{\Sigma}^{1/2})/\left\|\mathbf{P}_{j} \epsilon_{j\cdot}\right\|=O(1).$$

The normality of $S_{N}$ has yet to be proven. Define
$\sum_{j=2}^k Z_{j}$ where
$Z_{j}=\sum_{i=1}^{j-1}\sqrt{\frac{2}{N(N-1)}}\hat{\rho}_{i j},$
Let
$\mathcal{F}_{j}\doteq\sigma\{\epsilon_{1\cdot},\cdots,\epsilon_{j\cdot}\}$ be the $\sigma$-field
generated by $\{\epsilon_{i\cdot}, i\le j\}$, where $\epsilon_{i\cdot}=(\epsilon_{i1},\cdots,\epsilon_{iT})^{\prime}$.
Because $\epsilon_{1\cdot},\cdots,\epsilon_{N\cdot}$ are mutually independent under $H_{0}.$
Obviously, $\mE(Z_{j}\mid
\mathcal{F}_{j-1})=0$ and it follows that $\{\sum_{j=2}^k Z_{j},
\mathcal{F}_{k}; 2\le k\le N\}$ is a zero mean martingale.
According to %the Corollary 3.1 and Theorem 3.5 in
 the Martingale
central limit theorem in \cite{hall2014},  we only need to show
\begin{align}\label{clt1}
\frac{\sum_{j=2}^{N}\mE\left(Z_{j}^2\mid\mathcal{F}_{j-1}\right)}{\sigma^2_{S_{N}}}\cp
1,
\end{align}
and
\begin{align}\label{clt2}
\mE\left\{\sum_{j=2}^{N}\mE\left(Z_{j}^4|\mathcal{F}_{j-1}\right)\right\}=o\{\sigma^4_{S_{N}}\}.
\end{align}
We first focus on (\ref{clt1}).
It can be shown that
\begin{align*}
\sum_{j=2}^{N}\mE(Z_{j}^2|\mathcal{F}_{j-1})
=&\sum_{j=2}^{N}\mE\left[\left.\left\{\sqrt{\frac{2}{N(N-1)}}\sum_{i=1}^{j-1}\hat{\rho}_{i j}\right\}^2\right|\mathcal{F}_{j-1}\right]\\
=&\sum_{j=2}^{N}\frac{2}{N(N-1)}\mE\left\{\left.\sum_{i_1=1}^{j-1}\sum_{i_2=1}^{j-1}\hat{\rho}_{i_{1} j}\hat{\rho}_{i_{2} j}\right|\mathcal{F}_{n,j-1}\right\}\\
=&\sum_{j=2}^{N}\frac{2}{N(N-1)}\sum_{i_1=1}^{j-1}\sum_{i_2=1}^{j-1}\mE\left(\left.\hat{\rho}_{i_{1} j}\hat{\rho}_{i_{2} j}\right|\mathcal{F}_{n,j-1}\right)\\
=& A+B,
\end{align*}
where
\bse
A=\frac{2}{N(N-1)}\sum_{j=2}^{N}\sum_{i=1}^{j-1}\mE\left\{\left.\left(\hat{\rho}_{ij}\right)^{2}\right|\mathcal{F}_{n,j-1}\right\}\,\,\mbox{and}\,\,
B=\frac{4}{N(N-1)}\sum_{j=2}^{N}\underset{i_1<i_2}{\sum^{j-1}\sum^{j-1}}\mE\left(\left.\hat{\rho}_{i_{1} j}\hat{\rho}_{i_{2} j}\right|\mathcal{F}_{n,j-1}\right).
\ese
First, we consider $A$. By Lemmas \ref{lem:laplace approxi}, \ref{lem:matrix inequality} and \ref{le:moment of quadratic form}, we have
\begin{align*}
A=&\frac{2}{N(N-1)}\sum_{j=2}^{N}\sum_{i=1}^{j-1}
\Bigg\{\frac{\tr\left(\hat{\epsilon}_{i\cdot}^{\prime}\mathbf{M}_{j}\hat{\epsilon}_{i\cdot}\right)}{\tr\left(\mathbf{M}_{j}\right)
\hat{\epsilon}_{i\cdot}^{\prime}\hat{\epsilon}_{i\cdot}}+\frac{\mE\left({\epsilon}_{j\cdot}^{\prime}\mathbf{P}_{j}\hat{\epsilon}_{i\cdot}
\hat{\epsilon}_{i\cdot}^{\prime}\mathbf{P}_{j}\epsilon_{j\cdot}^{\prime}\right)\var\left(\epsilon_{j\cdot}^{\prime}
\mathbf{P}_{j}\epsilon_{j\cdot}\right)}{\tr^3\left(\mathbf{M}_{j}\right)\hat{\epsilon}_{i\cdot}^{\prime}\hat{\epsilon}_{i\cdot}}\\
&-\frac{\mE\left(\epsilon_{j\cdot}^{\prime}\mathbf{P}_{j}\hat{\epsilon}_{i\cdot}\hat{\epsilon}_{i\cdot}^{\prime}\mathbf{P}_{j}\epsilon_{j\cdot}
\epsilon_{j\cdot}^{\prime}\mathbf{P}_{j}\epsilon_{j\cdot}\right)-
\mE\left(\epsilon_{j\cdot}^{\prime}\mathbf{P}_{j}\hat{\epsilon}_{i\cdot}\hat{\epsilon}_{i\cdot}^{\prime}\mathbf{P}_{j}\epsilon_{j\cdot}\right)
\mE\left(\epsilon_{j\cdot}^{\prime}\mathbf{P}_{j}\epsilon_{j\cdot}\right)}{\tr^2\left(\mathbf{M}_{j}\right)\hat{\epsilon}_{i\cdot}^{\prime}\hat{\epsilon}_{i\cdot}}\Bigg\}+O\left(T^{-2}\right)\\
=&\frac{2}{N(N-1)}\sum_{j=2}^{N}\sum_{i=1}^{j-1}
\Bigg[\frac{\tr\left(\hat{\epsilon}_{i\cdot}^{\prime}\mathbf{M}_{j}\hat{\epsilon}_{i\cdot}\right)}{\tr\left(\mathbf{M}_{j}\right)
\hat{\epsilon}_{i\cdot}^{\prime}\hat{\epsilon}_{i\cdot}}+\frac{\tr\left(\hat{\epsilon}_{i\cdot}^{\prime}\mathbf{M}_{j}\hat{\epsilon}_{i\cdot}\right)
\{2\tr\left(\mathbf{M}_{j}^2
\right)+\gamma_{2}\tr(\mathbf{Q}_{jj}\circ\mathbf{Q}_{jj})\}}{\tr^3\left(\mathbf{M}_{j}\right)\hat{\epsilon}_{i\cdot}^{\prime}\hat{\epsilon}_{i\cdot}}\\
&-\frac{2\tr\left(\hat{\epsilon}_{i\cdot}^{\prime}\mathbf{M}^2_{j}\hat{\epsilon}_{i\cdot}\right)
}
{\tr^2\left(\mathbf{M}_{j}\right)\hat{\epsilon}_{i\cdot}^{\prime}\hat{\epsilon}_{i\cdot}}
-\frac{\gamma_{2}\tr[(\mathbf{\Sigma}^{1/2}\mathbf{P}_{j}\mathbf{P}_{i}
\epsilon_{i\cdot}\epsilon_{i\cdot}^{\prime}\mathbf{P}_{i}\mathbf{P}_{j}\mathbf{\Sigma}^{1/2})\circ\mathbf{Q}_{jj}]}
{\tr^2\left(\mathbf{M}_{j}\right)\hat{\epsilon}_{i\cdot}^{\prime}\hat{\epsilon}_{i\cdot}}\Bigg]+O\left(T^{-2}\right)\\
=&A_{1}+A_{2}-A_{3}-A_{4}+O\left(T^{-2}\right).
\end{align*}
By Lemmas \ref{lem:laplace approxi} and \ref{lem:sigma}, we have
\begin{align*}
\mE\left(A_{1}\right)=&\frac{2}{N(N-1)}\sum_{j=2}^{N}\sum_{i=1}^{j-1}\mE\left(\frac{\tr\left(\hat{\epsilon}_{i\cdot}^{\prime}\mathbf{M}_{j}\hat{\epsilon}_{i\cdot}\right)}{\tr\left(\mathbf{M}_{j}\right)
\hat{\epsilon}_{i\cdot}^{\prime}\hat{\epsilon}_{i\cdot}}\right)\\
=&\frac{2}{N(N-1)}\sum_{j=2}^{N}\sum_{i=1}^{j-1}\frac{\tr\left(\mathbf{M}_{i}\mathbf{M}_{j}\right)}{\tr\left(\mathbf{M}_{i}\right)\tr\left(\mathbf{M}_{j}\right)}+O\left\{\frac{1}{T\tr\left(\mathbf{M}_{j}\right)}\right\}\\
=&\sigma^2_{S_{N}}\left\{1+o(1)\right\}
\end{align*}
and
\begin{align*}
\var\left(A_{1}\right)=&\mE\left(A^2_{1}\right)-\mE^2\left(A_{1}\right)\\
\leq&
\frac{4}{N^2(N-1)^2}\sum_{j=2}^{N}\sum_{i=1}^{j-1}\mE\left\{\frac{\tr^2\left(\hat{\epsilon}_{i\cdot}^{\prime}\mathbf{M}_{j}\hat{\epsilon}_{i\cdot}\right)}
{\tr^2\left(\mathbf{M}_{j}\right)
\left(\hat{\epsilon}_{i\cdot}^{\prime}\hat{\epsilon}_{i\cdot}\right)^2}\right\}\\
%&+\frac{8}{N^2(N-1)^2}\sum_{j=2}^{N}\sum_{i_{1}<i_{2}}^{j-1}\mE\left\{\frac{\tr\left(\hat{\epsilon}_{i_{1}}^{\prime}\mathbf{M}_{j}\hat{\epsilon}_{i_{1}}\right)
%\tr\left(\hat{\epsilon}_{i_{2}}^{\prime}\mathbf{M}_{j}\hat{\epsilon}_{i_{2}}\right)}{\tr\left(\mathbf{M}_{j}\right)
%\hat{\epsilon}_{i_{1}}^{\prime}\hat{\epsilon}_{i_{1}}\tr\left(\mathbf{M}_{j}\right)
%\hat{\epsilon}_{i_{2}}^{\prime}\hat{\epsilon}_{i_{2}}}\right\}\\
&+\frac{8}{N^2(N-1)^2}\sum_{2\leq j_{1}<j_{2}\leq N}\sum_{1\leq i\leq j_{1}-1}\mE\left\{\frac{\tr\left(\hat{\epsilon}_{i\cdot}^{\prime}\mathbf{M}_{j_{1}}\hat{\epsilon}_{i\cdot}\right)
\tr\left(\hat{\epsilon}_{i\cdot}^{\prime}\mathbf{M}_{j_{2}}\hat{\epsilon}_{i\cdot}\right)}{\tr\left(\mathbf{M}_{j_{1}}\right)
\hat{\epsilon}_{i\cdot}^{\prime}\hat{\epsilon}_{i\cdot}\tr\left(\mathbf{M}_{j_{2}}\right)
\hat{\epsilon}_{i\cdot}^{\prime}\hat{\epsilon}_{i\cdot}}\right\}\\
%&+\frac{8}{N^2(N-1)^2}\sum_{2\leq j_{1}<j_{2}\leq N}\sum_{1\leq i\leq j_{1}-1}\mE\left\{\frac{\tr\left(\hat{\epsilon}_{i\cdot}^{\prime}\mathbf{M}_{j_{1}}\hat{\epsilon}_{i\cdot}\right)
%\tr\left(\hat{\epsilon}_{j_{1}}^{\prime}\mathbf{M}_{j_{2}}\hat{\epsilon}_{j_{1}}\right)}{\tr\left(\mathbf{M}_{j_{1}}\right)
%\hat{\epsilon}_{i\cdot}^{\prime}\hat{\epsilon}_{i\cdot}\tr\left(\mathbf{M}_{j_{2}}\right)
%\hat{\epsilon}_{j_{1}}^{\prime}\hat{\epsilon}_{j_{1}}}\right\}
\leq&\frac{4}{N^2(N-1)^2}\sum_{j=2}^{N}\sum_{i=1}^{j-1}\mE\left\{\frac{\tr^2\left(\hat{\epsilon}_{i\cdot}^{\prime}\mathbf{M}_{j}\hat{\epsilon}_{i\cdot}\right)}
{\tr^2\left(\mathbf{M}_{j}\right)
\left(\hat{\epsilon}_{i\cdot}^{\prime}\hat{\epsilon}_{i\cdot}\right)^2}\right\}\\
&+\frac{8}{N^2(N-1)^2}\sum_{2\leq j_{1}<j_{2}\leq N}\sum_{1\leq i\leq j_{1}-1}\frac{
\sqrt{\mE\left\{\frac{\tr^2\left(\hat{\epsilon}_{i\cdot}^{\prime}\mathbf{M}_{j_{1}}\hat{\epsilon}_{i\cdot}\right)}
{\left(\hat{\epsilon}_{i\cdot}^{\prime}\hat{\epsilon}_{i\cdot}\right)^2}\right\}
\mE\left\{\frac{\tr^2\left(\hat{\epsilon}_{i\cdot}^{\prime}\mathbf{M}_{j_{2}}\hat{\epsilon}_{i\cdot}\right)
}{\left(\hat{\epsilon}_{i\cdot}^{\prime}\hat{\epsilon}_{i\cdot}\right)^2}\right\}}}
{\tr\left(\mathbf{M}_{j_{1}}\right)\tr\left(\mathbf{M}_{j_{2}}\right)}\\
=&O\left(N^{-2}T^{-2}\right)+O\left(N^{-1}T^{-2}\right)\\
=&o\left(\sigma^4_{S_{N}}\right),
\end{align*}
where the last second equality holds because \begin{align*}
\mE\left\{\frac{\tr^2\left(\hat{\epsilon}_{i\cdot}^{\prime}\mathbf{M}_{j}\hat{\epsilon}_{i\cdot}\right)}
{\tr^2\left(\mathbf{M}_{j}\right)
\left(\hat{\epsilon}_{i\cdot}^{\prime}\hat{\epsilon}_{i\cdot}\right)^2}\right\}=&
\frac{\tr^2\left(\mathbf{M}_{i}\mathbf{M}_{j}
\right)+2\tr\left(\mathbf{M}_{i}\mathbf{M}_{j}\mathbf{M}_{i}\mathbf{M}_{j}\right)}{\tr^2\left(\mathbf{M}_{j}\right)\tr^2\left(\mathbf{M}_{i}\right)}\\
&+\frac{\gamma_{2}\tr[(\mathbf{\Sigma}^{1/2}
\mathbf{P}_{i}\mathbf{M}_{j}\mathbf{P}_{i}\mathbf{\Sigma}^{1/2})\circ(\mathbf{\Sigma}^{1/2}
\mathbf{P}_{i}\mathbf{M}_{j}\mathbf{P}_{i}\mathbf{\Sigma}^{1/2})]}{\tr^2\left(\mathbf{M}_{j}\right)\tr^2\left(\mathbf{M}_{i}\right)}
+O\left\{T^{-1}\tr^{-2}\left(\mathbf{M}_{j}\right)\right\}\\
\leq&\frac{\tr^2\left(\mathbf{M}_{i}\mathbf{M}_{j}
\right)+(2+|\gamma_{2}|)\tr\left(\mathbf{M}_{i}\mathbf{M}_{j}\mathbf{M}_{i}\mathbf{M}_{j}\right)}
{\tr^2\left(\mathbf{M}_{j}\right)\tr^2\left(\mathbf{M}_{i}\right)}+O\left\{T^{-1}\tr^{-2}\left(\mathbf{M}_{j}\right)\right\}\\
=&O(T^{-2}).
\end{align*}
Hence, we can conclude that $A_{1}=\sigma^2_{S_{N}}+o_{p}\left(\sigma^2_{S_{N}}\right).$ Then, we have
\begin{align*}
\mE\left(|A_{2}|\right)\leq&\frac{2}{N(N-1)}\sum_{j=2}^{N}\sum_{i=1}^{j-1}\mE\left[\frac{
\tr\left(\hat{\epsilon}_{i\cdot}^{\prime}\mathbf{M}_{j}\hat{\epsilon}_{i\cdot}\right)
\{2\tr\left(\mathbf{M}_{j}^2\right)+|\gamma_{2}|\tr(\mathbf{Q}_{jj}\circ\mathbf{Q}_{jj})\}}{\tr^3\left(\mathbf{M}_{j}\right)\hat{\epsilon}_{i\cdot}^{\prime}\hat{\epsilon}_{i\cdot}}\right]\\
=&\frac{2}{N(N-1)}\sum_{j=2}^{N}\sum_{i=1}^{j-1}\frac{2\tr\left(\mathbf{M}_{j}^2\right)+|\gamma_{2}|\tr(\mathbf{Q}_{jj}\circ\mathbf{Q}_{jj})}{\tr^3\left(\mathbf{M}_{j}\right)}
\mE\left\{\frac{\tr\left(\hat{\epsilon}_{i\cdot}^{\prime}\mathbf{M}_{j}\hat{\epsilon}_{i\cdot}\right)}{\hat{\epsilon}_{i\cdot}^{\prime}\hat{\epsilon}_{i\cdot}}\right\}\\
=&\frac{2}{N(N-1)}\sum_{j=2}^{N}\sum_{i=1}^{j-1}\frac{2\tr\left(\mathbf{M}_{j}^2\right)+|\gamma_{2}|\tr(\mathbf{Q}_{jj}\circ\mathbf{Q}_{jj})}{\tr^3\left(\mathbf{M}_{j}\right)}
\left\{\frac{\tr\left(\mathbf{M}_{i}\mathbf{M}_{j}\right)}{\tr\left(\mathbf{M}_{i}\right)}+O\left(T^{-1}\right)
\right\}\\
=&O\left(T^{-2}\right)=o\left(\sigma^2_{S_{N}}\right),
\end{align*}
where the last two equality holds due to $\tr(\mathbf{Q}_{jj}\circ\mathbf{Q}_{jj})\leq \tr(\mathbf{M}_{j}\mathbf{M}_{j})=O(T).$
Hence, we have $A_{2}=o_{p}\left(\sigma^2_{S_{N}}\right)$ by Markov inequality.
Next, we have
\begin{align*}
\mE\left(A_{3}\right)=&\frac{2}{N(N-1)}\sum_{j=2}^{N}\sum_{i=1}^{j-1}\mE\left\{\frac{
2\tr\left(\hat{\epsilon}_{i\cdot}^{\prime}\mathbf{M}^2_{j}\hat{\epsilon}_{i\cdot}\right)}
{\tr^2\left(\mathbf{M}_{j}\right)\hat{\epsilon}_{i\cdot}^{\prime}\hat{\epsilon}_{i\cdot}}\right\}\\
=&\frac{4}{N(N-1)}\sum_{j=2}^{N}\sum_{i=1}^{j-1}\frac{1}{\tr^2\left(\mathbf{M}_{j}\right)}
\mE\left\{\frac{\tr\left(\hat{\epsilon}_{i\cdot}^{\prime}\mathbf{M}^2_{j}\hat{\epsilon}_{i\cdot}\right)}{\hat{\epsilon}_{i\cdot}^{\prime}\hat{\epsilon}_{i\cdot}}\right\}\\
=&\frac{4}{N(N-1)}\sum_{j=2}^{N}\sum_{i=1}^{j-1}\frac{1}{\tr^2\left(\mathbf{M}_{j}\right)}
\left\{\frac{\tr\left(\mathbf{M}_{i}\mathbf{M}^2_{j}\right)}{\tr\left(\mathbf{M}_{i}\right)}+O\left(T^{-1}\right)
\right\}\\
=&O\left(T^{-2}\right)=o\left(\sigma^2_{S_{N}}\right),
\end{align*}
and
\begin{align*}
\mE(A_{4}/\gamma_{2})=&\frac{2}{N(N-1)}\sum_{j=2}^{N}\sum_{i=1}^{j-1}\mE\left\{\frac{\tr[(\mathbf{\Sigma}^{1/2}\mathbf{P}_{j}\mathbf{P}_{i}
\epsilon_{i\cdot}\epsilon_{i\cdot}^{\prime}\mathbf{P}_{i}\mathbf{P}_{j}\mathbf{\Sigma}^{1/2})\circ\mathbf{Q}_{jj}]}
{\tr^2\left(\mathbf{M}_{j}\right)\hat{\epsilon}_{i\cdot}^{\prime}\hat{\epsilon}_{i\cdot}}\right\}\\
\leq&\frac{2C}{N(N-1)}\sum_{j=2}^{N}\sum_{i=1}^{j-1}\mE\left\{\frac{\tr(\mathbf{\Sigma}^{1/2}\mathbf{P}_{j}\mathbf{P}_{i}
\epsilon_{i\cdot}\epsilon_{i\cdot}^{\prime}\mathbf{P}_{i}\mathbf{P}_{j}\mathbf{\Sigma}^{1/2})}
{\tr^2\left(\mathbf{M}_{j}\right)\hat{\epsilon}_{i\cdot}^{\prime}\hat{\epsilon}_{i\cdot}}\right\}\\
=&\frac{2C}{N(N-1)}\sum_{j=2}^{N}\sum_{i=1}^{j-1}\mE\left\{\frac{\tr(\epsilon_{i\cdot}^{\prime}\mathbf{P}_{i}\mathbf{M}_{j}\mathbf{P}_{i}
\epsilon_{i\cdot})}
{\tr^2\left(\mathbf{M}_{j}\right)\hat{\epsilon}_{i\cdot}^{\prime}\hat{\epsilon}_{i\cdot}}\right\}\\
=&\frac{2C}{N(N-1)}\sum_{j=2}^{N}\sum_{i=1}^{j-1}\left\{\frac{\tr(\mathbf{M}_{j}\mathbf{M}_{i})}
{\tr^2\left(\mathbf{M}_{j}\right)\tr(\mathbf{M}_{i})}+O\left(\frac{1}{T\tr^2\left(\mathbf{M}_{j}\right)}\right)\right\}\\
=&O\left(T^{-2}\right)=o\left(\sigma^2_{S_{N}}\right).
\end{align*}
Because $A_{3}$ and $A_{4}/\gamma_{2}$ are non-negative, we have $A_{3}=o_{p}\left(\sigma^2_{S_{N}}\right)$ and $A_{4}/\gamma_{2}=o_{p}\left(\sigma^2_{S_{N}}\right).$
Second, we focus on $B.$
\begin{align*}
B=&\frac{4}{N(N-1)}\sum_{j=2}^{N}\underset{i_1<i_2}{\sum^{j-1}\sum^{j-1}}\mE\left(\left.\hat{\rho}_{i_{1} j}\hat{\rho}_{i_{2} j}\right|\mathcal{F}_{n,j-1}\right)\\
=&\frac{4}{N(N-1)}\sum_{j=2}^{N}\left\{\underset{i_1<i_2}{\sum^{j-1}\sum^{j-1}}\mE\left(\left.\hat{\rho}_{i_{1} j}\hat{\rho}_{i_{2} j}\right|\mathcal{F}_{n,j-1}\right)\right\}\\
=&\sum_{j=2}^{N}\left\{\frac{4}{N(N-1)}\underset{i_1<i_2}{\sum^{j-1}\sum^{j-1}}\delta_{i_{1}i_{2}}\right\},
\end{align*}
where we denote that $\delta_{i_{1}i_{2}}\doteq\mE\left(\left.\hat{\rho}_{i_{1} j}\hat{\rho}_{i_{2} j}\right|\mathcal{F}_{n,j-1}\right).$
Obviously, we have $\mE\left(B\right)=0 $ due to the law of total expectation.
By Lemmas \ref{lem:laplace approxi} and \ref{lem:matrix inequality}, we have
\begin{align*}
\delta_{i_{1}i_{2}}
=&\frac{\hat{\epsilon}_{i_{1}\cdot}^{\prime}\mathbf{M}_{j}
\hat{\epsilon}_{i_{2}\cdot}}{\sqrt{\hat{\epsilon}_{i_{1}\cdot}^{\prime}\hat{\epsilon}_{i_{1}\cdot}
\hat{\epsilon}_{i_{2}\cdot}^{\prime}\hat{\epsilon}_{i_{2}\cdot}}\tr\left(\mathbf{M}_{j}\right)}
+\frac{\hat{\epsilon}_{i_{1}\cdot}^{\prime}\mathbf{M}_{j}
\hat{\epsilon}_{i_{2}\cdot}
\{2\tr(\mathbf{M}^2_{j})+\gamma_{2}\tr(\mathbf{Q}_{jj}\circ\mathbf{Q}_{jj})\}}
{\sqrt{\hat{\epsilon}_{i_{1}\cdot}^{\prime}\hat{\epsilon}_{i_{1}\cdot}
\hat{\epsilon}_{i_{2}\cdot}^{\prime}\hat{\epsilon}_{i_{2}\cdot}}\tr^3\left(\mathbf{M}_{j}\right)}\\
&-\frac{2\hat{\epsilon}_{i_{1}\cdot}^{\prime}\mathbf{M}^2_{j}
\hat{\epsilon}_{i_{2}\cdot}+\gamma_{2}\tr[(\mathbf{\Sigma}^{1/2}\mathbf{P}_{j}\hat{\epsilon}_{i_{1}\cdot}
\hat{\epsilon}_{i_{2}\cdot}^{\prime}\mathbf{P}_{j}\mathbf{\Sigma}^{1/2})\circ\mathbf{Q}_{jj}]}{\sqrt{\hat{\epsilon}_{i_{1}\cdot}^{\prime}\hat{\epsilon}_{i_{1}\cdot}
\hat{\epsilon}_{i_{2}\cdot}^{\prime}\hat{\epsilon}_{i_{2}\cdot}}\tr^2\left(\mathbf{M}_{j}\right)}+O\left(T^{-2}\right)
\end{align*}
for any $1\leq i_{1}<i_{2}<j\leq N.$
Moreover, we have
\begin{align*}
\mE\left(\delta^2_{i_{1}i_{2}}\right)\leq&5\mE\left\{\frac{\left(\hat{\epsilon}_{i_{1}\cdot}^{\prime}\mathbf{M}_{j}
\hat{\epsilon}_{i_{2}\cdot}\right)^2}{\hat{\epsilon}_{i_{1}\cdot}^{\prime}\hat{\epsilon}_{i_{1}\cdot}
\hat{\epsilon}_{i_{2}\cdot}^{\prime}\hat{\epsilon}_{i_{2}\cdot}\tr^2\left(\mathbf{M}_{j}\right)}\right\}
+5\mE\left\{\frac{(2+|\gamma_{2}|)^2\left(
\hat{\epsilon}_{i_{2}\cdot}^{\prime}\mathbf{M}_{j}\hat{\epsilon}_{i_{1}\cdot}\right)^2
\tr^2\left(\mathbf{M}^2_{j}\right)}
{\hat{\epsilon}_{i_{1}\cdot}^{\prime}\hat{\epsilon}_{i_{1}\cdot}
\hat{\epsilon}_{i_{2}\cdot}^{\prime}\hat{\epsilon}_{i_{2}\cdot}\tr^6\left(\mathbf{M}_{j}\right)}\right\}\\
&+5\mE\left\{\frac{4\left(\hat{\epsilon}_{i_{2}\cdot}^{\prime}\mathbf{M}^2_{j}\hat{\epsilon}_{i_{1}\cdot}\right)^2}
{\hat{\epsilon}_{i_{1}\cdot}^{\prime}\hat{\epsilon}_{i_{1}\cdot}
\hat{\epsilon}_{i_{2}\cdot}^{\prime}\hat{\epsilon}_{i_{2}\cdot}\tr^4\left(\mathbf{M}_{j}\right)}\right\}+
5\mE\left\{\frac{\gamma_{2}^2\tr^2[(\mathbf{\Sigma}^{1/2}\mathbf{P}_{j}\hat{\epsilon}_{i_{1}\cdot}
\hat{\epsilon}_{i_{2}\cdot}^{\prime}\mathbf{P}_{j}\mathbf{\Sigma}^{1/2})\circ\mathbf{Q}_{jj}]}
{\hat{\epsilon}_{i_{1}\cdot}^{\prime}\hat{\epsilon}_{i_{1}\cdot}
\hat{\epsilon}_{i_{2}\cdot}^{\prime}\hat{\epsilon}_{i_{2}\cdot}\tr^4\left(\mathbf{M}_{j}\right)}\right\}
+O\left(T^{-4}\right)\\
\leq&\frac{5\tr\left(\mathbf{M}_{i_{1}}\mathbf{M}_{j}\mathbf{M}_{i_{2}}\mathbf{M}_{j}\right)}
{\tr^2\left(\mathbf{M}_{j}\right)\tr\left(\mathbf{M}_{i_{1}}\right)\tr\left(\mathbf{M}_{i_{2}}\right)}
+\frac{5(2+|\gamma_{2}|)^2\tr\left(\mathbf{M}_{i_{1}}\mathbf{M}_{j}\mathbf{M}_{i_{2}}\mathbf{M}_{j}\right)\tr^2\left(\mathbf{M}^2_{j}\right)}
{\tr\left(\mathbf{M}_{i_{1}}\right)\tr\left(\mathbf{M}_{i_{2}}\right)\tr^6\left(\mathbf{M}_{j}\right)}\\
&+\frac{20\tr\left(\mathbf{M}_{i_{1}}\mathbf{M}^2_{j}\mathbf{M}_{i_{2}}\mathbf{M}^2_{j}\right)}
{\tr\left(\mathbf{M}_{i_{1}}\right)\tr\left(\mathbf{M}_{i_{2}}\right)\tr^4\left(\mathbf{M}_{j}\right)}
+\frac{5C^2\gamma^2_{2}T\tr\left(\mathbf{M}_{i_{1}}\mathbf{M}_{j}\right)\tr\left(\mathbf{M}_{i_{2}}\mathbf{M}_{j}\right)}
{\tr\left(\mathbf{M}_{i_{1}}\right)\tr\left(\mathbf{M}_{i_{2}}\right)\tr^4\left(\mathbf{M}_{j}\right)}
+O\left(T^{-3}\right)\\
=&O\left(T^{-3}\right),
\end{align*}
where the last two inequality holds due to $$0\leq \tr^{2}[(\mathbf{\Sigma}^{1/2}\mathbf{P}_{j}
\epsilon_{i_{1}}\epsilon_{i_{2}}^{\prime}\mathbf{P}_{j}\mathbf{\Sigma}^{1/2})
\circ \mathbf{Q}_{ii}]\leq TC^{2} (\epsilon_{i_{2}}^{\prime}\mathbf{M}_{j}\epsilon_{i_{2}})(\epsilon_{i_{1}}^{\prime}\mathbf{M}_{j}\epsilon_{i_{1}}),$$
by Lemma \ref{lem:matrix inequality}, for some constant $C\geq0.$
So, we can calculate $\mE(B^2),$
\begin{align*}
\mE\left(B^2\right)\leq & N\sum_{j=2}^{N}\frac{16}{N^2(N-2)^2}\mE\left(\underset{i_1<i_2}{\sum^{j-1}\sum^{j-1}}
\underset{i_3<i_4}{\sum^{j-1}\sum^{j-1}}\delta_{i_{1}i_{2}}\delta_{i_{3}i_{4}}\right)\\
\leq&\frac{16}{N(N-2)^2}\sum_{j=2}^{N}\underset{i_1<i_2}{\sum^{j-1}\sum^{j-1}}
\mE\left(\delta^2_{i_{1}i_{2}}\right)\\
=&O\left(T^{-3}\right)=o\left(\sigma^4_{S_{N}}\right),
\end{align*}
then, we obtain that $B=o_{p}\left(\sigma^2_{S_{N}}\right).$
Next, we will focus on (\ref{clt2}) and prove that
\begin{align*}
\sum_{j=2}^{N}\mE\left(Z_{j}^4\right)=o\{\sigma^4_{S_{N}}\}
\end{align*}
due to $\mE\left\{\sum_{j=2}^{N}\mE\left(Z_{j}^4|\mathcal{F}_{j-1}\right)\right\}=\sum_{j=2}^{N}\mE\left(Z_{j}^4\right).$
We divide $\sum_{j=2}^{N}\mE\left(Z_{j}^4\right)$ into two parts:
\begin{align*}
\sum_{j=2}^{N}\mE\left(Z_{j}^4\right)=&\frac{4}{N^2(N-1)^2}\sum_{j=2}^{N}
\sum_{i_{1}=1}^{j-1}\sum_{i_{2}=1}^{j-1}\sum_{i_{3}=1}^{j-1}\sum_{i_{4}=1}^{j-1}
\mE\left(\hat{\rho}_{i_{1} j}\hat{\rho}_{i_{2} j}\hat{\rho}_{i_{3}j}\hat{\rho}_{i_{4}j}\right)\\
=&\frac{4}{N^2(N-1)^2}\sum_{j=2}^{N}\sum_{i=1}^{j-1}\mE\left(\hat{\rho}^4_{i j}\right)
+\frac{12}{N^2(N-1)^2}\sum_{j=2}^{N}\underset{i_{1}\neq i_{2}}{\sum^{j-1}\sum^{j-1}}
\mE\left(\hat{\rho}^2_{i_{1}j}\hat{\rho}^2_{i_{2}j}\right)\\
=&C+D.
\end{align*}
By Lemma \ref{lem:laplace approxi}, we have
\begin{align*}
\mE\left(\hat{\rho}^4_{i j}\right)
=&\mE\left\{\mE\left(\left.\hat{\rho}^4_{i j}\right|\epsilon_{j\cdot}\right)\right\}\\
=&\mE\Bigg[\frac{2\tr\left(\hat{\epsilon}_{j\cdot}^{\prime}\mathbf{M}_{i}\hat{\epsilon}_{j\cdot}
\hat{\epsilon}_{j\cdot}^{\prime}\mathbf{M}_{i}\hat{\epsilon}_{j\cdot}\right)+
\tr^2\left(\hat{\epsilon}_{j\cdot}^{\prime}\mathbf{M}_{i}\hat{\epsilon}_{j\cdot}\right)
+\gamma_{2}\tr[(\mathbf{\Sigma}^{1/2}\mathbf{P}_{i}\hat{\epsilon}_{j\cdot}\hat{\epsilon}_{j\cdot}^{\prime}
\mathbf{P}_{i}\mathbf{\Sigma}^{1/2})\circ(\mathbf{\Sigma}^{1/2}\mathbf{P}_{i}\hat{\epsilon}_{j\cdot}\hat{\epsilon}_{j\cdot}^{\prime}
\mathbf{P}_{i}\mathbf{\Sigma}^{1/2})]}
{\left(\hat{\epsilon}_{j\cdot}^{\prime}\hat{\epsilon}_{j\cdot}\right)^2\tr^2\left(\mathbf{M}_{i}\right)}\\
&+
\frac{3\mE\left\{\left({\epsilon}_{i\cdot}^{\prime}\mathbf{P}_{i}
\hat{\epsilon}_{j\cdot}\hat{\epsilon}_{j\cdot}^{\prime}\mathbf{P}_{i}{\epsilon}_{i\cdot}\right)^2|\epsilon_{j\cdot}\right\}
\{2\tr\left(\mathbf{M}^2_{i}\right)+\gamma_{2}\tr(\mathbf{Q}_{ii}\circ\mathbf{Q}_{ii})\}}
{\left(\hat{\epsilon}_{j\cdot}^{\prime}\hat{\epsilon}_{j\cdot}\right)^2\tr^4\left(\mathbf{M}_{i}\right)}\\
&-\frac{2\mE\left\{\left({\epsilon}_{i\cdot}^{\prime}\mathbf{P}_{i}
\hat{\epsilon}_{j\cdot}\hat{\epsilon}_{j\cdot}^{\prime}\mathbf{P}_{i}{\epsilon}_{i\cdot}\right)^2
\left({\epsilon}_{i\cdot}^{\prime}\mathbf{P}_{i}
{\epsilon}_{i\cdot}\right)|\epsilon_{j\cdot}\right\}-2\mE\left\{\left({\epsilon}_{i\cdot}^{\prime}\mathbf{P}_{i}
\hat{\epsilon}_{j\cdot}\hat{\epsilon}_{j\cdot}^{\prime}\mathbf{P}_{i}{\epsilon}_{i\cdot}\right)^2|\epsilon_{j\cdot}\right\}
\mE\left({\epsilon}_{i\cdot}^{\prime}\mathbf{P}_{i}
{\epsilon}_{i\cdot}\right)}
{\left(\hat{\epsilon}_{j\cdot}^{\prime}\hat{\epsilon}_{j\cdot}\right)^2\tr^3\left(\mathbf{M}_{i}\right)}+O\left(T^{-2}\right)
\Bigg].
\end{align*}
Then, by Lemma \ref{le:moment of quadratic form},
we have \begin{align*}
&\mE\left\{\left({\epsilon}_{i\cdot}^{\prime}\mathbf{P}_{i}
\hat{\epsilon}_{j\cdot}\hat{\epsilon}_{j\cdot}^{\prime}\mathbf{P}_{i}{\epsilon}_{i\cdot}\right)^2
\left({\epsilon}_{i\cdot}^{\prime}\mathbf{P}_{i}
{\epsilon}_{i\cdot}\right)|\epsilon_{j\cdot}\right\}-\mE\left\{\left({\epsilon}_{i\cdot}^{\prime}\mathbf{P}_{i}
\hat{\epsilon}_{j\cdot}\hat{\epsilon}_{j\cdot}^{\prime}\mathbf{P}_{i}{\epsilon}_{i\cdot}\right)^2|\epsilon_{j\cdot}\right\}
\mE\left({\epsilon}_{i\cdot}^{\prime}\mathbf{P}_{i}
{\epsilon}_{i\cdot}\right)\\
=&\gamma_{4} \tr\left[(\mathbf{\Sigma}^{1/2}\mathbf{P}_{i}\hat{\epsilon}_{j\cdot}\hat{\epsilon}_{j\cdot}^{\prime}\mathbf{P}_{i}\mathbf{\Sigma}^{1/2}) \circ (\mathbf{\Sigma}^{1/2}\mathbf{P}_{i}\hat{\epsilon}_{j\cdot}\hat{\epsilon}_{j\cdot}^{\prime}\mathbf{P}_{i}\mathbf{\Sigma}^{1/2} ) \circ ( \mathbf{\Sigma}^{1/2}\mathbf{P}_{i}\mathbf{\Sigma}^{1/2})\right]\\&+2\gamma_{2} \tr\left(\mathbf{\Sigma}^{1/2}\mathbf{P}_{i}\hat{\epsilon}_{j\cdot}\hat{\epsilon}_{j\cdot}^{\prime}\mathbf{P}_{i}\mathbf{\Sigma}^{1/2}\right) \tr\left[(\mathbf{\Sigma}^{1/2}\mathbf{P}_{i}\hat{\epsilon}_{j\cdot}\hat{\epsilon}_{j\cdot}^{\prime}\mathbf{P}_{i}\mathbf{\Sigma}^{1/2} )\circ(\mathbf{\Sigma}^{1/2}\mathbf{P}_{i}\mathbf{\Sigma}^{1/2})\right] \\
&+8 \gamma_{2} \tr\left[(\mathbf{\Sigma}^{1/2}\mathbf{P}_{i}\hat{\epsilon}_{j\cdot}\hat{\epsilon}_{j\cdot}^{\prime}\mathbf{P}_{i}\mathbf{\Sigma}^{1/2} ) \circ\left(\mathbf{\Sigma}^{1/2}\mathbf{P}_{i}\hat{\epsilon}_{j\cdot}\hat{\epsilon}_{j\cdot}^{\prime}\mathbf{P}_{i}\mathbf{\Sigma}^{1/2} \mathbf{\Sigma}^{1/2}\mathbf{P}_{i}\mathbf{\Sigma}^{1/2}\right)\right] \\
&+4 \gamma_{2} \tr\left[(\mathbf{\Sigma}^{1/2}\mathbf{P}_{i}\mathbf{\Sigma}^{1/2} ) \circ\left(\mathbf{\Sigma}^{1/2}\mathbf{P}_{i}\hat{\epsilon}_{j\cdot}\hat{\epsilon}_{j\cdot}^{\prime}\mathbf{P}_{i}\mathbf{\Sigma}^{1/2} \mathbf{\Sigma}^{1/2}\mathbf{P}_{i}\hat{\epsilon}_{j\cdot}\hat{\epsilon}_{j\cdot}^{\prime}\mathbf{P}_{i}\mathbf{\Sigma}^{1/2}\right)\right]\\
&+4 \gamma_{1}^{2}\left[{\tau}_{T}^{\prime}\left\{\mathbf{I}_{T} \circ (\mathbf{\Sigma}^{1/2}\mathbf{P}_{i}\hat{\epsilon}_{j\cdot}\hat{\epsilon}_{j\cdot}^{\prime}\mathbf{P}_{i}\mathbf{\Sigma}^{1/2})\right\} \mathbf{\Sigma}^{1/2}\mathbf{P}_{i}\hat{\epsilon}_{j\cdot}\hat{\epsilon}_{j\cdot}^{\prime}\mathbf{P}_{i}\mathbf{\Sigma}^{1/2}\left\{\mathbf{I}_{T} \circ(\mathbf{\Sigma}^{1/2}\mathbf{P}_{i}\mathbf{\Sigma}^{1/2})\right\} {\tau}_{T}\right] \\
&+2 \gamma_{1}^{2}\left[{\tau}_{T}^{\prime}\left\{\mathbf{I}_{T} \circ( \mathbf{\Sigma}^{1/2}\mathbf{P}_{i}\hat{\epsilon}_{j\cdot}\hat{\epsilon}_{j\cdot}^{\prime}\mathbf{P}_{i}\mathbf{\Sigma}^{1/2})\right\} \mathbf{\Sigma}^{1/2}\mathbf{P}_{i}\mathbf{\Sigma}^{1/2}\left\{\mathbf{I}_{T} \circ (\mathbf{\Sigma}^{1/2}\mathbf{P}_{i}\hat{\epsilon}_{j\cdot}\hat{\epsilon}_{j\cdot}^{\prime}\mathbf{P}_{i}\mathbf{\Sigma}^{1/2})\right\} {\tau}_{T}\right] \\
&+4 \gamma_{1}^{2}\left[{\tau}_{T}^{\prime}\left\{(\mathbf{\Sigma}^{1/2}\mathbf{P}_{i}\hat{\epsilon}_{j\cdot}\hat{\epsilon}_{j\cdot}^{\prime}\mathbf{P}_{i}\mathbf{\Sigma}^{1/2} )\circ ( \mathbf{\Sigma}^{1/2}\mathbf{P}_{i}\hat{\epsilon}_{j\cdot}\hat{\epsilon}_{j\cdot}^{\prime}\mathbf{P}_{i}\mathbf{\Sigma}^{1/2} )\circ(\mathbf{\Sigma}^{1/2}\mathbf{P}_{i}\mathbf{\Sigma}^{1/2})\right\} {\tau}_{T}\right]\\
&+4 \tr\left(\mathbf{\Sigma}^{1/2}\mathbf{P}_{i}\hat{\epsilon}_{j\cdot}\hat{\epsilon}_{j\cdot}^{\prime}\mathbf{P}_{i}\mathbf{\Sigma}^{1/2}\right) \tr\left(\mathbf{\Sigma}^{1/2}\mathbf{P}_{i}\hat{\epsilon}_{j\cdot}\hat{\epsilon}_{j\cdot}^{\prime}\mathbf{P}_{i}\mathbf{\Sigma}^{1/2} \mathbf{\Sigma}^{1/2}\mathbf{P}_{i}\mathbf{\Sigma}^{1/2}\right) \\
&+8 \tr\left(\mathbf{\Sigma}^{1/2}\mathbf{P}_{i}\hat{\epsilon}_{j\cdot}\hat{\epsilon}_{j\cdot}^{\prime}\mathbf{P}_{i}\mathbf{\Sigma}^{1/2} \mathbf{\Sigma}^{1/2}\mathbf{P}_{i}\hat{\epsilon}_{j\cdot}\hat{\epsilon}_{j\cdot}^{\prime}\mathbf{P}_{i}\mathbf{\Sigma}^{1/2} \mathbf{\Sigma}^{1/2}\mathbf{P}_{i}\mathbf{\Sigma}^{1/2}\right)\\
\geq&\gamma_{4} \tr\left\{(\mathbf{\Sigma}^{1/2}\mathbf{P}_{i}\hat{\epsilon}_{j\cdot}\hat{\epsilon}_{j\cdot}^{\prime}\mathbf{P}_{i}\mathbf{\Sigma}^{1/2} )\circ( \mathbf{\Sigma}^{1/2}\mathbf{P}_{i}\hat{\epsilon}_{j\cdot}\hat{\epsilon}_{j\cdot}^{\prime}\mathbf{P}_{i}\mathbf{\Sigma}^{1/2} )\circ ( \mathbf{\Sigma}^{1/2}\mathbf{P}_{i}\mathbf{\Sigma}^{1/2})\right\}\\&+2\gamma_{2} \tr\left(\mathbf{\Sigma}^{1/2}\mathbf{P}_{i}\hat{\epsilon}_{j\cdot}\hat{\epsilon}_{j\cdot}^{\prime}\mathbf{P}_{i}\mathbf{\Sigma}^{1/2}\right) \tr\left\{(\mathbf{\Sigma}^{1/2}\mathbf{P}_{i}\hat{\epsilon}_{j\cdot}\hat{\epsilon}_{j\cdot}^{\prime}\mathbf{P}_{i}\mathbf{\Sigma}^{1/2} )\circ(\mathbf{\Sigma}^{1/2}\mathbf{P}_{i}\mathbf{\Sigma}^{1/2})\right\} \\
&+8 \gamma_{2} \tr\left[(\mathbf{\Sigma}^{1/2}\mathbf{P}_{i}\hat{\epsilon}_{j\cdot}\hat{\epsilon}_{j\cdot}^{\prime}\mathbf{P}_{i}\mathbf{\Sigma}^{1/2} ) \circ\left(\mathbf{\Sigma}^{1/2}\mathbf{P}_{i}\hat{\epsilon}_{j\cdot}\hat{\epsilon}_{j\cdot}^{\prime}\mathbf{P}_{i}\mathbf{\Sigma}^{1/2} \mathbf{\Sigma}^{1/2}\mathbf{P}_{i}\mathbf{\Sigma}^{1/2}\right)\right] \\
&+4 \gamma_{2} \tr\left[(\mathbf{\Sigma}^{1/2}\mathbf{P}_{i}\mathbf{\Sigma}^{1/2} ) \circ\left(\mathbf{\Sigma}^{1/2}\mathbf{P}_{i}\hat{\epsilon}_{j\cdot}\hat{\epsilon}_{j\cdot}^{\prime}\mathbf{P}_{i}\mathbf{\Sigma}^{1/2} \mathbf{\Sigma}^{1/2}\mathbf{P}_{i}\hat{\epsilon}_{j\cdot}\hat{\epsilon}_{j\cdot}^{\prime}\mathbf{P}_{i}\mathbf{\Sigma}^{1/2}\right)\right]\\
&+4 \gamma_{1}^{2}\left[{\tau}_{T}^{\prime}\left\{\mathbf{I}_{T} \circ ( \mathbf{\Sigma}^{1/2}\mathbf{P}_{i}\hat{\epsilon}_{j\cdot}\hat{\epsilon}_{j\cdot}^{\prime}\mathbf{P}_{i}\mathbf{\Sigma}^{1/2})\right\} \mathbf{\Sigma}^{1/2}\mathbf{P}_{i}\hat{\epsilon}_{j\cdot}\hat{\epsilon}_{j\cdot}^{\prime}\mathbf{P}_{i}\mathbf{\Sigma}^{1/2}\left\{\mathbf{I}_{T} \circ(\mathbf{\Sigma}^{1/2}\mathbf{P}_{i}\mathbf{\Sigma}^{1/2})\right\} {\tau}_{T}\right]\\&+2 \gamma_{1}^{2}\left[{\tau}_{T}^{\prime}\left\{\mathbf{I}_{T} \circ (\mathbf{\Sigma}^{1/2}\mathbf{P}_{i}\hat{\epsilon}_{j\cdot}\hat{\epsilon}_{j\cdot}^{\prime}\mathbf{P}_{i}\mathbf{\Sigma}^{1/2})\right\} \mathbf{\Sigma}^{1/2}\mathbf{P}_{i}\mathbf{\Sigma}^{1/2}\left\{\mathbf{I}_{T} \circ (\mathbf{\Sigma}^{1/2}\mathbf{P}_{i}\hat{\epsilon}_{j\cdot}\hat{\epsilon}_{j\cdot}^{\prime}\mathbf{P}_{i}\mathbf{\Sigma}^{1/2})\right\} {\tau}_{T}\right] \\
&+4 \gamma_{1}^{2}\left[{\tau}_{T}^{\prime}\left\{(\mathbf{\Sigma}^{1/2}\mathbf{P}_{i}\hat{\epsilon}_{j\cdot}\hat{\epsilon}_{j\cdot}^{\prime}\mathbf{P}_{i}\mathbf{\Sigma}^{1/2} )\circ ( \mathbf{\Sigma}^{1/2}\mathbf{P}_{i}\hat{\epsilon}_{j\cdot}\hat{\epsilon}_{j\cdot}^{\prime}\mathbf{P}_{i}\mathbf{\Sigma}^{1/2} )\circ(\mathbf{\Sigma}^{1/2}\mathbf{P}_{i}\mathbf{\Sigma}^{1/2})\right\} {\tau}_{T}\right],
\end{align*}
where the last inequality holds because the last four terms on the right side of the first equation above are all greater than or equal to zero, according to Lemma \ref{lem:matrix inequality}.

Then, we have
\begin{align*}
\mE\left(\hat{\rho}^4_{i j}\right)
\leq&\mE\Bigg(\frac{(3+|\gamma_{2}|)\left(\hat{\epsilon}_{j\cdot}^{\prime}\mathbf{M}_{i}\hat{\epsilon}_{j\cdot}\right)^2}
{\left(\hat{\epsilon}_{j\cdot}^{\prime}\hat{\epsilon}_{j\cdot}\right)^2\tr^2\left(\mathbf{M}_{i}\right)}
+\frac{3(3+|\gamma_{2}|)(2+|\gamma_{2}|)\left(\hat{\epsilon}_{j\cdot}^{\prime}\mathbf{M}_{i}\hat{\epsilon}_{j\cdot}\right)^2\tr\left(\mathbf{M}^2_{i}\right)}
{\left(\hat{\epsilon}_{j\cdot}^{\prime}\hat{\epsilon}_{j\cdot}\right)^2\tr^4\left(\mathbf{M}_{i}\right)}\\
&+2\tr^{-3}(\mathbf{M}_{i})(\hat{\epsilon}_{j\cdot}^{\prime}\hat{\epsilon}_{j\cdot})^{-2}
\Big\{\gamma_{4} \tr\left\{(\mathbf{\Sigma}^{1/2}\mathbf{P}_{i}\hat{\epsilon}_{j\cdot}\hat{\epsilon}_{j\cdot}^{\prime}\mathbf{P}_{i}\mathbf{\Sigma}^{1/2} )\circ ( \mathbf{\Sigma}^{1/2}\mathbf{P}_{i}\hat{\epsilon}_{j\cdot}\hat{\epsilon}_{j\cdot}^{\prime}\mathbf{P}_{i}\mathbf{\Sigma}^{1/2} )\circ ( \mathbf{\Sigma}^{1/2}\mathbf{P}_{i}\mathbf{\Sigma}^{1/2})\right\}\\&+2\gamma_{2} \tr\left(\mathbf{\Sigma}^{1/2}\mathbf{P}_{i}\hat{\epsilon}_{j\cdot}\hat{\epsilon}_{j\cdot}^{\prime}\mathbf{P}_{i}\mathbf{\Sigma}^{1/2}\right) \tr\left\{(\mathbf{\Sigma}^{1/2}\mathbf{P}_{i}\hat{\epsilon}_{j\cdot}\hat{\epsilon}_{j\cdot}^{\prime}\mathbf{P}_{i}\mathbf{\Sigma}^{1/2} )\circ(\mathbf{\Sigma}^{1/2}\mathbf{P}_{i}\mathbf{\Sigma}^{1/2})\right\} \\
&+8 \gamma_{2} \tr\left[(\mathbf{\Sigma}^{1/2}\mathbf{P}_{i}\hat{\epsilon}_{j\cdot}\hat{\epsilon}_{j\cdot}^{\prime}\mathbf{P}_{i}\mathbf{\Sigma}^{1/2} )\circ\left(\mathbf{\Sigma}^{1/2}\mathbf{P}_{i}\hat{\epsilon}_{j\cdot}\hat{\epsilon}_{j\cdot}^{\prime}\mathbf{P}_{i}\mathbf{\Sigma}^{1/2} \mathbf{\Sigma}^{1/2}\mathbf{P}_{i}\mathbf{\Sigma}^{1/2}\right)\right] \\
&+4 \gamma_{2} \tr\left[(\mathbf{\Sigma}^{1/2}\mathbf{P}_{i}\mathbf{\Sigma}^{1/2} )\circ\left(\mathbf{\Sigma}^{1/2}\mathbf{P}_{i}\hat{\epsilon}_{j\cdot}\hat{\epsilon}_{j\cdot}^{\prime}\mathbf{P}_{i}\mathbf{\Sigma}^{1/2} \mathbf{\Sigma}^{1/2}\mathbf{P}_{i}\hat{\epsilon}_{j\cdot}\hat{\epsilon}_{j\cdot}^{\prime}\mathbf{P}_{i}\mathbf{\Sigma}^{1/2}\right)\right]\\
&+4 \gamma_{1}^{2}\left[{\tau}_{T}^{\prime}\left\{\mathbf{I}_{T} \circ ( \mathbf{\Sigma}^{1/2}\mathbf{P}_{i}\hat{\epsilon}_{j\cdot}\hat{\epsilon}_{j\cdot}^{\prime}\mathbf{P}_{i}\mathbf{\Sigma}^{1/2})\right\} \mathbf{\Sigma}^{1/2}\mathbf{P}_{i}\hat{\epsilon}_{j\cdot}\hat{\epsilon}_{j\cdot}^{\prime}\mathbf{P}_{i}\mathbf{\Sigma}^{1/2}\left\{\mathbf{I}_{T} \circ(\mathbf{\Sigma}^{1/2}\mathbf{P}_{i}\mathbf{\Sigma}^{1/2})\right\} {\tau}_{T}\right]\\
&+2 \gamma_{1}^{2}\left[{\tau}_{T}^{\prime}\left\{\mathbf{I}_{T} \circ ( \mathbf{\Sigma}^{1/2}\mathbf{P}_{i}\hat{\epsilon}_{j\cdot}\hat{\epsilon}_{j\cdot}^{\prime}\mathbf{P}_{i}\mathbf{\Sigma}^{1/2})\right\} \mathbf{\Sigma}^{1/2}\mathbf{P}_{i}\mathbf{\Sigma}^{1/2}\left\{\mathbf{I}_{T} \circ (\mathbf{\Sigma}^{1/2}\mathbf{P}_{i}\hat{\epsilon}_{j\cdot}\hat{\epsilon}_{j\cdot}^{\prime}\mathbf{P}_{i}\mathbf{\Sigma}^{1/2})\right\} {\tau}_{T}\right] \\
&+4 \gamma_{1}^{2}\left[{\tau}_{T}^{\prime}\left\{(\mathbf{\Sigma}^{1/2}\mathbf{P}_{i}\hat{\epsilon}_{j\cdot}\hat{\epsilon}_{j\cdot}^{\prime}\mathbf{P}_{i}\mathbf{\Sigma}^{1/2} )\circ ( \mathbf{\Sigma}^{1/2}\mathbf{P}_{i}\hat{\epsilon}_{j\cdot}\hat{\epsilon}_{j\cdot}^{\prime}\mathbf{P}_{i}\mathbf{\Sigma}^{1/2} )\circ(\mathbf{\Sigma}^{1/2}\mathbf{P}_{i}\mathbf{\Sigma}^{1/2})\right\} {\tau}_{T}\right]\Big\}+O\left(T^{-2}\right)
\Bigg)\\
\leq&\mE\Bigg(\frac{(3+|\gamma_{2}|)\left(\hat{\epsilon}_{j\cdot}^{\prime}\mathbf{M}_{i}\hat{\epsilon}_{j\cdot}\right)^2}
{\left(\hat{\epsilon}_{j\cdot}^{\prime}\hat{\epsilon}_{j\cdot}\right)^2\tr^2\left(\mathbf{M}_{i}\right)}
+\frac{3(3+|\gamma_{2}|)(2+|\gamma_{2}|)\left(\hat{\epsilon}_{j\cdot}^{\prime}\mathbf{M}_{i}\hat{\epsilon}_{j\cdot}\right)^2\tr\left(\mathbf{M}^2_{i}\right)}
{\left(\hat{\epsilon}_{j\cdot}^{\prime}\hat{\epsilon}_{j\cdot}\right)^2\tr^4\left(\mathbf{M}_{i}\right)}
\\
&+2\tr^{-3}(\mathbf{M}_{i})(\hat{\epsilon}_{j\cdot}^{\prime}\hat{\epsilon}_{j\cdot})^{-2}
\Big\{C|\gamma_{4}| \tr\left(\mathbf{\Sigma}^{1/2}\mathbf{P}_{i}\hat{\epsilon}_{j\cdot}\hat{\epsilon}_{j\cdot}^{\prime}\mathbf{P}_{i}\mathbf{\Sigma}^{1/2} \mathbf{\Sigma}^{1/2}\mathbf{P}_{i}\hat{\epsilon}_{j\cdot}\hat{\epsilon}_{j\cdot}^{\prime}\mathbf{P}_{i}\mathbf{\Sigma}^{1/2}\right)\\
&+2C|\gamma_{2}| \tr\left(\mathbf{\Sigma}^{1/2}\mathbf{P}_{i}\hat{\epsilon}_{j\cdot}\hat{\epsilon}_{j\cdot}^{\prime}\mathbf{P}_{i}\mathbf{\Sigma}^{1/2}\right) \tr\left(\mathbf{\Sigma}^{1/2}\mathbf{P}_{i}\hat{\epsilon}_{j\cdot}\hat{\epsilon}_{j\cdot}^{\prime}\mathbf{P}_{i}\mathbf{\Sigma}^{1/2}\right) \\
&+8 |\gamma_{2} | (\hat{\epsilon}_{j\cdot}^{\prime}\mathbf{M}_{i}\hat{\epsilon}_{j\cdot})^{3/2} (\hat{\epsilon}_{j\cdot}^{\prime}\mathbf{M}^3_{i}\hat{\epsilon}_{j\cdot})^{1/2}\\
&+4C |\gamma_{2} | \tr\left(\mathbf{\Sigma}^{1/2}\mathbf{P}_{i}\hat{\epsilon}_{j\cdot}\hat{\epsilon}_{j\cdot}^{\prime}\mathbf{P}_{i}\mathbf{\Sigma}^{1/2} \mathbf{\Sigma}^{1/2}\mathbf{P}_{i}\hat{\epsilon}_{j\cdot}\hat{\epsilon}_{j\cdot}^{\prime}\mathbf{P}_{i}\mathbf{\Sigma}^{1/2}\right)\\
&+4 \gamma_{1}^{2}\left[{\tau}_{T}^{\prime}\left\{\mathbf{I}_{T} \circ (\mathbf{\Sigma}^{1/2}\mathbf{P}_{i}\hat{\epsilon}_{j\cdot}\hat{\epsilon}_{j\cdot}^{\prime}\mathbf{P}_{i}\mathbf{\Sigma}^{1/2})\right\} \mathbf{\Sigma}^{1/2}\mathbf{P}_{i}\hat{\epsilon}_{j\cdot}\hat{\epsilon}_{j\cdot}^{\prime}\mathbf{P}_{i}\mathbf{\Sigma}^{1/2}\left\{\mathbf{I}_{T} \circ(\mathbf{\Sigma}^{1/2}\mathbf{P}_{i}\mathbf{\Sigma}^{1/2})\right\} {\tau}_{T}\right]\\
&+2 \gamma_{1}^{2}\left[{\tau}_{T}^{\prime}\left\{\mathbf{I}_{T} \circ (\mathbf{\Sigma}^{1/2}\mathbf{P}_{i}\hat{\epsilon}_{j\cdot}\hat{\epsilon}_{j\cdot}^{\prime}\mathbf{P}_{i}\mathbf{\Sigma}^{1/2})\right\} \mathbf{\Sigma}^{1/2}\mathbf{P}_{i}\mathbf{\Sigma}^{1/2}\left\{\mathbf{I}_{T} \circ (\mathbf{\Sigma}^{1/2}\mathbf{P}_{i}\hat{\epsilon}_{j\cdot}\hat{\epsilon}_{j\cdot}^{\prime}\mathbf{P}_{i}\mathbf{\Sigma}^{1/2})\right\} {\tau}_{T}\right] \\
&+4 \gamma_{1}^{2}\left[{\tau}_{T}^{\prime}\left\{(\mathbf{\Sigma}^{1/2}\mathbf{P}_{i}\hat{\epsilon}_{j\cdot}\hat{\epsilon}_{j\cdot}^{\prime}\mathbf{P}_{i}\mathbf{\Sigma}^{1/2} )\circ ( \mathbf{\Sigma}^{1/2}\mathbf{P}_{i}\hat{\epsilon}_{j\cdot}\hat{\epsilon}_{j\cdot}^{\prime}\mathbf{P}_{i}\mathbf{\Sigma}^{1/2} )\circ(\mathbf{\Sigma}^{1/2}\mathbf{P}_{i}\mathbf{\Sigma}^{1/2})\right\} {\tau}_{T}\right]\Big\}+O\left(T^{-2}\right)
\Bigg).
\end{align*}
Because $8 |\gamma_{2} | (\hat{\epsilon}_{j\cdot}^{\prime}\mathbf{M}_{i}\hat{\epsilon}_{j\cdot})^{3/2} (\hat{\epsilon}_{j\cdot}^{\prime}\mathbf{M}^3_{i}\hat{\epsilon}_{j\cdot})^{1/2}/(\hat{\epsilon}_{j\cdot}^{\prime}\hat{\epsilon}_{j\cdot})^{2}=O(1),$
\begin{align*}
&\frac{4 \gamma_{1}^{2}\left[{\tau}_{T}^{\prime}\left\{\mathbf{I}_{T} \circ (\mathbf{\Sigma}^{1/2}\mathbf{P}_{i}\hat{\epsilon}_{j\cdot}\hat{\epsilon}_{j\cdot}^{\prime}\mathbf{P}_{i}\mathbf{\Sigma}^{1/2})\right\} \mathbf{\Sigma}^{1/2}\mathbf{P}_{i}\hat{\epsilon}_{j\cdot}\hat{\epsilon}_{j\cdot}^{\prime}\mathbf{P}_{i}\mathbf{\Sigma}^{1/2}\left\{\mathbf{I}_{T} \circ(\mathbf{\Sigma}^{1/2}\mathbf{P}_{i}\mathbf{\Sigma}^{1/2})\right\} {\tau}_{T}\right]}{(\hat{\epsilon}_{j\cdot}^{\prime}\hat{\epsilon}_{j\cdot})^{2}}\\
\leq&4 \gamma_{1}^{2}\sum_{s=1}^{T}\sum_{k=1}^{T}\Big(\frac{\mathbf{\Sigma}^{1/2}
\mathbf{P}_{i}\hat{\epsilon}_{j\cdot}\hat{\epsilon}_{j\cdot}^{\prime}
\mathbf{P}_{i}\mathbf{\Sigma}^{1/2}}
{\hat{\epsilon}_{j\cdot}^{\prime}\hat{\epsilon}_{j\cdot}}\Big)_{sk}\Big(\frac{\mathbf{\Sigma}^{1/2}
\mathbf{P}_{i}\hat{\epsilon}_{j\cdot}\hat{\epsilon}_{j\cdot}^{\prime}
\mathbf{P}_{i}\mathbf{\Sigma}^{1/2}}
{\hat{\epsilon}_{j\cdot}^{\prime}\hat{\epsilon}_{j\cdot}}\Big)_{ss}
\big(\mathbf{\Sigma}^{1/2}\mathbf{P}_{i}\mathbf{\Sigma}^{1/2}\big)_{kk}\\
\leq&4\gamma_{1}^{2}O(1)\sqrt{T^{2}\tr\Big(\frac{\mathbf{\Sigma}^{1/2}
\mathbf{P}_{i}\hat{\epsilon}_{j\cdot}\hat{\epsilon}_{j\cdot}^{\prime}
\mathbf{P}_{i}\mathbf{\Sigma}^{1/2}}
{\hat{\epsilon}_{j\cdot}^{\prime}\hat{\epsilon}_{j\cdot}}\Big)^{2}}=O(T)
\end{align*}
and \begin{align*}
&4 \gamma_{1}^{2}\left[{\tau}_{T}^{\prime}\left\{(\mathbf{\Sigma}^{1/2}\mathbf{P}_{i}\hat{\epsilon}_{j\cdot}\hat{\epsilon}_{j\cdot}^{\prime}\mathbf{P}_{i}\mathbf{\Sigma}^{1/2} )\circ ( \mathbf{\Sigma}^{1/2}\mathbf{P}_{i}\hat{\epsilon}_{j\cdot}\hat{\epsilon}_{j\cdot}^{\prime}\mathbf{P}_{i}\mathbf{\Sigma}^{1/2} )\circ(\mathbf{\Sigma}^{1/2}\mathbf{P}_{i}\mathbf{\Sigma}^{1/2})\right\} {\tau}_{T}\right]\Big\}\\
\leq&4\gamma_{1}^{2}\sum_{s=1}^{T}\sum_{k=1}^{T}\Big(\frac{\mathbf{\Sigma}^{1/2}
\mathbf{P}_{i}\hat{\epsilon}_{j\cdot}\hat{\epsilon}_{j\cdot}^{\prime}
\mathbf{P}_{i}\mathbf{\Sigma}^{1/2}}
{\hat{\epsilon}_{j\cdot}^{\prime}\hat{\epsilon}_{j\cdot}}\Big)_{sk}^{2}\big(\mathbf{\Sigma}^{1/2}\mathbf{P}_{i}\mathbf{\Sigma}^{1/2}\big)_{sk}\\
\leq&4\gamma_{1}^{2}O(1)\tr\Big(\frac{\mathbf{\Sigma}^{1/2}
\mathbf{P}_{i}\hat{\epsilon}_{j\cdot}\hat{\epsilon}_{j\cdot}^{\prime}
\mathbf{P}_{i}\mathbf{\Sigma}^{1/2}}
{\hat{\epsilon}_{j\cdot}^{\prime}\hat{\epsilon}_{j\cdot}}\Big)^{2}\\
=&O(1),
\end{align*} we have
\begin{align*}
\mE\left(\hat{\rho}^4_{i j}\right)\leq&\left\{\frac{(3+|\gamma_{2}|)}{\tr^2\left(\mathbf{M}_{i}\right)}+
\frac{3(3+|\gamma_{2}|)(2+|\gamma_{2}|)\tr\left(\mathbf{M}^2_{i}\right)}{\tr^4\left(\mathbf{M}_{i}\right)}\right\}
\mE\left\{\frac{\left(\hat{\epsilon}_{j\cdot}^{\prime}\mathbf{M}_{i}\hat{\epsilon}_{j\cdot}\right)^2}
{\left(\hat{\epsilon}_{j\cdot}^{\prime}\hat{\epsilon}_{j\cdot}\right)^2}\right\}\\
&+\mE\Big[\tr^{-3}(\mathbf{M}_{i})(\hat{\epsilon}_{j\cdot}^{\prime}\hat{\epsilon}_{j\cdot})^{-2}
\Big\{(C|\gamma_{4}|+2C|\gamma_{2}|+4C |\gamma_{2} | ) (\hat{\epsilon}_{j\cdot}^{\prime}\mathbf{M}_{i}\hat{\epsilon}_{j\cdot})^2\\
&+2 \gamma_{1}^{2}\left[{\tau}_{T}^{\prime}\left\{\mathbf{I}_{T} \circ (\mathbf{\Sigma}^{1/2}\mathbf{P}_{i}\hat{\epsilon}_{j\cdot}\hat{\epsilon}_{j\cdot}^{\prime}\mathbf{P}_{i}\mathbf{\Sigma}^{1/2})\right\} \mathbf{\Sigma}^{1/2}\mathbf{P}_{i}\mathbf{\Sigma}^{1/2}\left\{\mathbf{I}_{T} \circ(\mathbf{\Sigma}^{1/2}\mathbf{P}_{i}\hat{\epsilon}_{j\cdot}\hat{\epsilon}_{j\cdot}^{\prime}\mathbf{P}_{i}\mathbf{\Sigma}^{1/2})\right\} {\tau}_{T}\right]\Big\}\Big]\\
&+O\left(T^{-2}\right)\\
\leq&\left\{\frac{(3+|\gamma_{2}|)}{\tr^2\left(\mathbf{M}_{i}\right)}+
\frac{3(3+|\gamma_{2}|)(2+|\gamma_{2}|)\tr\left(\mathbf{M}^2_{i}\right)}{\tr^4\left(\mathbf{M}_{i}\right)}
+\frac{C|\gamma_{4}|+2C|\gamma_{2}|+4C |\gamma_{2} |}{\tr^3(\mathbf{M}_{i})}\right\}\\
\,\,\,\,&\times\bigg[\frac{2\tr\left\{\left(\mathbf{M}_{i}\mathbf{M}_{j}\right)^2\right\}
+\tr^2\left(\mathbf{M}_{i}\mathbf{M}_{j}\right)+\gamma_{2}\tr\{(\mathbf{\Sigma}^{1/2}\mathbf{P}_{j}\mathbf{M}_{i}
\mathbf{P}_{j}\mathbf{\Sigma}^{1/2})\circ(\mathbf{\Sigma}^{1/2}\mathbf{P}_{j}\mathbf{M}_{i}
\mathbf{P}_{j}\mathbf{\Sigma}^{1/2})\}}{\tr^2\left(\mathbf{M}_{j}\right)}\\
&+O\left(T^{-1}\right)\bigg]\\
&+2 \gamma_{1}^{2}\mE\Big\{\frac{{\tau}_{T}^{\prime}\left\{\mathbf{I}_{T} \circ (\mathbf{\Sigma}^{1/2}\mathbf{P}_{i}\hat{\epsilon}_{j\cdot}\hat{\epsilon}_{j\cdot}^{\prime}\mathbf{P}_{i}\mathbf{\Sigma}^{1/2})\right\} \mathbf{\Sigma}^{1/2}\mathbf{P}_{i}\mathbf{\Sigma}^{1/2}\left\{\mathbf{I}_{T} \circ(\mathbf{\Sigma}^{1/2}\mathbf{P}_{i}\hat{\epsilon}_{j\cdot}\hat{\epsilon}_{j\cdot}^{\prime}\mathbf{P}_{i}\mathbf{\Sigma}^{1/2})\right\} {\tau}_{T}}{\tr^{3}(\mathbf{M}_{i})(\hat{\epsilon}_{j\cdot}^{\prime}\hat{\epsilon}_{j\cdot})^{2}
}\Big\}+O\left(T^{-2}\right)\\
=&O\left(T^{-2}\right),
\end{align*}
where the last equality holds because of the fact
\begin{align}\label{fact}
&\frac{2\gamma_{1}^{2}{\tau}_{T}^{\prime}\left\{\mathbf{I}_{T} \circ (\mathbf{\Sigma}^{1/2}\mathbf{P}_{i}\hat{\epsilon}_{j\cdot}\hat{\epsilon}_{j\cdot}^{\prime}\mathbf{P}_{i}\mathbf{\Sigma}^{1/2})\right\} \mathbf{\Sigma}^{1/2}\mathbf{P}_{i}\mathbf{\Sigma}^{1/2}\left\{\mathbf{I}_{T} \circ(\mathbf{\Sigma}^{1/2}\mathbf{P}_{i}\hat{\epsilon}_{j\cdot}\hat{\epsilon}_{j\cdot}^{\prime}\mathbf{P}_{i}\mathbf{\Sigma}^{1/2})\right\} {\tau}_{T}}{(\hat{\epsilon}_{j\cdot}^{\prime}\hat{\epsilon}_{j\cdot})^{2}
}\n\\
\leq&2\gamma_{1}^{2}\sum_{s=1}^{T}\sum_{k=1}^{T}\Big(\frac{\mathbf{\Sigma}^{1/2}
\mathbf{P}_{i}\hat{\epsilon}_{j\cdot}\hat{\epsilon}_{j\cdot}^{\prime}
\mathbf{P}_{i}\mathbf{\Sigma}^{1/2}}
{\hat{\epsilon}_{j\cdot}^{\prime}\hat{\epsilon}_{j\cdot}}\Big)_{ss}\big(\mathbf{\Sigma}^{1/2}\mathbf{P}_{i}\mathbf{\Sigma}^{1/2}\big)_{sk}
\Big(\frac{\mathbf{\Sigma}^{1/2}
\mathbf{P}_{i}\hat{\epsilon}_{j\cdot}\hat{\epsilon}_{j\cdot}^{\prime}
\mathbf{P}_{i}\mathbf{\Sigma}^{1/2}}
{\hat{\epsilon}_{j\cdot}^{\prime}\hat{\epsilon}_{j\cdot}}\Big)_{kk}\n\\
\leq&2\gamma_{1}^{2}O(1)\sum_{s=1}^{T}\Big(\frac{\mathbf{\Sigma}^{1/2}
\mathbf{P}_{i}\hat{\epsilon}_{j\cdot}\hat{\epsilon}_{j\cdot}^{\prime}
\mathbf{P}_{i}\mathbf{\Sigma}^{1/2}}
{\hat{\epsilon}_{j\cdot}^{\prime}\hat{\epsilon}_{j\cdot}}\Big)_{ss}\sum_{k=1}^{T}\Big(\frac{\mathbf{\Sigma}^{1/2}
\mathbf{P}_{i}\hat{\epsilon}_{j\cdot}\hat{\epsilon}_{j\cdot}^{\prime}
\mathbf{P}_{i}\mathbf{\Sigma}^{1/2}}
{\hat{\epsilon}_{j\cdot}^{\prime}\hat{\epsilon}_{j\cdot}}\Big)_{kk}\n\\
\leq&2\gamma_{1}^{2}O(1)\tr^{2}\Big(\frac{\mathbf{\Sigma}^{1/2}
\mathbf{P}_{i}\hat{\epsilon}_{j\cdot}\hat{\epsilon}_{j\cdot}^{\prime}
\mathbf{P}_{i}\mathbf{\Sigma}^{1/2}}
{\hat{\epsilon}_{j\cdot}^{\prime}\hat{\epsilon}_{j\cdot}}\Big)=O(1).
\end{align}
So, $C=O\left(N^{-2}T^{-2}\right)=o\left(\sigma^4_{S_{N}}\right).$ Next,
we consider $D.$
\begin{align*}
\mE\left(\hat{\rho}^2_{i_{1}j}\hat{\rho}^2_{i_{2}j}\right)
\leq\sqrt{\mE\left(\hat{\rho}^4_{i_{1}j}\right)\mE\left(\hat{\rho}^4_{i_{2}j}\right)}=O\left(T^{-2}\right).
\end{align*}
Then, $D=O\left(N^{-1}T^{-2}\right)=o\left(\sigma^4_{S_{N}}\right)$ and $\sum_{j=2}^{N}\mE\left(Z_{j}^4\right)=o\{\sigma^4_{S_{N}}\}.$
Then, we complete the proof. \hfill$\Box$
\subsection{Proof of Lemma \ref{th:sum cons}}
Recall that
\begin{align*}
\hat{\sigma}^2_{S_{N}}=\frac{2}{N(N-1)}\sum_{j=2}^{N} \sum_{i=1}^{j-1} v_{j}^{\prime}\left(v_{i}-\bar{v}_{ij}\right) v_{i}^{\prime}\left(v_{j}-\bar{v}_{i j}\right),
\end{align*}
where $\bar{v}_{ij}=\frac{1}{N-2} \sum_{1<\tau \neq i, j<N} v_{\tau}$ and $v_{\tau}=\hat{\epsilon}_{\tau}/\|\hat{\epsilon}_{\tau}\|.$
We have
\begin{align*}
\hat{\sigma}^2_{S_{N}}=&\frac{2}{N(N-1)}\sum_{j=2}^{N} \sum_{i=1}^{j-1} v_{i}^{\prime}\left(v_{j}-\bar{v}_{ij}\right) v_{j}^{\prime}\left(v_{i}-\bar{v}_{i j}\right)\\
=&\frac{2}{N(N-1)}\sum_{j=2}^{N} \sum_{i=1}^{j-1} \left(v_{i}^{\prime}v_{j}\right)^2
-\frac{2}{N(N-1)}\sum_{j=2}^{N} \sum_{i=1}^{j-1}v_{i}^{\prime}v_{j}v_{j}^{\prime}\bar{v}_{i j}\\
&-\frac{2}{N(N-1)}\sum_{j=2}^{N} \sum_{i=1}^{j-1}v_{i}^{\prime}\bar{v}_{i j}v_{j}^{\prime}v_{i}
+\frac{2}{N(N-1)}\sum_{j=2}^{N} \sum_{i=1}^{j-1}v_{i}^{\prime}\bar{v}_{i j}v_{j}^{\prime}\bar{v}_{i j}\\
=&F_{1}-F_{2}-F_{3}+F_{4}.
\end{align*}
Obviously, $\mE\left(F_{1}\right)=\sigma^2_{S_{N}}\left\{1+o(1)\right\}$ and
$\mE\left(F_{i}\right)=0$ for $i=2,3,4.$ Then, we just need to prove that $\var\left(F_{1}\right)=o\left(\sigma^4_{S_{N}}\right)$ and
$$\mE\left(F^2_{i}\right)=o\left(\sigma^4_{S_{N}}\right)$$
for $i=2,\cdots,4.$
\begin{align*}
\var\left(F^2_{1}\right)\leq&\frac{4}{N^2(N-1)^2}\sum_{j=2}^{N}\sum_{i=1}^{j-1}
\mE\left\{\left(v_{i}^{\prime}v_{j}\right)^4\right\}
+\frac{8}{N^2(N-1)^2}\underset{i_{1}<i_{2}<i_{3}}{\sum\sum\sum}\mE\left\{\left(v_{i_{1}}^{\prime}v_{i_{2}}\right)^2
\left(v_{i_{1}}^{\prime}v_{i_{3}}\right)^2\right\}\\
&+\frac{8}{N^2(N-1)^2}\underset{i_{1}<i_{2}<i_{3}}{\sum\sum\sum}\mE\left\{\left(v_{i_{1}}^{\prime}v_{i_{3}}\right)^2
\left(v_{i_{2}}^{\prime}v_{i_{3}}\right)^2\right\}\\&+
\frac{8}{N^2(N-1)^2}\underset{i_{1}<i_{2}<i_{3}}{\sum\sum\sum}\mE\left\{\left(v_{i_{1}}^{\prime}v_{i_{2}}\right)^2
\left(v_{i_{2}}^{\prime}v_{i_{3}}\right)^2\right\}.
\end{align*}
Because $\mE\left\{\left(v_{i_{1}}^{\prime}v_{i_{2}}\right)^2
\left(v_{i_{1}}^{\prime}v_{i_{3}}\right)^2\right\}\leq\sqrt{\mE\left\{\left(v_{i_{1}}^{\prime}v_{i_{2}}\right)^4\right\}
\mE\left\{\left(v_{i_{1}}^{\prime}v_{i_{3}}\right)^4\right\}}$ and $\mE\left\{\left(v_{i_{1}}^{\prime}v_{i_{2}}\right)^4\right\}=\mE\left(\hat{\rho}^4_{i_{1}i_{2}}\right)=O\left(T^{-2}\right),$
we have $\var\left(F^2_{1}\right)=O\left(N^{-1}T^{-2}\right)=o\left(\sigma^4_{S_{N}}\right).$
We next deal with $F_{2}.$
$$F_{2}=\frac{2}{N(N-1)}\sum_{j=2}^{N} \sum_{i=1}^{j-1}v_{i}^{\prime}v_{j}v_{j}^{\prime}\bar{v}_{i j}=\frac{2}{N(N-1)(N-2)}\sum_{j=2}^{N} \sum_{i=1}^{j-1}\sum_{\tau\neq i,j}v_{i}^{\prime}v_{j}v_{j}^{\prime}{v}_{\tau}.$$
So,
\begin{align*}
\mE\left(F^2_{2}\right)=&\frac{4}{N^2(N-1)^2(N-2)^2}\sum_{j_{1}=2}^{N} \sum_{i_{1}=1}^{j_{1}-1}\sum_{\tau_{1}\neq i_{1},j_{1}}\sum_{j_{2}=2}^{N} \sum_{i_{2}=1}^{j_{2}-1}\sum_{\tau_{2}\neq i_{2},j_{2}}\mE\left(v_{i_{1}}^{\prime}v_{j_{1}}v_{j_{1}}^{\prime}v_{\tau_{1}}
v_{i_{2}}^{\prime}v_{j_{2}}v_{j_{2}}^{\prime}v_{\tau_{2}}\right)\\
=&\frac{4}{N^2(N-1)^2(N-2)^2}\sum_{j=2}^{N} \overset{j-1}{\sum_{i=1}} \sum_{\tau\neq i,j}\mE\left(v_{i}^{\prime}v_{j}v_{j}^{\prime}v_{\tau}
v_{i}^{\prime}v_{j}v_{j}^{\prime}v_{\tau}\right)\\
+&\frac{8}{N^2(N-1)^2(N-2)^2}\underset{j_{1}<j_{2}}{\sum^{N}\sum^{N}} \overset{j_{1}-1}{\sum_{i=1}} \sum_{\tau\neq i,j_{1},j_{2}}\mE\left(v_{i}^{\prime}v_{j_{1}}v_{j_{1}}^{\prime}v_{\tau}
v_{i}^{\prime}v_{j_{2}}v_{j_{2}}^{\prime}v_{\tau}\right).
\end{align*}
Because
\begin{align*}
\mE\left(v_{i}^{\prime}v_{j}v_{j}^{\prime}v_{\tau}
v_{i}^{\prime}v_{j}v_{j}^{\prime}v_{\tau}\right)=&
\mE\left\{\left(v_{i}^{\prime}v_{j}v_{j}^{\prime}v_{\tau}\right)^2\right\}\\
=&\mE\left(\hat{\rho}^2_{ij}\hat{\rho}^2_{j\tau}\right)\\
\leq&\sqrt{\mE\left(\hat{\rho}^4_{ij}\right)\mE\left(\hat{\rho}^4_{j\tau}\right)}\\
=&O\left(T^{-2}\right).
\end{align*}
and
\begin{align*}
&\mE\left(v_{i}^{\prime}v_{j_{1}}v_{j_{1}}^{\prime}v_{\tau}
v_{i}^{\prime}v_{j_{2}}v_{j_{2}}^{\prime}v_{\tau}\right)\\
\leq&\sqrt{\mE\left\{\left(v_{i}^{\prime}v_{j_{1}}v_{j_{1}}^{\prime}v_{\tau}\right)^2\right\}
\mE\left\{\left(v_{i}^{\prime}v_{j_{2}}v_{j_{2}}^{\prime}v_{\tau}\right)^2\right\}}\\
=&O\left(T^{-2}\right),
\end{align*}
we have $\mE\left(F^2_{2}\right)=O\left(N^{-2}T^{-2}\right)=o\left(\sigma^4_{S_{N}}\right).$
Similarly, we have $F_3=o_{p}\left(\sigma^2_{S_{N}}\right).$ Now, we consider $F_{4},$
\begin{align*}
F_{4}=&\frac{2}{N(N-1)}\sum_{j=2}^{N} \sum_{i=1}^{j-1}v_{i}^{\prime}\bar{v}_{i j}v_{j}^{\prime}\bar{v}_{i j}\\
=&\frac{2}{N(N-1)(N-2)^2}\sum_{j=2}^{N} \sum_{i=1}^{j-1}\sum_{\tau_{1}\neq i,j}\sum_{\tau_{2}\neq i,j}v_{i}^{\prime}{v}_{\tau_{1}}v_{j}^{\prime}{v}_{\tau_{2}}\\
=&\frac{2}{N(N-1)(N-2)^2}\sum_{j=2}^{N} \sum_{i=1}^{j-1}\sum_{\tau\neq i,j}v_{i}^{\prime}{v}_{\tau}v_{j}^{\prime}{v}_{\tau}+\frac{2}{N(N-1)(N-2)^2}\sum_{j=2}^{N} \sum_{i=1}^{j-1}\sum_{\tau_{1}\neq\tau_{2}\neq i,j}v_{i}^{\prime}{v}_{\tau_{1}}v_{j}^{\prime}{v}_{\tau_{2}}\\
=&F_{41}+F_{42}.
\end{align*}
So,
\begin{align*}
\mE\left(F^2_{41}\right)=&\frac{4}{N^2(N-1)^2(N-2)^4}\sum_{j=2}^{N} \sum_{i=1}^{j-1}\sum_{\tau_{1},\tau_{2}\neq i,j}\mE\left(v_{i}^{\prime}{v}_{\tau_{1}}v_{j}^{\prime}{v}_{\tau_{1}}v_{i}^{\prime}{v}_{\tau_{2}}v_{j}^{\prime}{v}_{\tau_{2}}\right)\\
=&O\left(N^{-4}\right)\sqrt{\mE\left(v_{i}^{\prime}{v}_{\tau_{1}}v_{j}^{\prime}{v}_{\tau_{1}}\right)^2
\mE\left(v_{i}^{\prime}{v}_{\tau_{2}}v_{j}^{\prime}{v}_{\tau_{2}}\right)^2}\\
=&O\left(N^{-4}T^{-2}\right).
\end{align*}
Similarly, we have $\mE\left(F^2_{42}\right)=O\left(N^{-4}T^{-2}\right).$
%\begin{align*}
%\mE\left(F^2_{42}\right)=&\frac{4}{N^2(N-1)^2(N-2)^4}\sum_{j=2}^{N} \sum_{i=1}^{j-1}\sum_{\tau_{1}\neq\tau_{2}\neq i,j}\mE\left(v_{i}^{\prime}{v}_{\tau_{1}}v_{j}^{\prime}{v}_{\tau_{1}}v_{i}^{\prime}{v}_{\tau_{2}}v_{j}^{\prime}{v}_{\tau_{2}}\right)\\
%=&O\left(N^{-4}\right)\mE\left(v_{i}^{\prime}{v}_{\tau_{1}}v_{j}^{\prime}{v}_{\tau_{1}}v_{i}^{\prime}{v}_{\tau_{2}}v_{j}^{\prime}{v}_{\tau_{2}}\right)\\
%=&O\left(N^{-4}\right)\sqrt{\mE\left(v_{i}^{\prime}{v}_{\tau_{1}}v_{j}^{\prime}{v}_{\tau_{1}}\right)^2
%\mE\left(v_{i}^{\prime}{v}_{\tau_{2}}v_{j}^{\prime}{v}_{\tau_{2}}\right)^2}\\
%=&O\left(N^{-4}T^{-2}\right)
%\end{align*}
Next, we can conclude that $\mE\left(F^2_{4}\right)=o\left(\sigma^4_{S_{N}}\right)$ due to
$\mE\left(F^2_{4}\right)\leq2\mE\left(F^2_{41}+F^2_{42}\right),$
which leads to
$$\hat{\sigma}^2_{S_{N}}/{\sigma}^2_{S_{N}} \rightarrow1 ,$$ in probability.\hfill$\Box$

\subsection{Proof of Lemma \ref{indep hat}}
By Theorem \ref{th:max null} and Lemma \ref{th:sum null}, the following hold,
\begin{align}
&\frac{\tr^2(\tilde{\mathbf{\Sigma}})}
{\|\tilde{\mathbf{\Sigma}}\|^2_{\mathrm{F}}}L_{N}-4 \log N+\log \log  N\rightarrow G(y)  \,\,\text {in distribution; }\label{Tmax} \\
&S_{N}/\hat{\sigma}_{S_{N}} \rightarrow \mathcal{N}(0,1) \text { in distribution. }\label{Tsum}
\end{align}
First, we will prove that if $\tilde{T}_{max}-\log N+\log \log  N$ and $S_{N}/\sigma_{S_{N}}$ are asymptotically independent,
then $\frac{\tr^2(\tilde{\mathbf{\Sigma}})}
{\|\tilde{\mathbf{\Sigma}}\|^2_{\mathrm{F}}}L_{N}-\log N+\log \log  N$ and $S_{N}/{\sigma}_{S_{N}}$ are also asymptotically
independent.
Due to (\ref{eiej}), (\ref{dis-rho/}) and (\ref{max:cons}), we have
\begin{align}\label{diff of Lv and LVhat}
\frac{\tr^2(\tilde{\mathbf{\Sigma}})}
{\|\tilde{\mathbf{\Sigma}}\|^2_{\mathrm{F}}}L_{N}-\tilde{T}_{max}\rightarrow0,
\end{align}
in probability.
Then, to show asymptotic independence, it is enough to show
\begin{align}\label{indepe}
\lim _{\min(N,T) \rightarrow \infty} P\left(S_{N}/\sigma_{S_{N}} \leq x,\frac{\tr^2(\tilde{\mathbf{\Sigma}})}{\|\tilde{\mathbf{\Sigma}}\|^2_{\mathrm{F}}}L_{N}-4 \log N+\log \log  N\leq y\right)=\Phi(x) \cdot G(y),
\end{align}
for any $x \in \mathbb{R}$ and $y \in \mathbb{R},$ where $\Phi(x)=(2 \pi)^{-1 / 2} \int_{-\infty}^{x} e^{-t^{2} / 2} d t$ and $G(y)=\exp \left\{-\frac{1}{\sqrt{8 \pi}} \exp \left(-\frac{y}{2}\right)\right\}.$ Let $$a_{N}\doteq4\log N-\log \log  N+y.$$
Due to (\ref{Tmax}) and (\ref{Tsum}), we know (\ref{indepe}) is equivalent to that
\begin{align}\label{indepLV}
\lim _{\min(N,T) \rightarrow \infty} P\left(S_{N}/\sigma_{S_{N}}\leq x, \frac{\tr^2(\tilde{\mathbf{\Sigma}})}{\|\tilde{\mathbf{\Sigma}}\|^2_{\mathrm{F}}}L_{N}>a_{N}\right)=\Phi(x) \cdot\left\{1-G(y)\right\},
\end{align}
for any $x \in \mathbb{R}$ and $y \in \mathbb{R}$. By the assumption, we know that
\begin{align}\label{indepLVhat}
\lim _{\min(N,T) \rightarrow \infty} P\left(S_{N}/\sigma_{S_{N}} \leq x, \tilde{T}_{max}>a_{N}\right)=\Phi(x) \cdot\left\{1-G(y)\right\},
\end{align}
for any $x \in \mathbb{R}$ and $y \in \mathbb{R}$. We show next that (\ref{indepLVhat}) implies (\ref{indepLV}). By (\ref{diff of Lv and LVhat}),
$$
\left|\frac{\tr^2(\tilde{\mathbf{\Sigma}})}
{\|\tilde{\mathbf{\Sigma}}\|^2_{\mathrm{F}}}L_{N}-\tilde{T}_{max}\right| \rightarrow 0
$$
in probability. Given $\epsilon \in(0,1) .$ Set
$$
\Omega=\left\{\left|\frac{\tr^2(\tilde{\mathbf{\Sigma}})}
{\|\tilde{\mathbf{\Sigma}}\|^2_{\mathrm{F}}}L_{N}-\tilde{T}_{max}\right|<\epsilon\right\}.
$$
Then
\begin{align}\label{OmegaV=1}
\lim _{\min(N,T) \rightarrow \infty} P\left(\Omega\right)=1.
\end{align}
Now,
\begin{align}\label{OmgaV condition}
& P\left(S_{N}/\sigma_{S_{N}} \leq x, \frac{\tr^2(\tilde{\mathbf{\Sigma}})}{\|\tilde{\mathbf{\Sigma}}\|^2_{\mathrm{F}}}L_{N}>a_{N}\right) \n\\
\leq & P\left(S_{N}/\sigma_{S_{N}} \leq x, \frac{\tr^2(\tilde{\mathbf{\Sigma}})}{\|\tilde{\mathbf{\Sigma}}\|^2_{\mathrm{F}}}L_{N}>a_{N}, \Omega\right)+P\left(\Omega^{c}\right).
\end{align}
$\operatorname{On} \Omega$, if $\frac{\tr^2(\tilde{\mathbf{\Sigma}})}{\|\tilde{\mathbf{\Sigma}}\|^2_{\mathrm{F}}}L_{N}>a_{N}$ then
\begin{align}\label{LVhatqeg}
\tilde{T}_{max} \geq \frac{\tr^2(\tilde{\mathbf{\Sigma}})}
{\|\tilde{\mathbf{\Sigma}}\|^2_{\mathrm{F}}}L_{N}-\left|\frac{\tr^2(\tilde{\mathbf{\Sigma}})}
{\|\tilde{\mathbf{\Sigma}}\|^2_{\mathrm{F}}}L_{N}-\tilde{T}_{max}\right|>a_{N}-\epsilon.
\end{align}
Define
\begin{align*}
\tilde{a}_{N}\doteq4\log N-\log\log N+y-2\epsilon,
\end{align*}
which makes sense for large $T$. Thus,
$$
a_{N}-\tilde{a}_{N}>\epsilon,
$$
and (\ref{LVhatqeg}) conclude that
$$
\tilde{T}_{max} \geq \tilde{a}_{N},
$$
as $p$ is sufficiently large. Review (\ref{OmgaV condition}). We have
$$
\begin{aligned}
& P\left(S_{N}/\sigma_{S_{N}} \leq x, \frac{\tr^2(\tilde{\mathbf{\Sigma}})}{\|\tilde{\mathbf{\Sigma}}\|^2_{\mathrm{F}}}L_{N}>a_{N}\right) \\
\leq & P\left(S_{N}/\sigma_{S_{N}} \leq x, \tilde{T}_{max} \geq \tilde{a}_{N}\right)+P\left(\Omega^{c}\right).
\end{aligned}
$$
Immediately from (\ref{indepLVhat}) and (\ref{OmegaV=1}) we get
$$
\limsup _{\min(N,T) \rightarrow \infty} P\left(S_{N}/\sigma_{S_{N}} \leq x, \frac{\tr^2(\tilde{\mathbf{\Sigma}})}{\|\tilde{\mathbf{\Sigma}}\|^2_{\mathrm{F}}}L_{N}>a_{N}\right) \leq \Phi(x) \cdot\left\{1-G(y-2 \epsilon)\right\},
$$
for any $\epsilon \in(0,1)$. Inspect that the left-hand side of the above does not depend on $\epsilon$. Letting $\epsilon \downarrow 0$, we obtain
\begin{align}\label{96}
\limsup _{\min(N,T) \rightarrow \infty} P\left(S_{N}/\sigma_{S_{N}} \leq x, \frac{\tr^2(\tilde{\mathbf{\Sigma}})}{\|\tilde{\mathbf{\Sigma}}\|^2_{\mathrm{F}}}L_{N}>a_{N}\right) \leq \Phi(x) \cdot\left\{1-G(y)\right\},
\end{align}
for any $x \in \mathbb{R}$ and $y \in \mathbb{R}$. In the following we will show the lower limit. Evidently,
\begin{align}\label{97}
& P\left(S_{N}/\sigma_{S_{N}} \leq x, \frac{\tr^2(\tilde{\mathbf{\Sigma}})}{\|\tilde{\mathbf{\Sigma}}\|^2_{\mathrm{F}}}L_{N}>a_{N}\right) \n\\
\geq & P\left(S_{N}/\sigma_{S_{N}} \leq x, \frac{\tr^2(\tilde{\mathbf{\Sigma}})}{\|\tilde{\mathbf{\Sigma}}\|^2_{\mathrm{F}}}L_{N}>a_{N}, \Omega\right).
\end{align}
Set
$$
\tilde{a}_{N}^{\prime}\doteq4\log N-\log \log  N+y+2\epsilon,
$$
Therefore,
$
\tilde{a}_{N}^{\prime}>a_{N}+\epsilon.
$
It is straightforward to verify that
\begin{align}
\left\{\tilde{T}_{max}>\tilde{a}_{N}^{\prime}, \Omega\right\} \subset\left\{\tilde{T}_{max}>a_{N}+\epsilon, \Omega\right\} \subset\left\{\frac{\tr^2(\tilde{\mathbf{\Sigma}})}{\|\tilde{\mathbf{\Sigma}}\|^2_{\mathrm{F}}}L_{N}>a_{N}, \Omega\right\},
\end{align}
as $N$ is sufficiently large, where the last inclusion follows from the definition of $\Omega .$ By (\ref{97}),
\begin{align*}
& P\left(S_{N}/\sigma_{S_{N}} \leq x, \frac{\tr^2(\tilde{\mathbf{\Sigma}})}{\|\tilde{\mathbf{\Sigma}}\|^2_{\mathrm{F}}}L_{N}>a_{N}\right) \\
\geq & P\left(S_{N}/\sigma_{S_{N}} \leq x, \tilde{T}_{max}>\tilde{a}_{N}^{\prime}, \Omega\right).
\end{align*}
Thus, from (\ref{indepLVhat}) and (\ref{OmegaV=1}) we get
$$
\liminf _{\min(N,T) \rightarrow \infty} P\left(S_{N}/\sigma_{S_{N}} \leq x, \frac{\tr^2(\tilde{\mathbf{\Sigma}})}{\|\tilde{\mathbf{\Sigma}}\|^2_{\mathrm{F}}}L_{N}>a_{N}\right) \geq \Phi(x) \cdot\left\{1-G(y+2 \epsilon)\right\},
$$
for any $\epsilon \in(0,1)$. Sending $\epsilon \downarrow 0$ we see
$$
\liminf _{\min(N,T) \rightarrow \infty} P\left(S_{N}/\sigma_{S_{N}}\leq x, \frac{\tr^2(\tilde{\mathbf{\Sigma}})}{\|\tilde{\mathbf{\Sigma}}\|^2_{\mathrm{F}}}L_{N}>a_{N}\right) \geq \Phi(x) \cdot\left\{1-G(y)\right\},
$$
for any $x \in \mathbb{R}$ and $y \in \mathbb{R}$.

We have proved that $\frac{\tr^2(\tilde{\mathbf{\Sigma}})}
{\|\tilde{\mathbf{\Sigma}}\|^2_{\mathrm{F}}}L_{N}-4 \log N+\log \log N$ and $S_{N}/\sigma_{S_{N}}$ are asymptotically independent.
Similarly, it is easy to prove that $\frac{\tr^2(\tilde{\mathbf{\Sigma}})}
{\|\tilde{\mathbf{\Sigma}}\|^2_{\mathrm{F}}}L_{N}-4 \log N+\log N$ and $S_{N}/\hat{\sigma}_{S_{N}}$ are also asymptotically
independent. Consequently, we complete the proof.
\hfill$\Box$
\subsection{Proof of Lemma \ref{linear}}
For $I_{l}$ appeared in $H(N, k),$ write $I_{l}=\left(i_{l}, j_{l}\right)$ for $l=1, \cdots, k .$ Now we classify the indices $I_{1}<I_{2}<\cdots<I_{k} \in \Lambda_{N}$ in the definition of $H(N, k)$ into three cases. Let $\Gamma_{N, 1}$ be the set of indices $\left(I_{1}, \cdots, I_{k}\right)$ such that no two of the $2 k$ indices $\left\{i_{l}, j_{l} ; l=1, \cdots, k\right\}$ are identical. Let $\Gamma_{N, 2}$ be the set of indices $\left(I_{1}, \cdots, I_{k}\right)$ such that either $i_{1}=$ $\cdots=i_{k}$ or $j_{1}=\cdots=j_{k} .$ Let $\Gamma_{N, 3}$ be the set of indices $I_{1}<I_{2}<\cdots<I_{k} \in$
$\Lambda_{N}$ excluding $\Gamma_{N, 1} \cup \Gamma_{N, 2} .$ In the following we will estimate
$$
D_{j}:=\sum_{I_{1}<I_{2}<\cdots<I_{k} \in \Gamma_{N, j}} \mathbb{P}\left(B_{I_{1}} B_{I_{2}} \cdots B_{I_{k}}\right)
$$
for $j=1,2,3$ one by one. We will see $D_{1}$ contributes essentially the sum in the expression of $H(N, k)$ by an easy argument; the term $D_{2}$ is negligible
and its computation is trivial; the term $D_{3}$ is also negligible but its estimate is most involved.

Step 1: the estimate of $D_{1}$. Recall $B_{I}=\left\{\left|\epsilon_{i\cdot}^{\prime}\epsilon_{j\cdot} \right|\geq l_{N}\right\}$ if $I=(i, j) \in$ $\Lambda_{N}.$ By the definition of $\Gamma_{N, 1},$ we know that $B_{I_{1}}, B_{I_{2}} \cdots, B_{I_{k}}$ are independent. For large $N$,
$$
\max _{I \in \Lambda_{N}} \mathbb{P}\left(B_{I}\right)=\mathbb{P}\left\{\left|\epsilon_{i\cdot}^{\prime}\epsilon_{j\cdot} \right|\geq l_{N}\right\} \leq \frac{C}{N^{2}}
$$
where the inequality holds due to (\ref{T12}). Then,
\begin{align}\label{O1}
D_{1} \leq \frac{C}{N^{2 k}} \cdot\binom{\frac{N(N-1)}{2}}{k}\leq \frac{C}{k !}
\end{align}
Step
2: the estimate of $D_{2}$. Evidently, the size of $\Gamma_{N, 2}$ is no more than $\binom{N}{1}\cdot\binom{N}{k}\cdot2 \leq 2 N^{k+1} .$
Recall that $z_{i}$ is the $i$-th row vector of matrix $\mathbf{Z}.$
Similarly, there exists some $C_{3} > 0$ such that for fixed $k>0$ and $a_{l}\in\{-1,1\},$ $1\leq l\leq k,  $
that
$$
\mathbb{P}\left(\frac{\Big(\sum_{l=1}^{k}a_{l}z_{{l}}\Big)^{\prime}\mathbf{\Sigma}^2\Big(\sum_{l=1}^{k}a_{l}z_{{l}}\Big)}
{\tr(\mathbf{\Sigma}^{2})}>k(1+\varepsilon_{3})\right)
\leq2N^{-3k},
$$
where $\varepsilon_{3}=C_{3}\sqrt{\log N/\tr(\mathbf{\Sigma}^2)}.$
Again, by Cram\'er type moderate deviation results in \cite{statulevivcius1966} and Theorem 1.1 in \cite{rudelson2013}, for $j_{1}\neq j_{2}\neq \cdots\neq j_{k},$
\begin{align}\label{step2}
&\mathbb{P}\left(B_{I_{1}} B_{I_{2}} \cdots B_{I_{k}}\right)\n\\
=&\mathbb{P}\left(|\epsilon_{i_{1}\cdot}^{\prime}\epsilon_{j_{1}\cdot}|>l_{N},\cdots, |\epsilon_{i_{1}\cdot}^{\prime}\epsilon_{j_{k}\cdot}|>l_{N}\right)\n\\
\leq&\sum_{a_{l}\in\{-1,1\}}\mathbb{P}\bigg\{\Big|\epsilon_{i_{1}\cdot}^{\prime}\Big(\sum_{l=1}^{k}a_{l}\epsilon_{j_{l}\cdot}
\Big)\Big|>k\cdot l_{N}\bigg\}\n\\
\leq&\sum_{a_{l}\in\{-1,1\}}\mathbb{P}\bigg\{\Big|z_{i1}^{\prime}\mathbf{\Sigma}\Big(\sum_{l=1}^{k}a_{l}z_{j_{l}}\Big)
\Big|>k\cdot \sqrt{a_{N}\tr(\mathbf{\Sigma}^2)}\bigg\}\n\\
\leq&\sum_{a_{l}\in\{-1,1\}}\mathbb{P}\bigg\{\frac{\Big|z_{i1}^{\prime}\mathbf{\Sigma}\Big(\sum_{l=1}^{k}a_{l}z_{j_{l}}\Big)
\Big|}{\sqrt{\Big(\sum_{l=1}^{k}a_{l}z_{j_{l}}\Big)^{\prime}\mathbf{\Sigma}^2\Big(\sum_{l=1}^{k}a_{l}z_{j_{l}}\Big)}}>\frac{\sqrt{ka_{N}}}{\sqrt{1+\varepsilon_{3}}}\bigg\}\n\\
&+\sum_{a_{l}\in\{-1,1\}}
\mathbb{P}\bigg\{\Big(\sum_{l=1}^{k}a_{l}z_{j_{l}}\Big)^{\prime}\mathbf{\Sigma}^2\Big(\sum_{l=1}^{k}a_{l}z_{j_{l}}\Big)
>k(1+\varepsilon_{3})\tr(\mathbf{\Sigma}^2)\bigg\}\n\\
\leq&\sum_{a_{l}\in\{-1,1\}}\mE\Bigg[\mE\bigg\{\frac{\Big|z_{i1}^{\prime}\mathbf{\Sigma}\Big(\sum_{l=1}^{k}a_{l}z_{j_{l}}\Big)
\Big|}{\sqrt{\Big(\sum_{l=1}^{k}a_{l}z_{j_{l}}\Big)^{\prime}\mathbf{\Sigma}^2\Big(\sum_{l=1}^{k}a_{l}z_{j_{l}}\Big)}}>\frac{ \sqrt{ka_{N}}}{\sqrt{1+\varepsilon_{3}}}|z_{j_{1}},\cdots,z_{j_{k}}\bigg\}\Bigg]+2^{k+1}N^{-3k}\n\\
\leq&\sum_{a_{l}\in\{-1,1\}}\left\{1+o(1)\right\}\frac{2}{\sqrt{ka_{N}2\pi}}\exp\left(-ka_{N}/2\right)+2^{k+1}N^{-3k}\n\\
\leq&\frac{2^{k+1}(2\pi)^{-1/2}}{\sqrt{4k\log N}}\exp\left[-2k\log N+k\log \log N /2-ky/2\right]+2^{k+1}N^{-3k}\n\\
\leq&2^{k+1}\left(\log N\right)^{k/2-1/2}N^{-2k}\exp(-ky/2)+2^{k+1}N^{-3k},
\end{align}
where $ka_{N}=4k\log N-k\log \log N +ky.$ So, for $k\geq 2,$ we have
\begin{align}\label{O2}
D_{2} & \leq 2 N^{k+1} \cdot \mathbb{P}\left(B_{I_{1}} B_{I_{2}} \cdots B_{I_{k}}\right)\n\\
&=2 N^{k+1} \cdot 2^{k+1}\left(\log N\right)^{k/2-1/2}N^{-2k}\exp(-ky/2) \rightarrow 0.
\end{align}
Step 3 : the estimate of $D_{3}$. Fix a tuple $\left(I_{1}, I_{2}, \cdots, I_{l}\right) \in \Gamma_{N, 3}$. By the ordering imposed on $\Lambda_{N},$ we see that $i_{1} \leq i_{2} \leq \cdots \leq i_{k} .$ Let's assume that
$$i_{1}=i_{2}=\cdots=i_{n_{1}},$$
$$i_{n_{1}+1}=i_{n_{1}+2}=\cdots=i_{n_{1}+n_{2}},$$
$$\cdots\cdots\cdots\cdots\cdots\cdots\cdots\cdots\cdots$$
$$i_{n_{1}+\cdots+n_{r-1}+1}=i_{n_{1}+\cdots+n_{r-1}+2}=\cdots=i_{n_{1}+\cdots+n_{r}},$$
where $n_{1}+\cdots+n_{r}=k.$
Let $\mathcal{F}_{1}$ be the set of random vectors $\left\{\epsilon_{j_{1}\cdot}, \epsilon_{i_{l}\cdot}, \epsilon_{j_{l}\cdot} ; 2 \leq l \leq k\right\}$.

Similar to (\ref{step2}),
for sufficiently large constant $C>0,$ we have
$$
\begin{aligned}
&\P\left(B_{I_{1}} B_{I_{2}} \cdots B_{I_{k}}\right) \\ =&\mE\left\{\P\left(B_{I_{1}} B_{I_{2}} \cdots B_{I_{k}} \mid \mathcal{F}_{1}\right)\right\} \\
=&\mE\left[\P\left\{\min_{1\leq l\leq n_{1}}|\epsilon_{i_{1}\cdot}^{\prime}\epsilon_{j_{l}\cdot}|\geq l_{N} \mid \mathcal{F}_{1}\right\} \cdot \prod_{l=n_{1}+1}^{k} I\left(B_{I_{l}}\right)\right]\\
\leq &\mE\left[\sum_{a_{l}\in\{-1,1\}}\P\left\{\Big|\epsilon_{i_{1}\cdot}^{\prime}(\sum_{l=1}^{n_{1}}a_{l}\epsilon_{j_{l}\cdot})\Big|\geq l_{N} \mid \mathcal{F}_{1}\right\} \cdot \prod_{l=n_{1}+1}^{k} I\left(B_{I_{l}}\right)\right]\\
\leq&\mE\left[\sum_{a_{l}\in\{-1,1\}}\P\left\{\Big|\epsilon_{i_{1}\cdot}^{\prime}(\sum_{l=1}^{n_{1}}a_{l}\epsilon_{j_{l}\cdot})\Big|\geq l_{N} \mid \mathcal{F}_{1}\right\}\right. \\
&\left.\quad\quad\cdot I\Big\{\frac{\big(\sum_{l=1}^{n_{1}}a_{l}z_{j_{l}}\big)^{\prime}\mathbf{\Sigma}^{2}\big(\sum_{l=1}^{n_{1}}a_{l}z_{j_{l}}\big)}
{n_{1}\tr(\mathbf{\Sigma}^{2})}>1+C\sqrt{\frac{\log N}{\tr(\mathbf{\Sigma}^{2})}}\Big\}\cdot \prod_{l=n_{1}+1}^{k} I\left(B_{I_{l}}\right)\right]\\
&+\mE\left[I\Big\{\frac{\big(\sum_{l=1}^{n_{1}}a_{l}z_{j_{l}}\big)^{\prime}\mathbf{\Sigma}^{2}\big(\sum_{l=1}^{n_{1}}a_{l}z_{j_{l}}\big)}
{n_{1}\tr(\mathbf{\Sigma}^{2})}>1+C\sqrt{\frac{\log N}{\tr(\mathbf{\Sigma}^{2})}}\Big\}\cdot \prod_{l=n_{1}+1}^{k} I\left(B_{I_{l}}\right)\right]\\
\leq&O(N^{-2n_{1}})\mE\left[\prod_{l=n_{1}+1}^{k} I\left(B_{I_{l}}\right)\right]+2N^{-3k}\\
\leq &O(N^{-2n_{1}})\P\left( B_{I_{n_{1}+1}} \cdots B_{I_{k}}\right)+2N^{-3k}.
\end{aligned}
$$
In the same way, we finally obtain
\begin{align*}
\P\left(B_{I_{1}} B_{I_{2}} \cdots B_{I_{k}}\right)=&\P\Big(\prod_{l=1}^{n_{1}}B_{I_{l}}\Big)\cdots\P\Big(\prod_{l=n_{r-1}+1}^{n_{r}}B_{I_{l}}\Big)\\
\leq& \frac{C}{N^{2n_{1}+\cdots+2n_{r}}}\\
\leq&\frac{C}{N^{2k}},
\end{align*}
for some constant $C$ and sufficiently large $N$.
Recall $I_{l}=\left(i_{l}, j_{l}\right)$ for each $1 \leq l \leq k .$ In view of the definition of $\Gamma_{N, 3},$ there are at least two of the $2 k$ indices from $\left\{\left(i_{l}, j_{l}\right) ; 1 \leq l \leq k\right\}$ are identical for any $\left(I_{1}, \cdots, I_{k}\right) \in \Gamma_{N, 3} .$ Let $\kappa=\left|\left\{i_{l}, j_{l} ; 1 \leq l \leq k\right\}\right|$ for such $\left(I_{1}, I_{2}, \cdots, I_{k}\right).$
Easily, $k+1 \leq \kappa \leq 2 k-1 .$ To see how many such $\left(I_{1}, \cdots, I_{k}\right)$ with $\mid\left\{i_{l}, j_{l} ; 1 \leq\right.$ $l \leq k\} \mid=\kappa,$ first pick $\kappa$ many indices from $\{1,2, \cdots, N\},$ which has the total number of ways $\binom{N}{\kappa}\leq N^{\kappa},$ then use the $\kappa$ many indices to make a $\left(I_{1}, \cdots, I_{k}\right) \in \Gamma_{N, 3} .$ The total number of ways to do so is no more than $\kappa^{2 k}$. Therefore,
$$
\left|\Gamma_{N, 3}\right| \leq \sum_{\kappa=k+1}^{2 k-1} N^{\kappa} \cdot \kappa^{2 k} \leq(2 k)^{2 k} \cdot N^{2 k-1}.
$$
Similar to the (\ref{step2}), for fixed $k$,
we can have $\mathbb{P}\left(B_{I_{1}} B_{I_{2}} \cdots B_{I_{k}}\right)=O(N^{-2k})$ and
\begin{align*}
D_{3} \leq&(2 k)^{2 k} \cdot2^k\left(\log N\right)^{k/2-1/2}N^{-2k}\exp(-ky/2)\\
\leq&(2 k)^{2 k} \cdot2^k\exp(-ky/2)\frac{\left(\log N\right)^{k/2-1/2}}{N}\rightarrow 0
\end{align*}
as $N$ is sufficiently large.  \hfill$\Box$

\subsection{Proof of Lemma \ref{linear2}}
For $I_{1}<I_{2}<\cdots<I_{k} \in \Lambda_{N},$ write $I_{l}=\left(i_{l}, j_{l}\right)$ for $l=1,2, \cdots, k .$ Set
$$
\Lambda_{N, k}=\left\{\left(i_{l}, j\right) ; i_{l}<j \leq N, 1 \leq l \leq k\right\} \bigcup\left\{\left(i, j_{l}\right) ; 1 \leq i<j_{l}, 1 \leq l \leq k\right\}
$$
for $k \geq 1 .$ It is easy to check that $\left|\Lambda_{N, k}\right|=\sum_{l=1}^{k}\left(N-i_{l}+j_{l}-2\right) .$ since $i_{l}<j_{l}$ for each $l,$ we see that
$$
k(N-1) \leq\left|\Lambda_{N, k}\right| \leq \sum_{l=1}^{k}\left(N+j_{l}\right) \leq 2 k N.
$$
Recall
$$
A_{N}=\left\{S_{N}/\sqrt{\sigma^2_{S_{N}}} \leq x\right\}  \quad x \in \mathbb{R},
$$
for $N\geq 3$ and
$$
S_{N}^{k}=\sqrt{\frac{2}{N(N-1)}}\sum_{(i, j) \in \Lambda_{N, k}}\hat{\rho}_{ij},
$$
for $N \geq k \geq 1$.
Observe that $B_{I_{1}} B_{I_{2}} \cdots B_{I_{k}}$ is an event generated by random vectors $\left\{\hat{\rho}_{ij};(i, j) \in \Lambda_{N, k}\right\} .$ A crucial observation is that $S_{N}-S_{N}^{k}$ is independent of $B_{I_{1}} B_{I_{2}} \cdots B_{I_{k}} .$ It is easy to see that
\begin{align*}
S_{N}^{k} =&\sqrt{\frac{2}{N(N-1)}}{\sum\sum}_{\left\{\left(i_{l}, j\right) ; i_{l}<j \leq N, 1 \leq l \leq k\right\}}\hat{\rho}_{i_{l}j}\\
&+\sqrt{\frac{2}{N(N-1)}}{\sum\sum}_{\left\{\left(i, j_{l}\right) ; 1 \leq i<j_{l}, 1 \leq l \leq k\right\}} \hat{\rho}_{ij_{l}}\\
&-\sqrt{\frac{2}{N(N-1)}}\sum_{s=1}^{k}\sum_{t=1}^{k} \hat{\rho}_{i_{s} j_{t}}\\
\doteq&D_{N, 1}+D_{N, 2}-D_{N, 3}.
\end{align*}
Fix $\epsilon \in(0,1),$ for even $\tau$ and $\tau>2$,
we have
\begin{align*}
\mE\left|D_{N, 1}\right|^{\tau}\leq k^{\tau-1}\frac{2^{\tau/2}}{N^{\tau/2}(N-1)^{\tau/2}}\sum_{l=1}^{k}\mE\left(\left|\sum_{j=i_{l}+1}^{N}\hat{\rho}_{i_{l}j}\right|^{\tau}\right),
\end{align*}
where
$$
\mE\left\{\left(\sum_{j=i_{l}+1}^{N}\hat{\rho}_{i_{l}j}\right)^{\tau}\right\}
=\mE\left[\mE\left\{\left.\left(\sum_{j=i_{l}+1}^{N}\hat{\rho}_{i_{l}j}\right)^{\tau}\right|\epsilon_{i_{l}\cdot}\right\}\right].
$$
By Lemma 2 in \cite{feng2020}, there exists a constant $K_{\tau}> 0$ depending on $\tau$ only such that
$$\mE\left\{\left.\left(\sum_{j=i_{l}+1}^{N}\hat{\rho}_{i_{l}j}\right)^{\tau}\right|\epsilon_{i_{l}\cdot}\right\}
\leq K_{\tau}(N-i_{l})^{\tau/2-1}\sum_{j=i_{l}+1}^{N}\mE\left\{\left.\left(\hat{\rho}_{i_{l}j}\right)^{\tau}\right|\epsilon_{i_{l}\cdot}\right\}.$$
So, we have
$$\mE\left|D_{N, 1}\right|^{\tau}\leq k^{\tau-1}\frac{2^{\tau/2}}{N^{\tau/2}(N-1)^{\tau/2}}\sum_{l=1}^{k}
K_{\tau}(N-i_{l})^{\tau/2-1}\sum_{j=i_{l}+1}^{N}\mE\left[\mE\left\{\left.\left(
\hat{\rho}_{i_{l}j}\right)^{\tau}\right|\epsilon_{i_{l}\cdot}\right\}\right].
$$
By Khintchine's inequality (Exercise 2.6.5 in \cite{vershynin2018}) and Lemma \ref{lem:laplace approxi}, it's easy to obtain that
$\mE\left[\mE\left\{\left.\left(
\hat{\rho}_{i_{l}j}\right)^{\tau}\right|\epsilon_{i_{l}\cdot}\right\}\right]\leq (\tau/2)^{\tau/4}O\left(T^{-\tau/2}\right).$
Hence, together with $\sigma_{S_{N}}=O\left(T^{-1/2}\right),$ we obtain that
\begin{align*}
\mathbb{P}\left(\left|D_{N,1}\right|>\sigma_{S_{N}}\epsilon\right) & \leq \frac{\mE\left|D_{N,1}\right|^{\tau}}{\sigma^{\tau}_{S_{N}}\epsilon^{\tau}}\\
&=\frac{K_{\tau}k^{\tau}\tau^{\tau/4}2^{\tau/4}}{\epsilon^{\tau}}\cdot O\left(N^{-\tau/2}\right),
\end{align*}
where the last equality holds due to Assumption \ref{assum:x_it}.
Similarly,
$$
\mathbb{P}\left(\left|D_{N,2}\right|>\sigma_{S_{N}}\epsilon\right) \leq \frac{k^{\tau}\tau^{\tau/4}2^{\tau/4}}{\epsilon^{\tau}}\cdot O\left(N^{-\tau/2}\right).
$$
Lastly, for even $\tau$ and $\tau>2$,
\begin{align*}
\mE\left(\left|D_{N, 3}\right|^{\tau}\right) & \leq \frac{2^{\tau/2}}{N^{\tau/2}(N-1)^{\tau/2}} \cdot k^{2(\tau-1)} \cdot \sum_{s=1}^{k} \sum_{t=1}^{k} \mE\left(\left|\hat{\rho}_{i_{s} j_{t}}\right|^{\tau}\right) \\
& \leq \frac{2^{\tau/2}}{N^{\tau/2}(N-1)^{\tau/2}} \cdot k^{2\tau}\mE\left(\left|\hat{\rho}_{i_{s} j_{t}}\right|^{\tau}\right).
\end{align*}
So, we have
$$
\begin{aligned}
\mathbb{P}\left\{\frac{|S_{N}^k|}{\sigma_{S_{N}}} \geq \epsilon\right\} &\leq C^{\prime} \cdot \frac{k^{2\tau}\tau^{\tau/4}}{N^{\tau/2}},
\end{aligned}
$$
for large $N$, where $C^{\prime}$ is a constant depending on $\epsilon$ but free of $N.$

Fix $I_{1}<I_{2}<\cdots<I_{k} \in \Lambda_{N} .$ By the definition of $A_{N}$,
$$
\begin{aligned}
& \mathbb{P}\left\{A_{N}(x) B_{I_{1}} B_{I_{2}} \cdots B_{I_{k}}\right\} \\
\leq & \mathbb{P}\left\{A_{N}(x) B_{I_{1}} B_{I_{2}} \cdots B_{I_{k}}, \frac{\left|S_{N}^{k} \right|}{\sigma_{S_{N}}}  <\epsilon\right\}+C^{\prime} \cdot \frac{k^{2\tau}\tau^{\tau/4}}{N^{\tau/2}} \\
\leq & \mathbb{P}\left\{\frac{S_{N}-S_{N}^{k}}{\sigma_{S_{N}}} \leq x+\epsilon, B_{I_{1}} B_{I_{2}} \cdots B_{I_{k}}\right\}+C^{\prime} \cdot \frac{k^{2\tau}\tau^{\tau/4}}{N^{\tau/2}} \\
=& \mathbb{P}\left\{\frac{S_{N}-S_{N}^{k}}{\sigma_{S_{N}}} \leq x+\epsilon\right\} \cdot \mathbb{P}\left(B_{I_{1}} B_{I_{2}} \cdots B_{I_{k}}\right)+C^{\prime} \cdot \frac{k^{2\tau}\tau^{\tau/4}}{N^{\tau/2}},
\end{aligned}
$$
by the independence between $S_{N}-S_{N}^{k}$ and $B_{I_{1}} B_{I_{2}} \cdots B_{I_{k}} .$ Now
$$
\begin{aligned}
& \mathbb{P}\left\{\frac{S_{N}-S_{N}^{k}}{\sigma_{S_{N}}} \leq x+\epsilon\right\} \\
\leq & \mathbb{P}\left\{\frac{S_{N}-S_{N}^{k}}{\sigma_{S_{N}}}  \leq x+\epsilon,\frac{\left|S_{N}^{k}\right|}{\sigma_{S_{N}}} <\epsilon\right\}+C^{\prime} \cdot \frac{k^{2\tau}\tau^{\tau/4}}{N^{\tau/2}} \\
\leq & \mathbb{P}\left\{\frac{S_{N}}{\sigma_{S_{N}}} \leq x+2 \epsilon\right\}+C^{\prime} \cdot \frac{k^{2\tau}\tau^{\tau/4}}{N^{\tau/2}} \\
\leq & \mathbb{P}\left\{A_{N}(x+2 \epsilon)\right\}+C^{\prime} \cdot \frac{k^{2\tau}\tau^{\tau/4}}{N^{\tau/2}}.
\end{aligned}
$$
Combing the two inequalities to get
\begin{align}\label{leq1}
& \mathbb{P}\left\{A_{N}(x) B_{I_{1}} B_{I_{2}} \cdots B_{I_{k}}\right\}\n \\
\leq & \mathbb{P}\left\{A_{N}(x+2 \epsilon)\right\} \cdot \mathbb{P}\left(B_{I_{1}} B_{I_{2}} \cdots B_{I_{k}}\right)+2 C^{\prime} \cdot \frac{k^{2\tau}\tau^{\tau/4}}{N^{\tau/2}}.
\end{align}
Similarly,
$$
\begin{aligned}
& \mathbb{P}\left\{\frac{S_{N}-S_{N}^k}{\sigma_{S_{N}}} \leq x-\epsilon, B_{I_{1}} B_{I_{2}} \cdots B_{I_{k}}\right\} \\
\leq & \mathbb{P}\left\{\frac{S_{N}-S_{N}^k}{\sigma_{S_{N}}}  \leq x-\epsilon, B_{I_{1}} B_{I_{2}} \cdots B_{I_{k}},\right.\\
&\left.\frac{\left|S_{N}^k\right|}{\sigma_{S_{N}}} <\epsilon\right\}+C^{\prime} \cdot \frac{k^{2\tau}\tau^{\tau/4}}{N^{\tau/2}} \\
\leq & \mathbb{P}\left\{\frac{S_{N}}{\sigma_{S_{N}}}  \leq x, B_{I_{1}} B_{I_{2}} \cdots B_{I_{k}}\right\}+C^{\prime} \cdot \frac{k^{2\tau}\tau^{\tau/4}}{N^{\tau/2}}.
\end{aligned}
$$
In other words, by independence,
\begin{align*}
&\mathbb{P}\left\{A_{N}(x) B_{I_{1}} B_{I_{2}} \cdots B_{I_{k}}\right\} \\
\geq&\mathbb{P}\left\{\frac{S_{N}-S_{N}^k}{\sigma_{S_{N}}}  \leq x-\epsilon\right\}\cdot \mathbb{P}\left(B_{I_{1}} B_{I_{2}} \cdots B_{I_{k}}\right)-C^{\prime} \cdot \frac{k^{2\tau}\tau^{\tau/4}}{N^{\tau/2}}.
\end{align*}
Furthermore,
\begin{align}
&\mathbb{P}\left\{\frac{S_{N}}{\sigma_{S_{N}}} \leq x-2 \epsilon\right\}\n\\
\leq &\mathbb{P}\left\{\frac{S_{N}}{\sigma_{S_{N}}} \leq x-2 \epsilon, \frac{\left|S_{N}^k\right|}{\sigma_{S_{N}}} <\epsilon\right\}+C^{\prime} \cdot \frac{k^{2\tau}\tau^{\tau/4}}{N^{\tau/2}} \n\\
\leq& \mathbb{P}\left\{\frac{S_{N}-S_{N}^k}{\sigma_{S_{N}}}  \leq x-\epsilon\right\}+C^{\prime} \cdot \frac{k^{2\tau}\tau^{\tau/4}}{N^{\tau/2}}.
\end{align}
The above two strings of inequalities imply
$$
\begin{aligned}
& \mathbb{P}\left\{A_{N}(x) B_{I_{1}} B_{I_{2}} \cdots B_{I_{k}}\right\} \\
\geq & \mathbb{P}\left\{\frac{S_{N}}{\sigma_{S_{N}}} \leq x-2 \epsilon\right\} \cdot \mathbb{P}\left(B_{I_{1}} B_{I_{2}} \cdots B_{I_{k}}\right)-2 C^{\prime} \cdot \frac{k^{2\tau}\tau^{\tau/4}}{N^{\tau/2}},
\end{aligned}
$$
which joining with (\ref{leq1}) yields
$$
\begin{aligned}
&\left|\mathbb{P}\left\{A_{N}(x) B_{I_{1}} B_{I_{2}} \cdots B_{I_{k}}\right\}-\mathbb{P}\left\{A_{N}(x)\right\} \cdot \mathbb{P}\left(B_{I_{1}} B_{I_{2}} \cdots B_{I_{k}}\right)\right| \\
\leq & \Delta_{N, \epsilon} \cdot \mathbb{P}\left(B_{I_{1}} B_{I_{2}} \cdots B_{I_{k}}\right)+4 C^{\prime} \cdot \frac{k^{2\tau}\tau^{\tau/4}}{N^{\tau/2}},
\end{aligned}
$$
where
$$
\Delta_{N, \epsilon}\doteq\left|\mathbb{P}\left\{A_{N}(x)\right\}-\mathbb{P}\left\{A_{N}(x+2 \epsilon)\right\}\right|+\left| \mathbb{P}\left\{A_{N}(x)\right\}-\mathbb{P}\left\{A_{N}(x-2 \epsilon) \right\}\right|.
$$
In particular,
\begin{align}\label{diff of phi}
\Delta_{N, \epsilon} \rightarrow|\Phi(x+2 \epsilon)-\Phi(x)|+|\Phi(x-2 \epsilon)-\Phi(x)|,
\end{align}
as $\min(N,T) \rightarrow \infty$ by Theorem \ref{th:max null}. As a consequence,
$$
\begin{aligned}
\zeta(N, k)\doteq& \sum_{I_{1}<I_{2}<\cdots<I_{k} \in \Lambda_{N}}\left[\mathbb{P}\left(A_{N}(x) B_{I_{1}} B_{I_{2}} \cdots B_{I_{k}}\right)-\right.\\
&\left.\mathbb{P}\left\{A_{N}(x)\right\} \cdot \mathbb{P}\left(B_{I_{1}} B_{I_{2}} \cdots B_{I_{k}}\right)\right] \\
\leq & \sum_{I_{1}<I_{2}<\cdots<I_{k} \in \Lambda_{N}}\left\{\Delta_{N, \epsilon} \cdot \mathbb{P}\left(B_{I_{1}} B_{I_{2}} \cdots B_{I_{k}}\right)+4 C^{\prime} \cdot \frac{k^{2\tau}\tau^{\tau/4}}{N^{\tau/2}}\right\} \\
\leq & \Delta_{N, \epsilon} \cdot H(N, k)+\left(4 C^{\prime}\right) \cdot\binom{\frac{1}{2}N(N-1)}{k} \cdot \frac{k^{2\tau}\tau^{\tau/4}}{N^{\tau/2}},
\end{aligned}
$$
where
$$
H(N, k)=\sum_{I_{1}<I_{2}<\cdots<I_{k} \in \Lambda_{k}} \mathbb{P}\left(B_{I_{1}} B_{I_{2}} \cdots B_{I_{k}}\right)
$$
as defined in Lemmas \ref{linear} and \ref{linear2}, we know $\lim \sup _{\min(N,T) \rightarrow \infty} H(N, k) \leq C / k !$, where $C$ is a universal constant. Picking $\tau=6k,$ and using the trivial fact $\binom{r}{i} \leq r^{i}$ for any integers $1 \leq i \leq r,$ we have that
$$
 \dbinom{\frac{1}{2}N(N-1)}{k}\cdot \frac{k^{2\tau}\tau^{\tau/4}}{N^{\tau/2 }} \leq N^{2 k} \cdot \frac{k^{2\tau}\tau^{\tau/4}}{N^{\tau/2 }} \leq \frac{k^{2\tau}\tau^{\tau/4}}{N^{k}}\rightarrow 0.
$$
Hence, from (\ref{diff of phi})
$$
\begin{aligned}
\limsup _{\min(N,T) \rightarrow \infty} \zeta(N, k) & \leq \frac{C}{k !} \cdot \limsup _{\min(N,T) \rightarrow \infty} \Delta_{N, \epsilon} \\
&=\frac{C}{k !} \cdot\left\{|\Phi(x+2 \epsilon)-\Phi(x)|+|\Phi(x-2 \epsilon)-\Phi(x)|\right\},
\end{aligned}
$$
for any $\epsilon>0 .$ The desired result follows by sending $\epsilon \downarrow 0$.
\hfill$\Box$

%\bibliographystyle{asa}
%\begingroup
%\baselineskip=16.5pt
%\bibliography{reference}
%\endgroup

\end{document}